\documentclass[12pt]{article}
\pdfoutput=1
\usepackage[utf8x]{inputenc}
\usepackage[colorlinks,linkcolor=blue,citecolor=blue,bookmarks,bookmarksnumbered]{hyperref}
\usepackage[scaled=0.85]{helvet}
\usepackage{amsmath,amssymb,accents,mathrsfs,XoohmE}
\usepackage{graphicx,color}
\usepackage{booktabs}
\usepackage{multirow}
\usepackage{placeins}
\usepackage{amsmath}
\usepackage{tikz}

\usepackage{enumitem}

\usepackage[aligntableaux=center,boxsize=1.2em]{ytableau}

\usepackage{XoohmE}

\definecolor{Green}  {rgb}{0.10,0.70,0.10} 
\definecolor{Orange} {rgb}{1.00,0.50,0.15} 
\definecolor{Red}    {rgb}{0.90,0.00,0.12} 
\definecolor{Purple} {rgb}{0.50,0.25,0.55} 
\definecolor{Turque} {rgb}{0.00,0.65,0.85} 
\definecolor{Blue}   {rgb}{0.00,0.00,1.00} 
\definecolor{Magenta}{rgb}{1.00,0.00,1.00} 
\definecolor{Gold}   {rgb}{1.00,0.75,0.25} 
\definecolor{Seaweed}{rgb}{0.01,0.24,0.09} 
\definecolor{Brown}  {rgb}{0.43,0.26,0.32} 
\definecolor{grey1}  {rgb}{0.20,0.20,0.20} 
\definecolor{grey2}  {rgb}{0.40,0.40,0.40} 
\definecolor{grey3}  {rgb}{0.60,0.60,0.60} 
\definecolor{grey4}  {rgb}{0.80,0.80,0.80} 
\definecolor{grey5}  {rgb}{0.90,0.90,0.90} 
\def\C#1#2{{\ifcase#1\or
             \color{Green}\or \color{Orange}\or \color{Red}\or
              \color{Purple}\or \color{Turque}\or \color{Blue}\or
               \color{Magenta}\or \color{Gold}\or \color{Seaweed}\or
                \color{Brown}\or\color{grey1}\or\color{grey2}\or
                 \color{grey3}\else\color{grey4}\fi#2}}

\definecolor{Slate} {rgb}{0.00,0.45,0.55}

\usepackage[enableskew,vcentermath]{youngtab}
\let\TC=\textcolor
\definecolor{Hey}{rgb}{.9,.05,.4}
\definecolor{orange}{rgb}{1,.5,0}
\definecolor{plum}{rgb}{.4,0,.6}
\definecolor{R}{rgb}{1,0,0}
\definecolor{G}{rgb}{0,1,0}
\definecolor{B}{rgb}{0,0,1}

\long\def\CMTred#1{\leavevmode\TC{red}{\sf#1}}

\long\def\CMTR#1{\leavevmode\TC{R}{\sf#1}}

\long\def\CMTB#1{\leavevmode\TC{B}{\sf#1}}

\usepackage{lipsum}
\usepackage{listings}
\definecolor{MyDarkGreen}{rgb}{0.0,0.4,0.0} 
\def\rI{{\rm I}}
\def\rJ{{\rm J}}

\def\hi{{\hat\imath}}
\def\hj{{\hat\jmath}}
\def\hk{{\hat{k}}}

\def\fracm#1#2{\hbox{\large{${\frac{{#1}}{{#2}}}$}}}

\def\be{\begin{equation}}
\def\ee{\end{equation}}
\newcommand{\bea}{\begin{eqnarray}}
\newcommand{\eea}{\end{eqnarray}}
\newcommand{\ena}{\end{eqnarray}}

\def\pp{{\mathchoice
          {
              \kern 1pt%
              \raise 1pt
              \vbox{\hrule width5pt height0.4pt depth0pt
                    \kern -2pt
                    \hbox{\kern 2.3pt
                          \vrule width0.4pt height6pt depth0pt
                          }
                    \kern -2pt
                    \hrule width5pt height0.4pt depth0pt}%
                    \kern 1pt
           }
            {
              \kern 1pt%
              \raise 1pt
              \vbox{\hrule width4.3pt height0.4pt depth0pt
                    \kern -1.8pt
                    \hbox{\kern 1.95pt
                          \vrule width0.4pt height5.4pt depth0pt
                          }
                    \kern -1.8pt
                    \hrule width4.3pt height0.4pt depth0pt}%
                    \kern 1pt
            }
            {
              \kern 0.5pt%
              \raise 1pt
              \vbox{\hrule width4.0pt height0.3pt depth0pt
                    \kern -1.9pt  
                    \hbox{\kern 1.85pt
                          \vrule width0.3pt height5.7pt depth0pt
                          }
                    \kern -1.9pt
                    \hrule width4.0pt height0.3pt depth0pt}%
                    \kern 0.5pt
            }
            {
              \kern 0.5pt%
              \raise 1pt
              \vbox{\hrule width3.6pt height0.3pt depth0pt
                    \kern -1.5pt
                    \hbox{\kern 1.65pt
                          \vrule width0.3pt height4.5pt depth0pt
                          }
                    \kern -1.5pt
                    \hrule width3.6pt height0.3pt depth0pt}%
                    \kern 0.5pt
            }
        }}

\def\mm{{\mathchoice
                       {
                             \kern 1pt
               \raise 1pt    \vbox{\hrule width5pt height0.4pt depth0pt
                                  \kern 2pt
                                  \hrule width5pt height0.4pt depth0pt}
                             \kern 1pt}
                       {
                            \kern 1pt
               \raise 1pt \vbox{\hrule width4.3pt height0.4pt depth0pt
                                  \kern 1.8pt
                                  \hrule width4.3pt height0.4pt depth0pt}
                             \kern 1pt}
                       {
                            \kern 0.5pt
               \raise 1pt
                            \vbox{\hrule width4.0pt height0.3pt depth0pt
                                  \kern 1.9pt
                                  \hrule width4.0pt height0.3pt depth0pt}
                            \kern 1pt}
                       {
                           \kern 0.5pt
             \raise 1pt  \vbox{\hrule width3.6pt height0.3pt depth0pt
                                  \kern 1.5pt
                                  \hrule width3.6pt height0.3pt depth0pt}
                           \kern 0.5pt}
                       }}

\def\ad{{\kern0.5pt
                   \alpha \kern-5.05pt \raise5.8pt\hbox{$\textstyle.$}\kern
0.5pt}}

\def\bd{{\kern0.5pt
                   \beta \kern-5.05pt \raise5.8pt\hbox{$\textstyle.$}\kern
0.5pt}}

\def\qd{{\kern0.5pt
                   q \kern-5.05pt \raise5.8pt\hbox{$\textstyle.$}\kern
0.5pt}}
\def\Dot#1{{\kern0.5pt
     {#1} \kern-5.05pt \raise5.8pt\hbox{$\textstyle.$}\kern
0.5pt}}

\catcode`@=11
\def\un#1{\relax\ifmmode\@@underline#1\else
        $\@@underline{\hbox{#1}}$\relax\fi}
\catcode`@=12

\def\a{\alpha}
\def\b{\beta}
\def\c{\chi}
\def\d{\delta}
\def\e{\epsilon}

\def\g{\gamma}

\def\i{\iota}
\def\j{\psi}
\def\k{\kappa}
\def\l{\lambda}

\def\n{\nu}
\def\o{\omega}

\def\r{\rho}
\def\s{\sigma}
\def\t{\tau}

\def\D{\Delta}

\def\L{\Lambda}
\def\O{\Omega}
\def\P{\Pi}

\def\S{\Sigma}

\def\ca{{\cal A}}

\def\cf{{\cal F}}

\def\dslash{\not{\hbox{\kern-2pt $\partial$}}}
\def\Dslash{\not{\hbox{\kern-4pt $D$}}}
\def\pslash{\not{\hbox{\kern-2.3pt $p$}}}
 \newtoks\slashfraction
 \slashfraction={.13}
 \def\slash#1{\setbox0\hbox{$ #1 $}
 \setbox0\hbox to \the\slashfraction\wd0{\hss \box0}/\box0 }
 
\def\kcr{{\hbox{\ro \char'170}}}                
\def\ktl{{\hbox{\ro \char'170}}}        
\def\ktr{{\hbox{\ro \char'170}}}        
\def\kbl{{\hbox{\ro \char'170}}}        
\def\kbr{{\hbox{\ro \char'170}}}        

\def\plpl{\raise-2pt\hbox{$\raise3pt\hbox{$_+$}\hskip-6.67pt\raise0.0pt
\hbox{$^+$}\hskip 0.01pt$}}
\def\mimi{\raise-2pt\hbox{$\raise3pt\hbox{$_-$}\hskip-6.67pt\raise0.0pt
\hbox{$^-$}\hskip 0.01pt$}}

\def\bo{{\raise.15ex\hbox{\large$\Box$}}}               
\def\pa{\partial}                                       
\def\TH{{\raise.2ex\hbox{$\displaystyle \bigodot$}\mskip-4.7mu \llap H \;}}
\def\face{{\raise.2ex\hbox{$\displaystyle \bigodot$}\mskip-2.2mu \llap {$\ddot
        \smile$}}}                                      

\def\dt#1{\on{\hbox{\bf .}}{#1}}                
\def\Dot#1{\dt{#1}}

\def\Tilde#1{\widetilde{#1}}                    
\def\VEV#1{\left\langle #1\right\rangle}        
\def\leftrightarrowfill{$\mathsurround=0pt \mathord\leftarrow \mkern-6mu
        \cleaders\hbox{$\mkern-2mu \mathord- \mkern-2mu$}\hfill
        \mkern-6mu \mathord\rightarrow$}
\def\dvec#1{\vbox{\ialign{##\crcr
        \leftrightarrowfill\crcr\noalign{\kern-1pt\nointerlineskip}
        $\hfil\displaystyle{#1}\hfil$\crcr}}}           
\def\dt#1{{\buildrel {\hbox{\LARGE .}} \over {#1}}}     

\def\fracm#1#2{\hbox{\large{${\frac{{#1}}{{#2}}}$}}}
\def\sfrac#1#2{{\vphantom1\smash{\lower.5ex\hbox{\small$#1$}}\over
        \vphantom1\smash{\raise.4ex\hbox{\small$#2$}}}} 
\def\bfrac#1#2{{\vphantom1\smash{\lower.5ex\hbox{$#1$}}\over
        \vphantom1\smash{\raise.3ex\hbox{$#2$}}}}       
\def\afrac#1#2{{\vphantom1\smash{\lower.5ex\hbox{$#1$}}\over#2}}    

\def\pa{\partial}      
\let\bm\relax
\newcommand{\bm}[1]{{\boldsymbol{#1}}}

\def\ad{{\Dot{\alpha}}}
\def\bd{{\Dot{\beta}}}

 \font\rOpe=cmsy10                        
 \def\ktl{{\hbox{\rOpe\char'170}}}        
 \def\kbl{{\hbox{\rOpe\char'170}}}        
 \def\kcr{{\reflectbox{\rOpe\char'170}}}        
 \def\ktr{{\reflectbox{\rOpe\char'170}}}        
 \def\kbr{{\reflectbox{\rOpe\char'170}}}        
 \def\Border{\vbox{\hsize0pt
        \setlength{\unitlength}{1mm}
        \newcount\xco
        \newcount\yco
        \xco=-21
        \yco=12
        \begin{picture}(0,0)(-7.5,0)
        \put(\xco,\yco){$\ktl$}
        \advance\yco by-1
        {\loop
        \put(\xco,\yco){$\kcr$}
        \advance\yco by-2
        \ifnum\yco>-240
        \repeat
        \put(\xco,\yco){$\kbl$}}
        \xco=170
        \yco=12
        \put(\xco,\yco){$\ktr$}
        \advance\yco by-1
        {\loop
        \put(\xco,\yco){$\kcr$}
        \advance\yco by-2
        \ifnum\yco>-240
        \repeat
        \put(\xco,\yco){$\kbr$}}
        \put(-19.5,13){\scalebox{.6065}{%
         University of Maryland Center for String and Particle  Theory \&\ Physics Department%
        |University of Maryland Center for String and Particle  Theory \&\ Physics Department}}
        \put(-19.5,-241.5){\scalebox{.5835}{%
         ****University of Maryland * Center for String and
         Particle  Theory* Physics Department****University of Maryland *Center
        for String and Particle  Theory* Physics Department}}
        \end{picture}
        \par\vskip-8mm}}
\definecolor{UMred}{rgb}{.9,.05,.2}
\definecolor{HUblue}{rgb}{.0,.3,.7}

\definecolor{Red}    {rgb}{0.90,0.00,0.12} 
\definecolor{Blue}   {rgb}{0.00,0.00,1.00} 
\definecolor{Green}  {rgb}{0.10,0.70,0.10} 
\definecolor{Turque} {rgb}{0.00,0.65,0.85} 
\definecolor{Orange} {rgb}{1.00,0.50,0.15} 
\definecolor{Magenta}{rgb}{1.00,0.00,1.00} 
\definecolor{Gold}   {rgb}{1.00,0.75,0.25} 
\definecolor{Seaweed}{rgb}{0.01,0.24,0.09} 
\definecolor{Purple} {rgb}{0.50,0.25,0.55} 
\definecolor{Brown}  {rgb}{0.43,0.26,0.32} 
\definecolor{grey1}  {rgb}{0.20,0.20,0.20} 
\definecolor{grey2}  {rgb}{0.40,0.40,0.40} 
\definecolor{grey3}  {rgb}{0.60,0.60,0.60} 
\definecolor{grey4}  {rgb}{0.80,0.80,0.80} 
\definecolor{grey5}  {rgb}{0.90,0.90,0.90} 
\def\C#1#2{{\ifcase#1\or
             \color{Red}\or \color{Green}\or \color{Blue}\or\
              \color{Turque}\or \color{Orange}\or \color{Magenta}\or
               \color{Gold}\or \color{Seaweed}\or \color{Purple}\or
                \color{Brown}\or\color{grey1}\or\color{grey2}\or
                 \color{grey3}\else\color{grey4}\fi#2}}

\definecolor{Slate} {rgb}{0.00,0.45,0.55}

\newdimen\parshift\parshift=\parindent
\catcode`@=11
 \long\def\@footnotetext#1{\insert\footins{\reset@font\footnotesize
           \interlinepenalty\interfootnotelinepenalty\splittopskip%
            \footnotesep\splitmaxdepth\dp\strutbox\floatingpenalty\@MM%
             \hsize\columnwidth\addtolength{\hsize}{-2\parindent}
              \@parboxrestore\protected@edef\@currentlabel%
              {\csname p@footnote\endcsname\@thefnmark}%
                \color@begingroup%
                 \@makefntext{\rule\z@\footnotesep\ignorespaces#1%
                  \@finalstrut\strutbox}%
                \color@endgroup}}
 \long\def\@makefntext#1{\hglue\parshift%
           \vbox{\noindent\baselineskip=11pt plus.5pt minus.5pt\hb@xt@0em{\hss\@makefnmark\kern1pt}#1}}
\catcode`@=12

\newskip\humongous \humongous=0pt plus 1000pt minus 1000pt
\def\caja{\mathsurround=0pt}
\def\eqalign#1{\,\vcenter{\openup2\jot \caja
        \ialign{\strut \hfil$\displaystyle{##}$&$
        \displaystyle{{}##}$\hfil\crcr#1\crcr}}\,}
\newif\ifdtup

\makeatletter
\def\section{\@startsection{section}{1}{\z@}
        {3ex plus-1ex minus-.2ex}{1pt plus1pt}{\large\sf\bfseries\boldmath}}
\def\subsection{\@startsection{subsection}{2}{\z@}
         {1.5ex plus-1ex minus-.2ex}{0.01pt plus1pt}{\sf\slshape}}
\def\subsubsection{\@startsection{subsubsection}{3}{\z@}
          {1.5ex plus-1ex minus-.2ex}{0.01pt plus0.2pt}{\sf\boldmath}}
\def\paragraph{\@startsection{paragraph}{4}{\z@}
           {.75ex \@plus.5ex \@minus.2ex}{-2mm}{\sf\bfseries\boldmath}}
\makeatother

 \allowdisplaybreaks
 \seceq

\newcommand{\aone}{{\un a}_1}
\newcommand{\bone}{{\un b}_1}
\newcommand{\cone}{{\un c}_1}
\newcommand{\done}{{\un d}_1}
\newcommand{\eone}{{\un e}_1}

\newcommand{\atwo}{{\un a}_2}
\newcommand{\btwo}{{\un b}_2}
\newcommand{\ctwo}{{\un c}_2}
\newcommand{\dtwo}{{\un d}_2}

\newcommand{\athree}{{\un a}_3}
\newcommand{\bthree}{{\un b}_3}
\newcommand{\cthree}{{\un c}_3}
\newcommand{\dthree}{{\un d}_3}

\newcommand{\afour}{{\un a}_4}
\newcommand{\bfour}{{\un b}_4}
\newcommand{\cfour}{{\un c}_4}
\newcommand{\dfour}{{\un d}_4}

\newcommand{\dotline}{\,\,\,\bullet\hspace{-0.35em}-}
\newcommand{\linedot}{-\hspace{-0.35em}\bullet\,\,\,}

\definecolor{skyblue}{rgb}{0.12, 0.46, 1.00}
\definecolor{brightpink}{rgb}{1.0, 0.0, 0.5}
\definecolor{darkgreen}{rgb}{0.10, 0.75, 0.24}

\newcommand{\sym}{\color{skyblue}{\bm \circ}}
\newcommand{\antisym}{\color{brightpink}{\ast}}
\newcommand{\nosym}{\color{darkgreen}{\thicksim}}
\newcommand{\symone}{\color{skyblue}{1}}
\newcommand{\symtwo}{\color{skyblue}{2}}

\newcommand{\nosymone}{\color{darkgreen}{1}}
\newcommand{\nosymtwo}{\color{darkgreen}{2}}
\newcommand{\nosymthree}{\color{darkgreen}{3}}
\newcommand{\antisymone}{\color{brightpink}{1}}

\newcommand{\antisymthree}{\color{brightpink}{3}}
\newcommand{\singleone}{\color{black}{1}}
\newcommand{\singletwo}{\color{black}{2}}
\newcommand{\singlethree}{\color{black}{3}}
\newcommand{\singlefour}{\color{black}{4}}

\newcommand{\dotlineup}{\raisebox{0.15em}{$\dotline$}}
\newcommand{\linedotup}{\raisebox{0.15em}{$\linedot$}}

\newcommand{\linedotdown}{\raisebox{-0.15em}{$\linedot$}}
\newcommand{\dotlineupdot}{\raisebox{0.15em}{$\dotline$}\hspace{-1.35em}\raisebox{-0.15em}{$\cdots$}}
\newcommand{\linedotupdot}{\raisebox{0.15em}{$\linedot$}\hspace{-1.35em}\raisebox{-0.15em}{$\cdots$}}
\newcommand{\dotlinedowndot}{\raisebox{-0.15em}{$\dotline$}\hspace{-1.35em}\raisebox{0.15em}{$\cdots$}}
\newcommand{\linedotdowndot}{\raisebox{-0.15em}{$\linedot$}\hspace{-1.35em}\raisebox{0.15em}{$\cdots$}}
\newcommand{\dotlinebb}{\,\,\,{\color{skyblue}{\bullet}}\hspace{-0.35em}-}
\newcommand{\linedotbb}{-\hspace{-0.35em}{\color{skyblue}{\bullet}}\,\,\,}
\newcommand{\dotlineupbb}{\raisebox{0.15em}{$\dotlinebb$}}

\newcommand{\dotlinedownbb}{\raisebox{-0.15em}{$\dotlinebb$}}

\newcommand{\dotlineupdotbb}{\raisebox{0.15em}{$\dotlinebb$}\hspace{-1.35em}\raisebox{-0.15em}{$\cdots$}}
\newcommand{\linedotupdotbb}{\raisebox{0.15em}{$\linedotbb$}\hspace{-1.35em}\raisebox{-0.15em}{$\cdots$}}
\newcommand{\dotlinedowndotbb}{\raisebox{-0.15em}{$\dotlinebb$}\hspace{-1.35em}\raisebox{0.15em}{$\cdots$}}
\newcommand{\linedotdowndotbb}{\raisebox{-0.15em}{$\linedotbb$}\hspace{-1.35em}\raisebox{0.15em}{$\cdots$}}

\newcommand{\sixteen}{{\raisebox{-0.1em}{$16$}}}
\newcommand{\sixteenbar}{{\raisebox{-0.1em}{$\overline{16}$}}}
\newcommand{\tinysixteen}{{\raisebox{-0.03em}{\scriptsize $16$}}}
\newcommand{\tinysixteenbar}{{\raisebox{-0.03em}{\scriptsize $\overline{16}$}}}
\newcommand{\indexsixteen}{\scalebox{.5}{$16$}}
\newcommand{\indexsixteenbar}{\scalebox{.5}{$\overline{16}$}}
\newcommand{\tinysixteenaltbar}{{\raisebox{-0.1em}{\tiny $\overset{(-)^{n}}{16~~}$}}}
\newcommand{\tinysixteenaltbarm}{{\raisebox{-0.1em}{\tiny $\overset{(-)^{m}}{16~~}$}}}
\newcommand{\SYTtri}{\mbox{\begin{picture}(4,4)
\put(1,0){\line(1,0){4}}
\put(1,0){\line(1,1){4}}
\put(5,0){\line(0,1){4}}
\end{picture}}}

\newcommand{\YThdots}[1]{\ydiagram{1}\ytableausetup{tabloids}\ydiagram[*(white) \cdots]{{#1}}\ytableausetup{notabloids}\ydiagram{1}}
\newcommand{\YTvdots}[1]{\renewcommand{\arraystretch}{0} \begin{tabular}{l} \ydiagram{1} \\ \ytableausetup{vtabloids}\ydiagram[*(white) \vdots]{#1}\ytableausetup{novtabloids} \\ \ydiagram{1} \end{tabular}}

\newcommand{\generalBYTpqrst}{{\overbrace{\CMTB{\ytableaushort{{}\cdots{},{}\cdots{},{
}\cdots{},{}\cdots{},{}\cdots{}}}}^{t} \hspace{-0.2em} \overbrace{\CMTB{\ytableaushort{{}\cdots{
},{}\cdots{},{}\cdots{},{}\cdots{},\none}}}^{s} \hspace{-0.2em} \overbrace{\CMTB{\ytableaushort{{
}\cdots{},{}\cdots{},{}\cdots{},\none,\none}}}^{r} \hspace{-0.2em} \overbrace{\CMTB{\ytableaushort{{
}\cdots{},{}\cdots{},\none,\none,\none}}}^{q} \hspace{-0.2em} \overbrace{\CMTB{\ytableaushort{{
}\cdots{},\none,\none,\none,\none}}}^{p}}}

\newcommand{\drop}{\CMTB{\ytableaushort{\dotline{\none[-\hspace{-0.3em}\partial\,\,\,]}}}}

\newcommand{\Diop}{ \partial_{\ell} ~+~ i \, \ell \, \drop }

\begin{document}

\thispagestyle{empty}
\noindent{\small
\hfill{$~~$}  \\ 
{}
}
\begin{center}
{\large \bf

Advening to Adynkrafields:\\ Young Tableaux to Component Fields
\vskip0.02in
of the 10D, $\mathcal{N} = 1$ Scalar Superfield
}   \\   [8mm]
{\large {
S.\ James Gates, Jr.\footnote{sylvester$_-$gates@brown.edu}${}^{,a, b}$,
Yangrui Hu\footnote{yangrui$_-$hu@brown.edu}${}^{,a,b}$, and
S.-N. Hazel Mak\footnote{sze$_-$ning$_-$mak@brown.edu}${}^{,a,b}$
}}
\\*[6mm]
\emph{
\centering
$^{a}$Brown Theoretical Physics Center,
\\[1pt]
Box S, 340 Brook Street, Barus Hall,
Providence, RI 02912, USA
\\[10pt]
$^{b}$Department of Physics, Brown University,
\\[1pt]
Box 1843, 182 Hope Street, Barus \& Holley,
Providence, RI 02912, USA
}
 \\*[20mm]
{ ABSTRACT}\\[5mm]
\parbox{142mm}{\parindent=2pc\indent\baselineskip=14pt plus1pt
Starting from higher dimensional adinkras constructed with nodes referenced by Dynkin 
Labels, we  define ``adynkras.''  These suggest a computationally direct way to
describe the component fields contained within supermultiplets in all superspaces.  We
explicitly discuss the cases of ten dimensional superspaces.  We show this is
possible by replacing conventional $\theta$-expansions by expansions over Young Tableaux
and component fields by Dynkin Labels.  Without the need to introduce $\s$-matrices, this
permits rapid passages from Adynkras $\to$ Young Tableaux $\to$ Component Field
Index Structures for both bosonic and fermionic fields while increasing computational
efficiency compared to the starting point that uses
superfields. In order to reach our goal, this work introduces a new graphical method,
``tying rules,'' that provides an alternative to Littlewood's 1950 mathematical results which 
proved branching rules result from using a specific Schur function series.
The ultimate point of this line of reasoning is the introduction of mathematical expansions based on Young Tableaux and that are algorithmically 
superior to superfields. The expansions are given the name of ``adynkrafields" as
they combine the concepts of adinkras and Dynkin Labels.
} \end{center}
\vfill
\vfill
\clearpage

\newpage
{\hypersetup{linkcolor=black}
\tableofcontents
}


\newpage
\section{Introduction}

Papert \cite{CT} introduced the phrase ``computational thinking" in 1980.  A 
search on-line can be found to lead to the following comment.

$~~~~~$ {\it {Computational Thinking (CT)... is essential to the development of 
computer}}  \newline $~~~~~~~~\,~~$
{\it {applications, but it can also be used to support problem solving across 
all }}  \newline  $~~~~~~~~~\,~$
{\it {disciplines, including math, science, and the humanities.}}

\noindent
One point this approach emphasizes is focused attention on the formulation of algorithms.  
In recent works \cite{HyDSG2,HyDSG3}, we have been exploring emerging opportunities 
created by the adinkra-based framework, enhanced algorithmic architectures, and computational 
applications to study superspace\footnote{It is an often overlooked historical fact that the
concept of ``superspace'' \cite{A-V} was introduced independently and separately from the concept 
of ``superfields''\cite{SandS} and we recognize S.\ Kuzenko for discussion.} supergravity in the 
ten and eleven dimensional geometrical limits of the heterotic string, superstrings, and M-Theory. 
These efforts have shown success as they permitted the {\it {complete}} deciphering of the 
Lorentz spectra in $\mathfrak{so}(1,10)$ and $\mathfrak{so}(1,9)$, respectively, for all 
component fields contained in scalar superfields.  We posit this as a notable advance against 
the benchmark established by Bergshoeff and de Roo \cite{BdR} since our results cover Type-II 
superspaces and 11D superspace.

Adinkras appropriate for 10D and 11D superfields, involving billions of degrees of freedom, 
have been successfully constructed by use of Young Tableaux \cite{HyDSG3}, Dynkin Labels 
\cite{yamatsu2015,Susyno,LieART}, and Plethysm \cite{Susyno,Plethysm,LiE}.  Successful algorithms 
based on the information held {\it {solely}} in these adinkras, as opposed to that in 
traditional $\theta$-expansions of superfields, thus emerged due to their increased 
calculational and computational efficiency.  

The efficacy of this approach can be understood if we analogize a superfield to a 
biological body where the adinkra plays the role of a genome.  By the study of genes 
and knowing their expressions, one can deduce information about structures. This is the reason 
why using the foundation of the adinkra concept, we were able to analyze
\cite{HyDSG3} {\it {all}} the $2^{31}$ (= 2,147,483,648) bosonic degrees of freedom
and {\it {all}} the $2^{31}$ (= 2,147,483,648) fermionic degrees of freedom in the
11D, $\cal N$ = 1 scalar superfield.  Scalar superfields act as gateways to the similar 
deciphering the component field spectra of superfields in all spin representations.  
Using this fact, we have begun the task of identifying superfields that contain the 
conformal graviton in these contexts.  

By this means we discovered, a bit surprisingly, the 11D, $\cal N$ = 1 scalar
superfield contains:
\vskip0.01in \indent
(a.) the symmetrical conformal graviton at the 16-th order of the $\theta$-expansion,  
\newline \indent
(b.) a 3-form
at the 16-th order of the $\theta$-expansion, 
\newline \indent
(c.) a conformal gravitino
at the 17-th order of the $\theta$-expansion, and
\newline \indent
(d.) 1,494 bosonic fields and 1,186 fermionic fields in general,
\vskip0.01in \noindent
... facts unknown from the time this theory was introduced into the literature.  
 
Furthermore, we found the 11D, $\cal N$ = 1 scalar superfield does not possess
the antisymmetrical part of the component level vielbein at the sixteenth level.  
Based on past experience with supergravity in superspace \cite{4DN2SG}, combined
with the results from the study of the 11D, $\cal N$ = 1 scalar superfield, the simplest
proposal for the 11D, $\cal N$ = 1 supergravity prepotential is a spinor superfield
$\Psi{}_{\a}$\footnote{Here, $\Psi{}_{\a}$ is an 11D, $\cal N$ = 1 superfield, not to be 
confused with the 10D, $\cal N$ = 1 superfield that we denote by the same symbol 
later in this work.}  where the complete component fields of Poincar\' e vielbein are contained 
at the 17-th level, along with the gauge 3-form, and the complete component fields of 
Poincar\' e gravitino are contained at the 18-th level in the $\theta$-expansion.

Another surprise uncovered was the prolific presence of the component graviton among 11D, $\cal N$ = 1 candidates for the SG prepotential.  
Our scan reveals this particular presence of supergravity on-shell component 
occurs in every superfield up to and including the 
$\Psi_{\CMTB{\{255255\}}} = \Psi_{\CMTB{[2,0,2,0,0]}} = \Psi_{\{ \athree, \afour | \aone\bone\cone, \atwo\btwo\ctwo \}}$\footnote{This translation of notations between dimensions, Dynkin labels and the index notation of an irrep will be explicitly explained in the following sections.}.

It may not be obvious if one begins with higher dimensional adinkras described by Dynkin 
Labels, that there is a path to component fields.   It is the purpose of this work to provide 
an end-to-end demonstration showing how this is carried out.  We will marshal the lessons 
learned in the 11D, $\cal N$ = 1 theory and apply them to the 10D, $\cal N$ = 1 scalar 
superfield as it is the basis for gaining a complete understanding of off-shell 
supersymmetical theories in this arena.

In this work, we will introduce a concept which shows some potential for becoming a 
computationally superior complement for superfields.  We call it the ``adynkrafield
formulation'' which appears as a natural consequence of the path we have explored.  The 
importance of two distinct sets of Young Tableaux, each associated with Dynkin Labels 
in our discussions, points toward the use of the Young Tableaux together with the 
introduction of a ``level parameter'' $\ell$ as a basis for expansions.  Initial 
evidence is given that these are sufficient to accurately investigate the domain where
heretofore traditional Grassmann coordinates in superfields, as defined by Salam and Strathdee,
provided the sole means enabling investigations.

In Chapter two, we present the adinkra for the 10D, $\cal N$ = 1 scalar superfield that
provides the starting point for our construction.  All nodes of the adinkra are described
by Dynkin Labels.  We review how computational efficiency is gained from this view
point.  Since adinkras utilizing Dynkin Labels play a key role, the new term ``adynkras''
is introduced to describe this particular form of adinkras.

In Chapter three, we turn to adapting the well known technology of Young Tableaux to 
the task of representing the irreducible bosonic representations of $\mathfrak{so}(10)$.  
It should be noted the challenge here comes about because there is a well accepted 
method for using Young Tableaux to represent the irreducible representations for 
$\mathfrak{su}(10)$, but not for $\mathfrak{so}(10)$.  Adaptations are necessary to allow 
a set of decorated Young Tableaux to accomplish the latter goal. 

In the literature, a number of other works have offered proposals for such adaptations to achieve this purpose \cite{fischler1981,hurni1987,KingES1983,BS1991,GSP1982}. Several of them use a very similar approach to ours \cite{fischler1981,hurni1987,KingES1983,BS1991}, which translates Dynkin labels to Young Tableaux, and puts bosonic and spinorial Young Tableaux side by side. Others associate irreducible representations with skew Young Tableaux which involves ``negative boxes'' for tensor product calculations \cite{GSP1982}.

Our discussion starts with the presentation of a logical path for the construction of a
projection matrix for $\mathfrak{su}(10)\supset \mathfrak{so}(10)$.  Here we set our
conventions for the embedding.  We next adopt a set of conventions for how the Dynkin
Labels are to be translated into conventional $\mathfrak{so}(10)$ Young Tableaux.  
This is followed by a brief discussion of how these $\mathfrak{so}(10)$ Young Tableaux
are reducible with respect to a set of ``adapted" $\mathfrak{so}(10)$ Young Tableaux
we introduce in this work and the extraction of irreducible tableaux.

Chapter four is devoted to the graphical rules for the $\mathfrak{su}(10)\supset\mathfrak{so}(10)$ branching rules.
Littlewood's rule is reviewed. Then we turn to the
introduction of a set of graphical rules, we call ``tying rules,'' 
that allow the $\mathfrak{su}(10)$ Young Tableaux to be decorated in such a way so 
as to produce irreducible $\mathfrak{so}(10)$ Young Tableaux.  This is illustrated in 
some examples.  However, this discussion is limited to bosonic Young Tableaux, i.e. 
those Tableaux that are associated with bosonic representations.

Chapter five aims to construct Young Tableau representations for the spinorial irreducible
representations of $\mathfrak{so}(10)$. Mixed Young Tableaux as well as the graphical rules 
that lead to correct dimensions are introduced. 
Mixed Young Tableaux highlight the facts that we are using two distinct types of Young Tableaux.  
Blue Tableaux are associated with bosonic indices and representations while red Tableaux are associated with spinor indices and representations.
This is accompanied by the concomitant task 
of describing a corresponding set of Dynkin Labels.  Illustrations are given to show how the 
issue of irreducibility is handled.

Chapter six presents the general graphical rules to get tensor product decompositions of a bosonic irrep with the basic spinor representation of $\mathfrak{so}(10)$.
The inverse relation between this tensor product rule and the dimension rule in last chapter is presented and demonstrated through examples.

Chapter seven brings all the strands of the previous chapters together with the explicit presentation of the field variables showing all their various types of indices associated with each node of the adynkra introduced in Chapter two.  
The derivation begins from the adynkra in chapter two in which the nodes are expressed in terms of Dynkin Labels. 
It is shown the Dynkin Labels contain sufficient data to derived a complete description of the Lorentz structure of all the component fields.  A complete description of the irreducibility conditions
is presented.

Chapter eight presents a new concept to which we give the name ``adynkrafields."  Adinkras 
and adynkras do not necessarily depend on field variables. In such settings, the adinkras 
and adynkras play a role similar to matrices in, for example, the study of $\mathfrak{su}(3)$ representations.

Chapter nine presents our conclusions.

There are three appendices included in this work.  
Appendix \ref{appen:example-notation} is devoted to present a dictionary between bosonic indices of field variables, irreducible Young
Tableaux, and Dynkin Labels.  
Appendix \ref{appen:SYTdimex} gives explicit examples of the dimension formula for Mixed Young Tableaux defined in Chapter five.
Appendix \ref{appen:spinorial-tensorproduct} is devoted to explicit examples of the graphical tensor product rules stated in Chapter six as well as providing demonstrations of the subject of
multiplication of the fundamental SYT by BYT's to obtain SYT's.  It thus covers the same 
topics as Chapter \ref{sec:spinorial-irreps} but now translates spinorial representations 
into field language by only considering tensor product decomposition.

The final portion of this paper includes our references.


\newpage
\section{The 10D, $\cal N$ = 1 Scalar Superfield Adynkra}
\label{sec:adynkra}

\begin{figure}[htp!]
\centering
\includegraphics[width=0.35\textwidth]{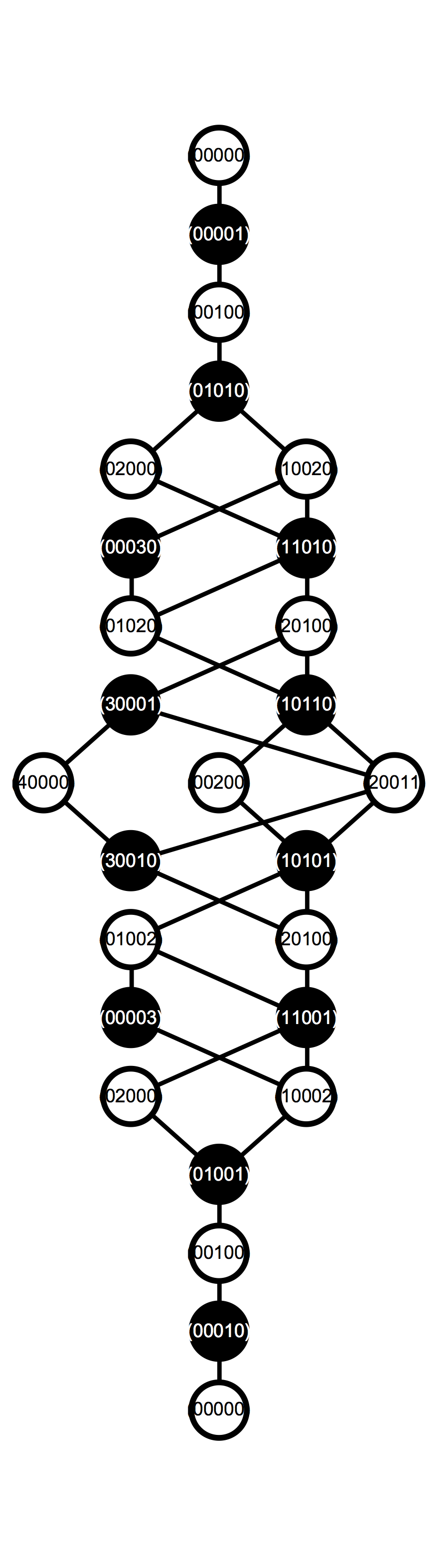}
\caption{Adinkra Diagram for 10D, $\mathcal{N} = 1$ Scalar Superfield}
\label{Fig:10DTypeI_Dynkin}
\end{figure}

\indent
In the work of \cite{HyDSG3} the adinkra for the 10D, $\cal N$ = 1 scalar
superfield, with nodes expressed in terms of Dynkin Labels, was
shown as it appears in Figure \ref{Fig:10DTypeI_Dynkin}.
Counting the number of open and closed nodes respectively implies this is a
superfield with 15 bosonic component fields and 12 fermionic component fields. The Dynkin Labels
on each node carry the information about the SO(1,9) Lorentz representation
of each field.  However, the image signifies another possibility to
which we will return shortly.

In most of our initial investigations of higher dimensional adinkras, the bulk of 
the discussions was carried out in cases where the dimensionality 
of the representations of the nodes was illustrated. There is an ambiguity in
such labeling. We can see this by considering the dimensionality formula in
the simpler case of $\mathfrak{su}(3)$.  In the next chapter shown in equation 
(\ref{dimnum}), there appears the relation between the dimensionality of a
representation in $\mathfrak{su}(3)$ specified by the integers $p$ and $q$ which 
occur in a Dynkin Label [$p$, $q$].  The form of $d(p, \, q)$ shows that the case 
where  $p$ = M and $q$ = N has the same dimensionality as the case where $p$ = N 
and $q$ = M for any positive integers M and N.  So for a fixed value of $d(p, \, q)$, the 
solutions for $p$ and $q$ are not unique.  Thus, labeling the nodes of the higher 
dimensional adinkra with Dynkin Labels removes the ambiguity.

The image in Figure \ref{Fig:10DTypeI_Dynkin} uses Dynkin Label to describe
the Lorentz representation associated with each node.  In the following, it will
be shown that this realization has a computationally superior property.  The
knowledge of the integers that appear in the Dynkin Label suffices to derive
complete component level field variables with their varied Lorentz index structures.  
Henceforth, we will refer to adinkras where their nodes are described by Dynkin
Labels as ``adynkras,'' replacing the letters ``ink'' in ``adinkra'' by the letters
``ynk'' from ``Dynkin.''

The ``platform'' of this current work is within the context of linearized Nordstr\" om supergravity in 10D, $\cal N$ = 1 superspace, as we established the foundation for this 
in the work of \cite{HyDSG1}.  One reason for performing this is the component level construction of the linearized 10D, $\cal N$ = 1 Nordstr\" om supergravity theory yields 
the simplest context in which the derivation faces all the same problems present in the  
general class of models described previously \cite{HyDSG2,HyDSG3,HyDSG1}.

In order to accomplish our task we need to develop: \newline \indent $~~~~~$
(a.) a set of new concepts based on direct graphical manipulations of Young
\newline \indent $~~~~~~~~~~~$
Tableaux allowing them to generate
$\mathfrak{su}(10)\supset \mathfrak{so}(10)$
branching rules, and
\newline \indent $~~~~~$
(b.) a ``translation dictionary'' for Dynkin Labels into indices
on field variables. \vskip.01in
\noindent
In particular for task (a.), to our knowledge, these will be new concepts and 
techniques introduced into the literature.

We need a well defined methodology for converting Dynkin Labels into indices on a 
set of field variables. We now turn to this in the less complicated context of the
10D, $\cal N$ = 1 system \cite{HyDSG1}.

Almost since their introduction \cite{Adnk1} and in one form or another, numbers
of  physicists have posed the question, ``Is the purpose of adinkras to replace
superfields?''  We have always responded, ``No, the purpose of adinkras is to
augment Salam-Strathdee superfields'' \cite{SandS}.  

From the time of the discovery of supersymmetry at the level of on-shell component
representations \cite{{GrVSak},GLKTMN,LKTMN}, the description's incompleteness 
was obvious.  To remedy this, Salam and Strathdee invented superfields.  However,
superfields are ill-posed as a convenient platform from which to study the structure
of the representation theory associated with spacetime supersymmetry.  A most forceful
demonstration of this was given in our work of \cite{HyDSG3}, where we demonstrated
the unwieldiness associated with the $\theta$-expansion of the scalar 11D, $\cal
N$ = 1 superfield itself acts as a computational impediment to deriving results. The 
problem arises due to the fact that the {\it {actual derivation}} of component results require
the subsidiary derivation of large numbers of Fierz identities.  In the context of this 
system, the totality of these derivations has never been shown in the literature.  We 
suspect the reason is the daunting numbers of these.

The primary purpose of adinkras is to provide a ``Goldilocks'' solution by avoiding
the incompleteness of the component-level approach while simultaneously avoiding
the unwieldiness of the superfield approach.  This is accomplished by banishing the
need for $\g$-matrices (actually $\s$-matrices in 10D) in deriving supersymmetrical
representation theory results.

\newpage
\section{Bosonic Irreps of $\mathfrak{so}(10)$ \& Irreducible Bosonic Young Tableaux}
\label{sec:so10BYT}

\subsection{Preview: $\mathfrak{su}(3)$ \& Irreducible Bosonic Young Tableaux}
\label{subsec:BYTindexpq}

In the manner of a warm-up, we review a discussion from a previous paper
\cite{4DN2H0L0R} regarding the  $\mathfrak{su}(3)$ algebra.  As covered there, 
we have seen the most general $\mathfrak{su}(3)$ Young Tableaux takes the
form
\begin{equation} \label{equ:p_q_young}
    \overbrace{\ytableaushort{{}\cdots{},{}\cdots{}}}^{q}\hspace{-0.2em}
    \overbrace{\ytableaushort{{}\cdots{},\none}}^{p} ~~~~~.
\end{equation}
We next introduce the Dynkin Label $[p,q]$ as the highest weight vector for irreps in $\mathfrak{su}(3)$.  
The dimensionality of the irrep with the Dynkin Label $[p,q]$ is given by the Weyl 
dimension formula applied to $\mathfrak{su}(3)$ \cite{WEY}
\be {
d(p, \, q) ~=~ \fracm 12 \, \left( \, p \,+\, 1  \, \right) \, \left( \,  q \,+\, 1\, \right)  \, \left( \,
 p \,+\, q \,+\, 2 \, \right) ~~~.
 \label{dimnum}
} \ee

\subsection{Feature: Beginning $\mathfrak{su}(10)$ $\to$ $\mathfrak{so}(10)$ \& Irreducible
Bosonic Young Tableaux}
\label{subsec:BYTindexpqrst}

We now look at the bosonic Young Tableaux that correspond to bosonic representations
of $\mathfrak{so}(10)$. Generally speaking, these are not necessarily irreducible. In
order to establish a graphical language to describe bosonic irreducible representations
in $\mathfrak{so}(10)$, we must define {\em irreducible} bosonic Young Tableaux.

Consider the projection matrix for $\mathfrak{su}(10)\supset \mathfrak{so}(10)$~\cite{yamatsu2015},
\begin{equation}
P_{\mathfrak{su}(10)\supset \mathfrak{so}(10)} ~=~
\begin{pmatrix}
1 & 0 & 0 & 0 & 0 & 0 & 0 & 0 & 1\\
0 & 1 & 0 & 0 & 0 & 0 & 0 & 1 & 0\\
0 & 0 & 1 & 0 & 0 & 0 & 1 & 0 & 0\\
0 & 0 & 0 & 1 & 0 & 1 & 0 & 0 & 0\\
0 & 0 & 0 & 1 & 2 & 1 & 0 & 0 & 0\\
\end{pmatrix}~~~.
\end{equation}
The highest weight of a specified irrep in $\mathfrak{su}(10)$ is a row vector $[p,q,r,s,t,u,v,w,x]$,
where $p$ to $x$ are non-negative integers. Since the $\mathfrak{su}(10)$ YT with $n$ vertical 
boxes is the conjugate of the one with $10-n$ vertical boxes, we need only consider the $u=v=w=x=0$ case.

Starting from the weight vector $[p,q,r,s,t,0,0,0,0]$ in $\mathfrak{su}(10)$, we define its projected
weight vector $[p,q,r,s,s+2t]$ in $\mathfrak{so}(10)$ as the Dynkin Label of the corresponding
irreducible bosonic Young Tableau.
\begin{equation}
    [p,q,r,s,s+2t] ~=~ [p,q,r,s,t,0,0,0,0] \, P^T_{\mathfrak{su}(10)\supset\mathfrak{so}(10)} ~~~.
\end{equation}
Thus, given an irreducible bosonic Young Tableau, we obtain the definition of its corresponding
bosonic irreducible Dynkin Label representation. Moreover, we can reverse this process and show the
one-to-one correspondence between bosonic Dynkin Label irreps and irreducible bosonic Young Tableaux.
Namely, given a Dynkin Label $[a,b,c,d,e]$, write a set of linear equations:
\begin{equation}
\begin{split}
    a ~=&~ p ~~~, \\
    b ~=&~ q ~~~, \\
    c ~=&~ r ~~~, \\
    d ~=&~ s ~~~, \\
    e ~=&~ s+2t ~~~,
\end{split}
\end{equation}
and obtain
\begin{equation}
\begin{split}
    p ~=&~ a ~~~, \\
    q ~=&~ b ~~~, \\
    r ~=&~ c ~~~, \\
    s ~=&~ d ~~~, \\
    t ~=&~ \frac{e - d}{2} ~~~.
\end{split}
\end{equation}
The problem is that in order to obtain a valid YT, $p$, $q$, $r$, $s$, and $t$ have to be non-negative 
integers, meaning $e-d$ has to be an non-negative even integer. When $e>d$, we need to 
prove that $e-d$ is even and assign the corresponding YT with subscript $({\rm IR}, +)$.  When 
$e<d$, we assign the corresponding YT with subscript $({\rm IR}, -)$ which is described by the 
alternate Dynkin Label given by  $[a,b,c,e,d]$. In this case, $p = a$, $q = b$, $r = c$, $s = e$, 
and $t = \frac{d-e}{2}$, where $d-e$ has to be even.

The proof is simple. Consider the congruence classes of a representation with 
Dynkin Label $[a,b,c,d,e]$ in SO(10),
\begin{equation}
    \begin{split}
        C_{c1}(R) ~:=&~ d + e ~~ ({\rm mod} ~ 2) ~~~,\\
        C_{c2}(R) ~:=&~ 2a + 2c + 3d + 5e ~~ ({\rm mod} ~ 4) ~~~.
    \end{split}
\end{equation}
A congruence class is an equivalent class of irreps. Based on
the above equations, there are totally four congruence classes in SO(10),
\begin{equation}
    \Big[ C_{c1}, C_{c2} \Big](R) ~=~
    \begin{cases}
    ~[0,0]\\
    ~[0,2]\\
    ~[1,1]\\
    ~[1,3]
    \end{cases} ~~~.
\end{equation}
The quantity $\Big[ C_{c1}, C_{c2} \Big]$ can be treated as a vector and it satisfies
\begin{equation}
    \Big[ C_{c1}, C_{c2} \Big](R_1\otimes R_2) ~=~ \Big[ C_{c1}(R_1)+C_{c1}(R_2)
    ~~ ({\rm mod} ~ 2) ~,~ C_{c2}(R_1)+C_{c2}(R_2) ~~ ({\rm mod} ~ 4)  \Big]~~.
\end{equation}
We know that
\begin{equation}
    \begin{cases}
    ~\text{bosonic irrep} \otimes \text{bosonic irrep} ~=~ \text{bosonic irrep}~~~,\\
    ~\text{bosonic irrep} \otimes \text{spinorial irrep} ~=~ \text{spinorial irrep}~~~,\\
    ~\text{spinorial irrep} \otimes \text{spinorial irrep} ~=~ \text{bosonic irrep}~~~.\\
    \end{cases}
\end{equation}
One can quickly check that $C_{c1}(R)$ actually classifies the bosonic irreps and spinorial
irreps:  $C_{c1}(R) = 0$ is bosonic and  $C_{c1}(R)=1$ is spinorial.  Consequently, a
bosonic irrep satisfies $d+e = 0 ~~ ({\rm mod} ~ 2)$ and consequently $d-e
= 0 ~~ ({\rm mod} ~ 2)$.

Summarizing, given an irreducible bosonic Young Tableau with $p$ columns of one box,
$q$ columns of two vertical boxes, $r$ columns of three vertical boxes, $s$ columns of
four vertical boxes, and $t$ columns of five vertical boxes, the Dynkin Label of its
corresponding bosonic irrep is $[p,q,r,s,s+2t]$ or $[p,q,r,s+2t,s]$ depending on its self-duality.  
Given a bosonic irrep with Dynkin Label $[a,b,c,d,e]$, its corresponding irreducible 
bosonic Young Tableau is composed of $a$ columns of one box, $b$ columns of two 
vertical boxes, $c$ columns of three vertical boxes, $d$ columns of four vertical boxes, 
and $|e-d|/2$ columns of five vertical boxes. Duality properties depend on the sign of 
$e-d$ which has been discussed. From the discussion above, we know that a bijection 
is established between Dynkin Labels and BYTs, that there's no ambiguity in the 
translation from Dynkin Labels to BYTs. When $d=e$ (even), it would mean $d$ 
columns of four vertical boxes, instead of sticking two sets of $e/2$ columns of five 
boxes of opposite dualities.

The simplest examples,
also the fundamental building blocks of a BYT, are given below.
\begin{equation}
\begin{gathered}
    {\CMTB{\ydiagram{1}}}_{{\rm IR}} ~\equiv~ \CMTB{[1,0,0,0,0]}  ~~~,~~~
    {\CMTB{\ydiagram{1,1}}}_{{\rm IR}} ~\equiv~ \CMTB{[0,1,0,0,0]} ~~~,~~~
    {\CMTB{\ydiagram{1,1,1}}}_{{\rm IR}} ~\equiv~  \CMTB{[0,0,1,0,0]} ~~~,  \\
    {\CMTB{\ydiagram{1,1,1,1}}}_{{\rm IR}} ~\equiv~ \CMTB{ [0,0,0,1,1] } ~~~,~~~
    {\CMTB{\ydiagram{1,1,1,1,1}}}_{{\rm IR},+} \hspace{-0.4em}\equiv~
    \CMTB{ [0,0,0,0,2] } ~~~,~~~
    {\CMTB{\ydiagram{1,1,1,1,1}}}_{{\rm IR},-} \hspace{-0.4em}\equiv~
    \CMTB{ [0,0,0,2,0] } ~~~.
\end{gathered} \label{equ:BYTbasic}
\end{equation}
We put ``IR'' as subscripts to indicate that these are irreducible representations\footnote{This convention is also adopted by \cite{hurni1987}.}.
Putting together these columns corresponds to adding their Dynkin Labels. All the
BYT with one or more columns of 5 boxes can be either self-dual or anti-self-dual.
Here we impose a rule that if there's no $+$ or $-$ subscript put at the corner, it is
assumed as the direct sum of the two irreps.
\begin{equation}
    {\CMTB{\ydiagram{1,1,1,1,1}}}_{\rm IR} ~=~
    {\CMTB{\ydiagram{1,1,1,1,1}}}_{{\rm IR},+} \hspace{-0.8em}\oplus~
    {\CMTB{\ydiagram{1,1,1,1,1}}}_{{\rm IR},-} ~.  \label{eqn:col5}
\end{equation}
With these basic elements, we can build different examples of BYTs,
\begin{align}
    {\CMTB{\ydiagram{3,1,1}}}_{\rm IR} \hspace{0.75em}~\equiv&~
    \CMTB{[2,0,1,0,0]} ~~~, \\
    {\CMTB{\ydiagram{3,2,2,2,1}}}_{{\rm IR},+} ~\equiv&~
     \CMTB{[1,0,0,1,3]} ~~~, \label{equ:E+}  \\
    {\CMTB{\ydiagram{3,2,2,2,1}}}_{{\rm IR},-} ~\equiv&~
    \CMTB{[1,0,0,3,1]} ~~~. \label{equ:E-}
\end{align}
These last two images illustrate the meaning of ``duality'' in the present context.  The
Tableaux shown in (\ref{equ:E+}) and (\ref{equ:E-}) each corresponds to tensors that
possess ten indices, five of which are totally antisymmetric as signified by the length
of the first column.  Thus, with respect to these indices the tensor are five-forms.  It is
a well recognized fact that a five-form in the context of a ten dimensional manifold can
either be dual or anti-dual.  This distinction is captured by the $\pm$ subscript shown
at the bottom of the tableaux.

\subsection{Indices Corresponding To Irreducible Bosonic Young Tableaux}
\label{subsec:BYTindex}

When translating the irreducible bosonic Young Tableaux into field representations,
Young Tableaux tell us the index structure of the field. 
In some literature \cite{curtright1982,dauriafre1982}, 
for the efficiency in expressing an index structure, 
an entire Young Tableau is drawn in the subscript of the field 
in replacement of a bunch of overlapping $(~~)$ and $[~~]$.
Here we develop the notation further such that it becomes compact and typable. 
We introduce the following notational conventions.
We put all the bosonic indices in a pair of curly braces ``$\{~\}$''.
We use ``$|$'' to separate indices in column(s) of YT with different heights
and ``,'' to separate indices in column(s) of YT with the same heights. 
It should be noted that the $\{~\}$-indices, irreducible bosonic Young
Tableaux, and Dynkin Labels are equivalent and have one-to-one correspondence.
The general expression is given below in Figure \ref{fig:index-notation0} and Figure \ref{fig:index-notation},
\begin{figure}[htp!]
    \centering
    \includegraphics[width=5.4in]{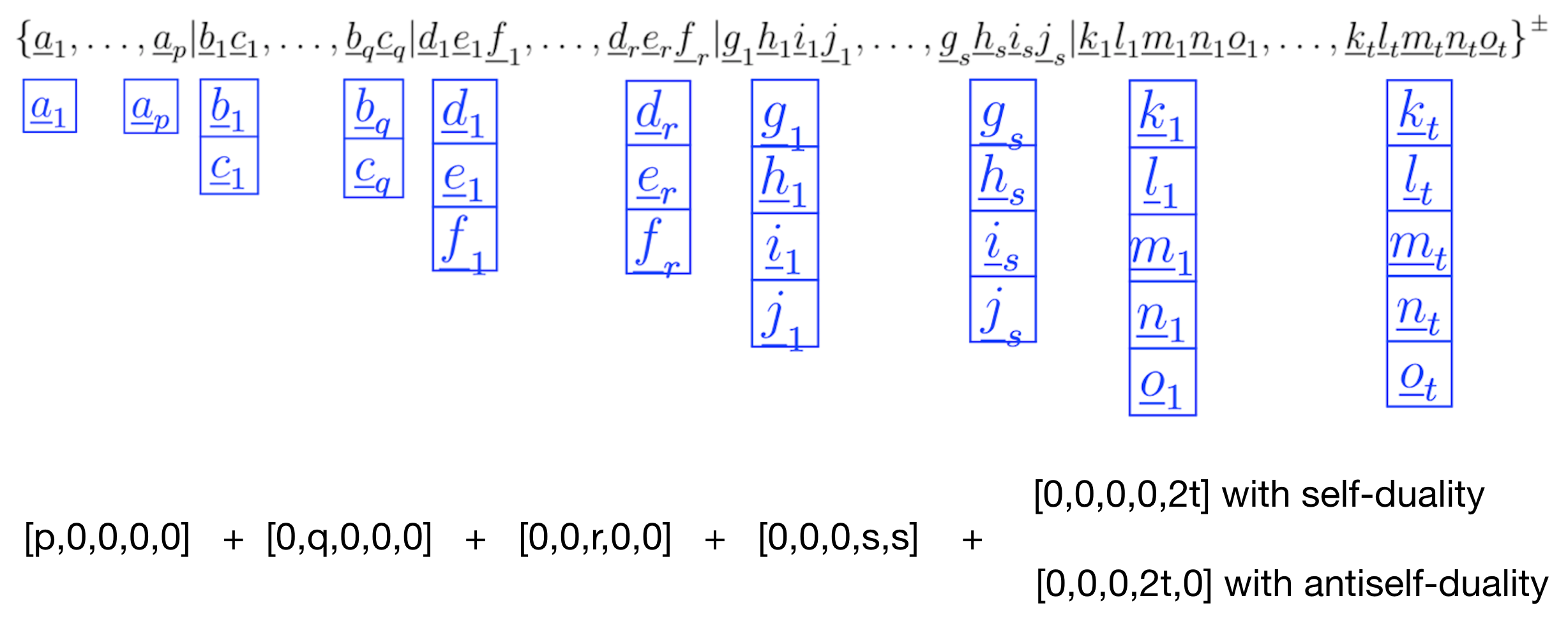}
    \caption{Young Tableaux-Index Structure Notation \& Conventions\, \# 1}
    \label{fig:index-notation0}
\end{figure}
\vskip0.1pt  \noindent
where in Figure \ref{fig:index-notation0} we have ``disassembled'' the YT to show how each column is affiliated with each type of
subscript structure. In Figure \ref{fig:index-notation}, we have assembled all the column into a proper YT.
\begin{figure}[htp!]
    \centering
    \includegraphics[width=3.8in]{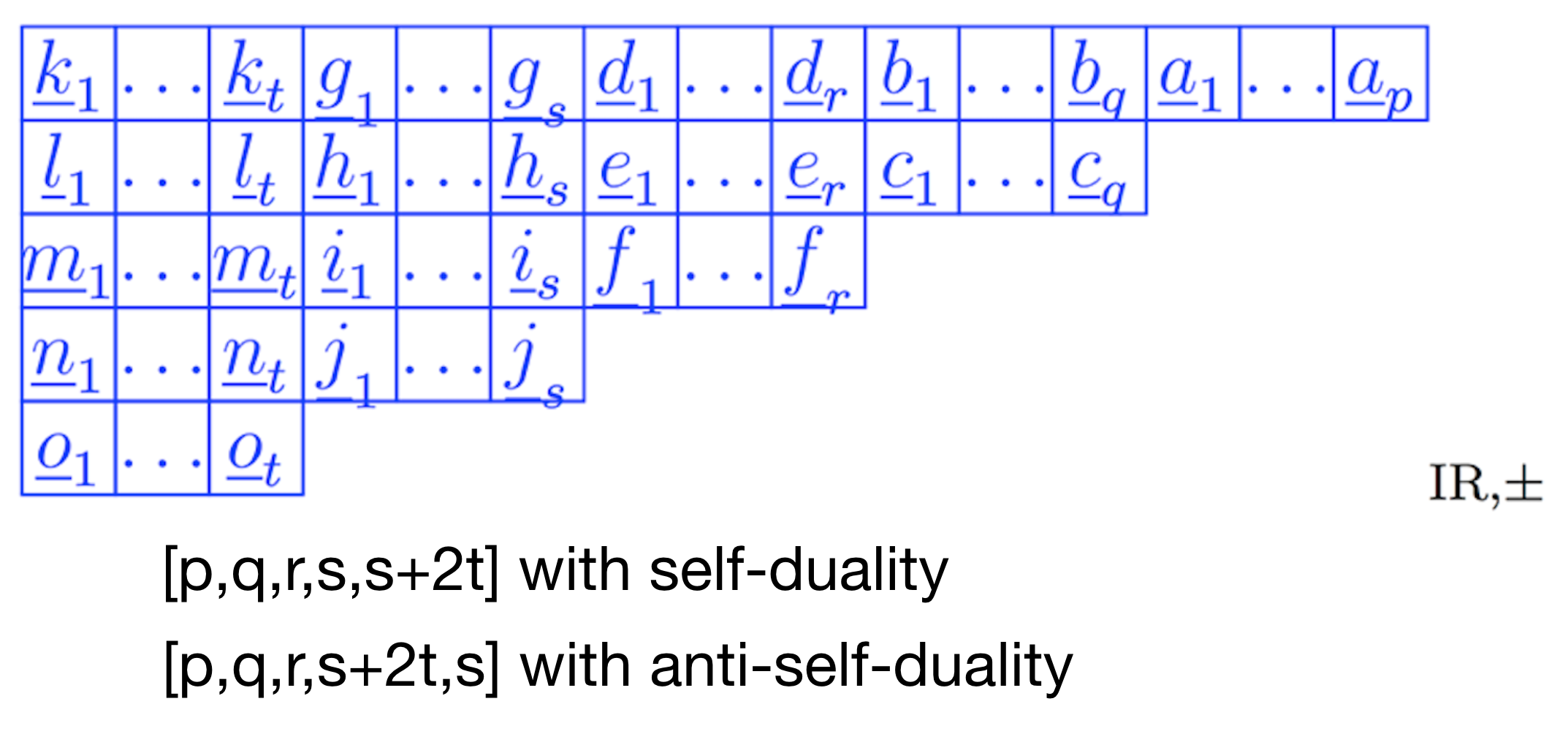}
    \caption{Young Tableaux-Index Structure Notation \& Conventions\, \# 2}
    \label{fig:index-notation}
\end{figure}

As one moves from the YT's shown in Figure~\ref{fig:index-notation0} to
Figure~\ref{fig:index-notation}, it is clear that the number of vertical boxes is tabulating the number of
1-forms, 2-forms, 3-forms, 4-forms, and 5-forms in the YT's.  These are the entries between
the vertical $|$ bars. These precisely correspond to the Dynkin Labels $p$, $q$, $r$, $s$,
and $t$.  A first example of the correspondence between the subscript conventions and
the affiliated YT and Dynkin Label is shown in (\ref{equ:index-notation1}). More examples are provided in Appendix~\ref{appen:example-notation}.
\begin{equation}
    \{{\un a}_2 , {\un a}_3| {\un a}_1  {\un b}_1   {\un c}_1  {\un d}_1\} ~~\equiv~~
    { \CMTB{{\ytableaushort{\aone \atwo \athree,\bone,\cone,\done}} } }_{{\rm IR}}
    ~~\equiv~~ \CMTB{[2,0,0,1,1]} ~~~.
    \label{equ:index-notation1}
\end{equation}

One remark is that the $\{~\}$-indices that include five vertical boxes within them are
separated into anti-self-dual and self-dual components that satisfy the equations
\begin{equation}
   \{{\un a}_1   {\un b}_1   {\un c}_1  {\un d}_1  {\un e}_1  \}{}^{\pm} ~=~ \pm
\fracm 1{5!} \, \epsilon{}_{{\un a}_1   {\un b}_1   {\un c}_1  {\un d}_1  {\un e}_1}
{}^{{\un a}_2   {\un b}_2   {\un c}_2  {\un d}_2  {\un e}_2}
\{ {\un a}_2   {\un b}_2   {\un c}_2  {\un d}_2  {\un e}_2    \}{}^{\pm}  ~~~.
\end{equation}

Figure~\ref{fig:index-notation} is the
$\mathfrak{so}(10)$ generalization of the one shown in Equation (\ref{equ:p_q_young}) for 
$\mathfrak{su}(3)$.  However, there is one important difference.  The YT itself shown in 
Equation (\ref{equ:p_q_young}) is related to irreducible representations of $\mathfrak{su}
(3)$, while in Figure~\ref{fig:index-notation} the YT with ``IR" subscript subject to certain conditions refers 
to irreducible representations of $\mathfrak{so}(10)$.
YT's without this subscript are reducible with respect to $\mathfrak{so}(10)$.
In the next chapter, we will deal with 
extracting irreducible representations of $\mathfrak{so}(10)$ from YT's without ``IR" subscript by a graphical means.

\subsection{Irreducibility Conditions}
\label{subsec:bosonicirrcond}

Irreducible bosonic Young Tableaux only tell us the index structures of the
fields when we translate the irrep descriptions into field variable language. If we want 
the correct d.o.f.\footnote{We use d.o.f. as the abbreviation for "degrees of freedom."} 
of the fields, we have to include the irreducibility conditions or constraints. The 
irreducibility conditions are effectuated by the branching rules for $\mathfrak{su}(10) 
\supset \mathfrak{so}(10)$.
Examples are
\begin{align}
\begin{split}
    \ydiagram{1}  ~=&~ { \CMTB{\ydiagram{1}} }_{{\rm IR}}  \\
    \{10\} ~=&~ \CMTB{\{10\}}
\end{split}  ~~~~~, \\
\begin{split}
    \ydiagram{2}  ~=&~ { \CMTB{\ydiagram{2}} }_{{\rm IR}} ~\oplus~  \CMTB{\cdot} \\
    \{55\} ~=&~ ~~\CMTB{\{54\}}~~  ~\oplus~  \CMTB{\{1\}}
\end{split} ~~~~~, \\
\begin{split}
    \ydiagram{2,1}  ~=&~ { \CMTB{\ydiagram{2,1}} }_{{\rm IR}}~\oplus~
    { \CMTB{\ydiagram{1}} }_{{\rm IR}}   \\
    \{330\} ~=&~ ~\CMTB{\{320\}}~\, ~\oplus~ \CMTB{\{10\}}
\end{split} ~~~~~, \\
\begin{split} \label{equ:2310}
    \ydiagram{2,1,1,1,1}  ~=&~ { \CMTB{\ydiagram{2,1,1,1,1}} }_{{\rm IR},-}
    \hspace{-0.5em}\oplus~ { \CMTB{\ydiagram{2,1,1,1,1}} }_{{\rm IR},+}
    \hspace{-0.5em}\oplus~ { \CMTB{\ydiagram{1,1,1,1}} }_{{\rm IR}}  \\
    \{2310\} ~=&~ \,\CMTB{\{1050\}}\, ~\oplus~ \CMTB{\{\overline{1050}\}}
    \, ~\oplus~ \CMTB{\{210\}}
\end{split} ~~~~~.
\end{align}
In each above equation, the leading term tells us the index structure corresponding to
the irrep and the remaining terms tell us the irreducible conditions or constraints. 
The degrees of freedom contributed by each term are also presented below the YT. They 
can be translated into the field language respectively,
\begin{align}
    \Phi{}_{\{{\un a}_1\}} ~:&~~~ {\rm N/A}  ~~~, \\
    \Phi{}_{\{{\un a}_1, {\un a}_2 \}} ~:&~~~ \eta^{{\un a}_1 {\un a}_2}\, \Phi{}_{\{{\un a}_1
    , {\un a}_2 \}} = 0  ~~~, \\
    \Phi{}_{\{{\un a}_2| {\un a}_1 {\un b}_1 \}} ~:&~~~ \eta^{{\un a}_1 {\un a}_2}\, \Phi{}_{
    \{{\un a}_2| {\un a}_1 {\un b}_1 \}} = 0  ~~~, \\
    \Phi{}_{\{{\un a}_2| {\un a}_1 {\un b}_1 {\un c}_1 {\un d}_1 {\un e}_1 \}} ~:&~~
    \begin{cases}
    ~ \eta^{{\un a}_1 {\un a}_2}\,\Phi{}_{\{{\un a}_2| {\un a}_1    {\un b}_1   {\un c}_1  
    {\un d}_1  {\un e}_1  \}} = 0  ~~~, \\
    ~ \Phi{}_{\{{\un a}_2| {\un a}_1 {\un b}_1 {\un c}_1 {\un d}_1 {\un e}_1 \}{}^{+}} ~=~
    \frac{1}{5!} \epsilon_{{\un a}_1  {\un b}_1 {\un c}_1 {\un d}_1 {\un e}_1}{}^{{\un f}_1
    {\un g}_1 {\un h}_1 {\un i}_1 {\un j}_1}\Phi{}_{\{{\un a}_2| {\un f}_1 {\un g}_1 {\un
    h}_1    {\un i}_1    {\un j}_1 \}{}^{+}}  ~~~, \\
    ~ \Phi{}_{\{{\un a}_2| {\un a}_1 {\un b}_1 {\un c}_1 {\un d}_1 {\un e}_1 \}{}^{-}}~=~-
    \frac{1}{5!}\epsilon_{{\un a}_1 {\un b}_1 {\un c}_1 {\un d}_1  {\un e}_1}{}^{{\un f}_1
    {\un g}_1 {\un h}_1    {\un i}_1 {\un j}_1}\Phi{}_{\{{\un a}_2| {\un f}_1   {\un g}_1 {\un
    h}_1 {\un i}_1 {\un j}_1 \}{}^{-}} ~~~.
    \end{cases}
\end{align}
where
\begin{equation}
    \Phi_{\{{\un a}_2 | {\un a}_1 {\un b}_1 {\un c}_1 {\un d}_1 {\un e}_1 \}} ~=~
    \Phi_{\{{\un a}_2 | {\un a}_1 {\un b}_1 {\un c}_1 {\un d}_1 {\un e}_1  \}{}^{+}}
    ~+~ \Phi_{\{{\un a}_2 | {\un a}_1 {\un b}_1 {\un c}_1 {\un d}_1 {\un e}_1  \}
    {}^{-}}  ~~~.
\end{equation}
as indicated in the text around Equation (\ref{eqn:col5}).

In the next chapter, we will see how these branching rules can be obtained by graphical rules.

\newpage
\section{10D Bosonic Young Tableau Tying Rules}
\label{sec:ties}

From Chapter \ref{sec:so10BYT}, we know irreducible bosonic Young Tableaux for 
$\mathfrak{so}(10)$ can be drawn. In the $A_{n-1} = \mathfrak{su}(n)$ algebra, one can 
calculate the dimension of an irrep by the well-known graphical device involving the use 
of the ``hook rule.'' This leads to the question whether there exists a {\em diagrammatic 
method} to calculate the dimension of an irrep from the $\mathfrak{so}(10)$ BYT directly, 
instead of translating it to a Dynkin Label $[a_1,a_2,a_3,a_4,a_5]$ constructed from the
five integers $a_1$, $a_2$, $a_3$, $a_4$, and $a_5$ followed by substituting 
the integers into the Weyl dimension formula for the $D_{5}$ algebra~\cite{yamatsu2015}
\begin{equation}\label{equ:Weyl}
\begin{split}
d(a_1,a_2,a_3,a_4,a_5)~=&~ \Big[ \prod_{k=1}^{3}\prod_{i=1}^{k} \Big( \frac{\sum_{j=i
}^{k}a_j}{k-i+1}+1\Big) \Big] \Big[ \prod_{k=0}^{3}\prod_{l=4}^{5} \Big( \frac{(\sum_{j=1}^{
k}a_{4-j})+a_l}{k+1}+1\Big) \Big]\\
&~\times~\Big[ \prod_{k=0}^{2}\prod_{i=k+1}^{3} \Big( \frac{(\sum_{j=k+1}^{i}a_{4-j})+
(\sum_{j=1}^{k}2a_{4-j})+a_4+a_5}{i+k+2}+1\Big) \Big] \\
~=&~  \left[ \, \frac{1}{7!\times 5!\times 4!\times 3!} \, \right]
(a_{1} + 1) (a_{2} + 1) (a_{3} + 1) (a_{4} + 1) 
(a_{5} + 1) \\
&~~~~~ (a_{1} + a_{2} + 2) (a_{2} + a_{3} + 2) (a_{3} + a_{4} + 2) (a_{3} + a_{5} + 2) \\
&~~~~~ (a_{1} + a_{2} + a_{3} + 3) (a_{2} + a_{3} + a_{4} + 3) (a_{2} + a_{3} + a_{5} + 3) 
(a_{3} + a_{4} + a_{5} + 3) \\
&~~~~~ (a_{1} + a_{2} + a_{3} + a_{4} + 4) (a_{1} + a_{2} + a_{3} + a_{5} + 4) (a_{2} + 
a_{3} + a_{4} + a_{5} + 4) \\
&~~~~~ (a_{2} + 2 a_{3} + a_{4} + a_{5} + 5) (a_{1} + a_{2} + a_{3} + a_{4} + a_{5} + 5) \\
&~~~~~ (a_{1} + a_{2} + 2 a_{3} + a_{4} + a_{5} + 6) (a_{1} + 2 a_{2} + 2 a_{3} + 
a_{4} + a_{5} + 7) ~~~.
\end{split}
\end{equation}

On the other hand, we already know there exist branching rules $\mathfrak{su}(10) \supset 
\mathfrak{so}(10)$ such that a $\mathfrak{su}(10)$ irrep can be projected to a direct sum 
of irreps in $\mathfrak{so}(10)$. In a branching rule, one of the irreps in $\mathfrak{so}(10)$ 
must have the same YT shape with that in $\mathfrak{su}(10)$. Also, the dimensions of the 
totality of representations obtained from the branching, when added together, should match 
that of the ``unbranched'' representation.  These properties manifest in the examples of the 
analytical expressions of irreducible conditions as mentioned in Section 
\ref{subsec:bosonicirrcond}.

This inspires us to invent a diagrammatic method to obtain the branching rules of 
$\mathfrak{su}(10) \supset \mathfrak{so}(10)$. Meanwhile, requiring the dimensions to 
match before and after projections would give us dimensions of BYT in $\mathfrak{so}(10)$.

This diagrammatic method is called tying rule.
The origin of the rule's name comes from the notion of ``ties''. Given two symmetric 
boxes in a BYT, one can``tie'' them by putting a node $\CMTB{\bullet}$ in each box and a 
line between them. In a field theoretical formulation using index notation, this means 
contracting the two symmetric indices with the flat metric $\eta^{\un{a} \un{b}}$, i.e. 
taking a ``trace''. The degrees of freedom are thus equivalent to those in the diagram 
where those two boxes are ``eliminated''. This is illustrated in Equation (\ref{eqn:tiedef}), 
where on the left the YT has two symmetric boxes, and on the right the ``tying'' action is 
applied on them.
\begin{equation}
\begin{aligned}
&\CMTB{\ytableaushort{{\un{a}}{\un{b}}}}  ~~~~~~&~~~~~~
\CMTB{\ytableaushort{\dotline\linedot}} ~\equiv&~ \CMTB{\cdot} \\
&\Phi_{\{\un{a},\un{b}\}} ~~~~~~&~~~~~~ \eta^{\un{a}\un{b}} \,
\Phi_{\{\un{a},\un{b}\}} ~\equiv&~ \varphi
\end{aligned} \label{eqn:tiedef}
\end{equation}
In the following, we will express everything about the tying rule graphically without 
mentioning the $\eta$-contractions, but readers should keep this in mind.

For completely antisymmetric BYTs, since no tie can be drawn (no two symmetric boxes), 
the rules for calculating the dimensions are exactly the same as the YTs in $\mathfrak{su}
(10)$. We will introduce the tying rules in the order of completely symmetric and two-column 
(Section \ref{subsec:tie2col}), three-column (Section \ref{subsec:tie3col}), and $n$-column 
$(n\geq4)$ BYTs (Section \ref{subsec:tiencol}), where the latter ones in the list have more 
complications and the former ones are specific cases of them. While the tying rules up till 3-column BYTs are complete, 
tying rules for 4-column and above remain incomplete as only several specific cases are verified. 
This requires further investigation beyond this paper. 

A remark is tying rule also works for general $\mathfrak{su}(N) \supset \mathfrak{so}(N)$, 
i.e. $A_{N-1} \supset D_{N/2}$ for even $N$ 
or $A_{N-1} \supset B_{(N-1)/2}$ for odd $N$\footnote{It seems likely that the proposal 
could be extended, with appropriate modifications to cases relating branchings of the 
$\mathfrak{su}(N)$ into the $\mathfrak{usp}(N)$ algebras
(i.e. $A_{2N-1}$ into $C_N$) as the latter possess a quadratic symplectic invariant. In 
this case, the tying would take place vertically not horizontally. In \cite{king1971}, 
King introduced two set of step-by-step graphical rules to describe $\mathfrak{su}(N) \supset \mathfrak{so}(N)$ and $\mathfrak{su}(N) \supset \mathfrak{usp}(N)$ branching rules respectively. 
The former is an interpretation of Littlewood's rule (that would be described later in Section \ref{subsec:BYTschur}), and one would see later how similar in spirit it is to the tying rule. We propose that the latter set of graphical rules by King is in fact in the same vein with the vertical tying rule, with the exact same mapping between Littlewood's rule and the horizontal tying rule.}. 
The only difference between branching into a $D$-series algebra and a $B$-series algebra is that the BYTs 
containing column(s) of $N/2$ boxes in the $D_{N/2}$ algebra can be split into a self-dual 
irrep and an anti-self-dual irrep (and thus contain two irreps), but none of the BYTs in the 
$B$-series algebra have these self-duality properties (so they will never be split into two 
irreps).
A little Mathematica program available in \url{https://github.com/SNHazelMak/TyingRule_SUnSOnBranching}
is written to verify the tying rules for all the BYTs up to 3-columns for $N > 6$.

Another remark is that based on tying rules we can also calculate the dimensionalities of bosonic irreps which will be presented in the following sections. 
One can quickly check the consistency of the Weyl dimension formula (\ref{equ:Weyl}), the $\mathfrak{su}(10)$ algebra hook rule, and our graphical tying rules. A simple python code to calculate Weyl dimension formula by inputting the Dynkin Label can be found in \url{https://github.com/1211890120/HigherDimlCounting/blob/master/Weyl_dimension_SO10.py}. 

But before we dive into the tying rule, we want to review a rule first proposed by Littlewood \cite{littlewood,king1971} and we propose as a conjecture\footnote{A 
rigorous mathematical proof is currently lacking, but in the following sections we present supporting
evidence.} that our tying rule (up to 3-columns) must be equivalent. The rule in question is carried out by 
utilizing a specific Schur function series
and a ``division operator'' of Young Tableaux. 
We will present the rule in Section \ref{subsec:BYTschur}.

\subsection{Littlewood's Rules by a Schur Function Series}
\label{subsec:BYTschur}

Schur functions were introduced by Issai Schur (1875 - 1941) in his doctoral dissertation 
in 1901 \cite{schur}. They are symmetric functions that are more general than elementary 
symmetric functions, which their symmetry properties can be represented by Young Tableaux. 
In 1934, Littlewood and Richardson published the famous 
Littlewood-Richardson rule \cite{littlewoodrichardson} that gives the combinatorial 
description of the coefficients in the decompositions of the product of two Schur functions 
as a linear combination of other Schur functions. It was then broadly applied in the area 
of classical Lie algebras to understand tensor products between two irreducible 
representations. For physicists, there are also familiar books \cite{SLnk,GrGI} where 
such introductions and discussions appear.

If we denote an irreducible representation of one of the $A_{n}$ algebras by $\{ \lambda \}$, where 
$\{ ~ \}$ means $A_{n}$ and $\lambda$ indicates the Young Tableau shape, we can 
write the Littlewood-Richardson rule as
\begin{equation}
\{\lambda\} \otimes \{\mu\} ~=~ \sum_{\nu} m^{\nu}_{\lambda\mu} \{\nu\} ~~~. 
\label{eqn:lrproduct}
\end{equation}
where $m^{\nu}_{\lambda\mu}$ is a number. The way to obtain the tensor product is 
familiar: put ``$a$'' labels in the first row of $\{\mu\}$, ``$b$'' labels in the second, ``$c$'' 
labels in the third, etc., and exploit all combinations of sticking these boxes containing 
these letters to the YT $\{\lambda\}$ such that they obey the rules,
\begin{enumerate}
\item The final Young diagram is regular (has a standard shape, not skew a shape);
\item No two of the same letters sit in the same column (so that symmetry is not violated); 
and
\item When going from left to right, top to bottom, the accumulated number of ``$a$'' 
is always greater than that of ``$b$'', and that of ``$b$'' is always greater than that of 
``$c$'', etc..
\end{enumerate}
The coefficients $m^{\nu}_{\lambda\mu}$ can then be obtained by counting the 
numbers of $\{\nu\}$ in the final decomposition.

We can then turn to a related operation, called quotient operation ``/''\footnote{Note that it is not the inverse operation of the Kronecker product $\otimes$.}, which is defined 
by
\begin{equation}
\{\nu\} / \{\mu\} ~=~ \sum_{\lambda} m^{\nu}_{\lambda\mu} \{\lambda\} ~~~, 
\label{eqn:lrquotient}
\end{equation}
where $m^{\nu}_{\lambda\mu}$ are exactly the same as the coefficients of the 
Littlewood-Richardson rule for products \cite{stanley,schutzenberger,lascoux}; 
and $\{\nu\} / \{\mu\}$ is called a skew Schur function. 

The coefficients above immediately suggest that the same rules 1 - 3 stated above 
apply to the quotient operation, but now we put letters from $\{\mu\}$ into 
$\{\nu\}$ and eliminate those boxes with letters.
An example of the quotient operation would be
\begin{equation}
\begin{split}
& {\CMTB{\ydiagram{4,2,1}}}_{\rm IR} ~\Bigg/~ {\CMTB{\ytableaushort{aa,b}}}_{\rm 
IR} \\
& ~=~ {\CMTB{ \ytableaushort{{}{}{\none[a]}{\none[a]},{}{\none[b]},{}} }}_{\rm IR} 
~\oplus~ {\CMTB{ \ytableaushort{{}{}{\none[a]}{\none[a]},{}{},{\none[b]}} }}_{\rm IR} 
~\oplus~ {\CMTB{\ytableaushort{{}{}{}{\none[a]},{}{\none[a]},{\none[b]}} }}_{\rm IR} 
~\oplus~ {\CMTB{\ytableaushort{{}{}{}{\none[a]},{}{\none[b]},{\none[a]}} }}_{\rm IR} 
~\oplus~ {\CMTB{\ytableaushort{{}{}{}{},{\none[b]}{\none[a]},{\none[a]}} }}_{\rm IR} 
~~~~~.
\end{split} \label{eqn:quotient}
\end{equation}
The meaning of these coefficients for quotient are the same as those for product, 
illustrated by this example, is
\begin{equation}
\begin{aligned}
{\CMTB{\ydiagram{2,1,1}}}_{\rm IR} ~\otimes~ {\CMTB{\ytableaushort{aa,b}}}_{\rm IR} 
~=&~~~ {\CMTB{ \ytableaushort{{}{}aa,{}b,{}} }}_{\rm IR} ~\oplus~ \cdots ~~~~~, \\
{\CMTB{\ydiagram{2,2}}}_{\rm IR} ~\otimes~ {\CMTB{\ytableaushort{aa,b}}}_{\rm IR} 
~=&~~~ {\CMTB{ \ytableaushort{{}{}aa,{}{},b} }}_{\rm IR} ~\oplus~ \cdots ~~~~~, \\
{\CMTB{\ydiagram{3,1}}}_{\rm IR} ~\otimes~ {\CMTB{\ytableaushort{aa,b}}}_{\rm IR} 
~=&~~~ {\CMTB{\ytableaushort{{}{}{}a,{}a,b} }}_{\rm IR} 
~\oplus~ {\CMTB{\ytableaushort{{}{}{}a,{}b,a} }}_{\rm IR} ~\oplus~ \cdots ~~~~~, \\
{\CMTB{\ydiagram{4}}}_{\rm IR} ~\otimes~ {\CMTB{\ytableaushort{aa,b}}}_{\rm IR} 
~=&~~~ {\CMTB{\ytableaushort{{}{}{}{},ba,a} }}_{\rm IR} ~\oplus~ \cdots ~~~~~,
\end{aligned}
\end{equation}
More explicitly, if we let
\begin{equation}
\begin{gathered}
\nu ~=~ {\CMTB{\ydiagram{4,2,1}}}_{\rm IR} ~~~,~~~ \mu ~=~ 
{\CMTB{\ydiagram{2,1}}}_{\rm IR} ~~~, \\ 
\lambda_{1} ~=~ {\CMTB{\ydiagram{2,1,1}}}_{\rm IR} ~~~,~~~ \lambda_{2} ~=~ 
{\CMTB{\ydiagram{2,2}}}_{\rm IR} ~~~,~~~ \lambda_{3} ~=~ {\CMTB{\ydiagram{3,1}}}_{\rm IR} 
~~~,~~~ \lambda_{4} ~=~ {\CMTB{\ydiagram{4}}}_{\rm IR} ~~~,
\end{gathered}
\end{equation}
according to the notations in Equations (\ref{eqn:lrproduct}) and (\ref{eqn:lrquotient}), 
then we have the coefficients
\begin{equation}
m^{\nu}_{\lambda_{1}\mu} ~=~ 1 ~~~,~~~ m^{\nu}_{\lambda_{2}\mu} ~=~ 1 ~~~,~~~ 
m^{\nu}_{\lambda_{3}\mu} ~=~ 2 ~~~,~~~ m^{\nu}_{\lambda_{4}\mu} ~=~ 1 ~~~.
\end{equation}

With these understandings of the quotient operation, we can proceed to the statement 
of the algorithm of calculating the branching rules of $\mathfrak{su}(n)\supset\mathfrak{so}(n)$.
By manipulating Schur function series, Littlewood proved  \cite{littlewood}
\begin{equation}
    \mathfrak{su}(n)\supset\mathfrak{so}(n) ~:~~ \{\lambda\} \supset [\lambda / D] ~~~, \label{eqn:littlewoodDrule}
\end{equation}
where $[ ~ ]$ means a representation in $\mathfrak{so}(n)$ algebra, and
\begin{equation}
D ~=~ {\CMTB{\cdot}} ~\oplus~ {\CMTB{\ydiagram{2}}}_{\rm IR} ~\oplus~ 
{\CMTB{\ydiagram{4}}}_{\rm IR} ~\oplus~ {\CMTB{\ydiagram{2,2}}}_{\rm IR} ~\oplus~ 
{\CMTB{\ydiagram{6}}}_{\rm IR} ~\oplus~ {\CMTB{\ydiagram{4,2}}}_{\rm IR} ~\oplus~ 
{\CMTB{\ydiagram{2,2,2}}}_{\rm IR} ~\oplus~ \cdots
\label{eqn:SchurD}
\end{equation}
is a Schur function series\footnote{This Schur function series $D$ contains all possible YTs constructed from $\CMTB{\ydiagram{2}}_{\rm IR}$. Therefore, the number of terms containing $2n$ box(es) is the integer partition of $n$. This is why the Littlewood's rule is very similar to tying rule - that ``$/D$'' means removing all the combinations of $\CMTB{\ydiagram{2}}_{\rm IR}$, while tying rules involve removing all the combinations of tied boxes $\CMTB{\ytableaushort{\dotline\linedot}}_{\rm IR}$.} written in our notation, which contains all partitions with all parts 
even. King in his 1971 paper \cite{king1971} understood this rule diagrammatically through 
the rules of obtaining quotients, like those in Equation (\ref{eqn:quotient}), by quotient-ing 
all the terms in $D$. He described this intuition as ``step-by-step nature of the trace removal 
process of $\mathfrak{so}(n)$'', which echoes with our definition of ``tie'' in Equation 
(\ref{eqn:tiedef}) but we are very explicit in the symmetric property of the $\eta$-metric by 
putting the two nodes on the same row.

For clarity, let us apply Equation (\ref{eqn:littlewoodDrule}) to
\begin{equation}
    \lambda ~=~ \ydiagram{4,3,1} ~~~~~.
\label{eqn:littlewoodex}
\end{equation}
By implementing quotient operator with each term in $D$, only the terms below give non-vanishing results,
\begin{equation}
\begin{split}
    {\CMTB{\ydiagram{4,3,1}}}_{\rm IR} ~\Bigg/~ \CMTB{\cdot} \hspace{5.57em} =&~~ {\CMTB{\ydiagram{4,3,1}}}_{\rm IR}  ~~~~~,  \\
    {\CMTB{\ydiagram{4,3,1}}}_{\rm IR} ~\Bigg/~ {\CMTB{\ytableaushort{aa}}}_{\rm IR} \hspace{2.5em} =&~~ {\CMTB{\ytableaushort{{}{}{}{\none[a]},{}{}{\none[a]},{}}}}_{\rm IR} ~\oplus~ {\CMTB{\ytableaushort{{}{}{}{\none[a]},{}{}{},{\none[a]}}}}_{\rm IR} ~\oplus~ {\CMTB{\ytableaushort{{}{}{}{},{}{\none[a]}{\none[a]},{}}}}_{\rm IR} \\
    &~ ~\oplus~ {\CMTB{\ytableaushort{{}{}{}{},{}{}{\none[a]},{\none[a]}}}}_{\rm IR} ~~~~~, \\
    {\CMTB{\ydiagram{4,3,1}}}_{\rm IR} ~\Bigg/~ {\CMTB{\ytableaushort{aaaa}}}_{\rm IR} =&~~ {\CMTB{\ytableaushort{{}{}{}{\none[a]},{}{\none[a]}{\none[a]},{\none[a]}}}}_{\rm IR} ~~~~~, \\
    {\CMTB{\ydiagram{4,3,1}}}_{\rm IR} ~\Bigg/~ {\CMTB{\ytableaushort{aa,bb}}}_{\rm IR} \hspace{2.5em} =&~~ {\CMTB{\ytableaushort{{}{}{\none[b]}{\none[a]},{}{\none[b]}{\none[a]},{}}}}_{\rm IR} ~\oplus~ {\CMTB{\ytableaushort{{}{}{\none[b]}{\none[a]},{}{}{\none[a]},{\none[b]}}}}_{\rm IR} ~\oplus~ {\CMTB{\ytableaushort{{}{}{}{\none[a]},{}{\none[b]}{\none[a]},{\none[b]}}}}_{\rm IR} ~~~~~, \\
    {\CMTB{\ydiagram{4,3,1}}}_{\rm IR} ~\Bigg/~ {\CMTB{\ytableaushort{aaaa,bb}}}_{\rm IR} =&~~ {\CMTB{\ytableaushort{{}{\none[b]}{\none[b]}{\none[a]},{}{\none[a]}{\none[a]},{\none[a]}}}}_{\rm IR} ~\oplus~ {\CMTB{\ytableaushort{{}{}{\none[b]}{\none[a]},{\none[b]}{\none[a]}{\none[a]},{\none[a]}}}}_{\rm IR} ~~~~~, 
\end{split} \label{eqn:littlewoodexDterms}
\end{equation}
while other terms vanish. This leads to the result
\begin{align}
\begin{split}
    \ydiagram{4,3,1} ~=&~~ {\CMTB{\ydiagram{4,3,1}}}_{\rm IR} ~\oplus~~ {\CMTB{\ydiagram{3,2,1}}}_{\rm IR} ~\oplus~~ {\CMTB{\ydiagram{3,3}}}_{\rm IR} ~\oplus~~ {\CMTB{\ydiagram{4,1,1}}}_{\rm IR} ~\oplus~~ ~{\CMTB{\ydiagram{4,2}}}_{\rm IR}  \\
    &~~~ \CMTB{  [1,2,1,0,0]  } ~~~~~~~~ \CMTB{[1,1,1,0,0]} ~~~~~~ \CMTB{ [0,3,0,0,0]  } ~~~~~~~~ \CMTB{[3,0,1,0,0]} ~~~~~~~~~~ \CMTB{[2,2,0,0,0]}  \\
    \{235950\}~~ ~=&~ ~~~\CMTB{\{174636\}}~~~ ~\oplus~ \,~~\CMTB{\{17920\}}~~ ~\oplus~ ~~~\CMTB{\{7644\}}~~~  ~\oplus~ ~~~~\CMTB{\{14784\}}~~~~ ~\oplus~ ~~~~\CMTB{\{16380\}}~~~~
\end{split} \nonumber \\[10pt]
\begin{split}
    & ~\oplus~~ 2 ~~ {\CMTB{\ydiagram{3,1}}}_{\rm IR} ~\oplus~~ {\CMTB{\ydiagram{2,1,1}}}_{\rm IR} ~\oplus~~ {\CMTB{\ydiagram{2,2}}}_{\rm IR} ~\oplus~~ {\CMTB{\ydiagram{1,1}}}_{\rm IR}  ~\oplus~~ {\CMTB{\ydiagram{2}}}_{\rm IR}  \\
    &~~~~~~~~ \CMTB{[2,1,0,0,0]} ~~~~~ \CMTB{ [1,0,1,0,0] } ~~ \CMTB{[0,2,0,0,0]} ~ \CMTB{ [0,1,0,0,0] } ~ \CMTB{[2,0,0,0,0]}  \\
    & ~\oplus~ \,~~(2) \CMTB{\{1386\}}~~\, ~\oplus~ ~~\CMTB{\{945\}}~~ ~\oplus~ ~~\CMTB{\{770\}}~\, ~\oplus~ ~\CMTB{\{45\}}\, ~\oplus~ \,~\CMTB{\{54\}}~
\end{split}  ~~~~~~.  \label{eqn:littlewoodexresults}
\end{align}

\subsection{Tying Rule for Completely Symmetric and Two-Column BYTs}
\label{subsec:tie2col}

Now we turn to tying rule. 
For completely symmetric and two-column bosonic Young Tableaux, there are two main 
steps to find the branching rules. Given a YT in $\mathfrak{su}(10)$, we perform the following 
to find how it decomposes into a direct sum of irreps in $\mathfrak{so}(10)$. 
\begin{description}[leftmargin=4.1em]
\item[Step 1:] Draw all possible combination of ties (including no tie at all). 
Vertical position of ties does not matter.
Keep one copy for each of the equivalent ones.
\item[Step 2:] For each diagram, wipe out the boxes with ties. 
Then the decomposition in $\mathfrak{so}(10)$ would be the sum of all these BYTs.
\end{description}

A very detailed description of this algorithm is given below to foster a thorough 
understanding about the meaning of the above two main steps. Let $n$ be the number 
of boxes of the Young Tableau to 
be decomposed.
\begin{enumerate}
\item Draw a tie between two boxes in the same row. That is equivalent to contracting 
two vector indices with a metric. Next we erase those two boxes and create a new 
Young Tableau (with $n-2$ boxes) in the decomposition.
\item Repeat the above step until no tie can be drawn further. In each step we create a new 
Young Tableau with $n-2t$ boxes, where $t$ is the number of ties. Then the entire Young 
Tableau can be decomposed into direct sums of $t_{\max}+1$ irreducible Young Tableaux 
(with subscript $IR$ as we denote them).
\item Look at the Young Tableau with the maximum number of ties (or the minimum number 
of boxes without ties). Calculate the dimension using the usual Young Tableau rules and 
call it $d_{t_{\max}}$.
\item Look at the Young Tableau with the next maximum number of ties (or the next 
minimum number of boxes without ties). Calculate the dimension using the usual Young 
Tableau rules and call it $\tilde{d}_{t_{\max}-1}$. Then the dimension of this irreducible 
Young Tableau is $d_{t_{\max}-1} = \tilde{d}_{t_{\max}-1} - d_{t_{\max}}$.
\item Repeat the above step for all irreducible Young Tableaux until $t=0$. For a general 
$t$, the dimension of that irreducible Young Tableau is $d_{t} = \tilde{d}_{t} - \sum_{t' > t} 
d_{t'}$, where $\tilde{d}_{t}$ is the dimension calculated by the usual Young Tableau 
rules, and $d_{t'}$ are the dimensions of the irreducible representations.
\item Finally, we obtain a decomposition of the reducible Young Tableau into a direct 
sum of irreducible Young Tableaux. The dimensions satisfy $\tilde{d}_{0} = d_{0} + 
d_{1} + \cdots + d_{t_{\max}-1} + d_{t_{\max}}$.
\end{enumerate}

To make things as clear as possible, we discuss two explicit examples. The first example is 
a completely symmetric Young Tableau. Steps 1 - 2 allow us to draw the ties and the 
irreducible Young Tableaux as follows.
\begin{equation}
\ydiagram{5} ~=~ {\CMTB{\ydiagram{5}}}_{\rm IR} ~\oplus~ 
{\CMTB{\ytableaushort{{}{}{}\dotline\linedot}}}_{\rm IR} ~\oplus~ 
{\CMTB{\ytableaushort{{}\dotline\linedot\dotline\linedot}}}_{\rm IR} 
~~~. \label{eqn:totsymBYTex}
\end{equation}
This is the branching rule for $\ydiagram{5}$ in $\mathfrak{su}(10)$ to $\mathfrak{so}(10)$. 
So $t_{\max} = 2$, and thus the final decomposition has $t_{\max}+1 = 3$ 
irreducible Young Tableaux. In this example, $n=5$, and the three irreducible Young 
Tableaux have 5, 3, and 1 box(es) respectively. Steps 3 - 5 allow us to find the dimensions 
of the irreducible Young Tableaux as follows.
\begin{equation}
\begin{split}
{\CMTB{\ytableaushort{{}\dotline\linedot\dotline\linedot}}}_{\rm IR} & \qquad d_{2} 
= 10 ~~~, \\
{\CMTB{\ytableaushort{{}{}{}\dotline\linedot}}}_{\rm IR} & \qquad d_{1} = \tilde{d}_{1} 
- d_{2} = \frac{10 \times 11 \times 12}{3 \times 2 \times 1} - 10 = 210 ~~~, \\
{\CMTB{\ydiagram{5}}}_{\rm IR} & \qquad d_{0} = \tilde{d}_{0} - d_{1} - d_{2} = 
\frac{10 \times 11 \times 12 \times 13 \times 14}{5 \times 4 \times 3 \times 2 \times 
1} - 210 -10 = 1782 ~~~.
\end{split}
\label{equ:STP1}
\end{equation}
Therefore,
\begin{equation}
\{2002\} = \CMTB {\{ 1782 \}} \oplus \CMTB {\{ 210' \}} \oplus \CMTB {\{ 10 \}}~~~,
\end{equation}
where the irreducible representation $\{2002\}$ of $\mathfrak{su}(10)$ corresponds 
to the Dynkin Label [5,0,0,0,0,0,0,0,0], while $\CMTB {\{ 1782 \}}$, $\CMTB {\{ 210' \}}$ 
and $\CMTB {\{ 10 \}}$ of $\mathfrak{so}(10)$ corresponds to the Dynkin Labels 
$\CMTB{[5,0,0,0,0]}$, $\CMTB{[3,0,0,0,0]}$ and $\CMTB{[1,0,0,0,0]}$ respectively.

The second example is a two-column Young Tableau. Steps 1 - 2 allow us to draw 
the ties and the irreducible Young Tableaux as follows.
\begin{equation}
\ydiagram{2,2,1} ~=~ {\CMTB{\ydiagram{2,2,1}}}_{\rm IR} ~\oplus~ 
{\CMTB{\ytableaushort{\dotline\linedot,{}{},{}}}}_{\rm IR} ~\oplus~ 
{\CMTB{\ytableaushort{\dotline\linedot,\dotline\linedot,{}}}}_{\rm IR}~~~.
\end{equation}
This is the branching rule for $\ydiagram{2,2,1}$ in $\mathfrak{su}(10)$ to $\mathfrak{so
}(10)$. Again $t_{\max} = 2$, so we have 3 irreducible Young Tableaux in the decomposition. 
And again $n=5$, so we have three irreducible Young Tableaux with 5, 3, and 1 box(es) 
respectively. Steps 3 - 5 allow us to find the dimensions of the irreducible Young Tableaux 
as follows.
\begin{equation}
\begin{split}
{\CMTB{\ytableaushort{\dotline\linedot,\dotline\linedot,{}}}}_{\rm IR} & \qquad d_{2} = 
10 ~~~, \\
{\CMTB{\ytableaushort{\dotline\linedot,{}{},{}}}}_{\rm IR} & \qquad d_{1} = \tilde{d}_{1
} - d_{2} = \frac{10 \times 11 \times 9}{3 \times 1 \times 1} - 10 = 320 ~~~, \\
{\CMTB{\ydiagram{2,2,1}}}_{\rm IR} & \qquad d_{0} = \tilde{d}_{0} - d_{1} - d_{2} = 
\frac{10 \times 11 \times 9 \times 10 \times 8}{4 \times 2 \times 3 \times 1 \times 1} - 
320 -10 = 2970 ~~~.
\end{split}
\label{equ:STP2}
\end{equation}
Therefore,
\begin{equation}
    \{3300\} = \CMTB {\{ 2970 \}} \oplus\CMTB { \{ 320 \}} \oplus \CMTB {\{ 10 \}}~~~,
\end{equation}
where the irreducible representation $\{3300\}$ of $\mathfrak{su}(10)$ corresponds 
to the Dynkin Label [0,1,1,0,0,0,0,0,0], while $\CMTB {\{ 2970 \}}$, $\CMTB {\{ 320 \}}$ 
and $\CMTB {\{ 10 \}}$ of $\mathfrak{so}(10)$ corresponds to the Dynkin Labels 
$\CMTB{[0,1,1,0,0]}$, $\CMTB{[1,1,0,0,0]}$ and $\CMTB{[1,0,0,0,0]}$ respectively.

\subsection{Tying Rule for Three-Column BYTs}
\label{subsec:tie3col}

For a three-column BYT, there are two steps to figure out how a $\mathfrak{su}(10)$ irrep 
(which is reducible in $\mathfrak{so}(10)$) decomposes into $\mathfrak{so}(10)$ irreducible 
parts.
\begin{description}[leftmargin=4.1em]
\item[Step 1:] Draw all the possible combinations of ties (including no tie at all). They can skip a box (as long as they are on the same row and remain symmetric). Vertical position of ties does not matter. Keep one copy for the equivalent ones.
\item[Step 2:] For each diagram, wipe out the boxes with the ties and rearrange them vertically {\em according to symmetry properties} (we'll explain it shortly) such that they have standard YT shapes. If a standard YT shape cannot be obtained, or the rearrangements cannot be performed with symmetry preserved, throw that BYT away. Then the decomposition in $\mathfrak{so}(10)$ would be the sum of all the remaining BYTs.
\end{description}

Let us focus on step 1 first. To illustrate the idea, we will discuss two examples below. 
\begin{align}
\begin{split}
    \ydiagram{3,1} ~~:&~~~~~~ {\CMTB{\ydiagram{3,1}}}_{\rm IR} ~~~,~~~ {\CMTB{\ytableaushort{{}\dotline\linedot,{}}}}_{\rm IR} ~~~,~~~ {\CMTB{\ytableaushort{\dotline\cdots\linedot,{}}}}_{\rm IR} ~~~,~~~ {\CMTB{\ytableaushort{\dotline\linedot{},{}}}}_{\rm IR} ~~~.
\end{split} \label{eqn:ex1step1} \\[10pt] 
\begin{split}
    \ydiagram{3,3,2,1} ~~:&~~~~~~ {\CMTB{\ydiagram{3,3,2,1}}}_{\rm IR} ~~~,~~~ {\CMTB{\ytableaushort{\dotline\linedot{},{}{}{},{}{},{}}}}_{\rm IR} ~~~,~~~ {\CMTB{\ytableaushort{\dotline\linedot{},\dotline\linedot{},{}{},{}}}}_{\rm IR} ~~~,~~~ {\CMTB{\ytableaushort{\dotline\linedot{},\dotline\linedot{},\dotline\linedot,{}}}}_{\rm IR} ~~~,~~~ \\
    &~~~~~~ {\CMTB{\ytableaushort{{}\dotline\linedot,{}{}{},{}{},{}}}}_{\rm IR}  ~~~,~~~  {\CMTB{\ytableaushort{{}\dotline\linedot,{}\dotline\linedot,{}{},{}}}}_{\rm IR} ~~~,~~~ {\CMTB{\ytableaushort{{}\dotline\linedot,{}\dotline\linedot,\dotline\linedot,{}}}}_{\rm IR} ~~~,~~~ {\CMTB{\ytableaushort{{}\dotline\linedot,\dotline\linedot{},{}{},{}}}}_{\rm IR} ~~~,~~~ \\
    &~~~~~~ {\CMTB{\ytableaushort{{}\dotline\linedot,\dotline\linedot{},\dotline\linedot,{}}}}_{\rm IR}  ~~~,~~~  {\CMTB{\ytableaushort{\dotline\cdots\linedot,{}{}{},{}{},{}}}}_{\rm IR} ~~~,~~~  {\CMTB{\ytableaushort{\dotline\cdots\linedot,\dotline\linedot{},{}{},{}}}}_{\rm IR} ~~~,~~~ {\CMTB{\ytableaushort{\dotline\cdots\linedot,\dotline\linedot{},\dotline\linedot,{}}}}_{\rm IR} ~~~,~~~ \\
    &~~~~~~ {\CMTB{\ytableaushort{\dotline\cdots\linedot,{}\dotline\linedot,{}{},{}}}}_{\rm IR} ~~~,~~~ {\CMTB{\ytableaushort{\dotline\cdots\linedot,{}\dotline\linedot,\dotline\linedot,{}}}}_{\rm IR}  ~~~,~~~  {\CMTB{\ytableaushort{\dotline\cdots\linedot,\dotline\cdots\linedot,{}{},{}}}}_{\rm IR} ~~~,~~~ {\CMTB{\ytableaushort{\dotline\cdots\linedot,\dotline\cdots\linedot,\dotline\linedot,{}}}}_{\rm IR} ~~~.
\end{split} \label{eqn:ex2step1}
\end{align}

Now, let us consider step 2. Before we proceed with examples, let us explain in detail the meaning of the symmetry properties we mentioned above. In general, consider a YT containing two columns of unequal numbers of boxes, one with $M$ boxes and one with $N$ boxes. Without loss of generality, let $M<N$.
\newcommand{\curlyM}{\ensuremath\raisebox{1.2em}{$\Bigg\} \, M$}}
\begin{equation}
    N \, \left\{ \CMTB{\ytableaushort{{}{},\vdots\vdots,{}{},\vdots,{}}} \curlyM \color{white}{} \right\} \color{black}{}
\end{equation} 
The symmetry relations go as follows.
\begin{enumerate}
\item The $i$-th box in the $M$-column is {\em symmetric} with the $i$-th box in the $N$-column $(i = 1, \dots, M)$;
\item The $i$-th box in the $M$-column is {\em antisymmetric} with the $j$-th box in the $N$-column for $j \neq i$ and $j \leq M$ $(j = 1, \dots, M)$; 
\item The $i$-th box in the $M$-column has {\em no symmetry relation} with the $k$-th box in the $N$-column for $k > M$ $(k = M + 1, \dots, N)$.
\end{enumerate}
If we use the example in Equation (\ref{eqn:ex2step1}), we can illustrate these three conditions as follows.
\begin{equation}
\begin{split}
    & {\CMTB{ \ydiagram[*(white) \sym]{1+2}
    *[*(white)]{3,3,2,1} }}_{\rm IR}  ~~~~~~~~~~~~~~~~  {\CMTB{ \ydiagram[*(white) \antisym]{2+1,1+1}
    *[*(white)]{3,3,2,1} }}_{\rm IR}  ~~~~~~~~~~~~~~~~  {\CMTB{ \ydiagram[*(white) \nosym]{2+1,0,1+1}
    *[*(white)]{3,3,2,1} }}_{\rm IR}   \\
    & \text{symmetric} ~~~~~~~~~~~~~ \text{antisymmetric} ~~~~~~~ \text{no symmetry relation}
\end{split} \label{eqn:tiesym1}
\end{equation}
We have explained the definitions of the symmetry relations between any pair of boxes in a BYT. What does it mean by saying the symmetry properties are {\em preserved}? Since we perform all the ties horizontally, we are only allowed to move the boxes vertically. Then three scenarios below are possible after rearrangements. 
\begin{equation}
\begin{split}
    & \CMTB{ \ydiagram[*(white) \sym]{2} }  ~~~~~~~~~~~~~~~~~~~\,  \CMTB{ \ydiagram[*(white) \antisym]{2} }  ~~~~~~~~~~~~~~~~~~~~  \CMTB{ \ydiagram[*(white) \nosym]{2} }   \\
    & \,\text{must} ~~~~~~~~~~~~~~~~~~\, \text{must not} ~~~~~~~~~~~~~ \text{no restriction}
\end{split} \label{eqn:tiesym2}
\end{equation}
Symmetry properties are {\em preserved} if the final BYT abides by these rules, i.e. the originally symmetric boxes must remain symmetric, the originally antisymmetric boxes must not become symmetric, and the boxes that do not have any symmetry relation originally can end up being symmetric.

With these in mind, we can now erase the boxes that are tied up (with a bullet in the middle), and rearrange them vertically. Let us start with the first example in Equation (\ref{eqn:ex1step1}). The last diagram can not be vertically rearranged to be a regular YT shape, 
\begin{equation}
    {\CMTB{ \ytableaushort{\dotline\linedot{},{}} }}_{\rm IR} ~\longrightarrow~ \CMTB{ \ytableaushort{{}\none{}} }  ~~~~~.
\end{equation}
The remaining diagrams could be rearranged as follows,
\begin{equation}
\begin{split}
    {\CMTB{ \ytableaushort{{}\dotline\linedot,{}} }}_{\rm IR} ~\longrightarrow&~~ {\CMTB{ \ydiagram{1,1} }}_{\rm IR}  \hspace{1.25em} ~~~~~, \\
    {\CMTB{ \ydiagram[*(white) \cdot\hspace{-0.1em}\nosym\hspace{-0.1em}\CMTB{\cdot}]{1+1}
    *[*(white) \nosym]{0,1}
    *[*(white) \dotline]{1} *[*(white) \linedot]{2+1}
    *[*(white)]{3,1} }}_{\rm IR}  ~\longrightarrow&~~  {\CMTB{ \ydiagram[*(white) \nosym]{2} }}_{\rm IR} ~~~~~.
\end{split}
\end{equation}
Both of these diagrams are allowed as they have standard YT shapes and all the symmetry properties are preserved. Together with the diagram with no tie, we can write the branching rule
\begin{equation}
\begin{split}
    \ydiagram{3,1} ~=&~~ {\CMTB{\ydiagram{3,1}}}_{\rm IR} ~~\oplus~~~ {\CMTB{\ydiagram{1,1}}}_{\rm IR} ~~\oplus~~~ {\CMTB{\ydiagram{2}}}_{\rm IR}  \\
    & ~\CMTB{[2,1,0,0,0]} ~~~~ \CMTB{[0,1,0,0,0]} ~~~ \CMTB{[2,0,0,0,0]}  \\
    \{1485\}~ ~=&~ ~~~\CMTB{\{1386\}}~~\, ~~\oplus~~ ~\CMTB{\{45\}}\, ~~\oplus~~ ~~\CMTB{\{54\}}
\end{split}  ~~~~~~. 
\label{equ:Tying31}
\end{equation}
We turn to the more complicated second example in Equation (\ref{eqn:ex2step1}). Before we perform
delicate vertical rearrangements, we can discard diagrams that obviously could not be arranged to 
standard YT shapes, including
\begin{equation}
\begin{split}
    & {\CMTB{\ytableaushort{\dotline\linedot{},\dotline\linedot{},{}{},{}}}}_{\rm IR}  \longrightarrow~  {\CMTB{ \ytableaushort{{}{}{},{}\none{}} }}  ~~~~~,~~~~~ {\CMTB{\ytableaushort{\dotline\linedot{},\dotline\linedot{},\dotline\linedot,{}}}}_{\rm IR}  \longrightarrow~  {\CMTB{ \ytableaushort{{}\none{},\none\none{}} }} ~~~~~,~~~~~ {\CMTB{\ytableaushort{{}\dotline\linedot,\dotline\linedot{},\dotline\linedot,{}}}}_{\rm IR} \longrightarrow~ {\CMTB{ \ytableaushort{{}\none{},{}} }} ~~~~~, \\
    & {\CMTB{\ytableaushort{\dotline\cdots\linedot,\dotline\cdots\linedot,{}{},{}}}}_{\rm IR}  \longrightarrow~  {\CMTB{ \ytableaushort{{}{},{}{},\none{}} }} \hspace{1.25em}  ~~~~~,~~~~~ {\CMTB{\ytableaushort{\dotline\cdots\linedot,\dotline\cdots\linedot,\dotline\linedot,{}}}}_{\rm IR}   \longrightarrow~  {\CMTB{ \ytableaushort{{}{},\none{}} }} \hspace{1.25em} ~~~~~.
\end{split}
\end{equation}
We can now carry out the rearrangements of the remaining diagrams. To facilitate the presentation, in the rearrangement process, we will use the number ``1'' to indicate the boxes in the first row of the final diagram, and the number ``2'' to indicate those in the second row, etc.. We will use the colors in Equations (\ref{eqn:tiesym1}) and (\ref{eqn:tiesym2}) to indicate the conditions of \textcolor{skyblue}{symmetric}, \textcolor{brightpink}{antisymmetric}, and \textcolor{darkgreen}{no symmetry relation} respectively. For a row with a single box, we use black color. Since it's always about pairs of boxes, some boxes might need to be indicated twice. To avoid this, we will specify the conditions of \textcolor{skyblue}{symmetric} and \textcolor{brightpink}{antisymmetric} first, then if the box next to it is in green color, it has \textcolor{darkgreen}{no symmetry relation} with the other boxes on the same row. Here are the ones that are allowed according to symmetry rules.
\begin{equation}
\begin{split}
    {\CMTB{ \ydiagram[*(white) \dotline]{1} *[*(white) \linedot]{1+1} *[*(white) \nosymtwo]{2+1}
    *[*(white) \symone]{0,3}
    *[*(white) \symtwo]{0,0,2} 
    *[*(white) \singlethree]{0,0,0,1} }}_{\rm IR}  \longrightarrow&~  
    {\CMTB{ \ydiagram[*(white) \symone]{3} 
    *[*(white) \symtwo]{0,2} *[*(white) \nosymtwo]{0,2+1}
    *[*(white) \singlethree]{0,0,1} }}_{\rm IR}   ~~,~~~~
    {\CMTB{ \ydiagram[*(white) \dotline]{1+1} *[*(white) \linedot]{2+1} *[*(white) \singlefour]{1}
    *[*(white) \symone]{0,3}
    *[*(white) \symtwo]{0,0,2} 
    *[*(white) \singlethree]{0,0,0,1} }}_{\rm IR}  \longrightarrow~  
    {\CMTB{ \ydiagram[*(white) \symone]{3} 
    *[*(white) \symtwo]{0,2}
    *[*(white) \singlethree]{0,0,1}
    *[*(white) \singlefour]{0,0,0,1}}}_{\rm IR}  ~~,~~~~
    {\CMTB{ \ydiagram[*(white) \dotline]{1+1,1+1} *[*(white) \linedot]{2+1,2+1} *[*(white) \singlethree]{1}
    *[*(white) \singlefour]{0,1}
    *[*(white) \symone]{0,0,2} 
    *[*(white) \singletwo]{0,0,0,1} }}_{\rm IR}  \longrightarrow~  
    {\CMTB{ \ydiagram[*(white) \symone]{2} 
    *[*(white) \singletwo]{0,1}
    *[*(white) \singlethree]{0,0,1}
    *[*(white) \singlefour]{0,0,0,1} }}_{\rm IR} \hspace{1.25em}~~, \\
    {\CMTB{ \ydiagram[*(white) \dotline]{1+1,1+1,1} *[*(white) \linedot]{2+1,2+1,1+1} *[*(white) \singleone]{1}
    *[*(white) \singletwo]{0,1}
    *[*(white) \singlethree]{0,0,0,1} }}_{\rm IR}  \longrightarrow&~  
    {\CMTB{ \ydiagram[*(white) \singleone]{1} 
    *[*(white) \singletwo]{0,1}
    *[*(white) \singlethree]{0,0,1} }}_{\rm IR} \hspace{2.5em}~~,~~~~
    {\CMTB{ \ydiagram[*(white) \dotline]{1+1,1} *[*(white) \linedot]{2+1,1+1} *[*(white) \singlethree]{1}
    *[*(white) \nosymone]{0,2+1}
    *[*(white) \symone]{0,0,2}
    *[*(white) \singletwo]{0,0,0,1} }}_{\rm IR}  \longrightarrow~  
    {\CMTB{ \ydiagram[*(white) \symone]{2} *[*(white) \nosymone]{2+1}
    *[*(white) \singletwo]{0,1}
    *[*(white) \singlethree]{0,0,1} }}_{\rm IR} ~~,~~~~
    {\CMTB{ \ydiagram[*(white) \dotline]{1} *[*(white) \linedot]{2+1} *[*(white) \cdot\nosymthree\CMTB{\cdot}]{1+1} 
    *[*(white) \symone]{0,3}
    *[*(white) \symtwo]{0,0,2} 
    *[*(white) \nosymthree]{0,0,0,1} }}_{\rm IR}  \longrightarrow~  
    {\CMTB{ \ydiagram[*(white) \symone]{3} 
    *[*(white) \symtwo]{0,2} 
    *[*(white) \nosymthree]{0,0,2} }}_{\rm IR}   ~~,~~~~ \\
    {\CMTB{ \ydiagram[*(white) \dotline]{1,1+1} *[*(white) \linedot]{2+1,2+1} *[*(white) \singlethree]{1} *[*(white) \cdot\nosymtwo\CMTB{\cdot}]{1+1}
    *[*(white) \singlethree]{0,1}
    *[*(white) \symone]{0,0,2} 
    *[*(white) \nosymtwo]{0,0,0,1} }}_{\rm IR}  \longrightarrow&~  
    {\CMTB{ \ydiagram[*(white) \symone]{2} 
    *[*(white) \nosymtwo]{0,2}
    *[*(white) \singlethree]{0,0,1} }}_{\rm IR} \hspace{1.25em}~~,~~~~
    {\CMTB{ \ydiagram[*(white) \dotline]{1,1} *[*(white) \linedot]{2+1,1+1} *[*(white) \singlethree]{1} *[*(white) \cdot\nosymtwo\CMTB{\cdot}]{1+1}
    *[*(white) \nosymone]{0,2+1}
    *[*(white) \symone]{0,0,2} 
    *[*(white) \nosymtwo]{0,0,0,1} }}_{\rm IR}  \longrightarrow~  
    {\CMTB{ \ydiagram[*(white) \symone]{2} *[*(white) \nosymone]{2+1}
    *[*(white) \nosymtwo]{0,2} }}_{\rm IR} ~~,~~~~
    {\CMTB{ \ydiagram[*(white) \dotline]{1,1+1,1} *[*(white) \linedot]{2+1,2+1,1+1} *[*(white) \cdot\nosymone\CMTB{\cdot}]{1+1}
    *[*(white) \singletwo]{0,1} 
    *[*(white) \nosymone]{0,0,0,1} }}_{\rm IR}  \longrightarrow~  
    {\CMTB{ \ydiagram[*(white) \nosymone]{2} 
    *[*(white) \singletwo]{0,1} }}_{\rm IR} \hspace{1.25em} ~~.
\end{split}
\end{equation}
The remaining diagram is
\begin{equation}
    {\CMTB{ \ydiagram[*(white) \dotline]{1,1,1} *[*(white) \linedot]{2+1,1+1,1+1} *[*(white) \cdot\antisymone\CMTB{\cdot}]{1+1}
    *[*(white) \antisymone]{0,2+1} 
    *[*(white) \nosymone]{0,0,0,1} }}_{\rm IR}  \longrightarrow~  
    {\CMTB{ \ydiagram[*(white) \nosymone]{1} *[*(white) \antisymone]{1+2} }}_{\rm IR}  ~~~,
\end{equation}
which is not allowed by the symmetry rule in Equation (\ref{eqn:tiesym2}), and thus disregarded. Therefore,
\begin{align}
\begin{split}
    \ydiagram{3,3,2,1} ~=&~~ {\CMTB{\ydiagram{3,3,2,1}}}_{\rm IR} ~\oplus~~ {\CMTB{\ydiagram{3,3,1}}}_{\rm IR} ~\oplus~~ {\CMTB{\ydiagram{3,2,1,1}}}_{\rm IR} ~\oplus~~ {\CMTB{\ydiagram{2,1,1,1}}}_{\rm IR} ~\oplus~~ ~{\CMTB{\ydiagram{1,1,1}}}_{\rm IR}  \\
    &~ \CMTB{  [0,1,1,1,1]  } ~~~~~~ \CMTB{[0,2,1,0,0]} ~~~~~~ \CMTB{ [1,1,0,1,1]  } ~~~~ \CMTB{[1,0,0,1,1]} ~~ \CMTB{[0,0,1,0,0]}  \\
    \{304920\} ~=&~ \,~\CMTB{\{192192\}}~ ~\oplus~ \,~~\CMTB{\{34398\}}~~ ~\oplus~ ~~\CMTB{\{36750\}}~~  ~\oplus~ ~\CMTB{\{1728\}}~ ~\oplus~ ~\CMTB{\{120\}}~
\end{split} \nonumber \\[10pt]
\begin{split}
    & ~\oplus~~ {\CMTB{\ydiagram{3,1,1}}}_{\rm IR} ~\oplus~~ {\CMTB{\ydiagram{3,2,2}}}_{\rm IR} ~\oplus~~ {\CMTB{\ydiagram{2,2,1}}}_{\rm IR} ~\oplus~~ {\CMTB{\ydiagram{3,2}}}_{\rm IR}  ~\oplus~~ {\CMTB{\ydiagram{2,1}}}_{\rm IR}  \\
    &~~~~~~ \CMTB{[2,0,1,0,0]} ~~~~~~ \CMTB{ [1,0,2,0,0] } ~~~~ \CMTB{[0,1,1,0,0]} ~~~~ \CMTB{ [1,2,0,0,0] } ~~~~ \CMTB{[1,1,0,0,0]}  \\
    & ~\oplus~ ~~~\CMTB{\{4312\}}~~\, ~\oplus~ \,~~\CMTB{\{27720\}}~~ ~\oplus~ ~\CMTB{\{2970\}}~ ~\oplus~ ~~~\CMTB{\{4410\}}~~\, ~\oplus~ \,~\CMTB{\{320\}}~ 
\end{split}  ~~~~~~.  
\label{equ:data1}
\end{align}

To foster a better feeling for these rules, especially about the diagrams that violate the symmetry properties, i.e. contain the antisymmetric boxes on the same row in Equation (\ref{eqn:tiesym2}), we put a very enlightening example at the end. Since it's very easy to observe which BYT would not end up being a standard shape, those would be excluded right away. The equivalent ones would not be written twice also.
\begin{equation}
\begin{split}
    \ydiagram{3,3,3,3,1} ~~:&~~~~~~ {\CMTB{\ydiagram{3,3,3,3,1}}}_{\rm IR} ~~~,~~~ {\CMTB{\ytableaushort{{}\dotline\linedot,{}{}{},{}{}{},{}{}{},{}}}}_{\rm IR} ~~~,~~~ {\CMTB{\ytableaushort{{}\dotline\linedot,{}\dotline\linedot,{}{}{},{}{}{},{}}}}_{\rm IR} ~~~,~~~ {\CMTB{\ytableaushort{{}\dotline\linedot,{}\dotline\linedot,{}\dotline\linedot,{}{}{},{}}}}_{\rm IR} ~~~,~~~ {\CMTB{\ytableaushort{{}\dotline\linedot,{}\dotline\linedot,{}\dotline\linedot,{}\dotline\linedot,{}}}}_{\rm IR} ~~~, \\
    &~~~~~~ {\CMTB{\ytableaushort{\dotline\cdots\linedot,{}{}{},{}{}{},{}{}{},{}}}}_{\rm IR} ~~~,~~~ {\CMTB{\ytableaushort{\dotline\cdots\linedot,{}\dotline\linedot,{}{}{},{}{}{},{}}}}_{\rm IR} ~~~,~~~ {\CMTB{\ytableaushort{\dotline\cdots\linedot,{}\dotline\linedot,{}\dotline\linedot,{}{}{},{}}}}_{\rm IR} ~~~,~~~
    {\CMTB{\ytableaushort{\dotline\cdots\linedot,{}\dotline\linedot,{}\dotline\linedot,{}\dotline\linedot,{}}}}_{\rm IR} ~~~,~~~  
    {\CMTB{\ytableaushort{\dotline\cdots\linedot,\dotline\linedot{},{}{}{},{}{}{},{}}}}_{\rm IR} ~~~,  \\
    &~~~~~~ {\CMTB{\ytableaushort{\dotline\cdots\linedot,\dotline\linedot{},{}\dotline\linedot,{}{}{},{}}}}_{\rm IR} ~~~,~~~ {\CMTB{\ytableaushort{\dotline\cdots\linedot,\dotline\linedot{},{}\dotline\linedot,{}\dotline\linedot,{}}}}_{\rm IR} ~~~,~~~
    {\CMTB{\ytableaushort{\dotline\cdots\linedot,\dotline\cdots\linedot,{}\dotline\linedot,{}{}{},{}}}}_{\rm IR} ~~~,~~~
    {\CMTB{\ytableaushort{\dotline\cdots\linedot,\dotline\cdots\linedot,{}\dotline\linedot,{}\dotline\linedot,{}}}}_{\rm IR} ~~~,~~~
    {\CMTB{\ytableaushort{\dotline\cdots\linedot,\dotline\cdots\linedot,{}\dotline\linedot,\dotline\linedot{},{}}}}_{\rm IR} ~~~.
\end{split} \label{eqn:ex3step1}
\end{equation}
Now, let us just focus on those diagrams that would inevitably contain two antisymmetric boxes in one row, and thus must be discarded.
\begin{align}
    {\CMTB{ \ydiagram[*(white) \dotline]{1,1} *[*(white) \linedot]{2+1,1+1} *[*(white) \cdot\antisymthree\CMTB{\cdot}]{1+1}
    *[*(white) \antisymthree]{0,2+1} 
    *[*(white) \symone]{0,0,3}
    *[*(white) \symtwo]{0,0,0,3}
    *[*(white) \nosymthree]{0,0,0,0,1} }}_{\rm IR}  \longrightarrow~ 
    {\CMTB{ \ydiagram[*(white) \symone]{3} *[*(white) \symtwo]{0,3} 
    *[*(white) \nosymthree]{0,0,1} *[*(white) \antisymthree]{0,0,1+2} }}_{\rm IR}  ~~~, \label{eqn:ex3discard1} \\
    {\CMTB{ \ydiagram[*(white) \dotline]{1,1,1+1} *[*(white) \linedot]{2+1,2+1,2+1} *[*(white) \cdot\nosymtwo\CMTB{\cdot}]{1+1} *[*(white) \cdot\antisymthree\CMTB{\cdot}]{0,1+1} 
    *[*(white) \antisymthree]{0,0,1}
    *[*(white) \symone]{0,0,0,3}
    *[*(white) \nosymtwo]{0,0,0,0,1} }}_{\rm IR}  \longrightarrow~ 
    {\CMTB{ \ydiagram[*(white) \symone]{3} *[*(white) \nosymtwo]{0,2} *[*(white) \antisymthree]{0,0,2} }}_{\rm IR}  ~~~. \label{eqn:ex3discard2}
\end{align}
From the way these two diagrams vanish, we know there are 4 more diagrams that vanish in the same way,
\begin{align}
    & 
    {\CMTB{\ytableaushort{\dotline\cdots\linedot,\dotline\linedot{},{}\dotline\linedot,{}{}{},{}}}}_{\rm IR} ~~~,~~~ {\CMTB{\ytableaushort{\dotline\cdots\linedot,\dotline\linedot{},{}\dotline\linedot,{}\dotline\linedot,{}}}}_{\rm IR} ~~~, \label{eqn:ex3discard1de} \\
    & 
    {\CMTB{\ytableaushort{\dotline\cdots\linedot,\dotline\cdots\linedot,{}\dotline\linedot,{}\dotline\linedot,{}}}}_{\rm IR} ~~~,~~~
    {\CMTB{\ytableaushort{\dotline\cdots\linedot,\dotline\cdots\linedot,{}\dotline\linedot,\dotline\linedot{},{}}}}_{\rm IR} ~~~. \label{eqn:ex3discard2de}
\end{align}
We call the two diagrams in Equation (\ref{eqn:ex3discard1de}) the {\em descendants} of the diagram in Equation (\ref{eqn:ex3discard1}), as both diagrams contain the pair of ties (on the top two rows) of the (\ref{eqn:ex3discard1}) diagram which caused the diagram to contain antisymmetric boxes on the same row; and therefore the descendants would contain these exact same antisymmetric boxes and be discarded for the exact same reason. In a similar sense, the two diagrams in Equation (\ref{eqn:ex3discard2de}) are the descendants of the diagram in Equation (\ref{eqn:ex3discard2}) as they contain the same problematic triplet of ties (on the top three rows) of the (\ref{eqn:ex3discard2}) diagram. Thus the final decomposition is 
\begin{align}
\begin{split}
    \ydiagram{3,3,3,3,1} ~~=&~~~ {\CMTB{\ydiagram{3,3,3,3,1}}}_{\rm IR} ~~\oplus~~~ {\CMTB{\ydiagram{3,3,3,1,1}}}_{\rm IR} ~~\oplus~~~ {\CMTB{\ydiagram{3,3,1,1,1}}}_{\rm IR} ~~\oplus~~~ {\CMTB{\ydiagram{3,1,1,1,1}}}_{\rm IR} ~~\oplus~~~ ~{\CMTB{\ydiagram{1,1,1,1,1}}}_{\rm IR}  \\
    &~~ \CMTB{[0,0,0,4,2]} ~~~~~~~~ \CMTB{[0,0,2,2,0]} ~~~~~~~~ \CMTB{[0,2,0,2,0]} ~~~~~~~~ \CMTB{[2,0,0,2,0]} ~~~~~~ \CMTB{[0,0,0,2,0]}  \\
    & \oplus \CMTB{[0,0,0,2,4]} ~~~~ \oplus \CMTB{[0,0,2,0,2]} ~~~~ \oplus \CMTB{[0,2,0,0,2]} ~~~~ \oplus \CMTB{[2,0,0,0,2]} ~~~ \oplus \CMTB{[0,0,0,0,2]}  \\
    \{1176120\} ~=&~ \,\begin{matrix} \CMTB{\{141570\}} \\ \oplus~ \CMTB{\{\overline{141570}\}} \end{matrix}\,  ~\oplus~ \,\begin{matrix} \CMTB{\{144144'\}} \\ \oplus~ \CMTB{\{\overline{144144'}\}} \end{matrix}\, ~\oplus~ ~\,\begin{matrix} \CMTB{\{46800\}} \\ \oplus~ \CMTB{\{\overline{46800}\}} \end{matrix}\,~  ~\oplus~ ~~\begin{matrix} \CMTB{\{4950\}} \\ \oplus~ \CMTB{\{\overline{4950}\}} \end{matrix}~\, ~\oplus~ ~\begin{matrix} \CMTB{\{126\}} \\ \oplus~ \CMTB{\{\overline{126}\}} \end{matrix}~ 
\end{split} \nonumber \\[10pt]
\begin{split}
    & ~\oplus~~ {\CMTB{\ydiagram{3,3,3,2}}}_{\rm IR} ~\oplus~~ {\CMTB{\ydiagram{3,3,2,1}}}_{\rm IR} ~\oplus~~ {\CMTB{\ydiagram{3,2,1,1}}}_{\rm IR} ~\oplus~~ {\CMTB{\ydiagram{2,1,1,1}}}_{\rm IR}   \\
    &~~~~~~ \CMTB{[0,0,1,2,2]} ~~~~~~ \CMTB{  [0,1,1,1,1]  } ~~~~~~ \CMTB{ [1,1,0,1,1]  } ~~~~ \CMTB{[1,0,0,1,1]}   \\
    & ~\oplus~ \,~\CMTB{\{270270\}}~\, ~\oplus~ \,~\CMTB{\{192192\}}~ ~\oplus~ ~~\CMTB{\{36750\}}~~ ~\oplus~ ~\CMTB{\{1728\}}~
\end{split}  ~~~~~~.  
\label{equ:data2}
\end{align}

\subsection{Tying Rule for $n$-Column BYTs $(n\geq4)$}
\label{subsec:tiencol}

For a $n$-column BYT with $n\geq4$, the tying rule is as follows.
\begin{description}[leftmargin=4.1em]
\item[Step 1:] Draw all the possible combinations of ties (including no tie at all). They can skip any number of box(es) (as long as they are on the same row and remain symmetric). Vertical position of ties does not matter. {\em Keep one copy for each of the equivalent ones}.
\item[Step 2:] For each diagram, erase the boxes with the ties and rearrange them vertically according to symmetry properties such that they become a standard YT shape. If a standard YT shape cannot be obtained, or the rearrangements cannot be performed with symmetry preserved, throw that BYT away. Then the decomposition in $\mathfrak{so}(10)$ would be the sum of all the remaining BYTs.
\end{description}
In the $\mathfrak{su}(10)\supset\mathfrak{so}(10)$ branching rules, when a Young Tableau in $\mathfrak{su}(10)$ contains 4 or more columns, the irreps in $\mathfrak{so}(10)$ begin to have multiplicities larger than one. The key to understanding 4+ columns is therefore in Step 1: what does {\em equivalence} mean when ties are drawn?

Notice that when a YT has 4 or more columns, there is {\em more than one way to put two ties in a row}. For example, two ties can be put in a row of 4 boxes in the following 3 different ways, 
\begin{equation}
    \CMTB{\ytableaushort{\dotline\linedot\dotline\linedot}} ~~~~~,~~~~~ \CMTB{\ytableaushort{\dotlineup\dotlinedowndot\linedotupdot\linedotdown}}~~~~~,~~~~~ \CMTB{\ytableaushort{\dotlineup\dotlinedowndot\linedotdowndot\linedotup}} ~~~~~,
\label{eqn:4coltiescombo}
\end{equation}
{\em given that these 4 boxes are inequivalent}. The question now comes - when will they become equivalent? Stated another way, could we exchange two nodes that separately belong to two ties on the same row, such that the diagram after operation is equivalent to the original diagram? Let us remark that in these exchanges the specific boxes tied would not change, therefore they give the same irrep and it is for counting the multiplicity of each irrep in $\mathfrak{so}(10)$ after the application of branching rules.

To answer this question, let us study some special cases. Recall that in Section \ref{subsec:tie2col}, the totally symmetric BYT example (\ref{eqn:totsymBYTex}) does not contain 3 copies of ${\CMTB{\ydiagram{1}}}_{\rm IR}$ but only one, as all of the five boxes are equivalent - they are all exchangeable (totally symmetric). So let us propose a condition of exchanging two nodes on two ties that would leave the diagram invariant.
\begin{enumerate}
\item When two nodes are in two columns with the same number of boxes - so the two boxes that the two nodes posited are equivalent and exchangeable.
\end{enumerate}
For the special class of BYTs with all columns having the same number of box(es), this rule must apply. An example other than (\ref{eqn:totsymBYTex}) is
\begin{equation}
\begin{split}
    \ydiagram{4,4} ~~:&~~~~~ 1 ~~ {\CMTB{\ydiagram{4}}}_{\rm IR} ~~:~~~~~ {\CMTB{\ytableaushort{\dotline\linedot\dotline\linedot,{}{}{}{}}}}_{\rm IR} ~=~ {\CMTB{\ytableaushort{\dotlineup\dotlinedowndot\linedotupdot\linedotdown,{}{}{}{}}}}_{\rm IR} ~=~ {\CMTB{\ytableaushort{\dotlineup\dotlinedowndot\linedotdowndot\linedotup,{}{}{}{}}}}_{\rm IR} ~~~~~, \\[10pt]
    &~~~~~ 1 ~~ {\CMTB{\ydiagram{2}}}_{\rm IR} \hspace{2.5em}~~:~~~~~ {\CMTB{\ytableaushort{\dotline\linedot\dotline\linedot,{}{}\dotline\linedot}}}_{\rm IR} ~=~ {\CMTB{\ytableaushort{\dotlineup\dotlinedowndot\linedotupdot\linedotdown,{}{}\dotline\linedot}}}_{\rm IR} ~=~ {\CMTB{\ytableaushort{\dotlineup\dotlinedowndot\linedotdowndot\linedotup,{}{}\dotline\linedot}}}_{\rm IR} ~~~~~,  \\[10pt]
    &~~~~~ 1 ~~ \CMTB{\ytableaushort{{\none[\cdot]}}} \hspace{4.55em}~~:~~~~~ {\CMTB{\ytableaushort{\dotline\linedot\dotline\linedot,\dotline\linedot\dotline\linedot}}}_{\rm IR} ~=~ {\CMTB{\ytableaushort{\dotlineup\dotlinedowndot\linedotupdot\linedotdown,\dotlineup\dotlinedowndot\linedotupdot\linedotdown}}}_{\rm IR} ~=~ {\CMTB{\ytableaushort{\dotlineup\dotlinedowndot\linedotdowndot\linedotup,\dotlineup\dotlinedowndot\linedotdowndot\linedotup}}}_{\rm IR} \\[5pt]
    &~~~~~~~~~~ \hspace{5.8em}~~  ~=~ {\CMTB{\ytableaushort{\dotline\linedot\dotline\linedot,\dotlineup\dotlinedowndot\linedotupdot\linedotdown}}}_{\rm IR} ~=~ {\CMTB{\ytableaushort{\dotline\linedot\dotline\linedot,\dotlineup\dotlinedowndot\linedotdowndot\linedotup}}}_{\rm IR} ~=~ {\CMTB{\ytableaushort{\dotlineup\dotlinedowndot\linedotupdot\linedotdown,\dotlineup\dotlinedowndot\linedotdowndot\linedotup}}}_{\rm IR}  ~~~~~.
\end{split} 
\end{equation}
Therefore, 
\begin{align}
\begin{split}
    \ydiagram{4,4} ~=&~~ {\CMTB{\ydiagram{4,4}}}_{\rm IR} ~\oplus~~ {\CMTB{\ydiagram{4,2}}}_{\rm IR} ~\oplus~~ {\CMTB{\ydiagram{4}}}_{\rm IR}  \\
    &~~~ \CMTB{  [0,4,0,0,0]  } ~~~~~~~~~~ \CMTB{ [2,2,0,0,0]  } ~~~~~~~~~ \CMTB{[4,0,0,0,0]}   \\
    \{70785\}~~ ~=&~ ~~~~\CMTB{\{52920\}}~~~~ ~\oplus~ ~~~~\CMTB{\{16380\}}~~~~  ~\oplus~ ~~~~~\CMTB{\{660\}}~~~~~ 
\end{split} \nonumber \\[10pt]
\begin{split}
    & ~\oplus~~ ~{\CMTB{\ydiagram{2,2}}}_{\rm IR} ~\oplus~~ ~{\CMTB{\ydiagram{2}}}_{\rm IR} ~\oplus~~ ~\CMTB{\ytableaushort{{\none[\cdot]}}}  \\
    &~~~~ \CMTB{[0,2,0,0,0]} ~~~ \CMTB{[2,0,0,0,0]} ~~ \CMTB{[0,0,0,0,0]}  \\
    & ~\oplus~ \,~~\CMTB{\{770\}}~~ ~\oplus~ ~~~\CMTB{\{54\}}~~~ ~\oplus~ ~~\CMTB{\{1\}}~ 
\end{split} ~~~~~. \label{eqn:4colex1}
\end{align}
In fact, for all BYTs with all columns having the same number of box(es), according to rule 1., all the irreps in the decompositions appear only once. 

Now, let us turn to another class of BYTs - those with all columns unequal. Rule 1. would not apply. Although all columns are inequivalent, the diagrams involving two ties on the same row do not naively have multiplicities 3. We thereby propose the second condition of exchanging two nodes that could make two diagrams equivalent. 
\begin{enumerate}\setcounter{enumi}{1}
\item When there is another tie connecting the two columns where the two nodes located - then this tie offers an additional symmetry.
\end{enumerate}
Let us demonstrate this rule with the following example.
\begin{equation}
\begin{split}
    \ydiagram{4,3,2,1} ~~:&~~~~~~ 3 ~~ {\CMTB{\ydiagram{3,2,1}}}_{\rm IR} ~~:~~~~~~ {\CMTB{\ytableaushort{\dotline\linedot\dotline\linedot,{}{}{},{}{},{}}}}_{\rm IR} ~\neq~ {\CMTB{\ytableaushort{\dotlineup\dotlinedowndot\linedotupdot\linedotdown,{}{}{},{}{},{}}}}_{\rm IR} ~\neq~ {\CMTB{\ytableaushort{\dotlineup\dotlinedowndot\linedotdowndot\linedotup,{}{}{},{}{},{}}}}_{\rm IR} ~~~~~, \\[10pt]
    &~~~~~~ 2 ~~ {\CMTB{\ydiagram{2,2}}}_{\rm IR} \hspace{1.25em}~~:~~~~~~ {\CMTB{\ytableaushort{\dotline\linedot\dotline\linedot,\dotline\cdots\linedot,{}{},{}}}}_{\rm IR} ~=~ {\CMTB{\ytableaushort{\dotlineupbb\dotlinedowndot\linedotdowndotbb\linedotup,\dotlinebb\cdots\linedotbb,{}{},{}}}}_{\rm IR} ~\neq~ {\CMTB{\ytableaushort{\dotlineup\dotlinedowndot\linedotupdot\linedotdown,\dotline\cdots\linedot,{}{},{}}}}_{\rm IR} ~~~~~,  \\[10pt]
    &~~~~~~ 2 ~~ {\CMTB{\ydiagram{2,1,1}}}_{\rm IR} \hspace{1.25em}~~:~~~~~~ {\CMTB{\ytableaushort{\dotline\linedot\dotline\linedot,{}\dotline\linedot,{}{},{}}}}_{\rm IR} ~=~ {\CMTB{\ytableaushort{\dotlineup\dotlinedowndotbb\linedotupdotbb\linedotdown,{}\dotlinebb\linedotbb,{}{},{}}}}_{\rm IR} ~\neq~ {\CMTB{\ytableaushort{\dotlineup\dotlinedowndot\linedotdowndot\linedotup,{}\dotline\linedot,{}{},{}}}}_{\rm IR} ~~~~~,  \\[10pt]
    &~~~~~~ 2 ~~ {\CMTB{\ydiagram{3,1}}}_{\rm IR} ~~:~~~~~~ {\CMTB{\ytableaushort{\dotline\linedot\dotline\linedot,\dotline\linedot{},{}{},{}}}}_{\rm IR} ~\neq~ {\CMTB{\ytableaushort{\dotlineup\dotlinedowndot\linedotupdot\linedotdown,\dotline\linedot{},{}{},{}}}}_{\rm IR} ~=~ {\CMTB{\ytableaushort{\dotlinedownbb\dotlineupdotbb\linedotupdot\linedotdown,\dotlinebb\linedotbb{},{}{},{}}}}_{\rm IR} ~~~~~.
\end{split} 
\end{equation}
These are the irreps with multiplicities not equal to one. Others have multiplicity one. Some examples are
\begin{equation}
\begin{split}
    & 1 ~~ {\CMTB{\ydiagram{2}}}_{\rm IR} \hspace{1.25em}~~:~~~~~~ {\CMTB{\ytableaushort{\dotline\linedot\dotline\linedot,\dotline\cdots\linedot,\dotline\linedot,{}}}}_{\rm IR} ~=~ {\CMTB{\ytableaushort{\dotlinedownbb\dotlineupdot\linedotupdotbb\linedotdown,\dotlinebb\cdots\linedotbb,\dotline\linedot,{}}}}_{\rm IR} ~=~ {\CMTB{\ytableaushort{\dotlineupbb\dotlinedowndotbb\linedotupdot\linedotdown,\dotline\cdots\linedot,\dotlinebb\linedotbb,{}}}}_{\rm IR} ~~~~~, \\
    & 1 ~~ {\CMTB{\ydiagram{1,1}}}_{\rm IR} \hspace{2.5em}~~:~~~~~~ {\CMTB{\ytableaushort{\dotline\linedot\dotline\linedot,{}\dotline\linedot,\dotline\linedot,{}}}}_{\rm IR} ~=~ {\CMTB{\ytableaushort{\dotlineup\dotlinedowndotbb\linedotupdotbb\linedotdown,{}\dotlinebb\linedotbb,\dotline\linedot,{}}}}_{\rm IR} ~=~ {\CMTB{\ytableaushort{\dotlinedownbb\dotlineupdotbb\linedotupdot\linedotdown,{}\dotline\linedot,\dotlinebb\linedotbb,{}}}}_{\rm IR} ~~~~~.
\end{split}
\end{equation}
Other symmetry rules in Step 2 are exactly the same as those stated in 3-column BYTs (Section \ref{subsec:tie3col}). Together we have
\begin{align}
\begin{split}
    \ydiagram{4,3,2,1} ~=&~~ {\CMTB{\ydiagram{4,3,2,1}}}_{\rm IR} ~\oplus~~ {\CMTB{\ydiagram{4,3,1}}}_{\rm IR} ~\oplus~~ {\CMTB{\ydiagram{4,2,1,1}}}_{\rm IR} ~\oplus~~ {\CMTB{\ydiagram{4,2,2}}}_{\rm IR} \\
    &~~~ \CMTB{  [1,1,1,1,1]  } ~~~~~~~~~~ \CMTB{  [1,2,1,0,0]  } ~~~~~~~~~~ \CMTB{  [2,1,0,1,1]  } ~~~~~~~~~ \CMTB{  [2,0,2,0,0]  }  \\
    \{1812096\}~~ ~=&~ ~~~\CMTB{\{1048576\}}~~\, ~\oplus~ ~~~\CMTB{\{174636\}}~~~ ~\oplus~ ~~~\CMTB{\{143000\}}~~~ ~\oplus~ ~~~\CMTB{\{112320\}}~~~
\end{split} \nonumber \\[10pt]
\begin{split}
    & ~\oplus~~ {\CMTB{\ydiagram{3,3,1,1}}}_{\rm IR} ~\oplus~~ {\CMTB{\ydiagram{3,2,2,1}}}_{\rm IR} ~\oplus~~ {\CMTB{\ydiagram{3,3,2}}}_{\rm IR}  \\
    &~~~~~~ \CMTB{[0,2,0,1,1]} ~~~~~~ \CMTB{ [1,0,1,1,1] } ~~~~~~ \CMTB{[0,1,2,0,0]}   \\
    & ~\oplus~ ~~\CMTB{\{73710\}}~~ ~\oplus~ \,~~\CMTB{\{72765\}}~~  ~\oplus~ ~~\CMTB{\{70070\}}~~ 
\end{split} \nonumber \\[10pt]
\begin{split}
    & ~\oplus~~ 3 ~~ {\CMTB{\ydiagram{3,2,1}}}_{\rm IR} ~\oplus~~ {\CMTB{\ydiagram{4,2}}}_{\rm IR} ~\oplus~~ {\CMTB{\ydiagram{4,1,1}}}_{\rm IR}  \\
    &~~~~~~~~ \CMTB{[1,1,1,0,0]} ~~~~~~~~~ \CMTB{ [2,2,0,0,0]  } ~~~~~~~~~~ \CMTB{[3,0,1,0,0]}   \\
    & ~\oplus~ ~~(3) \CMTB{\{17920\}}~~ ~\oplus~ ~~~~\CMTB{\{16380\}}~~~~  ~\oplus~ ~~~~\CMTB{\{14784\}}~~~~ 
\end{split} \nonumber \\[10pt]
\begin{split}
    & ~\oplus~~ ~{\CMTB{\ydiagram{3,1,1,1}}}_{\rm IR} ~\oplus~~ {\CMTB{\ydiagram{3,3}}}_{\rm IR} ~\oplus~~ {\CMTB{\ydiagram{2,2,1,1}}}_{\rm IR} ~\oplus~~ {\CMTB{\ydiagram{2,2,2}}}_{\rm IR}  \\
    &~~~~~~~ \CMTB{[2,0,0,1,1]} ~~~~~~ \CMTB{[0,3,0,0,0]} ~~~~ \CMTB{ [0,1,0,1,1]  } ~~ \CMTB{[0,0,2,0,0]}   \\
    & ~\oplus~ \,~~~\CMTB{\{8085\}}~~~\, ~\oplus~ ~~~\CMTB{\{7644\}}~~\, ~\oplus~ ~\CMTB{\{5940\}}~  ~\oplus~ ~\CMTB{\{4125\}}~ 
\end{split} \nonumber \\[10pt]
\begin{split}
    & ~\oplus~~ 2 ~~ {\CMTB{\ydiagram{3,1}}}_{\rm IR} ~\oplus~~ 2 ~~ {\CMTB{\ydiagram{2,1,1}}}_{\rm IR} ~\oplus~~ 2 ~~ {\CMTB{\ydiagram{2,2}}}_{\rm IR} ~\oplus~~ {\CMTB{\ydiagram{2}}}_{\rm IR}  ~\oplus~~ {\CMTB{\ydiagram{1,1}}}_{\rm IR}  \\
    &~~~~~~~~ \CMTB{[2,1,0,0,0]} ~~~~~~~~ \CMTB{ [1,0,1,0,0] } ~~~~~ \CMTB{[0,2,0,0,0]} ~~~~ \CMTB{ [2,0,0,0,0] } ~~ \CMTB{[0,1,0,0,0]}  \\
    & ~\oplus~ \,~~(2) \CMTB{\{1386\}}~~\, ~\oplus~ ~~(2) \CMTB{\{945\}}~\, ~\oplus~ \,~(2) \CMTB{\{770\}}~\, ~\oplus~ \,~~\CMTB{\{54\}}~~\, ~\oplus~ ~\CMTB{\{45\}}~ 
\end{split}  ~~~~~~.  \label{eqn:4colex2}
\end{align}

Now we report the status of those BYTs with some equal columns and some unequal columns. Although rules 1. and 2. apply to count most of the diagrams in the decompositions correctly, there are often one or two diagram(s) in each of those BYTs with multiplicities off by one. We hereby make a guess that there is a missing piece in these rules of exchanging nodes to count equivalences, and this is under investigation.

We close this section by restating the inspiration for our ``tying rules'' comes from the conventional 
physicist's approach to ``pulling out'' irreducible representations of tensors defined over orthogonal 
groups.  Namely, given a general symmetrical tensor that is reducible, and an invariant quadratic 
tensor, the physicist will typically ``pull out the irreps'' by contracting indices on the symmetric 
tensor with the invariant quadratic tensor. The contraction of indices ``ties'' a pair of indices 
together, hence ``tying rules.''

\newpage
\section{Spinorial Irreps of $\mathfrak{so}(10)$\label{sec:spinorial-irreps}}

In this chapter, we will exploit the graphical interpretation of Dynkin Labels of spinorial irreducible 
representations of $\mathfrak{so}(10)$ and its utility in translating an irrep to field-theoretical language. This 
can be applied to understand the following two aspects of a spinorial irrep:
\begin{enumerate}[label=(\alph*.)]
\item the proper index structure of the field; and
\item the irreducible constraints and the counting of the degrees of freedom.
\end{enumerate}
These understandings of spinorial irreps are tightly related to some tensor product rules. They are presented in Chapter \ref{sec:king-tensor}. 


Let us start with a general explanation for (a.). The basic elements of spinorial irreps are
\begin{equation}
\begin{split}
    \CMTred{ [0,0,0,0,1] } ~\equiv&~ \CMTred{\ytableaushort{\sixteen}} ~=~ \CMTred{\{16\}}  ~~~, \\
    \CMTred{ [0,0,0,1,0] } ~\equiv&~ \CMTred{\ytableaushort{\sixteenbar}} ~=~
    \CMTred{\{\overline{16}\}}  ~~~.
\end{split} \label{equ:SYTbasic}
\end{equation}
For convenience, we introduce a notation for mixed YT that is defined from the Dynkin Label. 
Since putting together the bosonic columns in Equation (\ref{equ:BYTbasic}) corresponds to 
adding Dynkin Labels, we can carry out a similar thing for these spinorial YT too. If we put together 
Equations (\ref{equ:SYTbasic}) and (\ref{equ:BYTbasic}), we can
draw a general mixed YT\footnote{
We represent the basic spinorial irrep by $\CMTred{\ydiagram{1}}$ and put it on the right of the BYT. 
In \cite{fischler1981,hurni1987}, they represent it by a long upward arrow $\uparrow$ and put it on the right of the BYT. 
In \cite{KingES1983,BS1991}, they represent it by a column of 5 $\SYTtri~$'s which means ``half'' of a column of 5 boxes (as $\CMTred{[0,0,0,0,1]}$ is ``half'' of $\CMTB{[0,0,0,0,2]}$ in terms of Dynkin labels), and put it on the right of the BYT. }
\begin{equation}
    {\generalBYTpqrst
    \CMTred{\ydiagram{1,0,0,0,0}}}_{{\rm IR},\pm} = \begin{cases}
        [\CMTB{p},\CMTB{q},\CMTB{r},\CMTB{s},\CMTB{s+2t}+\CMTred{1}], & \text{for}
        + \text{and } \CMTred{\ydiagram{1}} = \CMTred{\ytableaushort{\sixteen}} ~, \\
        [\CMTB{p},\CMTB{q},\CMTB{r},\CMTB{s+2t},\CMTB{s}+\CMTred{1}], & \text{for}
        - \text{and } \CMTred{\ydiagram{1}} = \CMTred{\ytableaushort{\sixteen}} ~, \\
        [\CMTB{p},\CMTB{q},\CMTB{r},\CMTB{s}+\CMTred{1},\CMTB{s+2t}], & \text{for}
        + \text{and } \CMTred{\ydiagram{1}} = \CMTred{\ytableaushort{\sixteenbar}}
        ~, \\
        [\CMTB{p},\CMTB{q},\CMTB{r},\CMTB{s+2t}+\CMTred{1},\CMTB{s}], & \text{for}
        - \text{and } \CMTred{\ydiagram{1}} = \CMTred{\ytableaushort{\sixteenbar}} ~.
    \end{cases}
\label{eqn:mixedYTtoDynkin}
\end{equation}
One point to note though, is $\CMTB{\ydiagram{1}}$ and $\CMTred{\ydiagram{1}}$ are two 
different types of YT, so when they are put together on a row, it {\em does not} imply that they 
are symmetric. Also, the above Dynkin Labels should all be red as they are spinorial irreps. 
The alternate coloring was just to illustrate the origin of those digits.
Two simple extra examples are
\begin{align}
    {\CMTB{\ydiagram{1}}\CMTred{\ytableaushort{\sixteen}}}_{{\rm IR}} \hspace{1.25em} ~=&~ \CMTred{ [1,0,0,0,1] } ~~~, \\
    {\CMTB{\ydiagram{2,2,1}}\CMTred{\ytableaushort{\sixteenbar,\none,\none}}}_{{\rm IR}} ~=&~ \CMTred{ [0,1,1,1,0] }  ~~~.
\end{align}

Note that although there is no ambiguity in translating from a mixed YT to a Dynkin Label for a spinorial 
irrep as in Equation (\ref{eqn:mixedYTtoDynkin}), there is an ambiguity for the converse translation. 
For a general bosonic irrep, the Dynkin Label $\CMTB{[a,b,c,d,e]}$ has 5 integers, while in the BYT there 
are 5 numbers $p$, $q$, $r$, $s$, and $t$ to describe the numbers of columns with 1, 2, 3, 4, and 5 
boxes respectively. So they already form a one-to-one mapping between Dynkin Labels and BYTs. Now the 
basic spinorial irreps in Equation (\ref{equ:SYTbasic}) are introduced. New YT notations $\CMTred{
\ytableaushort{\sixteen}}$ and $\CMTred{\ytableaushort{\sixteenbar}}$ are introduced, so there are 7 
fundamental objects for YTs. A general Dynkin Label $[a,b,c,d,e]$ remains to have only 5 integers, and 
there's one more degree of freedom that we can gain from the Dynkin Label: the parity of $|e-d|$. 
Therefore, we have 6 degrees of freedom for Dynkin Labels. So ambiguities arise. The simplest 
example is stated below.
\begin{equation}
\begin{split}
    ~ \CMTred{[0,0,0,1,2]} ~=&~ \CMTB{ [0,0,0,1,1] } + \CMTred{ [0,0,0,0,1] }  ~~~, \\
    ~ \CMTred{[0,0,0,1,2]} ~=&~ \CMTB{ [0,0,0,0,2] } + \CMTred{ [0,0,0,1,0] }  ~~~,
\end{split}
\label{eqn:ambiguity_ex}
\end{equation}
where the ``$+$'' here just means viewing each Dynkin Label as a vector and summing them 
integer by integer, but {\em not} any sort of direct sum of irreps. The way to resolve this is to eliminate 
one degree of freedom from the spinorial YT notation by making a choice - for the following 
dimension rules, we interpret {\em every} spinorial irrep (an irrep is spinorial when $|e-d|$ is 
odd) to be $\CMTred{ [0,0,0,0,1] }$ ``+'' some other bosonic Dynkin Labels, when $e>d$. When 
$e<d$, perform the conjugation first, apply the dimension rule, and carry out the conjugation again to obtain 
the final result. Interpreted in another way, that would mean for $e<d$, it would be $\CMTred{ 
[0,0,0,1,0] }$ ``+'' some other bosonic Dynkin Labels. With this condition, for the example in 
Equation (\ref{eqn:ambiguity_ex}), we {\em pick} 
\begin{equation}
~ \CMTred{[0,0,0,1,2]} ~=~ \CMTB{ [0,0,0,1,1] } + \CMTred{ [0,0,0,0,1] } ~=~ {\CMTB{
\ydiagram{1,1,1,1}}\CMTred{\ytableaushort{\sixteen,\none,\none,\none}}}_{{\rm IR}} ~~~,
\label{eqn:ambiguity_ex_choice}
\end{equation}
while for $\CMTred{[0,0,0,2,1]}$, since $e<d$, we carry out the conjugation first to find $\CMTred{
[0,0,0,1,2]}$, map it to YT notation according to Equation (\ref{eqn:ambiguity_ex_choice}), 
then perform conjugation again. Therefore, 
\begin{equation}
~ \CMTred{[0,0,0,2,1]} ~=~ \CMTB{ [0,0,0,1,1] } + \CMTred{ [0,0,0,1,0] } ~=~ {\CMTB{
\ydiagram{1,1,1,1}}\CMTred{\ytableaushort{\sixteenbar,\none,\none,\none}}}_{{\rm IR}} ~~~.
\end{equation}
Now the ambiguity has been cleared. This also brings us to two remarks.
\begin{enumerate}
\item Since we choose to translate a Dynkin Label $[a,b,c,d,e]$ with $e>d$ to YT notation, so in mixed YT notation, only self-dual BYTs remain - but of course if we 
perform a conjugation we obtain the anti-self-dual parts. Therefore, a general spinorial irrep (or its conjugate) can be represented by 
\begin{equation}
    {\generalBYTpqrst
    \CMTred{\ydiagram{1,0,0,0,0}}}_{{\rm IR},+} ~~~~~.
\end{equation}
\item While writing the irrep in dimension notation, it is $\CMTred{\{\dim\}}$ if the number of boxes in the BYT part is even, and $\CMTred{\{\overline{\dim}\}}$ if odd.
\end{enumerate}

For the following dimension rules, we would focus on the $e>d$ cases (and the readers can obtain the $e<d$ cases by conjugation). 
For convenience of presentation, from now on, $\CMTred{\ydiagram{1}}$ represents $
\CMTred{\ytableaushort{\sixteen}}$ unless specified otherwise. 
And the overall spinorial irrep would always be specified. 

\subsection{General Spinorial Irrep Graphical Dimension Rules}
\label{subsec:SYTdimrule}

How do we calculate the dimensions of these spinorial irreps graphically then? And what 
about their irreducible constraints? It turns out that these two questions are two sides of the 
same coin. In \cite{king1975}, the general formula for $D_{n}$ algebra spinorial irrep was 
first given,
\begin{equation}
    [\Delta;\lambda] ~=~ [\lambda/P] \otimes \Delta ~~~,
\label{eqn:kingrule}
\end{equation}
where $\Delta$ is the basic spinorial irrep, $\lambda$ is 
a BYT shape, and
\begin{equation}
    P ~=~ \sum_{m} (-1)^{m} \, {\overbrace{{\CMTB{\YThdots{1}}}}^{m}}_{\rm IR} ~~~,
\label{eqn:schurP}
\end{equation}
is a Schur function series.

A direct translation into our notation is
\begin{equation}
\begin{split}
    & \generalBYTpqrst
    {\CMTred{\ytableaushort{{},\none,\none,\none,\none}}}_{{\rm IR},+} \\
    &= \sum_{m=0}^{p+q+r+s+t} (-1)^{m} ~ \left(\, \generalBYTpqrst_{{\rm IR},+} \Bigg/ \, {\overbrace{\CMTB{\ytableaushort{{}\cdots{}}}}^{m}}_{\rm IR} \,\right)  ~\otimes~ \CMTred{\ytableaushort{\tinysixteenaltbarm}} ~~~~~,
\end{split} \label{eqn:kingrulegraphtensor}
\end{equation}
where on the right hand side 
\begin{equation}
    \CMTred{\ytableaushort{\tinysixteenaltbarm}} ~=~ \begin{cases}
    ~ \CMTred{\ytableaushort{\sixteen}} ~~~, & \text{for } m \text{ even} \\
    ~ \CMTred{\ytableaushort{\sixteenbar}} ~~~, & \text{for } m \text{ odd} \\
    \end{cases} ~~~~~.
\end{equation}
The SYT dimension formula is just taking the ``dim'' operator on both sides,
\begin{flalign*}
    \dim \left( \generalBYTpqrst
    {\CMTred{\ytableaushort{{},\none,\none,\none,\none}}}_{{\rm IR},+} \right) &&
\end{flalign*}
\begin{equation}
\begin{split}
    &= \left( \sum_{m=0}^{p+q+r+s+t} (-1)^{m} \dim \left(\, \generalBYTpqrst_{{\rm IR},+} \Bigg/ \, {\overbrace{\CMTB{\ytableaushort{{}\cdots{}}}}^{m}}_{\rm IR} \,\right) \right) \\
    & ~~~~~~~~~~~~~~~~~~~~~~~~~~~~~~~~~~~~~~~~~~~~~~~~~~~~~~~~~~~~~~~~~~~~~~~~~~~~~~~~~~~~~~~~~~~~~~~~~~~~~~ ~\times~ \dim\left(\, \CMTred{\ydiagram{1}} \,\right) ~~~~~.
\end{split} \label{eqn:kingrulegraph}
\end{equation}
If we look at the right hand side of Equation (\ref{eqn:kingrule}), we notice that the dimension of the basic spinorial irrep is pulled out. Let us focus on the bosonic part $[\lambda/P]$, i.e. the $\sum_{m=0}^{p+q+r+s+t}$ part in Equation (\ref{eqn:kingrulegraph}). If we look at it term-by-term, $m$ is the number of box(es) in a row removed in each term, and it ranges from $m=0$ (no box removed) to $m=p+q+r+s+t$ (entire row removed). Therefore, in the sum, 
\begin{equation}
\renewcommand{\arraystretch}{1.5}
\begin{tabular}{ccc}
$\overbrace{\CMTB{\ytableaushort{{}\cdots{},{}\cdots{},{}\cdots{},{}\cdots{},{}\cdots{}}}}^{t} 
\hspace{-0.2em} \overbrace{\CMTB{\ytableaushort{{}\cdots{},{}\cdots{},{}\cdots{},{}\cdots{},\none}
}}^{s} \hspace{-0.2em} \overbrace{\CMTB{\ytableaushort{{}\cdots{},{}\cdots{},{}\cdots{},\none,\none
}}}^{r} \hspace{-0.2em} \overbrace{\CMTB{\ytableaushort{{}\cdots{},{}\cdots{},\none,\none,\none}
}}^{q} \hspace{-0.2em} {\overbrace{\CMTB{\ytableaushort{{}\cdots{},\none,\none,\none,\none}}}^{
p}}_{\rm IR}$ & $\longrightarrow$ & $\overbrace{\CMTB{\ytableaushort{{}\cdots{},{}\cdots{},{}\cdots{
},{}\cdots{},\none}}}^{t} \hspace{-0.2em} \overbrace{\CMTB{\ytableaushort{{}\cdots{},{}\cdots{},{
}\cdots{},\none,\none}}}^{s} \hspace{-0.2em} \overbrace{\CMTB{\ytableaushort{{}\cdots{},{}\cdots{
},\none,\none,\none}}}^{r} \hspace{-0.2em} {\overbrace{\CMTB{\ytableaushort{{}\cdots{
},\none,\none,\none,\none}}}^{q}}_{\rm IR}$ \\
original diagram & $\longrightarrow$ & last diagram
\end{tabular} ~~~~~.
\end{equation}
This turns out to be an interesting combinatorics problem. Since the ``/'' operator would only count all the equivalent removals once, e.g. 
\begin{equation}
\begin{split}
    {\CMTB{\ytableaushort{{}{},\vdots\vdots,{}{},{}}}}_{\rm IR} \Bigg/ {\CMTB{\ydiagram{1}}}_{\rm IR} ~=&~~ {\CMTB{\ytableaushort{{}{},\vdots\vdots,{},{}}}}_{\rm IR} \oplus~ {\CMTB{\ytableaushort{{}{},\vdots\vdots,{}{}}}}_{\rm IR} ~~~~~, \\
    {\CMTB{\ytableaushort{{}{},\vdots\vdots,{}{}}}}_{\rm IR} \Bigg/ {\CMTB{\ydiagram{1}}}_{\rm IR} ~=&~~ {\CMTB{\ytableaushort{{}{},\vdots\vdots,{}}}}_{\rm IR} ~~~~~,
\end{split}
\label{eqn:quotientequiv}
\end{equation}
the quotient operator for each $m$
\begin{equation}
    \generalBYTpqrst_{{\rm IR},+} \Bigg/ \, {\overbrace{\CMTB{\ytableaushort{{}\cdots{}}}}^{m}}_{\rm IR} ~~~~~,
\end{equation}
can be thought as removing $m$ objects from the following $p+q+r+s+t$ objects, 
\begin{equation}
    \overbrace{\CMTB{\ytableaushort{t\cdots t}}}^{t}\hspace{-0.2em}
    \overbrace{\CMTB{\ytableaushort{s\cdots s}}}^{s}\hspace{-0.2em}
    \overbrace{\CMTB{\ytableaushort{r\cdots r}}}^{r}\hspace{-0.2em}
    \overbrace{\CMTB{\ytableaushort{q\cdots q}}}^{q}\hspace{-0.2em}
    \overbrace{\CMTB{\ytableaushort{p\cdots p}}}^{p} ~~~~~,
\end{equation}
with of course all $p$ objects are indistinguishable, all $q$ objects are indistinguishable, etc.. Therefore, for each $m$, the problem of how many terms appear there becomes:
\begin{quotation}
    \noindent
    Given a multiset $\{p,\cdots,p,q,\cdots,q,r,\cdots,r,s,\cdots,s,t,\cdots,t\}$, how many subsets with a fixed size $m$ are there?
\end{quotation}
Although we do not know the close form, we have an efficient algorithm to obtain the number\footnote{SNHM thanks Kevin Iga for this idea.}. If we write 
\begin{equation}
    ( 1 + x + \cdots + x^{p} ) ( 1 + x + \cdots + x^{q} ) ( 1 + x + \cdots + x^{r} ) ( 1 + x + \cdots + x^{s} ) ( 1 + x + \cdots + x^{t} ) ~=~ \sum_{m=0}^{p+q+r+s+t} c_{m} x^{m} ~~~,
\label{eqn:kevin}
\end{equation}
then $c_{m}$ equals the number of subsets of size $m$. To obtain all the subsets, just modify the above polynomial to 
\begin{equation}
    ( 1 + p + \cdots + p^{p} ) ( 1 + q + \cdots + q^{q} ) ( 1 + r + \cdots + r^{r} ) ( 1 + s + \cdots + s^{s} ) ( 1 + t + \cdots + t^{t} )  ~~~,
\label{eqn:kevinmodify}
\end{equation}
and take the part of the homogeneous polynomial of degree $m$ in it. Each term then corresponds to a subset, or a combination. 

To remove the abstractness, let us look at an example. Consider a mixed YT with the BYT part with $(p,q,r,s,t)=(1,2,1,0,0)$, i.e. 
\begin{equation}
    {\CMTB{\ydiagram{4,3,1}}\CMTred{\ydiagram{1,0,0}}}_{\rm IR} ~~~~~,
\end{equation}
and $m$ ranges from 0 to 4 $(=1+2+1)$. Application of Equations (\ref{eqn:kevin}) and (\ref{eqn:kevinmodify}) gives us the combinations in Table \ref{tab:SYTtermcount}.
\begin{table}[h!]
\centering
\begin{tabular}{|c|c|c|} \hline
    ~~$m$~~ & Number of combinations $c_{m}$ & Combinations \\ \hline
    0 & 1 & 1 \\ \hline
    1 & 3 & $p$, $q$, $r$ \\ \hline
    2 & 4 & $pq$, $pr$, $q^{2}$, $qr$ \\ \hline
    3 & 3 & $pq^{2}$, $pqr$, $q^{2}r$ \\ \hline
    4 & 1 & $pq^{2}r$ \\ \hline
\end{tabular}
\caption{All possible combinations of removing $m$ boxes for $(p,q,r,s,t)=(1,2,1,0,0)$}
\label{tab:SYTtermcount}
\end{table}

\noindent
Now let us translate back to the BYT notation and look at the terms in the sum one by one.
\begin{equation}
\begin{split}
    m=0 ~:&~~~ {\CMTB{\ytableaushort{{}{}{}{},{}{}{},{}}}}_{\rm IR} ~~~, \\
    m=1 ~:&~~~ {\CMTB{\ytableaushort{{}{}{}{\none[p]},{}{}{},{}}}}_{\rm IR} ~~~,~~~ {\CMTB{\ytableaushort{{}{}{}{},{}{}{\none[q]},{}}}}_{\rm IR} ~~~,~~~ {\CMTB{\ytableaushort{{}{}{}{},{}{}{},{\none[r]}}}}_{\rm IR} ~~~, \\
    m=2 ~:&~~~ {\CMTB{\ytableaushort{{}{}{}{\none[p]},{}{}{\none[q]},{}}}}_{\rm IR} ~~~,~~~ {\CMTB{\ytableaushort{{}{}{}{\none[p]},{}{}{},{\none[r]}}}}_{\rm IR} ~~~,~~~ {\CMTB{\ytableaushort{{}{}{}{},{}{\none[q]}{\none[q]},{}}}}_{\rm IR} ~~~,~~~ {\CMTB{\ytableaushort{{}{}{}{},{}{}{\none[q]},{\none[r]}}}}_{\rm IR} ~~~, \\
    m=3 ~:&~~~ {\CMTB{\ytableaushort{{}{}{}{\none[p]},{}{\none[q]}{\none[q]},{}}}}_{\rm IR} ~~~,~~~ {\CMTB{\ytableaushort{{}{}{}{\none[p]},{}{}{\none[q]},{\none[r]}}}}_{\rm IR} ~~~,~~~ {\CMTB{\ytableaushort{{}{}{}{},{}{\none[q]}{\none[q]},{\none[r]}}}}_{\rm IR} ~~~, \\
    m=4 ~:&~~~ {\CMTB{\ytableaushort{{}{}{}{\none[p]},{}{\none[q]}{\none[q]},{\none[r]}}}}_{\rm IR} ~~~.
\end{split}
\end{equation}
Another point to note is that each term is attached by a $(-1)^{m}$ factor. That means when even number of boxes are removed, we add the dimensions of those terms; when odd number of boxes are removed, we subtract their dimensions. Therefore, 
\begin{equation}
\begin{split}
    & \dim\left(\, {\CMTB{\ydiagram{4,3,1}}\CMTred{\ydiagram{1,0,0}}}_{\rm IR} \,\right) \\
    ~=&~ \Bigg(~ \dim\left(\, {\CMTB{\ydiagram{4,3,1}}}_{\rm IR} \,\right) \\
    &~~ - \dim\left(\, {\CMTB{\ydiagram{3,3,1}}}_{\rm IR} \,\right) - \dim\left(\, {\CMTB{\ydiagram{4,2,1}}}_{\rm IR} \,\right) - \dim\left(\, {\CMTB{\ydiagram{4,3}}}_{\rm IR} \,\right)  \\
    &~~ + \dim\left(\, {\CMTB{\ydiagram{3,2,1}}}_{\rm IR} \,\right) + \dim\left(\, {\CMTB{\ydiagram{3,3}}}_{\rm IR} \,\right) + \dim\left(\, {\CMTB{\ydiagram{4,1,1}}}_{\rm IR} \,\right) + \dim\left(\, {\CMTB{\ydiagram{4,2}}}_{\rm IR} \,\right) \\
    &~~ - \dim\left(\, {\CMTB{\ydiagram{3,1,1}}}_{\rm IR} \,\right) - \dim\left(\, {\CMTB{\ydiagram{3,2}}}_{\rm IR} \,\right) - \dim\left(\, {\CMTB{\ydiagram{4,1}}}_{\rm IR} \,\right) \\
    &~~ + \dim\left(\, {\CMTB{\ydiagram{3,1}}}_{\rm IR} \,\right) ~\Bigg)  ~\times~ \dim\left(\, \CMTred{\ydiagram{1}} \,\right) ~~~~~.
\end{split} \label{eqn:SYTdimExResult}
\end{equation}
Numerically, 
\begin{equation}
\begin{split}
    1260000 ~=&~ \big(~ 174636 \\
    &~~ ~-~ 34398 ~-~ 68640 ~-~ 37362 \\
    &~~ ~+~ 17920 ~+~ 7644 ~+~ 14784 ~+~ 16380 \\
    &~~ ~-~ 4312 ~-~ 4410 ~-~ 4608 \\
    &~~ ~+~ 1386  ~\big) ~\times~ 16 ~~~.
\end{split}
\end{equation}

In the following sections, we will apply the spinorial dimension rule (\ref{eqn:kingrulegraph}) 
for the spinorial irreps composed of completely antisymmetric, completely symmetric, 
two-equal-column, and two-unequal-column BYTs attached with a $\CMTred{\ytableaushort{\sixteen}}$. 
These are the types of spinorial irreps that would appear in the 10D, $\mathcal{N}=1$ scalar superfield. 
The general dimension formulas and the irreducible conditions will be presented in 
Sections \ref{subsec:SYTanti}, \ref{subsec:SYTsym}, \ref{subsec:SYTtwoeqcol} and \ref{subsec:SYTtwouneqcol}, 
while the explicit details of the corresponding examples will be presented in Appendix \ref{appen:SYTdimex}.

\subsection{Completely antisymmetric BYTs attached with $\CMTred{\{16\}}$}
\label{subsec:SYTanti}

First let us consider totally antisymmetric BYTs attached with a $\CMTred{\ytableaushort{\sixteen}}$. By applying (\ref{eqn:kingrulegraph}), one has
\begin{equation}
\begin{split}
    \dim \left(~ n \left\{ {\CMTB{\renewcommand{\arraystretch}{0} \begin{tabular}{l} \ydiagram{1} \CMTred{\ytableausetup{boxsize=1.4em}\ydiagram{1}} \\ \ytableausetup{boxsize=1.2em,vtabloids}\ydiagram[*(white) \vdots]{1,1}\ytableausetup{novtabloids} \\ \ydiagram{1} \end{tabular}}}_{\rm IR} ~\right) \color{white}{} \right\} \color{black}{} \hspace{-1em}=~ \left(~ \dim \left(~ n \left\{ {\CMTB{\YTvdots{1,1}}}_{\rm IR} ~\right) \color{white}{} \right\} \color{black}{} \hspace{-1em}-~ \dim \left(~ (n-1) \left\{ {\CMTB{\YTvdots{1}}}_{\rm IR} ~\right) \color{white}{} \right\} \color{black}{} \hspace{-1em}\right)  \\
    ~\times~  \dim \Big(~ \CMTred{\ydiagram{1}} ~\Big) ~~~~~.
\end{split} \label{eqn:SYTantisym}
\end{equation}
This formula for calculating the irrep dimensions (or the degrees of freedom of a field)
is very suggestive about how we should write out the irreducible conditions. 
First note that the removal of a box from the BYT part in the mixed YT is equivalent to the contraction by a $(\s^{\un{a}})_{\a\b}$ matrix. 
If we use the index notation we invented in Section \ref{subsec:BYTindex}, 
with the general field on the left and the general irreducible condition on the right, we 
find
\begin{equation}
    \Psi_{\{ \un{a}_{1} \cdots \un{a}_{n} \}}{}^{\a} ~~:~~~~~~~~~~~~ (\s^{\un{a}_{n}})^{\a\b}
    \Psi_{\{ \un{a}_{1} \cdots \un{a}_{n} \} \b } ~=~ 0  ~~~,
\end{equation}
where $n = 1, \dots, 5$. The degree of freedom of the irreducible condition is that of
$\text{d.o.f.}(\{ \un{a}_{1} \cdots \un{a}_{n-1} \}) \times \text{d.o.f.}(\a)$  as the sigma
matrix contracted one vector index out. Therefore, it is consistent with the dimension formula as written in Equation (\ref{eqn:SYTantisym}), or in index notation,
\begin{equation}
    \dim\left( \Psi_{\{ \un{a}_{1} \cdots \un{a}_{n} \}}{}^{\a} \right) ~=~ \text{d.o.f.}(\{
    \un{a}_{1} \cdots \un{a}_{n} \}) \times \text{d.o.f.}(\a) ~-~ \text{d.o.f.}(\{ \un{a}_{
    1} \cdots \un{a}_{n-1} \}) \times \text{d.o.f.}(\a) ~~~.
\label{eqn:SYTantisymIndex}
\end{equation}

\subsection{Completely symmetric BYTs attached with $\CMTred{\{16\}}$}
\label{subsec:SYTsym}

Now let us turn to totally symmetric BYTs attached with a $\CMTred{\ytableaushort{\sixteen}}$.
\begin{equation}
\begin{split}
    \dim \Big(~ \underbrace{\CMTB{\YThdots{2}}}_{n}\hspace{-0.18em}{\CMTred{\ydiagram{1}}}_{\rm IR} ~\Big) ~=&~ \Bigg(~ \dim \Big(~  \underbrace{\CMTB{\YThdots{2}}}_{n}\hspace{-1.43em}{\CMTB{\ydiagram{1}}}_{\rm IR} ~\Big) ~-~ \dim \Big(~  \underbrace{\CMTB{\YThdots{1}}}_{n-1}\hspace{-1.43em}{\CMTB{\ydiagram{1}}}_{\rm IR} ~\Big) ~+~ \cdots  \\
    &~+~ (-1)^{n-1} \dim \Big(~ {\CMTB{\ydiagram{1}}}_{\rm IR} ~\Big)  ~+~ (-1)^{n} \dim \big(~ \CMTB{\cdot} ~\big) ~\Bigg)  ~\times~ \dim \Big(~ \CMTred{\ydiagram{1}} ~\Big) ~~~~~.
\end{split}
\label{eqn:SYTsym}
\end{equation}
Again, we can write the analytical expressions of the fields on the left and
the corresponding set of irreducible constraints on the right as follows,
\begin{equation}
    \Psi_{\{ \un{a}_{1}, \cdots, \un{a}_{n} \}}{}^{\a} ~~:~~~~~~~~~~~~
    \begin{cases}
    (\s^{\un{a}_{n}})_{\a\b} \Psi_{\{ \un{a}_{1}, \cdots, \un{a}_{n} \}}{}^{ \a } ~\equiv~ \psi_{
    \{ \un{a}_{1}, \cdots, \un{a}_{n-1} \}\b} ~=~ 0  ~~~, \\
    (\s^{\un{a}_{n-1}})^{\g\b} \psi_{\{ \un{a}_{1}, \cdots, \un{a}_{n-1} \} \b } ~\equiv~ \psi_{
    \{ \un{a}_{1}, \cdots, \un{a}_{n-2} \}}{}^{\g} ~=~ 0  ~~~, \\
    ~~~~~~~~~~~~~~~~~~~ \vdots \\
    (\s^{\un{a}_{1}})^{\g\b} \psi_{\{ \un{a}_{1} \} \b } ~=~ 0 ~ (n \text{ odd}) ~~~,~~~ (\s^{
    \un{a}_{1}})_{\g\b} \psi_{\{ \un{a}_{1} \}}{}^{ \b } ~=~ 0 ~ (n \text{ even})~~~,
    \end{cases}
\end{equation}
where $n = 1, 2, \dots$. From these conditions, the dimension formula would be
\begin{equation}
\begin{split}
    \dim\left( \Psi_{\{ \un{a}_{1}, \cdots, \un{a}_{n} \}}{}^{\a} \right) ~=&~ \text{d.o.f.}(\{ \un{a}_{1}, \cdots, \un{a}_{n} \}) \times \text{d.o.f.}(\a) ~-~ \bigg(~ \text{d.o.f.}(\{ \un{a}_{1},\cdots, \un{a}_{n-1} \}) \times \text{d.o.f.}(\a) \\
    & ~-~ \Big(~ \text{d.o.f.}(\{ \un{a}_{1}, \cdots, \un{a}_{n-2} \}) \times \text{d.o.f.}(\a) ~-~ \cdots \\
    & ~-~ \big(~ \text{d.o.f.}(\{ \un{a}_{1} \}) \times \text{d.o.f.}(\a) ~-~ \text{d.o.f.}(\a) ~\big) ~ \Big) ~\bigg)  \\
    ~=&~ \Big(~ \text{d.o.f.}(\{ \un{a}_{1}, \cdots, \un{a}_{n} \}) ~-~ \text{d.o.f.}(\{ \un{a}_{1},\cdots, \un{a}_{n-1} \}) ~+~ \text{d.o.f.}(\{ \un{a}_{1}, \cdots, \un{a}_{n-2} \}) \\
    &~~~ ~-~ \cdots ~+~ (-1)^{n-1} \, \text{d.o.f.}(\{ \un{a}_{1} \}) ~+~ (-1)^{n} ~\Big) \times \text{d.o.f.}(\a)  ~~~,
\end{split}
\label{eqn:SYTsymIndex}
\end{equation}
which agrees with Equation (\ref{eqn:SYTsym}). One can see that for $n=1$, it recovers the $n=1$ case of 
the totally antisymmetric BYT attached with a $\CMTred{\ytableaushort{\sixteen}}$ in Equation (\ref{eqn:SYTantisymIndex}).

\subsection{Two-equal-column BYTs attached with $\CMTred{\{16\}}$}
\label{subsec:SYTtwoeqcol}

How about two column BYTs with same number of boxes in each column, attached with
a $\CMTred{\ytableaushort{\sixteen}}$? We have
\begin{equation}
\begin{split}
    \dim \left(~ {\CMTB{\ytableaushort{{}{},\vdots\vdots,{}{}}}\CMTred{\ydiagram{1,0,0}}}_{\rm IR} ~\right) ~=~ \left(~  \dim \left(~ {\CMTB{\ytableaushort{{}{},\vdots\vdots,{}{}}}}_{\rm IR} ~\right)  ~-~ \dim \left(~ {\CMTB{\ytableaushort{{}{},\vdots\vdots,{}}}}_{\rm IR} ~\right) ~+~ \dim \left(~ {\CMTB{\ytableaushort{{}{},\vdots\vdots}}}_{\rm IR} ~\right)  ~\right) \\
    ~\times~ \dim \Big(~ \CMTred{\ydiagram{1}} ~\Big) ~~~~~.
\end{split}
\label{eqn:SYTtwoeqcol}
\end{equation}
The general form of the irreps with index notations on the left and the set of irreducible
constraints on the right is
\begin{equation}
    \Psi_{\{ \un{a}_{1} \cdots \un{a}_{n}, \un{b}_{1} \cdots \un{b}_{n} \}}{}^{\a} ~~:~~~~~~~~~~~~
\begin{cases}
    (\s^{\un{b}_{n}})_{\a\b} \Psi_{\{ \un{a}_{1} \cdots \un{a}_{n}, \un{b}_{1} \cdots \un{b}_{n}\}}{}^{ \a } ~\equiv~ \psi_{\{ \un{a}_{1} \cdots \un{a}_{n}| \un{b}_{1} \cdots \un{b}_{n-1}\}\b} ~=~ 0  ~~~, \\
    (\s^{\un{a}_{n}})^{\g\b} \psi_{\{ \un{a}_{1} \cdots \un{a}_{n}| \un{b}_{1} \cdots \un{b}_{n-1} \} \b } ~=~ 0  ~~~,
\end{cases}
\end{equation}
where $n = 1, \dots, 5$. The dimension formula derived from these constraints is
\begin{equation}
\begin{split}
    \dim\left( \Psi_{\{ \un{a}_{1} \cdots \un{a}_{n}, \un{b}_{1} \cdots \un{b}_{n} \}}{}^{\a}\right) ~=&~ \text{d.o.f.}(\{ \un{a}_{1} \cdots \un{a}_{n}, \un{b}_{1} \cdots \un{b}_{n}\}) \times \text{d.o.f.}(\a)  \\
    & ~-~ \Big(~ \text{d.o.f.}(\{ \un{a}_{1} \cdots \un{a}_{n}| \un{b}_{1} \cdots \un{b}_{n-1} \}) \times \text{d.o.f.}(\a) \\
    &~~~ ~-~ \text{d.o.f.}(\{ \un{a}_{1} \cdots \un{a}_{n-1}, \un{b}_{1} \cdots \un{b}_{n-1} \}) \times \text{d.o.f.}(\a) ~\Big)  \\
    ~=&~ \Big(~ \text{d.o.f.}(\{ \un{a}_{1} \cdots \un{a}_{n}, \un{b}_{1} \cdots \un{b}_{n} \}) ~-~ \text{d.o.f.}(\{ \un{a}_{1} \cdots \un{a}_{n}| \un{b}_{1} \cdots \un{b}_{n-1} \}) \\
    &~~~ ~+~ \text{d.o.f.}(\{ \un{a}_{1} \cdots \un{a}_{n-1}, \un{b}_{1} \cdots \un{b}_{n-1} \}) ~\Big) \times \text{d.o.f.}(\a)  ~~~.
\end{split} 
\label{eqn:SYTtwoeqcolIndex}
\end{equation}
which corresponds to Equation (\ref{eqn:SYTtwoeqcol}). 
For $n=1$, this formula reduces to the dimension formula for the $n=2$ case of 
the totally symmetric BYT attached with a $\CMTred{\ytableaushort{\sixteen}}$ in Equation (\ref{eqn:SYTsymIndex}).

\subsection{Two-unequal-column BYTs attached with $\CMTred{\{16\}}$}
\label{subsec:SYTtwouneqcol}

Last but not least, one may wonder about the dimensions of the spinorial irreps
represented by two-column BYTs with different number of boxes in the two columns
attached with a $\CMTred{\ytableaushort{\sixteen}}$, 
\begin{equation}
\begin{split}
    \dim \left(~ {\CMTB{\ytableaushort{{}{},\vdots\vdots,{}{},\vdots,{}}}\CMTred{\ydiagram{1,0,0,0,0}}}_{\rm IR} ~\right) ~=&~ \left(~  \dim \left(~ {\CMTB{\ytableaushort{{}{},\vdots\vdots,{}{},\vdots,{}}}}_{\rm IR} ~\right)  ~-~ \dim \left(~ {\CMTB{\ytableaushort{{}{},\vdots\vdots,{},\vdots,{}}}}_{\rm IR} ~\right) \color{white}{}\right)\color{black}{} \\
    & \color{white}{}\left(\color{black}{} ~-~ \dim \left(~ {\CMTB{\ytableaushort{{}{},\vdots\vdots,{}{},\vdots}}}_{\rm IR} ~\right) ~+~ \dim \left(~ {\CMTB{\ytableaushort{{}{},\vdots\vdots,{},\vdots}}}_{\rm IR} ~\right) ~\right) \color{black}{}  ~\times~ \dim \Big(~ \CMTred{\ydiagram{1}} ~\Big) ~~~~~.
\end{split}
\label{eqn:SYTtwouneqcol}
\end{equation}
Now let us give the general expression for the fields and the general set of irreducible
conditions. There are two equivalent ways to write the set of irreducible conditions.
\begin{equation}
\begin{split}
\Psi_{\{ \un{a}_{1} \cdots \un{a}_{n} | \un{b}_{1} \cdots \un{b}_{m} \}}{}^{\a}  ~~ (m>n)
~~:~~~~~~~ &
\begin{cases}
(\s^{\un{b}_{m}})_{\a\b} \Psi_{\{ \un{a}_{1} \cdots \un{a}_{n}| \un{b}_{1} \cdots \un{b}_{m}
\}}{}^{ \a } ~\equiv~ \psi_{\{ \un{a}_{1} \cdots \un{a}_{n}| \un{b}_{1} \cdots \un{b}_{m-1} \}\b}
~=~ 0  ~~~, \\
(\s^{\un{a}_{n}})_{\a\b} \Psi_{\{ \un{a}_{1} \cdots \un{a}_{n}| \un{b}_{1} \cdots \un{b}_{m}
\}}{}^{ \a } ~=~ 0  ~~~, \\
(\s^{\un{a}_{n}})^{\g\b} \psi_{\{ \un{a}_{1} \cdots \un{a}_{n}| \un{b}_{1} \cdots \un{b}_{m-1}
\} \b } ~=~ 0  ~~~,
\end{cases} \\
& ~~~~~~~ \text{or} \\
& \begin{cases}
(\s^{\un{b}_{m}})_{\a\b} \Psi_{\{ \un{a}_{1} \cdots \un{a}_{n}| \un{b}_{1} \cdots \un{b}_{m}
\}}{}^{ \a } ~=~ 0  ~~~, \\
(\s^{\un{a}_{n}})_{\a\b} \Psi_{\{ \un{a}_{1} \cdots \un{a}_{n}| \un{b}_{1} \cdots \un{b}_{m}
\}}{}^{ \a } ~\equiv~ \psi_{\{ \un{a}_{1} \cdots \un{a}_{n-1}| \un{b}_{1} \cdots \un{b}_{m} \}
\b} ~=~ 0  ~~~, \\
(\s^{\un{b}_{m}})^{\g\b} \psi_{\{ \un{a}_{1} \cdots \un{a}_{n-1}| \un{b}_{1} \cdots \un{b}_{m}
\} \b } ~=~ 0  ~~~,
\end{cases}
\end{split}
\end{equation}
where $m = 2, \dots, 5$ and $n = 1, \dots, m-1$. The general dimension formula is
\begin{equation}
\begin{split}
\dim\left( \Psi_{\{ \un{a}_{1} \cdots \un{a}_{n}| \un{b}_{1} \cdots \un{b}_{m} \}}{}^{\a} \right)
~=&~ \text{d.o.f.}(\{ \un{a}_{1} \cdots \un{a}_{n}| \un{b}_{1} \cdots \un{b}_{m} \}) \times
\text{d.o.f.}(\a)  \\
& ~-~ \Big(~ \text{d.o.f.}(\{ \un{a}_{1} \cdots \un{a}_{n}| \un{b}_{1} \cdots \un{b}_{m-1} \})
\times \text{d.o.f.}(\a) \\
&~~~ ~-~ \text{d.o.f.}(\{ \un{a}_{1} \cdots \un{a}_{n-1}| \un{b}_{1} \cdots \un{b}_{m-1} \})
\times \text{d.o.f.}(\a) ~\Big)  \\
& ~-~ \text{d.o.f.}(\{ \un{a}_{1} \cdots \un{a}_{n-1}| \un{b}_{1} \cdots \un{b}_{m} \}) \times
\text{d.o.f.}(\a) \\
~=&~ \text{d.o.f.}(\{ \un{a}_{1} \cdots \un{a}_{n}| \un{b}_{1} \cdots \un{b}_{m} \}) \times
\text{d.o.f.}(\a)  \\
& ~-~ \Big(~ \text{d.o.f.}(\{ \un{a}_{1} \cdots \un{a}_{n-1}| \un{b}_{1} \cdots \un{b}_{m} \})
\times \text{d.o.f.}(\a) \\
&~~~ ~-~ \text{d.o.f.}(\{ \un{a}_{1} \cdots \un{a}_{n-1}| \un{b}_{1} \cdots \un{b}_{m-1} \})
\times \text{d.o.f.}(\a) ~\Big)  \\
& ~-~ \text{d.o.f.}(\{ \un{a}_{1} \cdots \un{a}_{n}| \un{b}_{1} \cdots \un{b}_{m-1} \}) \times
\text{d.o.f.}(\a) \\
~=&~ \Big(~ \text{d.o.f.}(\{ \un{a}_{1} \cdots \un{a}_{n}| \un{b}_{1} \cdots \un{b}_{m} \})
~-~ \text{d.o.f.}(\{ \un{a}_{1} \cdots \un{a}_{n}| \un{b}_{1} \cdots \un{b}_{m-1} \}) \\
&~~~ ~-~ \text{d.o.f.}(\{ \un{a}_{1} \cdots \un{a}_{n-1}| \un{b}_{1} \cdots \un{b}_{m} \}) \\
&~~~ ~+~ \text{d.o.f.}(\{ \un{a}_{1} \cdots \un{a}_{n-1}| \un{b}_{1} \cdots \un{b}_{m-1}
\}) ~\Big) \times \text{d.o.f.}(\a)  ~~~,
\end{split} \label{eqn:SYTtwouneqcolIndex}
\end{equation}
\sloppy
and one can check that it agrees with Equation (\ref{eqn:SYTtwouneqcol}). 
When one compares it to the dimension formula for two-equal-column BYT attached with 
a $\CMTred{\ytableaushort{\sixteen}}$ in Equation (\ref{eqn:SYTtwoeqcolIndex}), 
one quickly understands the relation between that formula
and this dimension formula (\ref{eqn:SYTtwouneqcolIndex}). 
Since for the case of two-equal-column mixed YT, $m=n$ and 
$\psi_{\{ \un{a}_{1} \cdots \un{a}_{n}| \un{b}_{1} \cdots \un{b}_{m-1} \} \b}
= \psi_{\{ \un{a}_{1} \cdots \un{a}_{n-1}| \un{b}_{1} \cdots \un{b}_{m} \} \b}$, 
so it was not counted twice and there was only one term 
$\text{d.o.f.}(\{ \un{a}_{1} \cdots \un{a}_{n}| \un{b}_{1} \cdots \un{b}_{n-1} \})$
in Equation (\ref{eqn:SYTtwoeqcolIndex}). For Equation (\ref{eqn:SYTtwouneqcolIndex}),
however, we have both $\text{d.o.f.}(\{ \un{a}_{1} \cdots \un{a}_{n}| \un{b}_{1} \cdots \un{b}_{m-1} \})$
and $\text{d.o.f.}(\{ \un{a}_{1} \cdots \un{a}_{n-1}| \un{b}_{1} \cdots \un{b}_{m} \})$ terms. 
This exactly reflects the meaning of ``/'' operator of BYT - that equivalent removals of one box will be only counted once, as indicated in Equation (\ref{eqn:quotientequiv}).

\newpage
\section{Tensor Product Rules of a Bosonic Irrep with the Basic Spinorial Irrep}
\label{sec:king-tensor}

For a $D_n$ algebra, the tensor product rule of a general BYT $[\lambda]$ with the basic spinorial irrep $\Delta = \CMTred{\{16\}}$ was first given by Murnaghan and Littlewood in 1938 and 1950 respectively \cite{murnaghan1938,littlewood1950b},
\begin{equation}
    [\lambda] \otimes \Delta ~=~ [\Delta;\lambda/Q] ~~~,
\label{eqn:SYTtensorrule}
\end{equation}
where $Q$ is a Schur function series defined by
\begin{equation}
    Q ~=~ \sum_{n} \color{white}{} \left\{ \color{black}{} \hspace{-0.5em} {\CMTB{\ytableaushort{{},\vdots,{}}}}_{\rm IR} \right\} \color{black}{} n ~~~.
\label{eqn:SchurQ}
\end{equation}
It turns out that the spinorial irrep dimension rule in Equation (\ref{eqn:kingrule}) was derived from this tensor product rule. In \cite{king1975}, King found the ``inverse'' of the Schur function series $Q$ to be $P$, as explicitly defined in Equation (\ref{eqn:schurP}). 

Following the footsteps of Section \ref{subsec:SYTdimrule}, we translate the tensor product rule to our notation.
\begin{equation}
\begin{split}
    & \generalBYTpqrst_{{\rm IR},+} ~\otimes~ \CMTred{\ydiagram{1}} \\
    &=~ \bigoplus_{n=0}^{h} \left( \left(\, h \left\{\generalBYTpqrst_{{\rm IR},+} \color{white}{} \right\} \color{black}{} \hspace{-1em} \Bigg/ \, \color{white}{} \left\{ \color{black}{} \hspace{-0.5em} {\CMTB{\ytableaushort{{},\vdots,{}}}}_{\rm IR} \right\} \color{black}{} n \,\right) \CMTred{\ytableaushort{\tinysixteenaltbar,\none,\none,\none,\none}} \right)  ~~~~~,
\end{split} \label{eqn:SYTtensorgeneral}
\end{equation}
where $\CMTred{\ytableaushort{\tinysixteenaltbar}}$ is attached to the BYT after the quotient operation; and 
\begin{equation}
    h ~=~ \begin{cases}
    ~ 5 ~~~, & \text{for } t \neq 0  \\
    ~ 4 ~~~, & \text{for } t = 0 \text{ and } s \neq 0 \\
    ~ 3 ~~~, & \text{for } t = s = 0 \text{ and } r \neq 0 \\
    ~ 2 ~~~, & \text{for } t = s = r = 0 \text{ and } q \neq 0 \\
    ~ 1 ~~~, & \text{for } t = s = r = q = 0 \text{ and } p \neq 0 \\
    ~ 0 ~~~, & \text{for } t = s = r = q = p = 0 \\
    \end{cases} ~~~~~,
\end{equation}
is the height of the BYT.
The tensor product rule is very similar to the SYT dimension rule in terms of combinatorics, and it's even simpler as we do not have the alternating signs in the final sum - we just take the direct sum of all possible objects we obtain. 

Let us show an example to illustrate the idea,
\begin{equation}
    {\CMTB{\ydiagram{4,3,1}}}_{\rm IR} ~\otimes~ \CMTred{\ydiagram{1}} ~~~~~.
\end{equation}
Since we have three rows of different numbers of boxes, we consider the three rows to be different. So $n$ ranges from 0 to $h=3$, and what we do is to remove $n$ objects from the following 3 objects,
\begin{equation}
    \CMTB{\ytableaushort{L,M,N}} ~~~.
\end{equation}
As these objects are all distinguishable, the combinatorics become very simple. The numbers of combinations are just the binomial coefficients $\begin{pmatrix}3\\n\end{pmatrix}$. If we look at the BYTs in the right hand side of Equation (\ref{eqn:SYTtensorgeneral}) without attaching the spinorial YT, term by term, we have
\begin{equation}
\begin{split}
    n=0 ~:&~~~ {\CMTB{\ytableaushort{{}{}{}{},{}{}{},{}}}}_{\rm IR} ~~~, \\
    n=1 ~:&~~~ {\CMTB{\ytableaushort{{}{}{}{\none[L]},{}{}{},{}}}}_{\rm IR} ~~~,~~~ {\CMTB{\ytableaushort{{}{}{}{},{}{}{\none[M]},{}}}}_{\rm IR} ~~~,~~~ {\CMTB{\ytableaushort{{}{}{}{},{}{}{},{\none[N]}}}}_{\rm IR} ~~~, \\
    n=2 ~:&~~~ {\CMTB{\ytableaushort{{}{}{}{\none[L]},{}{}{\none[M]},{}}}}_{\rm IR} ~~~,~~~ {\CMTB{\ytableaushort{{}{}{}{\none[L]},{}{}{},{\none[N]}}}}_{\rm IR} ~~~,~~~ {\CMTB{\ytableaushort{{}{}{}{},{}{}{\none[M]},{\none[N]}}}}_{\rm IR} ~~~, \\
    n=3 ~:&~~~ {\CMTB{\ytableaushort{{}{}{}{\none[L]},{}{}{\none[M]},{\none[N]}}}}_{\rm IR} ~~~.
\end{split}
\end{equation}
Therefore, 
\begin{equation}
\begin{split}
    {\CMTB{\ydiagram{4,3,1}}}_{\rm IR} ~\otimes~ \CMTred{\ydiagram{1}}
    ~~=&~~~~~\; {\CMTB{\ydiagram{4,3,1}}\CMTred{\ytableaushort{\sixteen,\none,\none}}}_{\rm IR} \\
    & ~\oplus~ {\CMTB{\ydiagram{3,3,1}}\CMTred{\ytableaushort{\sixteenbar,\none,\none}}}_{\rm IR} ~\oplus~ {\CMTB{\ydiagram{4,2,1}}\CMTred{\ytableaushort{\sixteenbar,\none,\none}}}_{\rm IR} ~\oplus~ {\CMTB{\ydiagram{4,3}}\CMTred{\ytableaushort{\sixteenbar,\none}}}_{\rm IR}  \\
    & ~\oplus~ {\CMTB{\ydiagram{3,2,1}}\CMTred{\ytableaushort{\sixteen,\none,\none}}}_{\rm IR} ~\oplus~ {\CMTB{\ydiagram{3,3}}\CMTred{\ytableaushort{\sixteen,\none}}}_{\rm IR} ~\oplus~ {\CMTB{\ydiagram{4,2}}\CMTred{\ytableaushort{\sixteen,\none}}}_{\rm IR} \\
    & ~\oplus~ {\CMTB{\ydiagram{3,2}}\CMTred{\ytableaushort{\sixteenbar,\none}}}_{\rm IR} ~~~~~.
\end{split} \label{eqn:SYTtensorExResult}
\end{equation}
Or in dimension notation, 
\begin{equation}
\begin{split}
    \CMTB{\{174636\}} ~\otimes~ \CMTred{\{16\}} ~=&~~~~~\; \CMTred{\{1260000\}} \\
    & ~\oplus~ \CMTred{\{258720\}} ~\oplus~ \CMTred{\{529200\}} ~\oplus~ \CMTred{\{332640\}} \\
    & ~\oplus~ \CMTred{\{144144\}} ~\oplus~ \CMTred{\{70560\}} ~\oplus~ \CMTred{\{155232\}} \\
    & ~\oplus~ \CMTred{\{43680\}} ~~~.
\end{split}
\end{equation}

To see how Schur function series $P$ and $Q$ as defined in Equations (\ref{eqn:schurP}) and (\ref{eqn:SchurQ}) are inverses of each other, let us use the above example to demonstrate the consistency of the SYT dimension rule (\ref{eqn:kingrule}) and the tensor product formula (\ref{eqn:SYTtensorrule}), i.e.
\begin{equation}
    \dim \left(~ {\CMTB{\ydiagram{4,3,1}}}_{\rm IR} ~\otimes~ \CMTred{\ydiagram{1}} ~\right) ~=~ \dim \left(~ {\CMTB{\ydiagram{4,3,1}}}_{\rm IR} ~\right) ~\times~ \dim \left(~ \CMTred{\ydiagram{1}} ~\right) ~~~~~.
\label{eqn:PQconsistency}
\end{equation}
In Equation (\ref{eqn:SYTdimExResult}), the dimension formula for ${\CMTB{\ydiagram{4,3,1}}\CMTred{\ydiagram{1,0,0}}}_{\rm IR}$ was shown. Next, we could write down the dimension formulas for each term on the right hand side of (\ref{eqn:SYTtensorExResult}). Note that the conjugation of an irrep would not change its dimension, so the same dimension formula applies. The dimension formula can be written in the notation (with even more details left out)
\begin{equation}
    \dim ([\Delta; \lambda]) ~=~ \Big(~ \sum_{i} \, b_{i} \, \dim(\lambda_{i}) ~\Big) ~\times~ \dim (\Delta) ~~~~~.
\label{eqn:dimsimple}
\end{equation}
where $b_{i}$ can take the values $+1$, $-1$ or $0$. 
Then we can conveniently put all the relevant dimension formulas in Table \ref{tab:PQinverse}. 
\begin{table}[h!]
\centering
\setlength{\tabcolsep}{5.3pt} 
\ytableausetup{boxsize=0.35em}
\begin{tabular}{|c|*{19}{c}|} \hline
    & \CMTB{\ydiagram{4,3,1}} & \CMTB{\ydiagram{3,3,1}} & \CMTB{\ydiagram{4,2,1}} & \CMTB{\ydiagram{4,3}} & \CMTB{\ydiagram{3,2,1}} & \CMTB{\ydiagram{3,3}} & \CMTB{\ydiagram{4,1,1}} & \CMTB{\ydiagram{4,2}} & \CMTB{\ydiagram{2,2,1}} & \CMTB{\ydiagram{3,1,1}} & \CMTB{\ydiagram{3,2}} & \CMTB{\ydiagram{4,1}} & \CMTB{\ydiagram{2,1,1}} & \CMTB{\ydiagram{2,2}} & \CMTB{\ydiagram{3,1}} & \CMTB{\ydiagram{4}} & \CMTB{\ydiagram{2,1}} & \CMTB{\ydiagram{3}} & \CMTB{\ydiagram{2}}  \\ \hline
    \CMTB{\ydiagram{4,3,1}}\CMTred{\ydiagram{1,0,0}} & 1 & -1 & -1 & -1 & 1 & 1 & 1 & 1 & 0 & -1 & -1 & -1 & 0 & 0 & 1 & 0 & 0 & 0 & 0 \\
    \CMTB{\ydiagram{3,3,1}}\CMTred{\ydiagram{1,0,0}} & 0 & 1 & 0 & 0 & -1 & -1 & 0 & 0 & 0 & 1 & 1 & 0 & 0 & 0 & -1 & 0 & 0 & 0 & 0 \\
    \CMTB{\ydiagram{4,2,1}}\CMTred{\ydiagram{1,0,0}} & 0 & 0 & 1 & 0 & -1 & 0 & -1 & -1 & 1 & 1 & 1 & 1 & -1 & -1 & -1 & 0 & 1 & 0 & 0 \\
    \CMTB{\ydiagram{4,3}}\CMTred{\ydiagram{1,0}} & 0 & 0 & 0 & 1 & 0 & -1 & 0 & -1 & 0 & 0 & 1 & 1 & 0 & 0 & -1 & -1 & 0 & 1 & 0 \\
    \CMTB{\ydiagram{3,2,1}}\CMTred{\ydiagram{1,0,0}} & 0 & 0 & 0 & 0 & 1 & 0 & 0 & 0 & -1 & -1 & -1 & 0 & 1 & 1 & 1 & 0 & -1 & 0 & 0 \\
    \CMTB{\ydiagram{3,3}}\CMTred{\ydiagram{1,0}} & 0 & 0 & 0 & 0 & 0 & 1 & 0 & 0 & 0 & 0 & -1 & 0 & 0 & 0 & 1 & 0 & 0 & -1 & 0 \\
    \CMTB{\ydiagram{4,2}}\CMTred{\ydiagram{1,0}} & 0 & 0 & 0 & 0 & 0 & 0 & 0 & 1 & 0 & 0 & -1 & -1 & 0 & 1 & 1 & 1 & -1 & -1 & 1 \\
    \CMTB{\ydiagram{3,2}}\CMTred{\ydiagram{1,0}} & 0 & 0 & 0 & 0 & 0 & 0 & 0 & 0 & 0 & 0 & 1 & 0 & 0 & -1 & -1 & 0 & 1 & 1 & -1 \\ \hline
    \CMTB{\ydiagram{4,3,1}}{\scriptsize{$\otimes$\,}}\CMTred{\ydiagram{1}} & 1 & 0 & 0 & 0 & 0 & 0 & 0 & 0 & 0 & 0 & 0 & 0 & 0 & 0 & 0 & 0 & 0 & 0 & 0 \\ \hline
\end{tabular}
\ytableausetup{boxsize=1.2em}
\caption{Consistency check of the tensor product formula with the dimension formula and the demonstration of $P$ as an inverse of $Q$}
\label{tab:PQinverse}
\end{table}

\noindent
The first column contains all the terms on the right hand side of (\ref{eqn:SYTtensorExResult}); and the first row contains all possible $\lambda_{i}$'s in Equation (\ref{eqn:dimsimple}) generated by them. The numbers in the middle of the table are $b_{i}$'s. Each row of $b_{i}$'s represents all the terms in the dimension formula for that irrep $[\Delta;\lambda]$ in the first entry of the row.
By the tensor product results in Equation (\ref{eqn:SYTtensorExResult}), adding up all the dimensions of the terms on the right hand side gives us the dimension of the tensor product. This operation corresponds to adding up each column of $b_{i}$'s in the middle of the table, which gives us the last row of the table - the dimension for the tensor product. As we can see, only the leading term remains, and all other terms sum up to zero, i.e. they cancel with each other. Therefore, we verify Equation (\ref{eqn:PQconsistency}). 

Moreover, we easily see that when summing up the terms in the Schur function series $Q$, the alternating signs generated by $P$ naturally allow non-leading terms cancel with each other. It is in the sense that the Schur function series $P$ is an inverse of the Schur function series $Q$.

Some more examples are shown in Appendix \ref{appen:spinorial-tensorproduct}, where we start from tensor product rules and then derive SYT dimension formulas. This approach is the converse of what we have just done.

\newpage
\section{Complete Descriptions of Component Fields in 10D, $\mathcal{N}=1$ Scalar Superfield\label{sec:comp-descrip}}

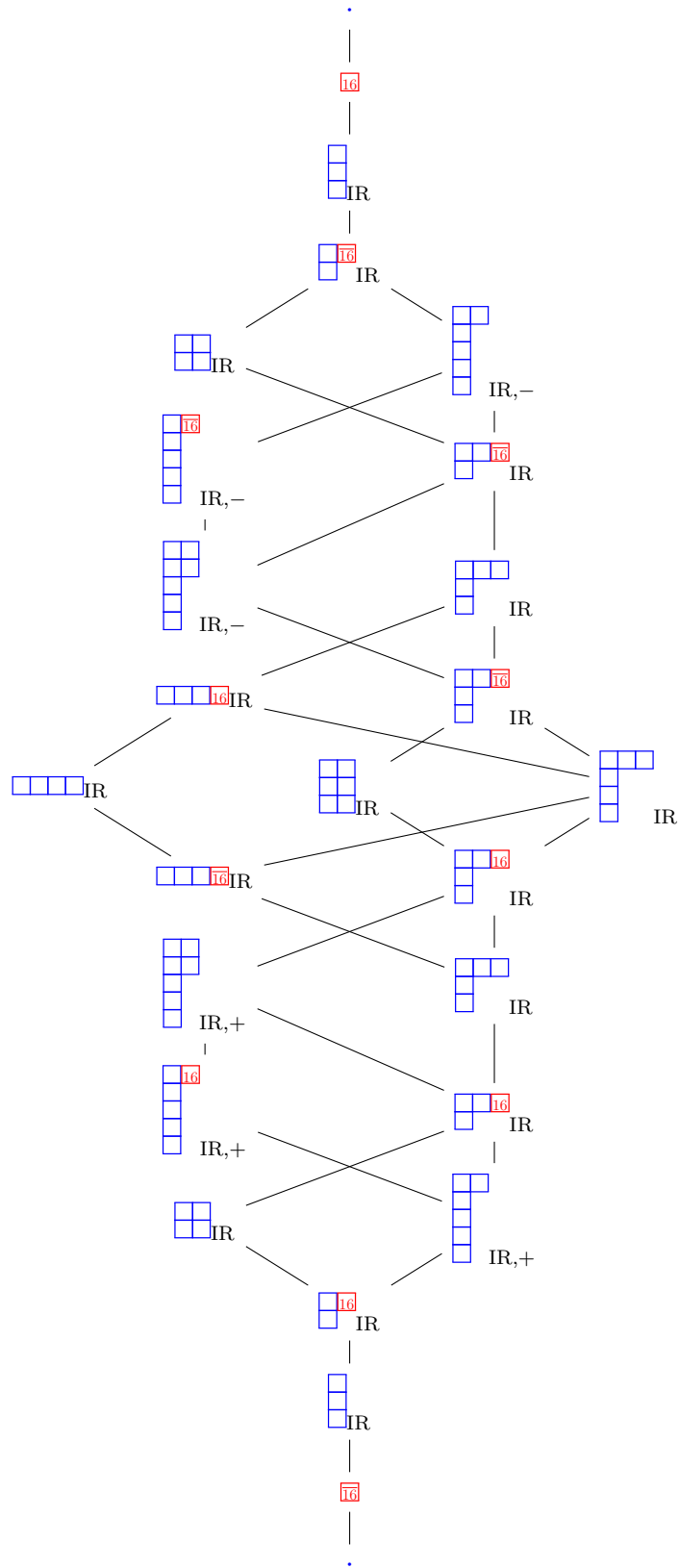
\begin{figure}[h!]
\centering
\ytableausetup{boxsize=0.55em}
\begin{tikzpicture}[state/.style={rectangle}]
\node[state] (0) at (0,-10.75) {$\CMTB{\ytableaushort{{\none[\cdot]}}}$};
\node[state] (1) at (0,-9.75) {$\CMTred{\ytableaushort{\indexsixteenbar}}$};
\node[state] (2) at (0,-8.5) {${\CMTB{\ydiagram{1,1,1}}}_{\rm IR}$};
\node[state] (3) at (0,-7.25) {${\CMTB{\ydiagram{1,1}}\CMTred{\ytableaushort{\indexsixteen,\none}}}_{\rm IR}$};
\node[state] (4a) at (-2,-6) {${\CMTB{\ydiagram{2,2}}}_{\rm IR}$};
\node[state] (4b) at (2,-6) {${\CMTB{\ydiagram{2,1,1,1,1}}}_{{\rm IR},+}$};
\node[state] (5a) at (-2,-4.5) {${\CMTB{\ydiagram{1,1,1,1,1}}\CMTred{\ytableaushort{\indexsixteen,\none,\none,\none,\none}}}_{{\rm IR},+}$};
\node[state] (5b) at (2,-4.5) {${\CMTB{\ydiagram{2,1}}\CMTred{\ytableaushort{\indexsixteen,\none}}}_{\rm IR}$};
\node[state] (6a) at (-2,-2.75) {${\CMTB{\ydiagram{2,2,1,1,1}}}_{{\rm IR},+}$};
\node[state] (6b) at (2,-2.75) {${\CMTB{\ydiagram{3,1,1}}}_{\rm IR}$};
\node[state] (7a) at (-2,-1.25) {${\CMTB{\ydiagram{3}}\CMTred{\ytableaushort{\indexsixteenbar}}}_{\rm IR}$};
\node[state] (7b) at (2,-1.25) {${\CMTB{\ydiagram{2,1,1}}\CMTred{\ytableaushort{\indexsixteen,\none,\none}}}_{\rm IR}$};
\node[state] (8a) at (-4,0) {${\CMTB{\ydiagram{4}}}_{\rm IR}$};
\node[state] (8b) at (0,0) {${\CMTB{\ydiagram{2,2,2}}}_{\rm IR}$};
\node[state] (8c) at (4,0) {${\CMTB{\ydiagram{3,1,1,1}}}_{\rm IR}$};
\node[state] (9a) at (-2,1.25) {${\CMTB{\ydiagram{3}}\CMTred{\ytableaushort{\indexsixteen}}}_{\rm IR}$};
\node[state] (9b) at (2,1.25) {${\CMTB{\ydiagram{2,1,1}}\CMTred{\ytableaushort{\indexsixteenbar,\none,\none}}}_{\rm IR}$};
\node[state] (10a) at (-2,2.75) {${\CMTB{\ydiagram{2,2,1,1,1}}}_{{\rm IR},-}$};
\node[state] (10b) at (2,2.75) {${\CMTB{\ydiagram{3,1,1}}}_{\rm IR}$};
\node[state] (11a) at (-2,4.5) {${\CMTB{\ydiagram{1,1,1,1,1}}\CMTred{\ytableaushort{\indexsixteenbar,\none,\none,\none,\none}}}_{{\rm IR},-}$};
\node[state] (11b) at (2,4.5) {${\CMTB{\ydiagram{2,1}}\CMTred{\ytableaushort{\indexsixteenbar,\none}}}_{\rm IR}$};
\node[state] (12a) at (-2,6) {${\CMTB{\ydiagram{2,2}}}_{\rm IR}$};
\node[state] (12b) at (2,6) {${\CMTB{\ydiagram{2,1,1,1,1}}}_{{\rm IR},-}$};
\node[state] (13) at (0,7.25) {${\CMTB{\ydiagram{1,1}}\CMTred{\ytableaushort{\indexsixteenbar,\none}}}_{\rm IR}$};
\node[state] (14) at (0,8.5) {${\CMTB{\ydiagram{1,1,1}}}_{\rm IR}$};
\node[state] (15) at (0,9.75) {$\CMTred{\ytableaushort{\indexsixteen}}$};
\node[state] (16) at (0,10.75) {$\CMTB{\ytableaushort{{\none[\cdot]}}}$};
\path [-](0) edge (1);
\path [-](1) edge (2);
\path [-](2) edge (3);
\path [-](3) edge (4a);
\path [-](3) edge (4b); 
\path [-](4a) edge (5b); 
\path [-](4b) edge (5a); 
\path [-](4b) edge (5b); 
\path [-](5a) edge (6a); 
\path [-](5b) edge (6a); 
\path [-](5b) edge (6b); 
\path [-](6a) edge (7b); 
\path [-](6b) edge (7a); 
\path [-](6b) edge (7b); 
\path [-](7a) edge (8a);
\path [-](7a) edge (8c);
\path [-](7b) edge (8b);
\path [-](7b) edge (8c);
\path [-](8a) edge (9a);
\path [-](8b) edge (9b);
\path [-](8c) edge (9a);
\path [-](8c) edge (9b);
\path [-](9a) edge (10b); 
\path [-](9b) edge (10a); 
\path [-](9b) edge (10b);
\path [-](10a) edge (11a); 
\path [-](10a) edge (11b); 
\path [-](10b) edge (11b); 
\path [-](11a) edge (12b); 
\path [-](11b) edge (12a); 
\path [-](11b) edge (12b);
\path [-](12a) edge (13); 
\path [-](12b) edge (13); 
\path [-](13) edge (14);
\path [-](14) edge (15); 
\path [-](15) edge (16); 
\end{tikzpicture}
\caption{10D, ${\cal N} = 1$ Adinkra in Young Tableaux}
\label{fig:adinkraYT}
\end{figure}

In this chapter, we present the complete descriptions of all component fields that occur in the 
10D, $\mathcal{N}=1$ scalar superfield decomposition. In \cite{HyDSG3} (and Section 
\ref{sec:adynkra} above) we presented the Lorentz descriptions of all the component fields. 
By using the techniques discussed in the previous sections, we can translate the Dynkin Labels of the component fields in Figure \ref{Fig:10DTypeI_Dynkin} to Young Tableaux. 
\begin{equation}
\ytableausetup{boxsize=0.8em}
{\cal V} ~=~ \begin{cases}
~~{\rm {Level}}-0 \,~~~~~~~~~~ \CMTB{\cdot} ~~~,  \\
~~{\rm {Level}}-1 \,~~~~~~~~~~ \CMTred{\ytableaushort{\tinysixteenbar}} ~~~,  \\
~~{\rm {Level}}-2 \,~~~~~~~~~~ {\CMTB{\ydiagram{1,1,1}}}_{\rm IR} ~~~,  \\[10pt]
~~{\rm {Level}}-3 \,~~~~~~~~~~ {\CMTB{\ydiagram{1,1}}\CMTred{\ytableaushort{\tinysixteen,\none}}}_{\rm IR}  ~~~,  \\
~~{\rm {Level}}-4 \,~~~~~~~~~~ {\CMTB{\ydiagram{2,2}}}_{\rm IR} ~\oplus~ {\CMTB{
\ydiagram{2,1,1,1,1}}}_{{\rm IR},+} ~~~,  \\
~~{\rm {Level}}-5 \,~~~~~~~~~~ {\CMTB{\ydiagram{1,1,1,1,1}}\CMTred{\ytableaushort{\tinysixteen,\none,\none,\none,\none}}}_{{\rm IR},+} ~\oplus~ {\CMTB{\ydiagram{2,1}}\CMTred{\ytableaushort{\tinysixteen,\none}}}_{\rm IR} ~~~,  \\[20pt]
~~{\rm {Level}}-6 \,~~~~~~~~~~ {\CMTB{\ydiagram{2,2,1,1,1}}}_{{\rm IR},+}  ~\oplus~ {\CMTB{\ydiagram{3,1,1}}}_{\rm IR} ~~~,  \\
~~{\rm {Level}}-7 \,~~~~~~~~~~ {\CMTB{\ydiagram{3}}\CMTred{\ytableaushort{\tinysixteenbar}}}_{\rm IR} ~\oplus
~ {\CMTB{\ydiagram{2,1,1}}\CMTred{\ytableaushort{\tinysixteen,\none,\none}}}_{\rm IR}  ~~~,  \\
~~{\rm {Level}}-8 \,~~~~~~~~~~ {\CMTB{\ydiagram{4}}}_{\rm IR} ~\oplus~ {\CMTB{\ydiagram{2,2,2}}}_{
\rm IR} ~\oplus~ {\CMTB{\ydiagram{3,1,1,1}}}_{\rm IR} ~~~,  \\
~~{\rm {Level}}-9 \,~~~~~~~~~~ {\CMTB{\ydiagram{3}}\CMTred{\ytableaushort{\tinysixteen}}}_{\rm IR} ~\oplus~ 
{\CMTB{\ydiagram{2,1,1}}\CMTred{\ytableaushort{\tinysixteenbar,\none,\none}}}_{\rm IR}  ~~~,  \\
~~{\rm {Level}}-10 ~~~~~~~~~ {\CMTB{\ydiagram{2,2,1,1,1}}}_{{\rm IR},-}  ~\oplus~ {\CMTB{\ydiagram{3,1,1}
}}_{\rm IR} ~~~,  \\[20pt]
~~{\rm {Level}}-11 ~~~~~~~~~ {\CMTB{\ydiagram{1,1,1,1,1}}\CMTred{\ytableaushort{\tinysixteenbar,\none,\none,\none,\none}}}_{{\rm IR},-} 
~\oplus~ {\CMTB{\ydiagram{2,1}}\CMTred{\ytableaushort{\tinysixteenbar,\none}}}_{\rm IR} ~~~,  \\
~~{\rm {Level}}-12 ~~~~~~~~~ {\CMTB{\ydiagram{2,2}}}_{\rm IR} ~\oplus~ {\CMTB{\ydiagram{2,1,1,1,1}}
}_{{\rm IR},-} ~~~,  \\
~~{\rm {Level}}-13 ~~~~~~~~~ {\CMTB{\ydiagram{1,1}}\CMTred{\ytableaushort{\tinysixteenbar,\none}}}_{\rm IR}  ~~~,  \\[5pt]
~~{\rm {Level}}-14 ~~~~~~~~~ {\CMTB{\ydiagram{1,1,1}}}_{\rm IR} ~~~,  \\
~~{\rm {Level}}-15 ~~~~~~~~~ \CMTred{\ytableaushort{\tinysixteen}} ~~~,  \\
~~{\rm {Level}}-16 ~~~~~~~~~ \CMTB{\cdot} ~~~.
\end{cases}
\label{equ:G}
\ytableausetup{boxsize=1.2em}
\end{equation}
Or we can directly translate the whole adynkra into Figure \ref{fig:adinkraYT}.

Given the data above, we can define a series ${\Tilde {\cal G}}(\ell)$  dependent on a real parameter 
``$\ell$'' via the equation
\be \ytableausetup{boxsize=0.8em}
\eqalign{
{\Tilde {\cal G}} &=  \CMTB{\cdot} \, \oplus \, \ell \,  \CMTred{\ytableaushort{\tinysixteen}}
\, \oplus \, \fracm 1{2} \,(\ell){}^{2} \, {\CMTB{\ydiagram{1,1,1}}}_{\rm IR} \, \oplus \,  \fracm 1{3!} \,(\ell){}^{3} \, 
{\CMTB{\ydiagram{1,1}}\CMTred{\ytableaushort{\tinysixteenbar,\none}}}_{\rm IR}
\, \oplus \,  \fracm 1{4!} \,(\ell){}^{4} \,  {\CMTB{\ydiagram{2,2}}}_{\rm IR} \, \oplus \,  \fracm 1{4!} \, (\ell){}^{4} \,  
{\CMTB{\ydiagram{2,1,1,1,1}}}_{{\rm IR},-}   \cr
& ~~~\, \oplus \, \fracm 1{5!} \, (\ell){}^{5} \,  {\CMTB{\ydiagram{1,1,1,1,1}}\CMTred{\ytableaushort{\tinysixteenbar,\none,\none,\none,\none}}}_{{\rm IR},-} 
\, \oplus \,  \fracm 1{5!} \, (\ell){}^{5} \,  {\CMTB{\ydiagram{2,1}}\CMTred{\ytableaushort{\tinysixteenbar,\none}}}_{\rm IR}  
\, \oplus \, \fracm 1{6!} \,  (\ell){}^{6} \,   {\CMTB{\ydiagram{2,2,1,1,1}}}_{{\rm IR},-}  \, \oplus \, \fracm 1{6!} \, ({\rm 
h}){}^{6} \,  {\CMTB{\ydiagram{3,1,1}}}_{\rm IR}    \cr
& ~~~\, \oplus \, \fracm 1{7!} \,  (\ell){}^{7} \,     {\CMTB{\ydiagram{3}}\CMTred{\ytableaushort{\tinysixteen}}}_{\rm IR} ~\oplus
~  \fracm 1{7!} \, (\ell){}^{7} \,  {\CMTB{\ydiagram{2,1,1}}\CMTred{\ytableaushort{\tinysixteenbar,\none,\none}}}_{\rm IR}       \cr
&  ~~~\, \oplus \, \fracm 1{8!} \, (\ell){}^{8} \,    {\CMTB{\ydiagram{4}}}_{\rm IR} \, \oplus \, \fracm 1{8!} \, (\ell){}^{8} \,  {\CMTB{\ydiagram{2,2,2}}}_{
\rm IR} \, \oplus \, \fracm 1{8!} \, (\ell){}^{8} \, {\CMTB{\ydiagram{3,1,1,1}}}_{\rm IR}        \cr
&  ~~~\, \oplus \, \fracm 1{9!} \, (\ell){}^{9} \,  {\CMTB{\ydiagram{3}}\CMTred{\ytableaushort{\tinysixteenbar}}}_{\rm IR} \, \oplus \, \fracm 1{9!} \, (\ell){}^{9} \, 
{\CMTB{\ydiagram{2,1,1}}\CMTred{\ytableaushort{\tinysixteen,\none,\none}}}_{\rm IR}         \cr
&  ~~~\, \oplus \, \fracm 1{10!} \, (\ell){}^{10} \,   {\CMTB{\ydiagram{2,2,1,1,1}}}_{{\rm IR},+}  \, \oplus \, \fracm 1{10!} \, (\ell){}^{10} \,  {\CMTB{\ydiagram{3,1,1}
}}_{\rm IR}  
\, \oplus \, \fracm 1{11!} \,  (\ell){}^{11} \,  {\CMTB{\ydiagram{1,1,1,1,1}}\CMTred{\ytableaushort{\tinysixteen,\none,\none,\none,\none}}}_{{\rm IR},+} 
\, \oplus \, \fracm 1{11!} \, (\ell){}^{11} \, 
{\CMTB{\ydiagram{2,1}}\CMTred{\ytableaushort{\tinysixteen,\none}}}_{\rm IR}         \cr
&  ~~~ \, \oplus \, \fracm 1{12!} \,  (\ell){}^{12} \,   {\CMTB{\ydiagram{2,2}}}_{\rm IR} \, \oplus \, \fracm 1{12!} \, (\ell){}^{12} \,  {\CMTB{\ydiagram{2,1,1,1,1}}
}_{{\rm IR},+}     \cr
& ~~~~ \, \oplus \,
\fracm 1{13!} \, (\ell){}^{13} \, {\CMTB{\ydiagram{1,1}}\CMTred{\ytableaushort{\tinysixteen,\none}}}_{\rm IR} 
\, \oplus \, \fracm 1{14!} \,  (\ell){}^{14} \, {\CMTB{\ydiagram{1,1,1}}}_{\rm IR} \, \oplus \,  \fracm 1{15!} \,  (\ell){}^{15} \,  \CMTred{\ytableaushort{\tinysixteenbar}} 
\, \oplus \, \fracm 1{16!}  \,(\ell){}^{16} \,  \CMTB{\cdot} ~~~~~.
} 
\ytableausetup{boxsize=1.2em}
\label{equ:G2}
\ee
We use the parameter ``$\ell$'' to track the level of the YT's that follow it, which will play an important role when we define the product of two superfields. Since the maximal number of the level is sixteen, the product of two YT's contributes to $({\ell})^n$ and when $n>16$ this piece will be ruled out. 
If we set ${\ell} = 1$, ${\Tilde {\cal G}}$ reduces to ${\cal G}$, which will be defined in (\ref{equ:calG}) in the next Chapter. They are alternate forms to each other. 

The index structures of all fifteen bosonic and twelve fermionic fields are identified below along 
with the level at which the fields occur in the adinkra of the superfield.
\begin{equation}
{\cal V} ~=~ \begin{cases}
~~{\rm {Level}}-0 \,~~~~~~~~~~ \Phi (x) ~~~,  \\
~~{\rm {Level}}-1 \,~~~~~~~~~~ \Psi_{\a} (x) ~~~,   \\
~~{\rm {Level}}-2 \,~~~~~~~~~~ \Phi_{\{ \aone \bone \cone \}} (x) ~~~,   \\
~~{\rm {Level}}-3 \,~~~~~~~~~~ \Psi_{\{ \aone \bone \}}{}^{\a} (x) ~~~,    \\
~~{\rm {Level}}-4 \,~~~~~~~~~~ \Phi_{\{ \aone \bone, \atwo \btwo \}} (x) ~~~,~~~ 
\Phi_{\{ \atwo | \aone \bone \cone \done \eone \}{}^{+}} (x) ~~~,   \\
~~{\rm {Level}}-5 \,~~~~~~~~~~ \Psi_{\{ \aone \bone \cone \done \eone \}{}^{+}}{}^{\a} (x) ~~~,~~~ 
\Psi_{\{ \atwo | \aone \bone \}}{}^{\a} (x) ~~~,   \\
~~{\rm {Level}}-6 \,~~~~~~~~~~ \Phi_{\{\atwo \btwo| \aone \bone \cone \done \eone \}{}^{+}} (x) ~~~,~~~ 
\Phi_{\{ \atwo , \athree | \aone \bone \cone \}} (x) ~~~,  \\
~~{\rm {Level}}-7 \,~~~~~~~~~~ \Psi_{\{ \aone, \atwo, \athree \}}{}_{\a} (x) ~~~,~~~ \Psi_{\{ \atwo | \aone 
\bone \cone \}}{}^{\a} (x) ~~~,   \\
~~{\rm {Level}}-8 \,~~~~~~~~~~ \Phi_{\{ \aone,  \atwo,  \athree, \afour \}} (x)  ~~~,~~~ \Phi_{\{ \aone \bone 
\cone , \atwo \btwo \ctwo \}} (x) ~~~,~~~ \Phi_{\{ \atwo, \athree | \aone \bone \cone \done \}} (x) ~~~,  \\
~~{\rm {Level}}-9 \,~~~~~~~~~~ \Psi_{\{ \aone , \atwo , \athree \}}{}^{\a}(x)  ~~~,~~~  \Psi_{\{ \atwo | \aone 
\bone \cone \}}{}_{\a}(x)  ~~~, \\
~~{\rm {Level}}-10 ~~~~~~~~~ \Phi_{\{ \atwo \btwo | \aone \bone \cone \done \eone \}{}^{-}}(x) ~~~,~~~  
{\widehat \Phi}_{\{ \atwo , \athree | \aone \bone \cone \}}(x)  ~~~, \\
~~{\rm {Level}}-11 ~~~~~~~~~ \Psi_{\{ \aone \bone \cone \done \eone \}{}^{-}}{}_{\a}(x)   ~~~,~~~  \Psi_{\{ 
\atwo | \aone \bone \}}{}_{\a}(x) ~~~, \\
~~{\rm {Level}}-12 ~~~~~~~~~ {\widehat \Phi}_{\{ \aone \bone , \atwo \btwo \}}(x)  ~~~,~~~  \Phi_{\{ \atwo | 
\aone \bone \cone \done \eone \}{}^{-}}(x) ~~~, \\
~~{\rm {Level}}-13 ~~~~~~~~~ \Psi_{\{ \aone \bone \}} {}_{\a}(x)  ~~~, \\
~~{\rm {Level}}-14 ~~~~~~~~~ {\widehat \Phi}{}_{\{ \aone \bone \cone \}}(x)  ~~~, \\
~~{\rm {Level}}-15 ~~~~~~~~~ \Psi^{\a}(x)  ~~~, \\
~~{\rm {Level}}-16 ~~~~~~~~~ {\widehat \Phi}(x)  ~~~.
\end{cases}
\label{equ:S1}
\end{equation}
These component fields are subject to irreducibility conditions listed below.
\begin{itemize}
\item Level-3:
\begin{equation}
    \Psi_{\{ \aone \bone \}} {}^{\a}(x) ~~:~~ 
    (\s^{ \aone })_{\a\b} \Psi_{\{ \aone \bone \}} {}^{\a}(x) = 0 ~~~.
\end{equation}
\item Level-4:
\begin{equation}
    \Phi_{\{ \aone \bone , \atwo \btwo \}}(x) ~~:~~ \begin{cases}
    \eta^{ \aone \atwo} \Phi_{\{ \aone \bone, \atwo \btwo \}}(x)~=~0~~~,\\
    \eta^{ \aone \atwo}\eta^{\bone \btwo} \Phi_{\{ \aone \bone, \atwo \btwo \}}(x) ~=~ 0~~~.
    \end{cases}
\end{equation}
\begin{equation}
    \Phi_{\{\atwo| \aone \bone \cone \done \eone \}{}^{+}}(x) ~~:~~ \begin{cases}
    \eta^{ \aone \atwo} \Phi_{\{\atwo| \aone \bone \cone \done \eone \}{}^{+}}(x)~=~0~~~,\\
    \Phi_{\{\atwo| \aone \bone \cone \done \eone \}{}^{+}}(x) ~=~
    +\frac{1}{5!}\epsilon_{ \aone \bone \cone \done \eone}{}^{{\un f}_1 
    {\un g}_1 {\un h}_1 {\un i}_1 {\un j}_1} \Phi_{\{\atwo| {\un f}_1 {\un g}_1
    {\un h}_1 {\un i}_1 {\un j}_1 \}{}^{+}}(x) ~~~.
    \end{cases}
\end{equation}
\item Level-5: 
\begin{equation}
    \Psi_{\{ \aone \bone \cone \done \eone \}{}^{+}}{}^{\a}(x) ~~:~~ \begin{cases}
    (\s^{\eone})_{\b\a} \Psi_{\{ \aone \bone \cone \done \eone \}{}^{+}}{}^{\a}(x) ~=~ 0  ~~~, \\
    \Psi_{\{ \aone \bone \cone \done \eone \}{}^{+}}{}^{\a}(x) ~=~ + \frac{1}{5!} \e_{\aone\bone\cone\done\eone}{}^{{\un f}_1 
    {\un g}_1 {\un h}_1 {\un i}_1 {\un j}_1} \Psi_{\{ {\un f}_1 
    {\un g}_1 {\un h}_1 {\un i}_1 {\un j}_1 \}{}^{+}}{}^{\a}(x) ~~~.
\end{cases}
\end{equation}
\begin{equation}
    \Psi_{\{{\un a}_2|  {\un a}_1 {\un b}_1  \}}{}^{\a}(x) ~~:~~ \begin{cases}
    (\s^{{\un b}_1})_{\a\d} \Psi_{\{{\un a}_2|   {\un a}_1 {\un b}_1  \}}{}^{\a}(x) ~\equiv~
    \psi_{\{{\un a}_1,  {\un a}_2  \}\d}(x)~=~0~~~,\\
    (\s^{{\un a}_2})_{\a\d} \Psi_{\{{\un a}_2|   {\un a}_1 {\un b}_1  \}}{}^{\a}(x) ~\equiv~
    \psi_{\{{\un a}_1  {\un b}_1  \}\d}(x)~=~0~~~,\\
    (\s^{{\un b}_1})^{\d\e} \psi_{\{{\un a}_1  {\un b}_1\}\d}(x)~=~0~~~.
\end{cases}
\end{equation}
\item Level-6: 
\begin{equation}
    \Phi{}_{\{{\un a}_2  {\un b}_2|  {\un a}_1 {\un b}_1 {\un c}_1  {\un d}_1 {\un e}_1 \}{}^{+}}(x) ~~:~~ \begin{cases}
    \eta^{{\un a}_1{\un a}_2} \Phi{}_{\{{\un a}_2  {\un b}_2|  {\un a}_1 {\un b}_1 {\un c}_1 {\un d}_1 {\un e}_1 \}{}^{+}}(x) = 0~~~, \\
    \eta^{{\un a}_1{\un a}_2}\eta^{{\un b}_1{\un b}_2} \Phi{}_{\{{\un a}_2  {\un b}_2|  {\un a}_1 {\un b}_1 {\un c}_1  {\un d}_1 {\un e}_1 \}{}^{+}}(x) = 0 ~~~, \\
    \Phi{}_{\{{\un a}_2  {\un b}_2|  {\un a}_1 {\un b}_1 {\un c}_1 {\un d}_1 {\un e}_1 \}{}^{+}}(x)
    = \frac{1}{5!}\epsilon_{{\un a}_1 {\un b}_1 {\un c}_1  {\un d}_1 {\un e}_1}{}^{{\un f}_1 {\un g}_1 {\un h}_1    {\un i}_1    {\un j}_1}\Phi{}_{\{{\un a}_2  {\un b}_2|  {\un f}_1 {\un g}_1 {\un h}_1  {\un i}_1 {\un j}_1 \}{}^{+}}(x)~~~.
\end{cases}
\end{equation}
\begin{equation}
    \Phi{}_{\{{\un a}_2 , {\un a}_3|   {\un a}_1  {\un b}_1     {\un c}_1  \}}(x) ~~:~~ \begin{cases}
    \eta^{{\un a}_2 {\un a}_3}\Phi{}_{\{{\un a}_2 , {\un a}_3|   {\un a}_1 {\un b}_1 {\un c}_1  \}}(x) = 0~~~,\\
    \eta^{{\un a}_1 {\un a}_3}\Phi{}_{\{{\un a}_2 , {\un a}_3|   {\un a}_1  {\un b}_1 {\un c}_1  \}}(x) = 0~~~.
    \end{cases}
\end{equation}
\item Level-7: 
\begin{equation}
    \Psi{}_{\{{\un a}_1,  {\un a}_2, {\un a}_3  \}}{}_{\a}(x) ~~:~~   \begin{cases}
    (\s^{{\un a}_3})^{\a\d}\Psi{}_{\{{\un a}_1,  {\un a}_2, {\un a}_3  \}}{}_{\a}(x) 
    ~\equiv~ \psi{}_{\{{\un a}_1,   {\un a}_2  \}}{}^{\d}(x)~=~0~~~,\\
    (\s^{{\un a}_2})_{\e\d}\psi{}_{\{{\un a}_1,   {\un a}_2 \}}{}^{\d}(x) 
    ~\equiv~\psi{}_{\{{\un a}_1 \}\e}(x)~=~0~~~,\\
    (\s^{{\un a}_1})^{\tau\e}\psi{}_{\{{\un a}_1  \}\e}(x)~=~0~~~.
\end{cases}
\end{equation}
\begin{equation}
    \Psi{}_{\{{\un a}_2| {\un a}_1 {\un b}_1 {\un c}_1  \}}{}^{\a}(x) ~~:~~ \begin{cases}
    (\s^{{\un a}_2})_{\a\d}\Psi{}_{\{{\un a}_2| {\un a}_1     {\un b}_1 {\un c}_1  \}}{}^{\a}(x) 
    ~\equiv~ \psi{}_{\{ {\un a}_1 {\un b}_1 {\un c}_1  \}}{}_{\d}(x)~=~0~~~,\\
    (\s^{{\un c}_1})_{\a\d}\Psi{}_{\{{\un a}_2| {\un a}_1  {\un b}_1 {\un c}_1  \}}{}^{\a}(x) 
    ~\equiv~ \psi{}_{\{{\un a}_2| {\un a}_1     {\un b}_1   \}}{}_{\d}(x)~=~0~~~,\\
    (\s^{{\un c}_1})^{\d\e}\psi{}_{\{ {\un a}_1 {\un b}_1 {\un c}_1  \}}{}_{\d}(x)~=~0~~~.
    \end{cases}
\end{equation}
\item Level-8: 
\begin{equation}
    \Phi_{\{{\un a}_1,  {\un a}_2,  {\un a}_3, {\un a}_4 \}}(x) ~~:~~   \begin{cases}
    \eta^{{\un a}_3 {\un a}_4} \Phi_{\{{\un a}_1,  {\un a}_2,  {\un a}_3, {\un a}_4 \}}(x)
    = 0~~~,\\
    \eta^{{\un a}_1 {\un a}_2}\eta^{{\un a}_3 {\un a}_4} \Phi_{\{{\un a}_1,  {\un a}_2,  
    {\un a}_3, {\un a}_4 \}}(x) = 0~~~.
    \end{cases}
\end{equation}
\begin{equation}
    \Phi_{\{{\un a}_1  {\un b}_1  {\un c}_1, {\un a}_2   {\un b}_2 {\un c}_2 \}}(x) ~~:~~
    \begin{cases}
    \eta^{{\un c}_1 {\un c}_2} \Phi_{\{{\un a}_1  {\un b}_1  {\un c}_1, {\un a}_2   {\un b}_2 {\un c}_2 \}}(x) = 0~~~,\\
    \eta^{{\un b}_1 {\un b}_2}\eta^{{\un c}_1 {\un c}_2} \Phi_{\{{\un a}_1  {\un b}_1  {\un c}_1, {\un a}_2   {\un b}_2 {\un c}_2 \}}(x) = 0~~~,\\
    \eta^{{\un a}_1 {\un a}_2}\eta^{{\un b}_1 {\un b}_2}\eta^{{\un c}_1 {\un c}_2} 
    \Phi_{\{{\un a}_1  {\un b}_1  {\un c}_1, {\un a}_2   {\un b}_2 {\un c}_2 \}}(x) = 0~~~.
    \end{cases}
\end{equation}
\begin{equation}
    \Phi_{\{{\un a}_2, {\un a}_3| {\un a}_1    {\un b}_1   {\un c}_1   {\un d}_1 \}}(x) ~~:~~
    \begin{cases}
    \eta^{{\un a}_2 {\un a}_3} \Phi_{\{{\un a}_2, {\un a}_3| {\un a}_1 {\un b}_1 {\un c}_1 {\un d}_1 \}}(x) = 0~~~,\\
    \eta^{{\un a}_1 {\un a}_3} \Phi_{\{{\un a}_2, {\un a}_3| {\un a}_1 {\un b}_1 {\un c}_1
    {\un d}_1 \}}(x) = 0~~~.
    \end{cases}
\end{equation}
\item Level-9:
\begin{equation}
    \Psi_{\{{\un a}_1,  {\un a}_2, {\un a}_3  \}}{}^{\a}(x) ~~:~~  \begin{cases}
    (\s^{{\un a}_3})_{\a\d}\Psi_{\{{\un a}_1,  {\un a}_2 {\un a}_3  \}}{}^{\a}(x) 
    ~\equiv~ \psi_{\{{\un a}_1,   {\un a}_2  \}}{}_{\d}(x)~=~0~~~,\\
    (\s^{{\un a}_2})^{\e\d}\psi_{\{{\un a}_1,   {\un a}_2  \}}{}_{\d}(x) ~\equiv~\psi_{\{{\un a}_1  \}}{}^{\e}(x)~=~0~~~,\\
    (\s^{{\un a}_1})_{\tau\e}\psi_{\{{\un a}_1  \}}{}^{\e}(x)~=~0~~~.
    \end{cases}
\end{equation}
\begin{equation}
    \Psi_{\{{\un a}_2| {\un a}_1 {\un b}_1 {\un c}_1 \}}{}_{\a}(x) ~~:~~ \begin{cases}
    (\s^{{\un a}_2})^{\a\d}\Psi_{\{{\un a}_2| {\un a}_1 {\un b}_1 {\un c}_1 \}}{}_{\a}(x) 
    ~\equiv~ \psi_{\{ {\un a}_1 {\un b}_1 {\un c}_1  \}}{}^{\d}(x)~=~0~~~,\\
    (\s^{{\un c}_1})^{\a\d}\Psi_{\{{\un a}_2| {\un a}_1 {\un b}_1 {\un c}_1\}}{}_{\a}(x) ~\equiv~
    \psi_{\{{\un a}_2| {\un a}_1  {\un b}_1 \}}{}^{\d}(x)~=~0~~~,\\
    (\s^{{\un c}_1})_{\d\e}\psi_{\{ {\un a}_1 {\un b}_1 {\un c}_1  \}}{}^{\d}(x)~=~0~~~.
    \end{cases}
\end{equation}
\item Level-10:
\begin{equation}
\Phi_{\{{\un a}_2 {\un b}_2| {\un a}_1 {\un b}_1 {\un c}_1 {\un d}_1 {\un e}_1 \}{}^{-}}(x) ~~:~~ \begin{cases}
\eta^{{\un a}_1{\un a}_2} \Phi_{\{{\un a}_2 {\un b}_2| {\un a}_1 {\un b}_1 {\un c}_1 {\un
d}_1 {\un e}_1 \}{}^{-}}(x) = 0~~~,\\
\eta^{{\un a}_1{\un a}_2}\eta^{{\un b}_1{\un b}_2} \Phi_{\{{\un a}_2 {\un b}_2| {\un a
}_1 {\un b}_1 {\un c}_1 {\un d}_1 {\un e}_1 \}{}^{-}}(x) = 0~~~,\\
\Phi_{\{{\un a}_2 {\un b}_2| {\un a}_1 {\un b}_1 {\un c}_1 {\un d}_1 {\un e}_1 \}{}^{-}}(x)
= -\frac{1}{5!}\epsilon_{{\un a}_1 {\un b}_1 {\un c}_1 {\un d}_1 {\un e}_1}{}^{{\un f}_1 {\un g}_1
{\un h}_1 {\un i}_1 {\un j}_1}\Phi_{\{{\un a}_2 {\un b}_2| {\un f}_1 {\un g}_1 {\un h}_1 
 {\un i}_1 {\un j}_1 \}{}^{-}}(x)~~~.
\end{cases}
\end{equation} 
\begin{equation}
{\widehat \Phi}_{\{{\un a}_2 , {\un a}_3| {\un a}_1 {\un b}_1 {\un c}_1 \}}(x) ~~:~~
 \begin{cases}
\eta^{{\un a}_2 {\un a}_3} {\widehat \Phi}_{\{{\un a}_2 , {\un a}_3| {\un a}_1 {\un b}_1 
{\un c}_1 \}}(x) = 0~~~,\\
\eta^{{\un a}_1 {\un a}_3} {\widehat \Phi}_{\{{\un a}_2 , {\un a}_3| {\un a}_1 {\un b}_1 
 {\un c}_1 \}}(x) = 0~~~.
\end{cases}
\end{equation} 
\item Level-11:
\begin{equation}
    \Psi_{\{ \aone \bone \cone \done \eone \}{}^{-}}{}_{\a}(x) ~~:~~ \begin{cases}
    (\s^{\eone})^{\b\a} \Psi_{\{ \aone \bone \cone \done \eone \}{}^{-}}{}_{\a}(x) ~=~ 0  ~~~, \\
    \Psi_{\{ \aone \bone \cone \done \eone \}{}^{-}}{}_{\a}(x) ~=~ - \frac{1}{5!} \e_{\aone\bone\cone\done\eone}{}^{{\un f}_1 
    {\un g}_1 {\un h}_1 {\un i}_1 {\un j}_1} \Psi_{\{ {\un f}_1 
    {\un g}_1 {\un h}_1 {\un i}_1 {\un j}_1 \}{}^{-}}{}_{\a}(x) ~~~.
\end{cases}
\end{equation}
\begin{equation}
\Psi_{\{{\un a}_2|   {\un a}_1 {\un b}_1  \}}{}_{\a}(x) ~~:~~  \begin{cases}
(\s^{{\un b}_1})^{\a\d}\Psi_{\{{\un a}_2|   {\un a}_1 {\un b}_1  \}\a}(x) ~\equiv~\psi_{\{{
\un a}_1,   {\un a}_2
\}}{}^{\d}(x)~=~0~~~,\\
(\s^{{\un a}_2})^{\a\d}\Psi_{\{{\un a}_2|   {\un a}_1 {\un b}_1  \}\a}(x) ~\equiv~\psi_{\{{\un
a}_1  {\un b}_1  \}}{}^{\d}(x)~=~0~~~,\\
(\s^{{\un b}_1})_{\d\e}\psi_{\{{\un a}_1  {\un b}_1 \}}{}^{\d}(x)~=~0~~~.
\end{cases} 
\end{equation}
\item Level-12:
\begin{equation}
{\widehat \Phi}{}_{\{{\un a}_1  {\un b}_1,   {\un a}_2 {\un b}_2 \}}(x) ~~:~~  
 \begin{cases}
\eta^{{\un a}_1  {\un a}_2}{\widehat \Phi}{}_{\{{\un a}_1  {\un b}_1,   {\un a}_2
{\un b}_2 \}}(x)~=~0~~~,\\
\eta^{{\un a}_1  {\un a}_2}\eta^{{\un b}_1  {\un b}_2}{\widehat \Phi}{}_{\{{\un a}_1  {\un b
}_1,   {\un a}_2 {\un b}_2 \}}(x)~=~0~~~.
\end{cases}
\end{equation}
\begin{equation}
\Phi{}_{\{{\un a}_2| {\un a}_1 {\un b}_1 {\un c}_1 {\un d}_1 {\un e}_1 \}{}^{-}}(x) ~~:~~
 \begin{cases}
\eta^{{\un a}_1  {\un a}_2}\Phi{}_{\{{\un a}_2| {\un a}_1 {\un b}_1 {\un c}_1  {\un d}_1    
{\un e}_1 \}{}^{-}}(x)~=~0~~~,\\
\Phi{}_{\{{\un a}_2| {\un a}_1   {\un b}_1 {\un c}_1    {\un d}_1    {\un e}_1 \}{}^{-}}(x)~=~-
\frac{1}{5!}\epsilon_{{\un a}_1   {\un b}_1 {\un c}_1    {\un d}_1    {\un e}_1}{}^{{\un f}_1  
{\un g}_1 {\un h}_1 {\un i}_1 {\un j}_1}\Phi{}_{\{{\un a}_2| {\un f}_1 {\un g}_1 {\un h}_1    
{\un i}_1    {\un j}_1 \}{}^{-}}(x)~~~.
\end{cases}
\end{equation}
\item Level-13:
\begin{equation}
\label{equ:S2}
    \Psi{}_{\{{\un a}_1  {\un b}_1 \}} {}_{\a}(x) ~~:~~ 
    (\s^{{\un a}_1})^{\a\b} \Psi{}_{\{{\un a}_1  {\un b}_1 \}} {}_{\a}(x) = 0 ~~~.
\end{equation}
\end{itemize}

The important point to note is this presentation of all the component fields in the 10D,
$\cal N$ = 1 scalar superfield, with their index structures showing a complete set of vector
and spinor indices, is achieved {\it {without}} ever introducing any $\s$-matrices.

Let us here consider the results in this chapter from a 
slightly different perspective.  We begin from the space of fields that we denote by
$ \{ {\cal F} \}$.  This space naturally is bisected $\{ {\cal F} \} ~=~ \{ {\cal F} \}_b ~
\oplus~ \{ {\cal F} \}_f $ according whether a specific field is a bosonic or 
fermionic representation of the Lorentz algebra.  Thus a representation of $\{ {\cal F} \}$
can be written in the form
\begin{equation}
\{ {\cal F} \} ~=~ \{ 
\begin{matrix} \text{scalar} \\ \phi(x) \\ \text{spin}-0 \end{matrix} ,~ 
\begin{matrix} \text{photon} \\ A_{\un{a}}(x) \\ \text{spin}-1 \end{matrix} ,~ 
\begin{matrix} \text{graviton} \\ h_{\un{a}\un{b}}(x) \\ \text{spin}-2 \end{matrix} ,
\dots \} ~\oplus~ \{
\begin{matrix} \text{spinor} \\ \l^{\a}(x) \\ \text{spin}-\frac{1}{2} \end{matrix} ,~ 
\begin{matrix} \text{spinor} \\ \l_{\a}(x) \\ \text{spin}-\frac{1}{2} \end{matrix} ,~ 
\begin{matrix} \text{gravitino} \\ \psi_{\un{a}}{}^{\b}(x) \\ \text{spin}-\frac{3}{2} \end{matrix} ,~ 
\begin{matrix} \text{gravitino} \\ \psi_{\un{a}}{}_{\b}(x) \\ \text{spin}-\frac{3}{2} \end{matrix} ,
\dots \} ~~~,
\label{equ:Fspace}
\end{equation}
and the ellipses in the equations denote the fact these are infinite dimensional
sets\footnote{
Above, for convenience only, we have used the spin nomenclature of the field space in 
4D to describe these fields.}.  One point the discussion in this chapter makes clear
is that the irreducible decomposition of the fields in $ \{ {\cal F} \}$ can be 
efficiently accomplished by the replacement of the various vector and spinor indices
with Dynkin Label as precisely the same information is conveyed. Moreover, since
the Dynkin Labels provide a specification of YT's, the latter can replace
space-time vector and spinor indices.
When this is done, (\ref{equ:Fspace}) is replaced by
\begin{equation}
\ytableausetup{boxsize=0.55em}
\{ {\cal F} \} ~=~ \{ 
\begin{matrix} \text{scalar} \\ \phi(x) \\ \text{spin}-0 \end{matrix} ,~ 
\begin{matrix} \text{photon} \\ A_{\CMTB{\ydiagram{1}}}(x) \\ \text{spin}-1 \end{matrix} ,~ 
\begin{matrix} \text{graviton} \\ h_{\CMTB{\ydiagram{2}}}(x) \\ \text{spin}-2 \end{matrix} ,
\dots \} ~\oplus~ \{
\begin{matrix} \text{spinor} \\ \l_{\CMTred{\ytableaushort{\indexsixteen}}}(x) \\ \text{spin}-\frac{1}{2} \end{matrix} ,~ 
\begin{matrix} \text{spinor} \\ \l_{\CMTred{\ytableaushort{\indexsixteenbar}}}(x) \\ \text{spin}-\frac{1}{2} \end{matrix} ,~ 
\begin{matrix} \text{gravitino} \\ \psi_{\CMTB{\ydiagram{1}}\CMTred{\ytableaushort{\indexsixteenbar}}}(x) \\ \text{spin}-\frac{3}{2} \end{matrix} ,~ 
\begin{matrix} \text{gravitino} \\ \psi_{\CMTB{\ydiagram{1}}\CMTred{\ytableaushort{\indexsixteen}}}(x) \\ \text{spin}-\frac{3}{2} \end{matrix} ,
\dots \} ~~~.
\label{equ:FspaceDynk}
\end{equation}

Finally, in view of the perspective of Wigner, who indicated that free elementary particles 
are irreducible representations of Poincar\' e group, one can rewrite (\ref{equ:FspaceDynk})
in the form
\begin{equation}
\ytableausetup{boxsize=0.8em}
\{ {\cal F} \} ~=~ \{ 
\begin{matrix} \text{scalar} \\ \CMTB{\cdot}~(x) \\ \text{spin}-0 \end{matrix} ,~ 
\begin{matrix} \text{photon} \\ \CMTB{\ydiagram{1}}(x) \\ \text{spin}-1 \end{matrix} ,~ 
\begin{matrix} \text{graviton} \\ \CMTB{\ydiagram{2}}(x) \\ \text{spin}-2 \end{matrix} ,
\dots \} ~\oplus~ \{
\begin{matrix} \text{spinor} \\ \CMTred{\ytableaushort{\tinysixteen}}(x) \\ \text{spin}-\frac{1}{2} \end{matrix} ,~ 
\begin{matrix} \text{spinor} \\ \CMTred{\ytableaushort{\tinysixteenbar}}(x) \\ \text{spin}-\frac{1}{2} \end{matrix} ,~ 
\begin{matrix} \text{gravitino} \\ \CMTB{\ydiagram{1}}\CMTred{\ytableaushort{\tinysixteenbar}}(x) \\ \text{spin}-\frac{3}{2} \end{matrix} ,~ 
\begin{matrix} \text{gravitino} \\ \CMTB{\ydiagram{1}}\CMTred{\ytableaushort{\tinysixteen}}(x) \\ \text{spin}-\frac{3}{2} \end{matrix} ,
\dots \} ~~~,
\ytableausetup{boxsize=1.2em}
\label{equ:FspaceWgN}
\end{equation}
where space-time dependent YT's replace fields.

The supersymmetry transformation is a map that acts between the spaces $\{ {\cal F} \}_b$ 
and $ \{ {\cal F} \}_f$.  Physicists have long assumed this map is homotopic to the identity 
map and thus assume the existence of an infinitesimal operator $\d_Q$ which depends on a 
parameter $\e^{\a}$ (also valued in the double cover) with the property
\be { \eqalign{ 
\d_Q(\e^{\a}) \, {\cal F} ~=~   &\{ \Tilde {\cal F} \}_f ~\oplus~ \{ \Tilde 
{\cal F} \}_b  }  ~~~.
} \ee
The elements of $\{ \Tilde {\cal F} \}_f$ are linear in $\e^{\a}$ and linear in the elements 
of $\{ {\cal F} \}_f$ and may involve tensors that are invariant under the action of isometries 
of the metric of the d-dimensional manifold and can involve first derivatives. The elements of 
$\{ \Tilde {\cal F} \}_b$ are linear in $\e^{\a}$ and linear in the elements of $\{ {\cal F} \}_b$ 
and may involve tensors that are invariant under the action of isometries of the metric of
the D-dimensional Lorentz algebra acting on the manifold and can involve first derivatives.

There are another set of infinitesimal variations, ``translations'' that can be defined on the 
space of fields.  These are denoted by $\d_P$ and depend on parameters $\xi^{\un a}$ where 
these parameters are valued in the tangent space to the manifold
\be { 
\eqalign{ 
\d_P(\xi^{\un a}) \, {\cal F} ~=~  (\xi^{\un a}\pa_{\un a} &\{  {\cal 
F} \}_b ) ~\oplus~ (\xi^{\un a}\pa_{\un a} 
\{  {\cal F} \}_f ) }   ~~~.
}  \ee
A system which consists of a subset of all the fields in $\{ {\cal F} \} $ is supersymmetric if 
the subset realizes the following equation
\be {  \eqalign{ 
\d_Q(\e_1^{\a}) \, \d_Q(\e_2^{\b}) ~-~ \d_Q(\e_2^{\a}) \, \d_Q(\e_1^{\b}) 
~=~ \d_P(\xi^{\un a}) 
~~~, 
 } 
 } \ee
where $ \xi^{\un a} ~=~i 2 < \e_1^{\a} (\g^{\un a}){}_{\a \, \b} \e_2^{\b} >$.  Systems satisfying 
this condition are said to be ``off-shell supersymmetric'' or to possess ``off-shell spacetime 
supersymmetry.''   The remarkable fact is that now decades after its first statement, the general 
{\it {irreducible}} solution to this problem is still {\em {not}} known.

What the physics community has quite effectively used is the fact that there is a related set 
of equations on the space of fields that is simpler to solve.

A hypersurface in field space may be defined by imposing some differential equations on 
the fields.  For example, the scalar field might be harmonic, satisfying the condition that its 
d'Alembertian vanishes.  In physics we call such a condition ``an equation of motion'' if it is 
derivable by the extremization of some function, typically denoted by $S$, that we call the 
action.  Let us denote such equations of  motion generically by the symbol $\pa S$.  Most 
of the discussions in the physics  literature involve representations such that
\be {
 \eqalign{ 
\d_Q(\e_1^{\a}) \, \d_Q(\e_2^{\b}) ~-~ \d_Q(\e_2^{\a}) \, \d_Q(\e_1^{\b}) 
~=~ \d_P(\xi^{\un a}) ~+~ \pa S
 }   
 } ~~~. \ee
Any formulation that is equivalent to superfields implies $\pa S \,=\, 0$ up
to gauge transformations.

\newpage
\section{From Superfields to Adynkrafields}

The path we have travelled began with the adynkra in Figure \ref{Fig:10DTypeI_Dynkin} that suggested a
reconceptualization of the superfield in terms of Young Tableaux which possess
a well understood algebra.  This replacement eliminates the need $\sigma$-matrices
in the expansion of the superfield.  As a consequence of branching rules, derivable
from decoration of the Young Tableaux, Fierz identities are eliminated with a huge
savings in terms of computational costs.  On the basis of these, far more efficient
algorithms for the explicit component-level examination of superfields are
achievable... as we have shown in the present work.

The adynkra shown in Figure \ref{Fig:10DTypeI_Dynkin} can be expressed
totally in a field-independent manner and purely in terms of group-theoretical
constructs mathematically in terms of $\cal G$ with the definition
\be{ \label{equ:calG}  \eqalign{
{\cal G} ~=~
&1 ~\oplus~ \ell \, \left\{  \left(\, \CMTred{\ydiagram{1}} \,\right) \, 
\times [a_1,b_1,c_1,d_1,e_1]  \right\} ~\oplus
~ \bigoplus_{p = 2}^{16}
 \, {\fracm 1{p!}} \, (\ell){}^p  \left\{  \left( \,
\CMTred{\ydiagram{1}} ~ ( \wedge \, \CMTred{\ydiagram{1}} \, )^{p - 2}  \wedge
 \CMTred{\ydiagram{1}} \, \right)  \, \times
[a_p,b_p,c_p,d_p,e_p]  \right\}    ~~~,
} }\ee
and where a number of definitions must be understood and these include:
\begin{enumerate}[label=(\alph*.)]
\item $\CMTred{\ydiagram{1}}$ denotes the SYT introduced in (\ref{equ:SYTbasic}),
\item the $\wedge$ product denotes the usual rule for multiplying two tableaux,\\
but restricted so that only single column resultants are kept,
\item $[a_p,b_p,c_p,d_p,e_p]$ denotes a Dynkin Label for an irrep in $\mathfrak{so}(10)$
where the \\
quantities $a_p$, $b_p$, $c_p$, $d_p$, and $e_p$ are a set of integers,
\item $\CMTred{{\cal A} } \, \times [a_p,b_p,c_p,d_p,e_p]$ = $[a_p,b_p,c_p,d_p,e_p]$ where
$\CMTred{{\cal A} }$ is a single column SYT \\
containing the irrep $[a_p,b_p,c_p,d_p,e_p]$ otherwise $\CMTred{{\cal A} }  \,
\times [a_p,b_p,c_p,d_p,e_p]$ = 0,
\item $\CMTred{{\cal A} } \, \times [a_p,b_p,c_p,d_p,e_p] = {\rm m} \, [a_p,b_p,c_p,d_p,e_p]$ if
instead $\CMTred{{\cal A} }$ contains the representation \\
$[a_p,b_p,c_p,d_p,e_p]$ {\rm m}-times, and finally
\item $ \left\{ \CMTred{{\cal A} }  \, \times [a_p,b_p,c_p,d_p,e_p] \right \}$ is a notation implying 
independent sums to be taken over \\
all possible values of $a_p$, $b_p$ ,$c_p$ ,$d_p$, and $e_p$.
\end{enumerate}

The mathematical object $\cal G$, as illustrated in the alternate form shown in 
(\ref{equ:G2}) for the case of 10D, $\cal N$ = 1 superfields, is the fundamental 
one we have been studying in the works of \cite{HyDSG2,HyDSG3} and that has allowed 
unprecedented clarity and access to the component field structures of high dimensional 
superfield theories.  

For arbitrary dimensions D, we can generalize the result of (\ref{equ:calG}) to be of the form
\be{ \label{equ:calGg}  \eqalign{
{\cal G} ~=~
&1 ~\oplus~ \ell \, \left\{  \left(\, \CMTred{\ydiagram{1}} \,\right) \, \times [Dk{}_{1}]  \right\}
~\oplus~   \bigoplus_{p = 2}^{d} \, {\fracm 1{p!}} \, (\ell){}^p  \left\{  \left( \,
\CMTred{\ydiagram{1}} ~ ( \wedge \, \CMTred{\ydiagram{1}} \, )^{p - 2}  \wedge
 \CMTred{\ydiagram{1}} \, \right)  \, \times
[Dk{}_{p}]   \right\}    ~~~,
} }\ee
where $[Dk{}_{1}] $ and $[Dk{}_{p}] $ denote Dynkin Labels appropriate for the
fields over the D-dimensional manifold and the quantity $d$ is the dimensionality
of the minimal spinor representation for the D-dimensional manifold.

We may call $\Tilde {\cal G}$ and $\cal G$ ``adynkra series,'' 
which appears to be an adapted and specialized series expansion in terms of Young
Tableaux.

$~~~$ {\it {All of these results strongly suggest adynkras
are pointing in the direction}}  \newline $~~~~~~\,~~$ 
{\it {of using series expansion in terms of YT's as a tool to 
gain the most}}  \newline $~~~~~~\,~~$ 
{\it {fundamental mathematical understanding of this class of problems.}}

\noindent
The results of (\ref{equ:G2}) can be combined with those in (\ref{equ:S1}) - (\ref{equ:S2})
to yield an ``adynkrafield.''
\begin{align}\label{equ:GHat}  \ytableausetup{boxsize=0.8em}
{\widehat {\cal G}}(x)
~=&~
\Phi(x)
\,+\, {\ell} \,  \CMTred{\ytableaushort{\tinysixteen}} \,
\Psi{}_{\a}(x)
~+~ \fracm 1{2} \,({\ell}){}^{2} \, {\CMTB{\ydiagram{1,1,1}}}_{\rm IR} \,
\Phi_{\{ \aone \bone \cone \}} (x)
~ + ~   \fracm 1{3!} \,({\ell}){}^{3} \,
{\CMTB{\ydiagram{1,1}}\CMTred{\ytableaushort{\tinysixteenbar,\none}}}_{\rm IR} \,
\Psi_{\{ \aone \bone \}}{}^{\a} (x)  {~~~~~~~~~~}  \nonumber\\
&
~+~  \fracm 1{4!} \,({\ell}){}^{4} \,  {\CMTB{\ydiagram{2,2}}}_{\rm IR} \,
\Phi_{\{ \aone \bone, \atwo \btwo \}} (x)
~+~  \fracm 1{4!} \, ({\ell}){}^{4} \,  {\CMTB{\ydiagram{2,1,1,1,1}}}_{{\rm IR},-} \,
\Phi_{\{ \atwo | \aone \bone \cone \done \eone \}{}^{+}} (x)  \nonumber\\
&~+~ \fracm 1{5!} \, ({\ell}){}^{5} \,
{\CMTB{\ydiagram{1,1,1,1,1}}\CMTred{\ytableaushort{\tinysixteenbar,\none,\none,\none,\none}}}_{{\rm IR},-} \,
\Psi_{\{ \aone \bone \cone \done \eone \}{}^{+}}{}^{\a} (x)
\, + \,  \fracm 1{5!} \, ({\ell}){}^{5} \,  {\CMTB{\ydiagram{2,1}}\CMTred{\ytableaushort{\tinysixteenbar,\none}}}_{\rm IR} \, \Psi_{\{ \atwo | \aone \bone \}}{}^{\a} (x)\nonumber\\
& \, + \, \fracm 1{6!} \,  ({\ell}){}^{6} \,   {\CMTB{\ydiagram{2,2,1,1,1}}}_{{\rm IR},-} \,
\Phi_{\{\atwo \btwo| \aone \bone \cone \done \eone \}{}^{+}} (x)
\, + \, \fracm 1{6!} \, ({\ell}){}^{6} \,  {\CMTB{\ydiagram{3,1,1}}}_{\rm IR} 
\, \Phi_{\{ \atwo , \athree | \aone \bone \cone \}} (x)   \nonumber\\
& \, + \, \fracm 1{7!} \,  ({\ell}){}^{7} \,     {\CMTB{\ydiagram{3}}\CMTred{\ytableaushort{\tinysixteen}}}_{\rm IR} \,
\Psi_{\{ \aone, \atwo, \athree \}}{}_{\a} (x)
~+~  \fracm 1{7!} \, ({\ell}){}^{7} \,  {\CMTB{\ydiagram{2,1,1}}
\CMTred{\ytableaushort{\tinysixteenbar,\none,\none}}}_{\rm IR}   \,
\Psi_{\{ \atwo | \aone \bone \cone \}}{}^{\a} (x)   \nonumber \\
& \, + \, \fracm 1{8!} \, ({\ell}){}^{8} \,    {\CMTB{\ydiagram{4}}}_{\rm IR} \,
\Phi_{\{ \aone,  \atwo,  \athree, \afour \}} (x)
\, + \, \fracm 1{8!} \, ({\ell}){}^{8} \,  {\CMTB{\ydiagram{2,2,2}}}_{\rm IR}  \,
\Phi_{\{ \aone \bone \cone , \atwo \btwo \ctwo \}} (x)   \nonumber\\
& \, + \, \fracm 1{8!} \, ({\ell}){}^{8} \,  {\CMTB{\ydiagram{3,1,1,1}}}_{\rm IR}  \,
\Phi_{\{\atwo,\athree | \aone \bone \cone \done \}} (x)   \nonumber\\
&  \, + \, \fracm 1{9!} \, ({\ell}){}^{9} \,  {\CMTB{\ydiagram{3}}\CMTred{\ytableaushort{\tinysixteenbar}}}_{\rm IR}
 \Psi_{\{ \aone , \atwo , \athree \}}{}^{\a}(x)
\, + \, \fracm 1{9!} \, ({\ell}){}^{9} \,
{\CMTB{\ydiagram{2,1,1}}\CMTred{\ytableaushort{\tinysixteen,\none,\none}}}_{\rm IR}    \,
 \Psi_{\{ \atwo | \aone \bone \cone \}}{}_{\a}(x)\nonumber \\
&  \, + \, \fracm 1{10!} \, ({\ell}){}^{10} \,   {\CMTB{\ydiagram{2,2,1,1,1}}}_{{\rm IR},+} \,
 \Phi_{\{ \atwo \btwo | \aone \bone \cone \done \eone \}{}^{-}}(x)
  \, + \, \fracm 1{10!} \, ({\ell}){}^{10} \,  {\CMTB{\ydiagram{3,1,1}
}}_{\rm IR} \, {\widehat \Phi}_{\{ \atwo , \athree | \aone \bone \cone \}}(x) {~~~~~}  \nonumber \\
& 
\, + \, \fracm 1{11!} \,  ({\ell}){}^{11} \,  {\CMTB{\ydiagram{1,1,1,1,1}}\CMTred{\ytableaushort{\tinysixteen,\none,\none,\none,\none}}}_{{\rm IR},+} \,  \Psi_{\{ \aone \bone \cone \done \eone \}{}^{-}}{}_{\a}(x)
\, + \, \fracm 1{11!} \, ({\ell}){}^{11} \,
{\CMTB{\ydiagram{2,1}}\CMTred{\ytableaushort{\tinysixteen,\none}}}_{\rm IR}    \,
 \Psi_{\{ \atwo | \aone \bone \}}{}_{\a}(x)   {~~} \nonumber \\
&  \, + \, \fracm 1{12!} \,  ({\ell}){}^{12} \,   {\CMTB{\ydiagram{2,2}}}_{\rm IR} \,
 {\widehat \Phi}_{\{ \aone \bone , \atwo \btwo \}}(x)
\, + \, \fracm 1{12!} \, ({\ell}){}^{12} \,  {\CMTB{\ydiagram{2,1,1,1,1}}
}_{{\rm IR},+}    \,  \Phi_{\{ \atwo |
\aone \bone \cone \done \eone \}{}^{-}}(x) \nonumber\\
& \, + \,
\fracm 1{13!} \, ({\ell}){}^{13} \, {\CMTB{\ydiagram{1,1}}\CMTred{\ytableaushort{\tinysixteen,\none}}}_{\rm IR}  \, \Psi_{\{ \aone \bone \}} {}_{\a}(x)
\, + \, \fracm 1{14!} \,  ({\ell}){}^{14} \, {\CMTB{\ydiagram{1,1,1}}}_{\rm IR} \, {\widehat \Phi}{}_{\{ \aone \bone \cone \}}(x)
\, + \,  \fracm 1{15!} \,  ({\ell}){}^{15} \,  \CMTred{\ytableaushort{\tinysixteenbar}}  \Psi^{\a}(x)\nonumber
\\
& 
\, + \, \fracm 1{16!}  \,({\ell}){}^{16} \,  {\widehat \Phi}(x) ~~~, 
\ytableausetup{boxsize=1.2em}
\end{align}
and it is to understand that each index on every field has all of it indices ``tied'' to indices in 
the YT that precedes it.  Bosonic indices are tied to blue boxes in the YT. Fermionic indices are 
tied to the red boxes. It should be noted that at most one red box appears for all terms.
The result in (\ref{equ:GHat}) arises from taking (\ref{equ:calG}) followed by the use of
branching rules in the cases of even numbers of red boxes to replace them by blue boxes.
Finally we have evaluated the ``inner product'' ${\cal G} \cdot \{ {\cal F} \}$.

Further with this definition, it seems reasonable to define 
a supercovariant derivative $\cal D$ via the formula
\be
{\cal D} ~=~ \Diop  
\ee
when acting on such an adynkrafield.  Here the $\CMTB {\pa}$ operator is the derivative
with respect to space-time coordinate.  Its index is tied to the index in the blue box that
precedes it.  Continuing investigation of such objects seem to possess the promise to unravel 
further long-standing mysteries about the structure of the representation theory 
of space-time supersymmetry.

The adynkra in (\ref{equ:G}) can be tensored with the $\CMTred{\ytableaushort{\sixteenbar}}$
representation.  This yields the following decomposition of $\cal{V}_{\alpha}$ at each level:
\begin{itemize}\ytableausetup{boxsize=0.8em}
    \item Level-0: $\CMTred{\ytableaushort{\tinysixteenbar}}$
    \item Level-1: ${\CMTB{\ydiagram{1}}}_{\rm IR}~\oplus~ {\CMTB{\ydiagram{1,1,1}}}_{\rm IR}~\oplus~ {\CMTB{\ydiagram{1,1,1,1,1}}}_{\rm IR,-}$
    \item Level-2: $\CMTred{\ytableaushort{\tinysixteen}}~\oplus~
    {\CMTB{\ydiagram{1}}\CMTred{\ytableaushort{\tinysixteenbar}}}_{\rm IR}~\oplus~
    {\CMTB{\ydiagram{1,1}}\CMTred{\ytableaushort{\tinysixteen,\none}}}_{\rm IR}~\oplus~{\CMTB{\ydiagram{1,1,1}}\CMTred{\ytableaushort{\tinysixteenbar,\none,\none}}}_{\rm IR}$
    \item Level-3: ${\CMTB{\ydiagram{1,1}}}_{\rm IR}~\oplus~ {\CMTB{\ydiagram{1,1,1,1}}}_{\rm IR}
    ~\oplus~ {\CMTB{\ydiagram{2,2}}}_{\rm IR}
    ~\oplus~ {\CMTB{\ydiagram{2,1,1}}}_{\rm IR}
    ~\oplus~ {\CMTB{\ydiagram{2,1,1,1,1}}}_{\rm IR,+}
    ~\oplus~ {\CMTB{\ydiagram{2,2,1,1}}}_{\rm IR}$
     \item Level-4: ${\CMTB{\ydiagram{1}}\CMTred{\ytableaushort{\tinysixteen}}}_{\rm IR}~\oplus~
    {\CMTB{\ydiagram{1,1}}\CMTred{\ytableaushort{\tinysixteenbar,\none}}}_{\rm IR}~\oplus~
    {\CMTB{\ydiagram{1,1,1,1,1}}\CMTred{\ytableaushort{\tinysixteen,\none,\none,\none,\none}}}_{\rm IR,+}~\oplus~
    {\CMTB{\ydiagram{1,1,1}}\CMTred{\ytableaushort{\tinysixteen,\none,\none}}}_{\rm IR}~\oplus~
    (2)\,{\CMTB{\ydiagram{2,1}}\CMTred{\ytableaushort{\tinysixteen,\none}}}_{\rm IR}~\oplus~
    {\CMTB{\ydiagram{2,2}}\CMTred{\ytableaushort{\tinysixteenbar,\none}}}_{\rm IR}
    ~\oplus~
    {\CMTB{\ydiagram{2,1,1,1}}\CMTred{\ytableaushort{\tinysixteen,\none,\none,\none}}}_{\rm IR} $
    \item Level-5: ${\CMTB{\ydiagram{1,1,1,1,1}}}_{\rm IR,+}
    ~\oplus~{\CMTB{\ydiagram{2,1}}}_{\rm IR}~\oplus~ {\CMTB{\ydiagram{2,1,1,1}}}_{\rm IR}
    ~\oplus~ {\CMTB{\ydiagram{2,2,1}}}_{\rm IR}
    ~\oplus~ (2)\,{\CMTB{\ydiagram{2,2,1,1,1}}}_{\rm IR,+}
    ~\oplus~ {\CMTB{\ydiagram{3,1,1}}}_{\rm IR}
    ~\oplus~ {\CMTB{\ydiagram{3,2}}}_{\rm IR}\\[10pt]
    ~\oplus~{\CMTB{\ydiagram{3,1,1,1,1}}}_{\rm IR,+}
    ~\oplus~
    {\CMTB{\ydiagram{2,2,2,2,1}}}_{\rm IR,+}
    ~\oplus~  {\CMTB{\ydiagram{3,2,1,1}}}_{\rm IR}$
     \item Level-6: ${\CMTB{\ydiagram{1,1}}\CMTred{\ytableaushort{\tinysixteen,\none}}}_{\rm IR}~\oplus~
    {\CMTB{\ydiagram{2}}\CMTred{\ytableaushort{\tinysixteen}}}_{\rm IR}~\oplus~
    {\CMTB{\ydiagram{1,1,1,1}}\CMTred{\ytableaushort{\tinysixteen,\none,\none,\none}}}_{\rm IR}
    ~\oplus~
    {\CMTB{\ydiagram{3}}\CMTred{\ytableaushort{\tinysixteenbar}}}_{\rm IR}~\oplus~{\CMTB{\ydiagram{2,1}}\CMTred{\ytableaushort{\tinysixteenbar,\none}}}_{\rm IR}~\oplus~
    {\CMTB{\ydiagram{2,1,1,1,1}}\CMTred{\ytableaushort{\tinysixteen,\none,\none,\none,\none}}}_{\rm IR,+}\\[10pt]
     ~\oplus~{\CMTB{\ydiagram{2,2}}\CMTred{\ytableaushort{\tinysixteen,\none}}}_{\rm IR}
     ~\oplus~(2)\, {\CMTB{\ydiagram{2,1,1}}\CMTred{\ytableaushort{\tinysixteen,\none,\none}}}_{\rm IR}
     ~\oplus~{\CMTB{\ydiagram{3,1}}\CMTred{\ytableaushort{\tinysixteen,\none}}}_{\rm IR}
     ~\oplus~
    {\CMTB{\ydiagram{2,2,1,1}}\CMTred{\ytableaushort{\tinysixteen,\none,\none,\none}}}_{\rm IR}
     ~\oplus~{\CMTB{\ydiagram{3,1,1}}\CMTred{\ytableaushort{\tinysixteenbar,\none,\none}}}_{\rm IR}
    $
    \item Level-7: ${\CMTB{\ydiagram{4}}}_{\rm IR}~\oplus~{\CMTB{\ydiagram{2,1,1}}}_{\rm IR}~\oplus~
    {\CMTB{\ydiagram{2,1,1,1,1}}}_{\rm IR,+}
    ~\oplus~{\CMTB{\ydiagram{3,1}}}_{\rm IR}~\oplus~{\CMTB{\ydiagram{2,2,2}}}_{\rm IR}~\oplus~{\CMTB{\ydiagram{2,2,1,1}}}_{\rm IR}~\oplus~{\CMTB{\ydiagram{2,2,2,1,1}}}_{\rm IR,+}\\[10pt]
    ~\oplus~(2)\,{\CMTB{\ydiagram{3,1,1,1}}}_{\rm IR}~\oplus~{\CMTB{\ydiagram{4,1,1}}}_{\rm IR}~\oplus~{\CMTB{\ydiagram{4,1,1,1,1}}}_{\rm IR,-}
    ~\oplus~{\CMTB{\ydiagram{3,2,1}}}_{\rm IR}~\oplus~{\CMTB{\ydiagram{3,2,1,1,1}}}_{\rm IR,+}
    ~\oplus~{\CMTB{\ydiagram{3,2,2,1}}}_{\rm IR}$
        \item Level-8: ${\CMTB{\ydiagram{2}}\CMTred{\ytableaushort{\tinysixteenbar}}}_{\rm IR}~\oplus~
    {\CMTB{\ydiagram{1,1,1}}\CMTred{\ytableaushort{\tinysixteen,\none,\none}}}_{\rm IR}~\oplus~
    (2)\,{\CMTB{\ydiagram{3}}\CMTred{\ytableaushort{\tinysixteen}}}_{\rm IR}~\oplus~
    {\CMTB{\ydiagram{2,1}}\CMTred{\ytableaushort{\tinysixteen,\none}}}_{\rm IR}~\oplus~
    {\CMTB{\ydiagram{4}}\CMTred{\ytableaushort{\tinysixteenbar}}}_{\rm IR}~\oplus~
    (2)\,{\CMTB{\ydiagram{2,1,1}}\CMTred{\ytableaushort{\tinysixteenbar,\none,\none}}}_{\rm IR}\\[10pt]
    ~\oplus~{\CMTB{\ydiagram{2,1,1,1}}\CMTred{\ytableaushort{\tinysixteen,\none,\none,\none}}}_{\rm IR} 
    ~\oplus~{\CMTB{\ydiagram{3,1}}\CMTred{\ytableaushort{\tinysixteenbar,\none}}}_{\rm IR}
    ~\oplus~
    {\CMTB{\ydiagram{2,2,1}}\CMTred{\ytableaushort{\tinysixteen,\none,\none}}}_{\rm IR}
    ~\oplus~
    {\CMTB{\ydiagram{2,2,2}}\CMTred{\ytableaushort{\tinysixteenbar,\none,\none}}}_{\rm IR}
    ~\oplus~
    {\CMTB{\ydiagram{3,1,1}}\CMTred{\ytableaushort{\tinysixteen,\none,\none}}}_{\rm IR}
    ~\oplus~{\CMTB{\ydiagram{3,1,1,1}}\CMTred{\ytableaushort{\tinysixteenbar,\none,\none,\none}}}_{\rm IR} 
    $
    \item Level-9: ${\CMTB{\ydiagram{3}}}_{\rm IR}~\oplus~{\CMTB{\ydiagram{2,1,1,1}}}_{\rm IR}~\oplus~{\CMTB{\ydiagram{2,2,1}}}_{\rm IR}~\oplus~{\CMTB{\ydiagram{2,2,1,1,1}}}_{\rm IR,-}
    ~\oplus~(2)\,{\CMTB{\ydiagram{3,1,1}}}_{\rm IR}~\oplus~{\CMTB{\ydiagram{4,1}}}_{\rm IR}~\oplus~{\CMTB{\ydiagram{3,1,1,1,1}}}_{\rm IR,-}\\[10pt]
    ~\oplus~{\CMTB{\ydiagram{3,1,1,1,1}}}_{\rm IR,+}
    ~\oplus~{\CMTB{\ydiagram{2,2,2,1}}}_{\rm IR}~\oplus~{\CMTB{\ydiagram{3,2,2}}}_{\rm IR}~\oplus~{\CMTB{\ydiagram{4,1,1,1}}}_{\rm IR}~\oplus~{\CMTB{\ydiagram{3,2,1,1}}}_{\rm IR}~\oplus~{\CMTB{\ydiagram{3,2,2,1,1}}}_{\rm IR,-}$
\end{itemize}

\begin{itemize}\ytableausetup{boxsize=0.8em}
    \item Level-10: ${\CMTB{\ydiagram{1,1,1,1,1}}\CMTred{\ytableaushort{\tinysixteenbar,\none,\none,\none,\none}}}_{\rm IR,-}~\oplus~
    {\CMTB{\ydiagram{2}}\CMTred{\ytableaushort{\tinysixteen}}}_{\rm IR}~\oplus~
    {\CMTB{\ydiagram{1,1,1}}\CMTred{\ytableaushort{\tinysixteenbar,\none,\none}}}_{\rm IR}~\oplus~
    {\CMTB{\ydiagram{3}}\CMTred{\ytableaushort{\tinysixteenbar}}}_{\rm IR}~\oplus~
    (2)\,{\CMTB{\ydiagram{2,1}}\CMTred{\ytableaushort{\tinysixteenbar,\none}}}_{\rm IR}~\oplus~
    {\CMTB{\ydiagram{2,1,1}}\CMTred{\ytableaushort{\tinysixteen,\none,\none}}}_{\rm IR}\\[10pt]
    ~\oplus~{\CMTB{\ydiagram{2,1,1,1}}\CMTred{\ytableaushort{\tinysixteenbar,\none,\none,\none}}}_{\rm IR}
    ~\oplus~{\CMTB{\ydiagram{3,1}}\CMTred{\ytableaushort{\tinysixteen,\none}}}_{\rm IR}
    ~\oplus~{\CMTB{\ydiagram{2,2,1,1,1}}\CMTred{\ytableaushort{\tinysixteenbar,\none,\none,\none,\none}}}_{\rm IR,-}
    ~\oplus~
    {\CMTB{\ydiagram{2,2,1}}\CMTred{\ytableaushort{\tinysixteenbar,\none,\none}}}_{\rm IR}
    ~\oplus~
    {\CMTB{\ydiagram{3,1,1}}\CMTred{\ytableaushort{\tinysixteenbar,\none,\none}}}_{\rm IR}
    $
    \item Level-11: ${\CMTB{\ydiagram{2,2}}}_{\rm IR}~\oplus~{\CMTB{\ydiagram{2,1,1}}}_{\rm IR}~\oplus~(2)\,{\CMTB{\ydiagram{2,1,1,1,1}}}_{\rm IR,-}~\oplus~{\CMTB{\ydiagram{3,1}}}_{\rm IR}~\oplus~{\CMTB{\ydiagram{2,2,2,2,2}}}_{\rm IR,-}~\oplus~{\CMTB{\ydiagram{2,2,1,1}}}_{\rm IR}~\oplus~{\CMTB{\ydiagram{2,2,2,1,1}}}_{\rm IR,-}\\[10pt]
    ~\oplus~{\CMTB{\ydiagram{3,1,1,1}}}_{\rm IR}~\oplus~{\CMTB{\ydiagram{3,2,1}}}_{\rm IR}~\oplus~{\CMTB{\ydiagram{3,2,1,1,1}}}_{\rm IR,-}$
    \item Level-12: $(2)\,{\CMTB{\ydiagram{1,1}}\CMTred{\ytableaushort{\tinysixteenbar,\none}}}_{\rm IR}~\oplus~{\CMTB{\ydiagram{2}}\CMTred{\ytableaushort{\tinysixteenbar}}}_{\rm IR}~\oplus~
    {\CMTB{\ydiagram{1,1,1,1,}}\CMTred{\ytableaushort{\tinysixteenbar,\none,\none,\none}}}_{\rm IR}
    ~\oplus~
    {\CMTB{\ydiagram{2,1}}\CMTred{\ytableaushort{\tinysixteen,\none}}}_{\rm IR}~\oplus~
    {\CMTB{\ydiagram{2,1,1,1,1}}\CMTred{\ytableaushort{\tinysixteenbar,\none,\none,\none,\none}}}_{\rm IR,-}\\[10pt]
    ~\oplus~
    {\CMTB{\ydiagram{2,2}}\CMTred{\ytableaushort{\tinysixteenbar,\none}}}_{\rm IR}~\oplus~{\CMTB{\ydiagram{2,1,1}}\CMTred{\ytableaushort{\tinysixteenbar,\none,\none}}}_{\rm IR}$
    \item Level-13: ${\CMTB{\ydiagram{1,1,1}}}_{\rm IR}~\oplus~{\CMTB{\ydiagram{1,1,1,1,1}}}_{\rm IR,-}~\oplus~{\CMTB{\ydiagram{2,1}}}_{\rm IR}~\oplus~{\CMTB{\ydiagram{2,1,1,1}}}_{\rm IR}~\oplus~{\CMTB{\ydiagram{2,2,1}}}_{\rm IR}~\oplus~{\CMTB{\ydiagram{2,2,1,1,1}}}_{\rm IR,-}$
    \item Level-14: $\CMTred{\ytableaushort{\tinysixteen}}~\oplus~
    {\CMTB{\ydiagram{1}}\CMTred{\ytableaushort{\tinysixteenbar}}}_{\rm IR}~\oplus~
    {\CMTB{\ydiagram{1,1}}\CMTred{\ytableaushort{\tinysixteen,\none}}}_{\rm IR}~\oplus~{\CMTB{\ydiagram{1,1,1}}\CMTred{\ytableaushort{\tinysixteenbar,\none,\none}}}_{\rm IR}$
    \item Level-15: $\CMTB{\cdot}~\oplus~{\CMTB{\ydiagram{1,1}}}_{\rm IR}~\oplus~{\CMTB{\ydiagram{1,1,1,1}}}_{\rm IR}$
    \item Level-16: $\CMTred{\ytableaushort{\tinysixteenbar}}$
\ytableausetup{boxsize=1.2em}
\end{itemize}

We can also apply $ {\cal D}$ to the adynkrafield in (\ref{equ:GHat}),
\begin{align}  \ytableausetup{boxsize=0.8em}
{\cal D} {\widehat {\cal G}}(x)
~=&~ 
\CMTred{\ytableaushort{\tinysixteen}} \,
\Psi{}_{\a}(x) 
~+~ ({\ell}) \left[ \,
{\CMTB{\ydiagram{1,1,1}}}_{\rm IR} \,
\Phi_{\{ \aone \bone \cone \}} (x) 
~+~
i\,  {\CMTB{\ydiagram{1}}}_{\rm IR}{\pa}_{{\un a}_{1}}
\Phi(x)
  \, \right]\nonumber\\
&~+~ \fracm 1{2!}\, ({\ell}){}^2 \, \left[ \, 
{\CMTB{\ydiagram{1,1}}\CMTred{\ytableaushort{\tinysixteenbar,\none}}}_{\rm IR} \,
\Psi_{\{ \aone \bone \}}{}^{\a} (x) 
~+~ 
i \, 2\,  ( \,\CMTred{\ytableaushort{\tinysixteenbar}}~+~
{\CMTB{\ydiagram{1}}\CMTred{\ytableaushort{\tinysixteen}}}_{\rm IR}  \, )\,  {\pa}_{{\un a}_{1}}\, \Psi{}_{\a}(x) 
\, \right] \nonumber\\ 
&~+~ \fracm 1{3!}\, ({\ell}){}^3 \, \left[ \, 
{\CMTB{\ydiagram{2,2}}}_{\rm IR} \,
\Phi_{\{ \aone \bone, \atwo \btwo \}} (x)
~+~{\CMTB{\ydiagram{2,1,1,1,1}}}_{{\rm IR},-} \,
\Phi_{\{ \atwo | \aone \bone \cone \done \eone \}{}^{+}} (x)\right. \nonumber\\
&\left. {~~~~~~~~~~~~~~~~} ~+~ i \, 3\,  \Bigg(\,
{\CMTB{\ydiagram{1,1}}}_{{\rm IR}}~+~{\CMTB{\ydiagram{1,1,1,1}}}_{{\rm IR}}~+~{\CMTB{\ydiagram{2,1,1}}}_{{\rm IR}}
 \,\Bigg) \,  {\pa}_{{\un a}_{2}}\, \Phi_{\{ \aone \bone \cone \}} (x)
\, \right]   \nonumber\\ 
&~+~ \fracm 1{4!}\, ({\ell}){}^4 \, \Bigg[ \,
{\CMTB{\ydiagram{1,1,1,1,1}}\CMTred{\ytableaushort{\tinysixteenbar,\none,\none,\none,\none}}}_{{\rm IR},-} \,
\Psi_{\{ \aone \bone \cone \done \eone \}{}^{+}}{}^{\a} (x)
~+~{\CMTB{\ydiagram{2,1}}\CMTred{\ytableaushort{\tinysixteenbar,\none}}}_{\rm IR} \, \Psi_{\{ \atwo | \aone \bone \}}{}^{\a} (x) \nonumber\\
& {~~~~~~~~~~~~~~~~}
~+~ i \, 4\,  \Big(\,
{\CMTB{\ydiagram{1}}\CMTred{\ytableaushort{\tinysixteenbar}}}_{\rm IR}~+~{\CMTB{\ydiagram{1,1}}\CMTred{\ytableaushort{\tinysixteen,\none}}}_{\rm IR}~+~{\CMTB{\ydiagram{1,1,1}}\CMTred{\ytableaushort{\tinysixteenbar,\none,\none}}}_{\rm IR}~+~{\CMTB{\ydiagram{2,1}}\CMTred{\ytableaushort{\tinysixteenbar,\none}}}_{\rm IR}
 \,\Big)\,  {\pa}_{{\un a}_{2}}\,\Psi_{\{ \aone \bone \}}{}^{\a} (x)
\, \Bigg]  \nonumber\\
&~+~ \fracm 1{5!}\, ({\ell}){}^5 \, \left[ \,
{\CMTB{\ydiagram{2,2,1,1,1}}}_{{\rm IR},-} \,
\Phi_{\{\atwo \btwo| \aone \bone \cone \done \eone \}{}^{+}} (x)
~+~ {\CMTB{\ydiagram{3,1,1}}}_{\rm IR} 
\, \Phi_{\{ \atwo , \athree | \aone \bone \cone \}} (x) \right. \nonumber\\
&\left. {~~~~~~~~~~~~~~~~}
~+~ i \, 5\,  \Big(\,
{\CMTB{\ydiagram{2,1}}}_{{\rm IR}}~+~{\CMTB{\ydiagram{2,2,1}}}_{{\rm IR}}~+~{\CMTB{\ydiagram{3,2}}}_{{\rm IR}}\,
 \Big)\,  {\pa}_{{\un a}_{3}}\,\Phi_{\{ \aone \bone, \atwo \btwo \}} (x)\right. \nonumber\\
&\left. {~~~~~~~~~~~~~~~~}
~+~ i \, 5\,  \left(\,
{\CMTB{\ydiagram{1,1,1,1,1}}}_{{\rm IR},-}~+~{\CMTB{\ydiagram{2,1,1,1}}}_{{\rm IR}}~+~{\CMTB{\ydiagram{2,2,1,1,1}}}_{{\rm IR},-}~+~{\CMTB{\ydiagram{3,1,1,1,1}}}_{{\rm IR},-}\,
 \right)\,  {\pa}_{{\un a}_{3}}\,\Phi_{\{ \atwo | \aone \bone \cone \done \eone \}{}^{+}} (x)
\, \right]  \nonumber\\ 
&~+~ \fracm 1{6!}\, ({\ell}){}^6 \, \left[ \, 
{\CMTB{\ydiagram{3}}\CMTred{\ytableaushort{\tinysixteen}}}_{\rm IR} \,
\Psi_{\{ \aone, \atwo, \athree \}}{}_{\a} (x)
~+~{\CMTB{\ydiagram{2,1,1}}
\CMTred{\ytableaushort{\tinysixteenbar,\none,\none}}}_{\rm IR}   \,
\Psi_{\{ \atwo | \aone \bone \cone \}}{}^{\a} (x)  \right. \nonumber\\
&\left.  {~~~~~~~~~~~~~~~}
~+~ i \, 6\,  \left(\,
{\CMTB{\ydiagram{1,1,1,1}}\CMTred{\ytableaushort{\tinysixteenbar,\none,\none,\none}}}_{\rm IR}~+~{\CMTB{\ydiagram{2,1,1,1,1}}\CMTred{\ytableaushort{\tinysixteenbar,\none,\none,\none,\none}}}_{\rm IR,-}
 \,\right)\,  {\pa}_{{\un a}_{2}}\,\Psi_{\{ \aone \bone \cone \done \eone \}{}^{+}}{}^{\a} (x)
 ~+~ i \, 6\,  \Big(\,
{\CMTB{\ydiagram{1,1}}\CMTred{\ytableaushort{\tinysixteenbar,\none}}}_{\rm IR}
~+~{\CMTB{\ydiagram{2}}\CMTred{\ytableaushort{\tinysixteenbar}}}_{\rm IR}
 \right. \nonumber\\
 &\left.  {~~~~~~~~~~~~~~~}
~+~{\CMTB{\ydiagram{2,1}}\CMTred{\ytableaushort{\tinysixteen,\none}}}_{\rm IR}
~+~{\CMTB{\ydiagram{2,2}}\CMTred{\ytableaushort{\tinysixteenbar,\none}}}_{\rm IR}~+~{\CMTB{\ydiagram{2,1,1}}\CMTred{\ytableaushort{\tinysixteenbar,\none,\none}}}_{\rm IR}~+~{\CMTB{\ydiagram{3,1}}\CMTred{\ytableaushort{\tinysixteenbar,\none}}}_{\rm IR}
 \,\Big)\,  {\pa}_{{\un a}_{3}}\,\Psi_{\{ \atwo | \aone \bone \}}{}^{\a} (x)
\, \right] 
\nonumber\\
&~+~ \fracm 1{7!}\, ({\ell}){}^7 \, \left[ \,
{\CMTB{\ydiagram{4}}}_{\rm IR} \,
\Phi_{\{ \aone,  \atwo,  \athree, \afour \}} (x)
~+~{\CMTB{\ydiagram{2,2,2}}}_{\rm IR}  \,
\Phi_{\{ \aone \bone \cone , \atwo \btwo \ctwo \}} (x)
\right.\nonumber\\
&\left. {~~~~~~~~~~~~~~~}
~+~{\CMTB{\ydiagram{3,1,1,1}}}_{\rm IR}  \,
\Phi_{\{\atwo,\athree | \aone \bone \cone \done \}} (x)
~+~ i \, 7\,  \Bigg(\,{\CMTB{\ydiagram{2,1,1,1,1}}}_{{\rm IR},-} 
~+~{\CMTB{\ydiagram{2,2,1,1}}}_{{\rm IR}}~+~{\CMTB{\ydiagram{2,2,2,1,1}}}_{{\rm IR},-}
\right. \nonumber\\
&\left. {~~~~~~~~~~~~~~~} ~+~{\CMTB{\ydiagram{3,2,1,1,1}}}_{{\rm IR},-}\,
 \Bigg)\,  {\pa}_{{\un a}_{3}}\,\Phi_{\{\atwo \btwo| \aone \bone \cone \done \eone \}{}^{+}} (x)
  ~+~ i \, 7\,  \Bigg(\,
{\CMTB{\ydiagram{2,1,1}}}_{{\rm IR}} 
~+~{\CMTB{\ydiagram{3,1}}}_{{\rm IR}}~+~{\CMTB{\ydiagram{3,1,1,1}}}_{{\rm IR}}
 \right. \nonumber\\
&\left.{~~~~~~~~~~~~~~~}  ~+~{\CMTB{\ydiagram{4,1,1}}}_{{\rm IR}}~+~{\CMTB{\ydiagram{3,2,1}}}_{{\rm IR}}\,
 \Bigg)\,  {\pa}_{{\un a}_{3}}\,\Phi_{\{ \atwo , \athree | \aone \bone \cone \}} (x)
\, \right] \nonumber\\
&~+~ \fracm 1{8!}\, ({\ell}){}^8 \, \left[ \,
{\CMTB{\ydiagram{3}}\CMTred{\ytableaushort{\tinysixteenbar}}}_{\rm IR}
 \Psi_{\{ \aone , \atwo , \athree \}}{}^{\a}(x)
 ~+~{\CMTB{\ydiagram{2,1,1}}\CMTred{\ytableaushort{\tinysixteen,\none,\none}}}_{\rm IR}    \,
 \Psi_{\{ \atwo | \aone \bone \cone \}}{}_{\a}(x)
 ~+~ i \, 8\,  \Big(\,
{\CMTB{\ydiagram{2}}\CMTred{\ytableaushort{\tinysixteen}}}_{\rm IR}
 \right. \nonumber\\
&\left. {~~~~~~~~~~~~~~~}
~+~{\CMTB{\ydiagram{3}}\CMTred{\ytableaushort{\tinysixteenbar}}}_{\rm IR}
~+~{\CMTB{\ydiagram{4}}\CMTred{\ytableaushort{\tinysixteen}}}_{\rm IR}~+~{\CMTB{\ydiagram{3,1}}\CMTred{\ytableaushort{\tinysixteen,\none}}}_{\rm IR}\,
 \Big)\,  {\pa}_{{\un a}_{4}}\,\Psi_{\{ \aone, \atwo, \athree \}}{}_{\a} (x)
 \right. \nonumber\\
&\left. {~~~~~~~~~~~~~~~}
~+~ i \, 8\,  \Bigg(\,
{\CMTB{\ydiagram{1,1,1}}\CMTred{\ytableaushort{\tinysixteenbar,\none,\none}}}_{\rm IR} ~+~{\CMTB{\ydiagram{2,1}}\CMTred{\ytableaushort{\tinysixteenbar,\none}}}_{\rm IR}~+~{\CMTB{\ydiagram{2,1,1}}\CMTred{\ytableaushort{\tinysixteen,\none,\none}}}_{\rm IR}~+~{\CMTB{\ydiagram{2,1,1,1}}\CMTred{\ytableaushort{\tinysixteenbar,\none,\none,\none}}}_{\rm IR}~+~{\CMTB{\ydiagram{2,2,1}}\CMTred{\ytableaushort{\tinysixteenbar,\none,\none}}}_{\rm IR}
\right. \nonumber\\
&\left. {~~~~~~~~~~~~~~~}
~+~{\CMTB{\ydiagram{3,1,1}}\CMTred{\ytableaushort{\tinysixteenbar,\none,\none}}}_{\rm IR}\,
 \Bigg)\,  {\pa}_{{\un a}_{3}}\,\Psi_{\{ \atwo | \aone 
\bone \cone \}}{}^{\a} (x)
\, \right] \nonumber \\
&~+~ \fracm 1{9!}\, ({\ell}){}^9 \, \left[ \,
{\CMTB{\ydiagram{2,2,1,1,1}}}_{{\rm IR},+} \,
 \Phi_{\{ \atwo \btwo | \aone \bone \cone \done \eone \}{}^{-}}(x)
 ~+~{\CMTB{\ydiagram{3,1,1}
}}_{\rm IR} \, {\widehat \Phi}_{\{ \atwo , \athree | \aone \bone \cone \}}(x)\right. \nonumber\\
&\left. {~~~~~~~~~~~~~~~} ~+~ i \, 9\,  \Big(\,
{\CMTB{\ydiagram{3}}}_{{\rm IR}}~+~{\CMTB{\ydiagram{5}}}_{{\rm IR}}~+~{\CMTB{\ydiagram{4,1}}}_{{\rm IR}}\,
 \Big)\,  {\pa}_{{\un a}_{5}}\, \Phi_{\{ \aone,  \atwo,  \athree, \afour \}} (x)\right. \nonumber\\
&\left.{~~~~~~~~~~~~~~~}  ~+~ i \, 9\,  \Bigg(\,
{\CMTB{\ydiagram{2,2,1}}}_{{\rm IR}}~+~{\CMTB{\ydiagram{2,2,2,1}}}_{{\rm IR}}~+~{\CMTB{\ydiagram{3,2,2}}}_{{\rm IR}}\,
 \Bigg)\,  {\pa}_{{\un a}_{3}}\, \Phi_{\{ \aone \bone 
\cone , \atwo \btwo \ctwo \}} (x)\right. \nonumber\\
&\left. {~~~~~~~~~~~~~~~} ~+~ i \, 9\,  \Bigg(\,
{\CMTB{\ydiagram{2,1,1,1}}}_{{\rm IR}}~+~{\CMTB{\ydiagram{3,1,1}}}_{{\rm IR}}~+~{\CMTB{\ydiagram{3,1,1,1,1}}}_{{\rm IR},-}~+~{\CMTB{\ydiagram{3,1,1,1,1}}}_{{\rm IR},+}~+~{\CMTB{\ydiagram{4,1,1,1}}}_{{\rm IR}}\right. \nonumber\\
&\left. {~~~~~~~~~~~~~~~} ~+~{\CMTB{\ydiagram{3,2,1,1}}}_{{\rm IR}}\,
 \Bigg)\,  {\pa}_{{\un a}_{4}}\, \Phi_{\{ \atwo, \athree | \aone \bone \cone \done \}} (x) 
\, \right]  \nonumber\\ 
&~+~ \fracm 1{10!}\, ({\ell}){}^{10} \, \left[ \, 
{\CMTB{\ydiagram{1,1,1,1,1}}\CMTred{\ytableaushort{\tinysixteen,\none,\none,\none,\none}}}_{{\rm IR},+} \,  \Psi_{\{ \aone \bone \cone \done \eone \}{}^{-}}{}_{\a}(x)
~+~{\CMTB{\ydiagram{2,1}}\CMTred{\ytableaushort{\tinysixteen,\none}}}_{\rm IR}    \,
 \Psi_{\{ \atwo | \aone \bone \}}{}_{\a}(x) \right. \nonumber\\
&\left. {~~~~~~~~~~~~~~~} ~+~ i \, 10\,  \Big(\,
{\CMTB{\ydiagram{2}}\CMTred{\ytableaushort{\tinysixteenbar}}}_{\rm IR}~+~{\CMTB{\ydiagram{3}}\CMTred{\ytableaushort{\tinysixteen}}}_{\rm IR}~+~{\CMTB{\ydiagram{4}}\CMTred{\ytableaushort{\tinysixteenbar}}}_{\rm IR}
 \right. \nonumber\\
&\left. {~~~~~~~~~~~~~~~}
~+~{\CMTB{\ydiagram{3,1}}\CMTred{\ytableaushort{\tinysixteenbar,\none}}}_{\rm IR}\,
 \Big)\,  {\pa}_{{\un a}_{4}}\,\Psi_{\{ \aone , \atwo , \athree \}}{}^{\a}(x)
 ~+~ i \, 10\,  \Bigg(\,
{\CMTB{\ydiagram{1,1,1}}\CMTred{\ytableaushort{\tinysixteen,\none,\none}}}_{\rm IR}~+~{\CMTB{\ydiagram{2,1}}\CMTred{\ytableaushort{\tinysixteen,\none}}}_{\rm IR}~+~{\CMTB{\ydiagram{2,1,1}}\CMTred{\ytableaushort{\tinysixteenbar,\none,\none}}}_{\rm IR}
 \right. \nonumber\\
&\left. {~~~~~~~~~~~~~~~} ~+~{\CMTB{\ydiagram{2,1,1,1}}\CMTred{\ytableaushort{\tinysixteen,\none,\none,\none}}}_{\rm IR}~+~{\CMTB{\ydiagram{2,2,1}}\CMTred{\ytableaushort{\tinysixteen,\none,\none}}}_{\rm IR}~+~{\CMTB{\ydiagram{3,1,1}}\CMTred{\ytableaushort{\tinysixteen,\none,\none}}}_{\rm IR}\,
 \Bigg)\,  {\pa}_{{\un a}_{3}}\,\Psi_{\{ \atwo | \aone 
\bone \cone \}}{}_{\a}(x) \, \right] 
\nonumber\\
&~+~ \fracm 1{11!}\, ({\ell}){}^{11} \, \Bigg[ \,
{\CMTB{\ydiagram{2,2}}}_{\rm IR} \,
 {\widehat \Phi}_{\{ \aone \bone , \atwo \btwo \}}(x)
 ~+~{\CMTB{\ydiagram{2,1,1,1,1}}
}_{{\rm IR},+}    \,  \Phi_{\{ \atwo |
\aone \bone \cone \done \eone \}{}^{-}}(x) \nonumber\\
& {~~~~~~~~~~~~~~~}
~+~ i \, 11\,  \left(\,
{\CMTB{\ydiagram{2,1,1,1,1}}}_{{\rm IR},+}~+~{\CMTB{\ydiagram{2,2,1,1}}}_{{\rm IR}}~+~{\CMTB{\ydiagram{2,2,2,1,1}}}_{{\rm IR},+}~+~{\CMTB{\ydiagram{3,2,1,1,1}}}_{{\rm IR},+}\,
 \right)\,  {\pa}_{{\un a}_{3}}\,\Phi_{\{ \atwo \btwo | \aone \bone \cone \done \eone \}{}^{-}}(x) \nonumber\\
& {~~~~~~~~~~~~~~~}
~+~ i \, 11\,  \Bigg(\,
{\CMTB{\ydiagram{2,1,1}}}_{{\rm IR}}~+~{\CMTB{\ydiagram{3,1}}}_{{\rm IR}}~+~{\CMTB{\ydiagram{3,1,1,1}}}_{{\rm IR}}~+~{\CMTB{\ydiagram{4,1,1}}}_{{\rm IR}}
 \nonumber\\
& {~~~~~~~~~~~~~~~}
~+~{\CMTB{\ydiagram{3,2,1}}}_{{\rm IR}}\,
 \Bigg)\,  {\pa}_{{\un a}_{4}}\,{\widehat \Phi}_{\{ \atwo , \athree | \aone \bone \cone \}}(x)
\, \Bigg]     \nonumber\\ 
& ~+~ \fracm 1{12!}\, ({\ell}){}^{12} \,   \Bigg[ \,
{\CMTB{\ydiagram{1,1}}\CMTred{\ytableaushort{\tinysixteen,\none}}}_{\rm IR}  \, \Psi_{\{ \aone \bone \}} {}_{\a}(x)
~+~ i \, 12\,  \left(\,
{\CMTB{\ydiagram{1,1,1,1}}\CMTred{\ytableaushort{\tinysixteen,\none,\none,\none}}}_{\rm IR}~+~{\CMTB{\ydiagram{2,1,1,1,1}}\CMTred{\ytableaushort{\tinysixteen,\none,\none,\none,\none}}}_{\rm IR,+ }
 \,\right)\,  {\pa}_{{\un a}_{2}}\,\Psi_{\{ \aone \bone \cone \done \eone \}{}^{-}}{}_{\a}(x)
  \nonumber\\
& {~~~~~~~~~~~~~~~}  ~+~ i \, 12\,  \Bigg(\,
{\CMTB{\ydiagram{1,1}}\CMTred{\ytableaushort{\tinysixteen,\none}}}_{\rm IR}~+~{\CMTB{\ydiagram{2}}\CMTred{\ytableaushort{\tinysixteen}}}_{\rm IR}~+~{\CMTB{\ydiagram{2,1}}\CMTred{\ytableaushort{\tinysixteenbar,\none}}}_{\rm IR}~+~{\CMTB{\ydiagram{2,2}}\CMTred{\ytableaushort{\tinysixteen,\none}}}_{\rm IR}~+~{\CMTB{\ydiagram{2,1,1}}\CMTred{\ytableaushort{\tinysixteen,\none,\none}}}_{\rm IR}
 \nonumber\\
& {~~~~~~~~~~~~~~~}
~+~{\CMTB{\ydiagram{3,1}}\CMTred{\ytableaushort{\tinysixteen,\none}}}_{\rm IR}
 \, \Bigg)\,  {\pa}_{{\un a}_{3}}\, \Psi_{\{ 
\atwo | \aone \bone \}}{}_{\a}(x) 
\, \Bigg] \nonumber\\
& ~+~ \fracm 1{13!}\, ({\ell}){}^{13} \, \Bigg[ \,
{\CMTB{\ydiagram{1,1,1}}}_{\rm IR} \, {\widehat \Phi}{}_{\{ \aone \bone \cone \}}(x)
~+~ i \, 13\,  \Bigg(\,
{\CMTB{\ydiagram{2,1}}}_{{\rm IR}}~+~{\CMTB{\ydiagram{2,2,1}}}_{{\rm IR}}
~+~{\CMTB{\ydiagram{3,2}}}_{{\rm IR}}\,
 \Bigg)\,  {\pa}_{{\un a}_{3}}\,{\widehat \Phi}_{\{ \aone \bone , \atwo \btwo \}}(x) \nonumber\\
&  {~~~~~~~~~~~~~~~}
~+~ i \, 13\,  \left(\,
{\CMTB{\ydiagram{1,1,1,1,1}}}_{{\rm IR},+}~+~{\CMTB{\ydiagram{2,1,1,1}}}_{{\rm IR}}~+~{\CMTB{\ydiagram{2,2,1,1,1}}}_{{\rm IR},+}~+~{\CMTB{\ydiagram{3,1,1,1,1}}}_{{\rm IR},+}\,
 \right)\,  {\pa}_{{\un a}_{3}}\,\Phi_{\{ \atwo | 
\aone \bone \cone \done \eone \}{}^{-}}(x)
\, \Bigg] \nonumber\\ 
&~+~ \fracm 1{14!}\, ({\ell}){}^{14} \, \left[ \, 
\CMTred{\ytableaushort{\tinysixteenbar}}  \Psi^{\a}(x)
~+~ i \, 14\,  \Big(\,
{\CMTB{\ydiagram{1}}\CMTred{\ytableaushort{\tinysixteen}}}_{\rm IR}~+~{\CMTB{\ydiagram{1,1}}\CMTred{\ytableaushort{\tinysixteenbar,\none}}}_{\rm IR}~+~{\CMTB{\ydiagram{1,1,1}}\CMTred{\ytableaushort{\tinysixteen,\none,\none}}}_{\rm IR}~+~{\CMTB{\ydiagram{2,1}}\CMTred{\ytableaushort{\tinysixteen,\none}}}_{\rm IR}
 \, \Big)\,  {\pa}_{{\un a}_{2}}\,\Psi_{\{ \aone \bone \}} {}_{\a}(x)
\, \right]  \nonumber\\
&~+~ \fracm 1{15!}\, ({\ell}){}^{15} \, \left[ \,
{\widehat \Phi}(x)
~+~ i \, 15\,  \Bigg(\,
{\CMTB{\ydiagram{1,1}}}_{{\rm IR}}~+~{\CMTB{\ydiagram{1,1,1,1}}}_{{\rm IR}}~+~{\CMTB{\ydiagram{2,1,1}}}_{{\rm IR}}
 \,
 \Bigg)\,  {\pa}_{{\un a}_{2}}\,{\widehat \Phi}{}_{\{ \aone \bone \cone \}}(x)
\, \right] \nonumber\\
&~+~  i \,\fracm 1{15!}\, ({\ell}){}^{16} \,  (\,\CMTred{\ytableaushort{\tinysixteen}}~+~
{\CMTB{\ydiagram{1}}\CMTred{\ytableaushort{\tinysixteenbar}}}_{\rm IR}  \,
 )\,  {\pa}_{{\un a}_{1}}\,\Psi^{\a}(x) ~~~.
\ytableausetup{boxsize=1.2em}
\label{equ:dGHat}
\end{align}


It can be seen the effect of the operator $\cal D$ upon calculating
$\cal D$ ${\widehat {\cal G}}(x)$ is similar to the result that is
found for the tensoring calculation 
$\CMTred{\ytableaushort{\sixteenbar}}$ ${\widehat {\cal G}}(x)$.
However, the two calculations yield different results.  One way to
quantify the difference is to define
\be  {\label{equ:DeltaWZg}
\D{}_{WZg}(n) ~=~ 
\frac{17!  \, \times \, n}{(n + 1)! \, (16-n)!} ~~~~~~~~~~~~~,
} \ee
so that at each value of $n$ the quantity $\D{}_{WZg}(n)$ counts the 
number of degrees of freedom in the difference 
[ $\CMTred{\ytableaushort{\sixteenbar}}$ 
- $\cal D$ ] ${\widehat {\cal G}}(x)$ at Level-$n$\footnote{The way to understand the reason for the form of (\ref{equ:DeltaWZg2}) is to recall the lowest component field of$~~~~~~~~~~~~~~~~~~~~~~~~$ $\cal G$ must describe the component level gauge parameter that is present even in the WZ gauge $~~~~~~~~~~~~~~~~~~$ and hence the mismatch between the naive equality of bosons versus fermions.
}.  A further 
calculation reveals
\be  {  \label{equ:DeltaWZg2}
{\sum_{n = 1}^{16}} \, \D{}_{WZg}(n) ~=~ 983,041 ~=~ \left[ (16 \, - \, 1) \, \times \,( 65,536)
\right]~+~ 1 
~~~.
} \ee
If we regard $\CMTred{\ytableaushort{\sixteenbar}}$ 
${\widehat {\cal G}}(x)$ as a connection adynkrafield and 
${\cal D} {\widehat {\cal G}}(x)$ as its gauge transformation,
then 491,521 is the number bosonic components which is not equal
to the 491,520 fermionic components contained in
the adynkrafield connection in a Wess-Zumino gauge with respect
this gauge transformation.

\subsection{From Symbolic Notation To Tensor Notation}

As we have mentioned, the fields that are shown in (\ref{equ:GHat}) have the
indices on the field contracted with the "invisible" indices on the YT's. 
For this equation, the contractions should be relatively straight forward to
surmise.  Instead with (\ref{equ:dGHat}) these contractions may not appear
so obvious.  Thus, in this subsection, time will be spent showing some
example in the hope the general case will be made clear.

Let us begin with the second order terms on the second line of (\ref{equ:dGHat})
\begin{align}  \ytableausetup{boxsize=0.8em}
\left[ \, {\cal D} {\widehat {\cal G}}(x) \, \right] {}_{({\ell}){}^2}
~&=
~+~ \fracm 1{2!}\, ({\ell}){}^2 \, \left[ \, 
{\CMTB{\ydiagram{1,1}}\CMTred{\ytableaushort{\tinysixteenbar,\none}}}_{\rm IR} \,
\Psi_{\{ \aone \bone \}}{}^{\a}
~+~ 
i \, 2\,  ( \,\CMTred{\ytableaushort{\tinysixteenbar}}~+~
{\CMTB{\ydiagram{1}}\CMTred{\ytableaushort{\tinysixteen}}}_{\rm IR}  \, )\,  {\pa}_{{\un a}_{1}}\, \Psi{}_{\a}
\, \right] ~~~.
\ytableausetup{boxsize=1.2em}
\label{equ:dGHatL-2}
\end{align}
On the leading term involving $\Psi_{\{ \aone \bone \}}{}^{\a} (x)$, the $\aone$-index, and the          
$\bone$-index are contracted with the two invisible bosonic indices of  $\ytableausetup{boxsize=0.8em} {\CMTB{\ydiagram{1,1}}}$ and the ${\a}$-index is contracted with the invisible fermionic index of $\CMTred{\ytableaushort{\tinysixteenbar}}$. For the remaining terms there are subtleties to consider.  Our deliberations begin by expanding the final two terms in (\ref{equ:dGHatL-2})
\be \eqalign{
( \,\CMTred{\ytableaushort{\tinysixteenbar}}~+~
{\CMTB{\ydiagram{1}}\CMTred{\ytableaushort{\tinysixteen}}}_{\rm IR}  \, )\,  {\pa}_{{\un a}_{1}}\, \Psi{}_{\a} 
~&=~
\CMTred{\ytableaushort{\tinysixteenbar}} \, ( \, {\pa}_{{\un a}_{1}}\, \Psi{}_{\a}  \, ) ~+~
{\CMTB{\ydiagram{1}}\CMTred{\ytableaushort{\tinysixteen}}}_{\rm IR} \, ( \, {\pa}_{{\un a}_{1}}\, \Psi{}_{\a}  \, )
~~~.
}  \label{equ:dGHatL-2a}
\ee
Next we introduce the relation
\be \eqalign{
 {\pa}_{{\un a}_{1}}\, \Psi{}_{\a}  ~=~  [ \, 
 {\pa}_{{\un a}_{1}}\, \Psi{}_{\a}  \,-\, k_1 \, ({\s}{}_{{\un a}_{1}}){}_{\a \d} ({\s}{}^{{\un b}_{1}}){}^{\d \g }
 \, {\pa}_{{\un b}_{1}}\,\Psi{}_{\g} \, ]  \,+\, k_1 \,  ({\s}{}_{{\un a}_{1}}){}_{\a \d} ({\s}{}^{{\un b}_{1}}){}^{\d \g }\, {\pa}_{{\un b}_{1}}\, \Psi{}_{\g} 
}  \label{equ:dGHatL-2b}
\ee
which implies the equations in (\ref{equ:dGHatL-2c}) and (\ref{equ:dGHatL-2d})
\be \eqalign{ {~~~~~~~~}
 \CMTred{\ytableaushort{\tinysixteenbar}} \,{\pa}_{{\un a}_{1}}\, \Psi{}_{\a}  ~=~  \CMTred{\ytableaushort{\tinysixteenbar}} \,
 [ \,  {\pa}_{{\un a}_{1}}\, \Psi{}_{\a}  \,-\, k_1 \, ({\s}{}_{{\un a}_{1}}){}_{\a \d} ({\s}{}^{{\un b}_{1}}){}^{\d \g }
 \,\, {\pa}_{{\un b}_{1}}\, \Psi{}_{\g} \, ]  \,+\, k_1 \, \CMTred{\ytableaushort{\tinysixteenbar}} \,
 ({\s}{}_{{\un a}_{1}}){}_{\a \d} ({\s}{}^{{\un b}_{1}}){}^{\d \g } 
 \, {\pa}_{{\un b}_{1}}\, \Psi{}_{\g}  ~~~,
}  \label{equ:dGHatL-2c}
\ee
and 
\be \eqalign{  
 {\CMTB{\ydiagram{1}}\CMTred{\ytableaushort{\tinysixteen}}}_{\rm IR}
 {\pa}_{{\un a}_{1}}\, \Psi{}_{\a}  ~=~ {\CMTB{\ydiagram{1}}\CMTred{\ytableaushort{\tinysixteen}}}_{\rm IR} [ \, 
 {\pa}_{{\un a}_{1}}\, \Psi{}_{\a}  \,-\, k_1 \, ({\s}{}_{{\un a}_{1}}){}_{\a \d} ({\s}{}^{{\un b}_{1}}){}^{\d \g }
 \, {\pa}_{{\un b}_{1}}\,
 \, \Psi{}_{\g} \, ]  \,+\, k_1 \, {\CMTB{\ydiagram{1}}\CMTred{\ytableaushort{\tinysixteen}}}_{\rm IR}
 ({\s}{}_{{\un a}_{1}}){}_{\a \d} ({\s}{}^{{\un b}_{1}}){}^{\d \g }\, {\pa}_{{\un b}_{1}}\, \Psi{}_{\g}  ~~~.
}  \label{equ:dGHatL-2d}
\ee
Our conventions for 10D, $\cal N$ = 1 superspace have previously been given in \cite{NORDSG}
where it was presented that we use a ``mostly plus'' Minkowski metric and define
a 10D set of Pauli matrices by
\be
(\sigma^{\un{a}})_{\alpha\beta}\, (\sigma^{\un{b}})^{\beta\gamma} + (\sigma^{
\un{b}})_{\alpha\beta} \, (\sigma^{\un{a}})^{\beta\gamma} = 2 \, \eta^{\un{a}\un{
b}} \, \delta_{\alpha}{}^{\gamma}~~~, 
\label{equ:NORDSG-1}
\ee
whereby this equation implies the following result
\begin{align}
(\sigma^{\un{a}})_{\alpha\beta} \, (\sigma^{\un{b}})^{\beta\gamma} ~=~
\eta^{\un{a}\un{b}} \, \delta_{\alpha}{}^{\gamma} ~+~
(\sigma^{\un{a}\un{b}})_{\alpha}{}^{\gamma}   ~~~, ~~~
\label{equ:NORDSG-3}
\end{align}
for our calculations. Contraction on the vector indices in (\ref{equ:NORDSG-3})
yields
\begin{align}
(\sigma^{\un{a}})_{\alpha\beta} (\sigma_{\un{a}})^{\beta\gamma} ~=~ 10\, \delta_{\alpha}{}^{\gamma} 
~~~.
\label{equ:NORDSG-4}
\end{align}
Thus, if we define $[ {\cal P}{}_{1} ]{}_{\un a}{}^{\un{b}} \,{}_{\alpha}{}^{\gamma} $ 
and $[ {\cal P}{}_{2} ]{}_{\un a}{}^{\un{b}} \,{}_{\alpha}{}^{\gamma} $ via the equations
\be{
\left[ {\cal P}{}_{1} \right]{}_{\un a}{}^{\un{b}} \,{}_{\alpha}{}^{\gamma} 
~=~   \left[ \delta{}_{\un a}{}^{\un{b}}  \, \delta{}_{\alpha}{}^{\gamma} 
\,-\, \fracm 1{10} \, ({\s}{}_{{\un a}}){}_{\a \d} ({\s}{}^{{\un b}}){}^{\d \g } \, \right]
~~~,~~~ \left[ {\cal P}{}_{2} \right]{}_{\un a}{}^{\un{b}} \,{}_{\alpha}{}^{\gamma}  
~=~  \fracm1{10} \, ({\s}{}_{{\un a}}){}_{\a \d} ({\s}{}^{{\un b}}){}^{\d \g }
} \label{equ:NORDSG-5}
\ee
we obtain
\be{
\left[ {\cal P}{}_{1} \right]{}_{\un a}{}^{\un{c}} \,{}_{\alpha}{}^{\d} \, 
\left[ {\cal P}{}_{1} \right]{}_{\un c}{}^{\un{b}} \,{}_{\delta}{}^{\gamma} 
~=~
\left[ {\cal P}{}_{1} \right]{}_{\un a}{}^{\un{b}} \,{}_{\alpha}{}^{\gamma} 
~~,~~
\left[ {\cal P}{}_{2} \right]{}_{\un a}{}^{\un{c}} \,{}_{\alpha}{}^{\d} \, 
\left[ {\cal P}{}_{2} \right]{}_{\un c}{}^{\un{b}} \,{}_{\delta}{}^{\gamma} 
~=~
\left[ {\cal P}{}_{2} \right]{}_{\un a}{}^{\un{b}} \,{}_{\alpha}{}^{\gamma} 
~~,~~
\left[ {\cal P}{}_{1} \right]{}_{\un a}{}^{\un{c}} \,{}_{\alpha}{}^{\d} \, 
\left[ {\cal P}{}_{2} \right]{}_{\un c}{}^{\un{b}} \,{}_{\delta}{}^{\gamma} 
~=~ 0 ~~.
} \label{equ:NORDSG-6}
\ee
Upon choosing $k_1$ = $\fracm 1{10}$, we can rewrite the results shown in
(\ref{equ:dGHatL-2c}) and (\ref{equ:dGHatL-2d}) in the forms
\be \eqalign{ 
 \CMTred{\ytableaushort{\tinysixteenbar}} \,  {\pa}_{{\un a}_{1}}\, \Psi{}_{\a}  ~=~  \CMTred{\ytableaushort{\tinysixteenbar}} \,
 \left[ {\cal P}{}_{1} \right]{}_{\un{a}_{1}}{}^{\un{c}_{1}} \,{}_{\alpha}{}^{\d} \,
 [ \,  {\pa}_{{\un c}_{1}}\, \Psi{}_{\d} \, ]  \,+\,  \CMTred{\ytableaushort{\tinysixteenbar}} \,
 \left[ {\cal P}{}_{2} \right]{}_{\un{a}_{1}}{}^{\un{c}_{1}} \,{}_{\alpha}{}^{\d} \,
 [ \,  {\pa}_{{\un c}_{1}}\, \Psi{}_{\d} \, ]   ~~~, {~~~~~}
}  \label{equ:dGHatL-2e}
\ee
and 
\be \eqalign{ 
 {\CMTB{\ydiagram{1}}\CMTred{\ytableaushort{\tinysixteen}}}_{\rm IR}
 {\pa}_{{\un a}_{1}}\, \Psi{}_{\a}  
  ~=~  {\CMTB{\ydiagram{1}}\CMTred{\ytableaushort{\tinysixteen}}}_{\rm IR} \,
 \left[ {\cal P}{}_{1} \right]{}_{\un{a}_{1}}{}^{\un{c}_{1}} \,{}_{\alpha}{}^{\d} \,
 [ \,  {\pa}_{{\un c}_{1}}\, \Psi{}_{\d} \, ]  \,+\,  
 {\CMTB{\ydiagram{1}}\CMTred{\ytableaushort{\tinysixteen}}}_{\rm IR} \,
 \left[ {\cal P}{}_{2} \right]{}_{\un{a}_{1}}{}^{\un{c}_{1}} \,{}_{\alpha}{}^{\d} \,
 [ \,  {\pa}_{{\un c}_{1}}\, \Psi{}_{\d} \, ]  
 ~~~.
}  \label{equ:dGHatL-2f}
\ee
The projection operators $\left[ {\cal P}{}_{1} \right]{}_{\un{a}}{}^{\un{b}} \,{}_{\a}{}^{\b}$
and $\left[ {\cal P}{}_{2} \right]{}_{\un{a}}{}^{\un{b}} \,{}_{\a}{}^{\b}$ are precisely
the ones that separate the irreducible projections of ${\pa}_{{\un b}}\, \Psi{}_{\b}$ into its
``sigma-traceless'' part and its ``pure sigma trace'' part.  Thus, it follows that (\ref{equ:dGHatL-2e})
and (\ref{equ:dGHatL-2f}) can be rewritten in the simpler forms
\be \eqalign{ 
 \CMTred{\ytableaushort{\tinysixteenbar}} \,  {\pa}_{{\un a}_{1}}\, \Psi{}_{\a}  ~=~   \CMTred{\ytableaushort{\tinysixteenbar}} \,
 \left[ {\cal P}{}_{2} \right]{}_{\un{a}_{1}}{}^{\un{b}_{1}} \,{}_{\alpha}{}^{\b} \,
 [ \,  {\pa}_{{\un b}_{1}}\, \Psi{}_{\b} \, ]   ~~~, \,
}  \label{equ:dGHatL-2g}
\ee
and 
\be \eqalign{ 
 {\CMTB{\ydiagram{1}}\CMTred{\ytableaushort{\tinysixteen}}}_{\rm IR}
 {\pa}_{{\un a}_{1}}\, \Psi{}_{\a}  
  ~=~  {\CMTB{\ydiagram{1}}\CMTred{\ytableaushort{\tinysixteen}}}_{\rm IR} \,
 \left[ {\cal P}{}_{1} \right]{}_{\un{a}_{1}}{}^{\un{b}_{1}} \,{}_{\alpha}{}^{\b} \,
 [ \,  {\pa}_{{\un b}_{1}}\, \Psi{}_{\b} \, ]    
 ~~~.
}  \label{equ:dGHatL-2h}
\ee

This set of calculations yields a very general rule when contracting the indices
on a YT with a following expression involving fermionic fields. Namely, when the 
number of blue boxes in the YT does not match the number of bosonic indices in 
the expression, this is due to the appearance of ``sigma traces'' in the 
fermionic expression.

We now turn to the third order terms on the second line of (\ref{equ:dGHat})
\begin{align}  \ytableausetup{boxsize=0.8em}
\left[ \, {\cal D} {\widehat {\cal G}}(x) \, \right] {}_{({\ell}){}^3}
~&=
~+~ \fracm 1{3!}\, ({\ell}){}^3 \, \left[ \, 
{\CMTB{\ydiagram{2,2}}}_{\rm IR} \,
\Phi_{\{ \aone \bone, \atwo \btwo \}} (x)
~+~{\CMTB{\ydiagram{2,1,1,1,1}}}_{{\rm IR},-} \,
\Phi_{\{ \atwo | \aone \bone \cone \done \eone \}{}^{+}} (x)\right. \nonumber\\
&\left. {~~~~~~~~~~~~~~~~} ~+~ i \, 3\,  \Bigg(\,
{\CMTB{\ydiagram{1,1}}}_{{\rm IR}}~+~{\CMTB{\ydiagram{1,1,1,1}}}_{{\rm IR}}~+~{\CMTB{\ydiagram{2,1,1}}}_{{\rm IR}}
 \,\Bigg)\,  {\pa}_{{\un a}_{2}}\, \Phi_{\{ \aone \bone \cone \}} (x)
\, \right] ~~~.
\ytableausetup{boxsize=1.2em}
\label{equ:dGHatL-3ex}
\end{align}

On the leading term involving $\Phi_{\{ \aone \bone, \atwo \btwo \}} (x)$, the $\aone$-index, and the $\bone$-index are contracted with the two invisible bosonic indices in the first column of the preceding YT, and the $\atwo$-index, and the $\btwo$-index are contracted with the two invisible bosonic indices in the second column of the preceding YT.

On the second term involving $\Phi_{\{ \atwo | \aone \bone \cone \done \eone \}{}^{+}} (x)$, 
the $\aone$, ..., $\eone$
indices are contracted with the five invisible bosonic indices in the first column of the preceding
YT, and the $\atwo$-index is contracted with the single box in the second column of the preceding YT.

The really interesting terms appear on the second line of (\ref{equ:dGHatL-3ex}).  Here our deliberations 
begin by expanding the final three terms,
\be \eqalign{  \ytableausetup{boxsize=0.8em}
 {~~~~} \Bigg(\,
{\CMTB{\ydiagram{1,1}}}_{{\rm IR}}~+~{\CMTB{\ydiagram{1,1,1,1}}}_{{\rm IR}}
~+~{\CMTB{\ydiagram{2,1,1}}}_{{\rm IR}}
 \,\Bigg) \,  {\pa}_{{\un a}_{2}}\, \Phi_{\{ \aone \bone \cone \}} (x)
~&=~
{\CMTB{\ydiagram{1,1}}}_{{\rm IR}} \,   {\pa}_{{\un a}_{2}}\, \Phi_{\{ \aone \bone \cone \}} (x)
~+~{\CMTB{\ydiagram{1,1,1,1}}}_{{\rm IR}}  \,  {\pa}_{{\un a}_{2}}\, \Phi_{\{ \aone \bone \cone \}} (x)
\cr
&{~~~~~~} 
~+~{\CMTB{\ydiagram{2,1,1}}}_{{\rm IR}} \,  {\pa}_{{\un a}_{2}}\, \Phi_{\{ \aone \bone \cone \}} (x)
~~~.
}  \label{equ:dGHatL-3a}
\ee
Here the only subtlety involves the first term.

On the leading term involving ${\pa}_{{\un a}_{2}}\, \Phi_{\{ \aone \bone \cone \}} (x)$, it is
immediately clear there are only two invisible bosonic indices on the YT, but there are four
bosonic indices on the spatial derivative acting on the field that follows the tableau.  The matching
of indices can only be reconciled if two of the indices are ``contracted away.''  This means
we must have
\be \eqalign{  \ytableausetup{boxsize=0.8em}
{\CMTB{\ydiagram{1,1}}}_{{\rm IR}} \,   {\pa}_{{\un a}_{2}}\, \Phi_{\{ \aone \bone \cone \}} (x)
~=~  {\CMTB{\ydiagram{1,1}}}_{{\rm IR}} \,  \eta{}^{\aone \atwo}   \, {\pa}_{{\un a}_{2}}\, 
\Phi_{\{ \aone \bone \cone \}} (x)  ~~~.
}  \label{equ:dGHatL-3b} 
\ee
with the $\bone$ and $\cone$ indices contracted with the preceding YT.  It should be 
noted that the antisymmetry of the the indices of the field  $\Phi_{\{ \aone \bone \cone \}} (x)$
implies that the choice of which index is contracted with the partial derivative is
immaterial.

For the second term involving ${\pa}_{{\un a}_{2}}\, \Phi_{\{ \aone \bone \cone \}} (x)$, the 
process is straightforward.  Each bosonic index in the expression ${\pa}_{{\un a}_{2}}\, 
\Phi_{\{ \aone \bone \cone \}} (x)$ is contracted with an invisible index of one of the
boxes in the YT.

Finally, for the third and last term involving ${\pa}_{{\un a}_{2}}\, \Phi_{\{ \aone \bone \cone \}} (x)$,
the index on the partial derivative is contracted with the invisible index of the box in the second
column.  The remaining indices $\aone$, $\bone$, and $\cone$ are each contracted with
one of the invisible indices associated with the boxes in the first column of the preceding
YT.

This set of calculations yields a very general rule, analogous to the one appearing 
below  (\ref{equ:dGHatL-2h}), when contracting the indices on a YT with a following 
expression involving bosonic fields. Namely, when the number of blue boxes in the 
YT does not match the number of bosonic indices in the expression, this is due to 
the appearance of  the Minkowski metric in the bosonic expression.

\subsection{From Adynkrafields Back To 1D Adinkras}

A final amusing matter is that having reached the introduction of 
adynkrafields, we can take a limit where the higher dimensional 
adynkrafields are forced into the form of 1D, $N$ = 16 valise
adinkras!  The is done by imposing the condition that all of the
field variables depend solely on a time-like coordinate $\t$ and
the imposition of the condition that $(\ell){}^2$ = 1.  This leads to
$$  \eqalign{  \ytableausetup{boxsize=0.8em}
{\widehat {\cal G}}{}_{Adnk}&(\t)
= ~  \cr
&{\bm {\Bigg \{ }} \Phi(\t)
\,+\,  \fracm 1{2} \, {\CMTB{\ydiagram{1,1,1}}}_{\rm IR} \,
\Phi_{\{ \aone \bone \cone \}} (\t) 
~+~  \fracm 1{4!}  \,  {\CMTB{\ydiagram{2,2}}}_{\rm IR} \,
\Phi_{\{ \aone \bone, \atwo \btwo \}} (\t)
~+~  \fracm 1{4!} \,  {\CMTB{\ydiagram{2,1,1,1,1}}}_{{\rm IR},-} \,
\Phi_{\{ \atwo | \aone \bone \cone \done \eone \}{}^{+}} (\t)  } $$
$$  \eqalign{ 
&{~~~~~~} + \, \fracm 1{6!} \,    {\CMTB{\ydiagram{2,2,1,1,1}}}_{{\rm IR},-} \,
\Phi_{\{\atwo \btwo| \aone \bone \cone \done \eone \}{}^{+}} (\t)
\, + \, \fracm 1{6!} \,  {\CMTB{\ydiagram{3,1,1}}}_{\rm IR} 
\, \Phi_{\{ \atwo , \athree | \aone \bone \cone \}} (\t)
{~~~~~~~~~~~~~~~~~~~~~~~~~~~~~~}
 } $$
 $$  \eqalign{ 
&  {~~~~~~~~~~~} + \, \fracm 1{8!}  \,    {\CMTB{\ydiagram{4}}}_{\rm IR} \,
\Phi_{\{ \aone,  \atwo,  \athree, \afour \}} (\t) \, + \, 
\fracm 1{8!} \,   {\CMTB{\ydiagram{2,2,2}}}_{\rm IR}  \,
\Phi_{\{ \aone \bone \cone , \atwo \btwo \ctwo \}} (\t)   \, + \, \fracm 1{8!} \, {\CMTB{\ydiagram{3,1,1,1}}}_{\rm IR}  \,
\Phi_{\{\atwo,\athree | \aone \bone \cone \done \}} (\t)
{~~~~~~~~~~~}   }  $$
$$  \eqalign{ 
 \, + \, \fracm 1{10!} \,  {\CMTB{\ydiagram{3,1,1}
}}_{\rm IR} \, {\widehat \Phi}_{\{ \atwo , \athree | \aone \bone \cone \}}(\t)  
~\, + \, \fracm 1{10!} \,   {\CMTB{\ydiagram{2,2,1,1,1}}}_{{\rm IR},+} \,
\Phi_{\{ \atwo \btwo | \aone \bone \cone \done \eone \}{}^{-}}(\t)
{~~~~~~~~~~~~~~~~~~~~~}  
} $$
$$  \eqalign{
&  ~~~ \, + \, \fracm 1{12!} \,    {\CMTB{\ydiagram{2,2}}}_{\rm IR} \,
{\widehat \Phi}_{\{ \aone \bone , \atwo \btwo \}}(\t)
\, + \, \fracm 1{12!} \,   {\CMTB{\ydiagram{2,1,1,1,1}}
}_{{\rm IR},+}    \,  \Phi_{\{ \atwo |
\aone \bone \cone \done \eone \}{}^{-}}(\t) 
\, + \, \fracm 1{14!} \,  {\CMTB{\ydiagram{1,1,1}}}_{\rm IR} \, {\widehat \Phi}{}_{\{ \aone \bone \cone \}}(\t){~~~}
}  $$
$$
{~~~~~~}  + \, \fracm 1{16!}  \,    {\widehat \Phi}(\t)   ~ {\bm {\Bigg \} }}  \,+\, 
{\ell} \, {\bm {\Bigg \{ }} ~ \CMTred{\ytableaushort{\tinysixteen}} \, \Psi{}_{\a}(\t) ~ + ~   \fracm 1{3!} \,
{\CMTB{\ydiagram{1,1}}\CMTred{\ytableaushort{\tinysixteenbar,\none}}}_{\rm IR} \,
\Psi_{\{ \aone \bone \}}{}^{\a} (\t)  
+~ \fracm 1{5!} \, 
{\CMTB{\ydiagram{1,1,1,1,1}}\CMTred{\ytableaushort{\tinysixteenbar,\none,\none,\none,\none}}}_{{\rm IR},-} \,
\Psi_{\{ \aone \bone \cone \done \eone \}{}^{+}}{}^{\a} (\t)
{~~}  
 $$
$$  \eqalign{ {~~~~~~~~}
& ~~~+~   \fracm 1{5!} \,   {\CMTB{\ydiagram{2,1}}\CMTred{\ytableaushort{\tinysixteenbar,\none}}}_{\rm IR} \, \Psi_{\{ \atwo | \aone \bone \}}{}^{\a} (\t)
\, + \, \fracm 1{7!} \,       {\CMTB{\ydiagram{3}}\CMTred{\ytableaushort{\tinysixteen}}}_{\rm IR} \,
\Psi_{\{ \aone, \atwo, \athree \}}{}_{\a} (\t)
~+~  \fracm 1{7!} \,   {\CMTB{\ydiagram{2,1,1}}
\CMTred{\ytableaushort{\tinysixteenbar,\none,\none}}}_{\rm IR}   \,
\Psi_{\{ \atwo | \aone \bone \cone \}}{}^{\a} (\t) {~~~~~~~~~~~~~~~~~}
 } $$
$$ \eqalign{
&  + \, \fracm 1{9!} \,   {\CMTB{\ydiagram{3}}\CMTred{\ytableaushort{\tinysixteenbar}}}_{\rm IR}
\Psi_{\{ \aone , \atwo , \athree \}}{}^{\a}(\t)  \, + \, \fracm 1{9!} \, 
{\CMTB{\ydiagram{2,1,1}}\CMTred{\ytableaushort{\tinysixteen,\none,\none}}}_{\rm IR}    \,
 \Psi_{\{ \atwo | \aone \bone \cone \}}{}_{\a}(\t) 
{~~~~~~~~~~~~~~~~~~~~~~~~~~~~~~}   
 }  $$
$$  \eqalign{ 
& ~~~~
\, + \, \fracm 1{11!} \,   {\CMTB{\ydiagram{1,1,1,1,1}}\CMTred{\ytableaushort{\tinysixteen,\none,\none,\none,\none}}}_{{\rm IR},+} \,  \Psi_{\{ \aone \bone \cone \done \eone \}{}^{-}}{}_{\a}(\t)
\, + \, \fracm 1{11!} \, 
{\CMTB{\ydiagram{2,1}}\CMTred{\ytableaushort{\tinysixteen,\none}}}_{\rm IR}    \,
\Psi_{\{ \atwo | \aone \bone \}}{}_{\a}(\t) 
\, + \,
\fracm 1{13!} \,{\CMTB{\ydiagram{1,1}}\CMTred{\ytableaushort{\tinysixteen,\none}}}_{\rm IR}  \, \Psi_{\{ \aone \bone \}} {}_{\a}(\t)
 }  $$
\be \label{equ:GHatad} {
{ \eqalign{ 
\, + \,  \fracm 1{15!} \,   \CMTred{\ytableaushort{\tinysixteenbar}}  \Psi^{\a}(\t)  ~ {\bm {\Bigg \} }}
 ~~~~,  {~~~~~~~~~~~~~~~~~~~~~~~~~~~~~~~~~~~~~~~~~~}
 {~~~~~~~~~~~~~~~~~~~~~~~~~~~~~~} 
}} \ytableausetup{boxsize=1.2em}
} 
\ee
which can be simplified further to eliminate all the factors involving the
factorial function by rescaling the field variables appropriately.  

This 1D, N = 16 valise adinkra system clearly contains 32,768 bosons and
32,768 fermions.  It also contains the information associated with the 
Lorentz representations (via the YT's) of the original 10D, $\cal N$ = 1
scalar supermultiplet for which it is the hologram.  Application of the
$\cal D$ operator to this expansion shown above while retaining terms 
only up to order $(\ell)$ will permit the derivations of the ${\cal {GR}}$
(d, $N$) matrices \cite{GRana1,GRana2} associated with this system.

In the work of \cite{G-1}, the inaugural discussion relating supermultiplets
in greater than 1D to those in 1D was given.  There was a portion of the
derivation that was not explicitly presented.  Let us recall from this
past work, valise adinkra systems were described in the following manner.

A set of bosonic fields $\Phi_i$ and fermionic fields $\Psi_{\hat k}$ (where
the index $i$ takes on values for one to any integer d and the index ${\hat k}$ 
ranges of the same values) describe valise adinkra systems.  Furthermore, two 
sets of matrices $\left( {\rm L}{}_{{}_{\rm I}}\right) {}_{i \, {\hat k}}$ and 
$\left( {\rm R}{}_{{}_{\rm I}}\right) {}_{{\hat k} \, i}$ are also introduced 
where the index ${\rm I}$ ranges over the integers, but its maximum value $N$ 
is not necessarily restricted to be the same as that of the indices $i$ and 
${\hat k}$.  The following conditions may be imposed on the matrices
\be \eqalign{
(\,{\rm L}_\rI\,)_i{}^\hj\>(\,{\rm R}_\rJ\,)_\hj{}^k + (\,{\rm L}_\rJ
\,)_i{}^\hj\>(\,{\rm R}_\rI\,)_\hj{}^k &= 2\,\d_{\rI\rJ}\,\d_i{}^k~~,\cr
(\,{\rm R}_\rJ\,)_\hi{}^j\>(\, {\rm L}_\rI\,)_j{}^\hk + (\,{\rm R}_\rI
\,)_\hi{}^j\>(\,{\rm L}_\rJ\,)_j{}^\hk &= 2\,\d_{\rI\rJ}\,\d_\hi{}^\hk~~,
}  \label{GarDNAlg1}
\ee
\be
~~~~
(\,{\rm R}_\rI\,)_\hj{}^k\,\d_{ik} = (\,{\rm L}_\rI\,)_i{}^\hk\,\d_{\hj\hk}~~,
\label{GarDNAlg2}
\end{equation}
along with the following differential equations being imposed on 
$\Phi_{i}$ and $\Psi_{\hat k}$
\be
{\rm D}{}_{{}_{\rm I}} \Phi_i ~=~ i \, \left( {\rm L}{}_{{}_{\rm I}}\right) {}_{i \, {\hat k}}  \,  \Psi_{\hat k}
~~~,~~~
{\rm D}{}_{{}_{\rm I}} \Psi_{\hat k} ~=~  \left( {\rm R}{}_{{}_{\rm I}}\right) {}_{{\hat k} \, i}  \, {{d ~} \over { d \t}} \, \Phi_{i}  ~~.
 \label{chiD0J}
\ee
so the equations (\ref{GarDNAlg1}), (\ref{GarDNAlg2}), and (\ref{chiD0J})
uniformly imply the operator equation is satisfied 
\be {
{\rm D}{}_{{}_{\rm I}} \, {\rm D}{}_{{}_{\rm J}} ~+~ 
{\rm D}{}_{{}_{\rm J}} \, {\rm D}{}_{{}_{\rm I}} ~=~ i \, 2 \, 
\d{}_{{\rm I} \,{\rm J}} {{d ~} \over { d \t}}  ~~~,
} \label{1Dsusyop}
\ee
on $\Phi_{i}$ and $\Psi_{\hat k}$. The fields $\Phi_{i}$ and $\Psi_{\hat k}$ 
can be related to the 1D fields in (\ref{equ:GHatad}) via
\be \eqalign{
{\Phi}_{i}(\t) ~&=~ \left\{ \, \Phi(\t) , \, \Phi_{\{ \aone \bone \cone \}} (\t) , \, \Phi_{\{ \aone \bone, \atwo \btwo \}} (\t), \, \dots ~ \right\}   ~~~~, \cr
\Psi_{\hat k}(\t) ~&=~ \left\{ \, \Psi{}_{\a}(\t)   , \,  \Psi_{\{ \aone \bone \}}{}^{\a} (\t)  , \, \Psi_{\{ \aone \bone \cone \done \eone \}{}^{+}}{}^{\a} (\t),  \, \dots ~  \right\}   ~~~~.
}  \label{adyadi}
\ee
Currently, it is not clear how to construct the ${\rm D}{}_{{}_{\rm I}}$ operators from $\cal D$.  
This will be the subject of a future investigation.

The algebra for the L-matrices and
R-matrices in (\ref{GarDNAlg1}) defines the 
${\cal {GR}}$(d, $N$) algebra or the ``Garden Algebra'' (d, $N$).  In the present
context d = 32,768 and $N$ = 16.
In future investigations (as it will be possible to study the case where the 
ranges of $i$ and $\hat k$ covers 1, $\dots$, 32768, and the range of $\rm 
I$ covers 1, $\dots$, 16), derivation will uncover how the 
$\left( {\rm L}{}_{{}_{\rm I}}\right)$ and $\left( {\rm R}{}_{{}_{\rm I}}\right)$ 
matrices holographically store the information of the YT's concerning the Lorentz 
representations of the fields in (\ref{equ:GHatad}).  This is the portion not 
undertaken in \cite{G-1}.

\subsection{Adynkras, and Links Between Nodes}

It is clear from the presentation in Equation (\ref{equ:G}) for $\cal V$ (as well as the one on pages
57 - 59 for ${ {
\CMTred{\ytableaushort{\sixteenbar}}}} \, {\cal V}$)
that each adinkra associates some Level numbers with sets of YT's.  Furthermore, the
YT's are partitioned into two classes. The BYT's possess no red boxes while the SYT's
possess one red box.  Let us introduce notational devices for this division.  We will
use the symbol ${\CMTB{\{ {\cal R}^{(i)}\}}}{}_{p}$ to denote the  ``i-th''  BYT at Level-$p$.
It should be noted that the range of the index ``i'' depends on the value of
$p$.  In a similar manner, we will use the symbol ${\CMTR{\{ {\cal R}^{(i)}\}}}{}_{p}$
to denote the  ``i-th''  SYT at Level-$p$.  Once more it should be noted that the range 
of the index ``i-th'' depends on the value of $p$.

Next, we introduce four coefficients $ {\CMTB{ c }}{}_{{\CMTR{\{ {\cal R}^{(j)}\}}}_{p+1}}^{+{\CMTB {(i)}}  }$,
$ {\CMTB{ c }}{}_{{\CMTR{\{ {\cal R}^{(j)}\}}}_{p-1}}^{-{\CMTB {(i)}}  }$, 
$ {\CMTB{ c }}{}_{{\CMTB{\{ {\cal R}^{(i)}\}}}_{p+1}}^{+{\CMTR {(j)}}  }$,
and ${\CMTB{ c }}{}_{{\CMTB{\{ {\cal R}^{(i)}\}}}_{p-1}}^{-{\CMTR {(j)}}  }$ which are determined
by examining properties of the adynkra. 
In particular, there are four
calculations to be implemented, and these are respectively
\be {
{\CMTB{ c }}{}_{{\CMTR{\{ {\cal R}^{(j)}\}}}_{p+1}}^{+{\CMTB {(i)}}  } ~=~ {\cal F}{}_{1}
\left[ \,  \left(  \, \CMTred{\ytableaushort{\sixteenbar}}  \otimes  {\CMTB{\{ {\cal R}^{(i)}\}}}{}_p 
\, \right)   \cap   \, {\CMTR{\{ {\cal R}^{(j)}\}}}{}_{p + 1}    \, \right]  ~~~,
}  \label{L1}
\ee
\be {
 {\CMTB{ c }}{}_{{\CMTR{\{ {\cal R}^{(j)}\}}}_{p-1}}^{-{\CMTB {(i)}}  } ~=~ {\cal F}{}_{2}
 \left[ \, \left(  \, \CMTred{\ytableaushort{\sixteenbar}}   \otimes  {\CMTR{\{ {\cal R}^{(j)}\}}}{}_{p - 1}
 \, \right) \cap  \, {\CMTB{\{ {\cal R}^{(i)}\}}}{}_p    \, \right]  ~~~,
}  \label{L2}
\ee
\be {
{\CMTB{ c }}{}_{{\CMTB{\{ {\cal R}^{(i)}\}}}_{p+1}}^{+{\CMTR {(j)}}  } ~=~{\cal F}{}_{3}
\left[ \, \left(  \,  \CMTred{\ytableaushort{\sixteenbar}} \otimes  {\CMTR{\{ {\cal R}^{(j)}\}}}{}_p  
\, \right)   \cap \, {\CMTB{\{ {\cal R}^{(i)}\}}}{}_{p + 1}    \, \right] ~~~,
}  \label{L3}
\ee
\be {
{\CMTB{ c }}{}_{{\CMTB{\{ {\cal R}^{(i)}\}}}_{p-1}}^{-{\CMTR {(j)}}  } ~=~ {\cal F}{}_{4}
\left[ \, \left( \, \CMTred{\ytableaushort{\sixteenbar}}   \,  \otimes  \,{\CMTB{\{ {\cal R}^{(i)}\}}}{}_{p - 1} 
 \right)   \cap  \, {\CMTR{\{ {\cal R}^{(j)}\}}}{}_p    \, \right] ~~~,
}  \label{L4}
\ee
where ${\cal F}{}_{1}$, ${\cal F}{}_{2}$, ${\cal F}{}_{3}$, and ${\cal F}{}_{4}$, are functions.  
All of these functions have the property that if the intersections indicated as their respective 
arguments vanish, then the functions output the value of zero.  The functions ${\cal F}{}_{1}$
and, ${\cal F}{}_{3}$ yield outputs of the value one if their respective intersections are 
non-vanishing.  The functions ${\cal F}{}_{2}$ and ${\cal F}{}_{4}$ yield outputs of the value 
of 0 or 1 if their respective intersections are non-vanishing.  
When the value of the $\CMTB{c}$-coefficient is 0, there is no link between those particular two nodes. When it is 1, there is a link.
For these the intersection 
principle can only tell us which links must be absent. However, the appearance of the links 
in the adinkra does not {\it {necessarily}} imply the corresponding normalization coefficients 
have to be non-vanishing.

\subsection{Adynkrafields, Expansion Basis Change, and Superfields}

The discussion in this chapter also points to a relation between the concepts 
in the adinkras, adynkrafields approach, and traditional description 
supermultiplets in terms of superfields.

Let us construct a quantity $\cal K$ from a superfield that can be viewed 
as an analog to $\cal G$.  We can ``strip'' a scalar superfield of all of
its field components which suggests the construct
\be{ \label{equ:calKg}  \eqalign{
{\cal K} ~=~ 1 ~\oplus~ \theta{}^{\a} ~\oplus~  ~ \bigoplus_{p = 2}^{16} ~
\theta{}^{\a{}_{1}} \,  \cdots \,  \theta{}^{\a{}_{p}}     ~~~.
} }\ee
Up to normalization factors it can be argued, initially, that the two expressions $\cal G$ 
and $\cal K$ are related to one another via a change of basis $\theta{}^{\a}$ 
$\to$ $\ell  \, \, \CMTred{\ydiagram{1}}\, $.  This transformation is consistent if there is an
understanding that the ``red box'' actually carries an ``invisible'' spinor index.  We have
used this convention in writing (\ref{equ:GHat}), (\ref{equ:dGHat}), and (\ref{equ:GHatad}).
This is also an assumption implicitly used through out this chapter. The ``blue boxes" carry
invisible vectorial indices, and the ``red boxes" carry invisible spinorial indices.

The exponential of the level parameter $\ell$ plays an important role.  It is seen that in 
(\ref{equ:GHat}) this parameter tracks the level of the YT's as they appear in the higher
dimensional adynkra.  So the exponents range from 0 to 16. On the other hand, within 
the equation (\ref{equ:GHatad}) which applies to only one dimensional valise systems, 
the exponent of $\ell$ only takes on values 0 and 1.  Within the context of 1D, $N=16$ theories,
there are many possible values of this exponent.  The exponents in (\ref{equ:GHatad})
correspond to a two-level adinkra while the ones in (\ref{equ:GHat}) correspond to
a full sixteen-level adinkra in 1D if we set $x{}^0=\tau$ and all spatial components
$x{}^i$ = 0.   There are many other choices for the values of the exponents of $\ell$
as explored in the work of \cite{GHS} in one class of examples.

In the context adynkras and adinkras, the exponents of the level parameter play 
another important role.  The exponents control the engineering dimension of the
fields that follow in the expansion in (\ref{equ:GHat}) and (\ref{equ:GHatad}).  So for
example, the fact that all the bosons in (\ref{equ:GHatad}) are associative with $(\ell){}^0$
implies all the bosons possess the same engineering dimensions.  Similarly, the
fact that all the fermions in (\ref{equ:GHatad}) are associate with $(\ell){}^1$
implies all the fermions possess the same engineering dimensions.  However,
the engineering dimensions of the fermions differs by a unit of (mass)${}^{- \frac 12}$
from the engineering dimensions of the bosons.   This follows from the interpretation
of adynkrafields arising from superfields via the change of basis $\theta{}^{\a}$ $\to$ $\ell  
\, \, \CMTred{\ydiagram{1}}\, $.

There is one implication about the substitution suggestion as the $\theta$-variable
is anti-commuting.  This demands that the product $\ell  \, \CMTred{\ydiagram{1}}$
should also be anti-commuting.  The most natural way to do this is to assume the
anti-commutivity of the red box.

Finally, the results earlier in this section, may be combined with the result in
(\ref{equ:FspaceDynk}) to gain an insight into how this formalism works.  We
first introduce an even more condense notation by writing the ``blue spacetime
derivative'' $
 {    {\CMTB {\pa}} }  ~=~ 
\drop
$.
Now some samples of component actions can be presented in the simplest cases of
actions for the spin-0, spin-1/2, and spin-1 fields from (\ref{equ:FspaceDynk}) as
illustractions.  The respective actions are simply
\be
{\cal S}{}_{\{0\}} ~=~ \int d^{10} x \,  \, \fracm 12 \, \VEV {   {    {\CMTB {\pa}} \, \phi}(x)  
 \, \Big|  \,  {    {\CMTB {\pa}} \, \phi} (x)} ~~~,
{~~~~~~~~~~} \ee

\be
\ytableausetup{boxsize=0.8em}
{\cal S}{}_{{\CMTred{\ytableaushort{\indexsixteen}}}} ~=~ \int d^{10} x \,  \,  i  \,  \fracm 12 \, \VEV {  
\l_{\CMTred{\ytableaushort{\indexsixteen}}}(x) \,  \Big|  \,  {\CMTB {\pa}}   \, \l_{\CMTred{\ytableaushort{\indexsixteen}}}(x)       }
~~~,
{~~~~~} \ee

\be
\ytableausetup{boxsize=0.8em}
{\cal S}{}_{{\CMTred{\ytableaushort{\indexsixteen}}}} ~=~ \int d^{10} x \,  \,  i  \,  \fracm 12 \, \VEV {  
\l_{\CMTred{\ytableaushort{\indexsixteenbar}}}(x) \,  \Big|  \,  {\CMTB {\pa}}   \, \l_{\CMTred{\ytableaushort{\indexsixteenbar}}}(x)       }
~~~,
{~~~~~}\ee

\be
{\cal S}{}_{\{1\}} ~=~ \int d^{10} x \,  \, \fracm 14 \, \VEV {  {    {\CMTB {\pa}} \, }  \wedge  A_{\CMTB{\ydiagram{1}}}(x)   \, \Big|  \,   {    {\CMTB {\pa}} \,} 
 \,   \wedge    \,  A_{\CMTB{\ydiagram{1}}}(x)} ~~~,
\ee
where notation $\VEV{~}$ indicates that intermediate calculations are to be done
in the space of the YT's and after these are completely evaluated, only the 
coefficient of the singlet Tableau is retained.  We are thus advocating an 
effectively index-free notation until the final stage of the derivation of results.  
As Tableaux and Dynkin Labels are interchangeable, this is the orgin of the gain 
in calculational efficiency based on the used of adynkrafields.

\newpage
\section{Conclusion}

In this work we have shown all the steps that allow one to begin with an adynkra of
the 10D, $\cal N$ = 1 scalar superfield and apply a well defined set of rules to
``tease'' from this starting point and finally obtain the field variables (together with their irreducibility
conditions) for which the Dynkin Labels provide descriptions.

There remain a few more steps before one obtains a complete component-level
description of 10D, $\cal N$ = 1 Nordstr\" om supergravity theory.  These include:
\vskip0.01in \indent
(a.) the use of the adynkra to provide a starting point for
ansatz\" e for the
\newline \indent $~~~~~$
component
level supersymmetry variations on each component field,
\newline \indent $~~~~~$
and
\newline \indent
(b.) substitute all of these results in the expressions that appear in \cite{HyDSG1} which
\newline \indent $~~~~~$
relate 10D, $\cal N$ = 1 Nordstr\" om
supergeometry to the 10D, $\cal N$ = 1 scalar
\newline \indent $~~~~~$
superfield.
\vskip0.01in \noindent
It should be noted the adynkra plays a role in efficiency in part (a.).  Namely only
representations connected by links in the adynkra are allowed to appear in the
component level supersymmetry variations.  This together with Lorentz invariance
fixes (up to a set of constants) the form of these variations.  These final constants
are fixed by the condition of closure of the SUSY algebra.  Using modern IT applications,
it should be possible to completely fix this final set of constants.

However, we should remind the reader that even if this is all explicitly carried out,
one still has a reducible construction.  That is a separate problem needing further
investigation of the properties of the quantities $\Tilde {\cal G}$ or $\cal G$.

We believe the explicit presentation in this work should be convincing to the skeptical
reader that there exists a well-defined set of steps that starts with the 10D, $\cal N$ = 1
adynkra shown in Figure \ref{Fig:10DTypeI_Dynkin} and leads to the complete component
field description given in Chapter \ref{sec:comp-descrip}. To our knowledge, this is the first such completely
explicit presentation at the component level in the literature.

To tie these results with the corresponding geometrical ones discussed in \cite{HyDSG1},
we note that $\Phi(x)$ at Level-0 is the scalar graviton, while $\Psi{}_{\a}(x)$ at Level-1 is
the non-conformal part of the gravitino.  Finally, the component field $\Phi{}_{  \{ {\un a}{}_{
1} {\un b}{}_{1} {\un c}{}_{1} \} }(x)$ corresponds to the lowest component of a superfield so
that $\Phi{}_{  \{ {\un a}{}_{1} {\un b}{}_{1} {\un c}{}_{1} \} }(x)$ = $G{}_{\{ {\un a}{}_{1} {\un b}
{}_{1} {\un c}{}_{1} \} }(x)$ $\propto$ $(\s{}_{  {\un a}{}_{1} {\un b}{}_{1}{\un c}{}_{1} }){}_{\a \b}$
$ \left( {\rm D}{}^{\a} {\rm D}{}^{\b} \, \cal{V} \right)|$.  From Eq (5.11) and Eq. (5.14) in the work
of \cite{HyDSG1}, we know this field $G{}_{  \{ {\un a}{}_{1} {\un b}{}_{1} {\un c}{}_{1} \} }$  is
the lowest component of a quantity that appears in the supertorsion $T{}_{\a \, {\un b}}{}^{\g}$
and the supercurvature $R{}_{\a \, \b}{}^{{\un c} \, {\un d}}$.  All the remaining component
fields seen at Levels 3 - 16 occur as the lowest components of superfields obtained by
applications of spinor  derivatives of orders 1 - 14 to $G{}_{ \{ {\un a}{}_{1} {\un b}{}_{1} {\un
c}{}_{1} \} }$ in the supertorsion $T{}_{\a \, {\un b}}{}^{\g}$ and the supercurvature $R{}_{
\a \, \b}{}^{{\un c} \, {\un d}}$.

For the skeptic, we present this as the most explicit evidence to date that, at least with respect
to a supergeometrical formulation, there exists a well-defined theory of 10D, $\cal N$ = 1
Nordstr\" om supergravity expressed in terms of the fifteen bosonic and twelve fermionic
component fields exhibited in Chapter \ref{sec:comp-descrip}.

In Chapter \ref{sec:ties}, we have checked the consistency of branching rules of $\mathfrak{su}(10)\supset\mathfrak{so}(10)$ as well as the Weyl dimension 
formula, the  $\mathfrak{su}(10)$ algebra hook rule, and our graphical tying rules 
for all BYTs up to 3-columns. We are thus able to show our graphical rules for branching rules and the 
dimensionality of irreps actually have a substantial support. So we cast this into the form of a conjecture.

\begin{quotation}
 \noindent Conjecture: \vspace{0.5em}\newline
 {\it{The calculation of the branching rules for general $\mathfrak{su}(N) \supset \mathfrak{so}(N)$}} 
 \newline \noindent
 {\it{where $A_{N-1} \supset D_{N/2}$ for even $N$, 
 or $A_{N-1} \supset B_{(N-1)/2}$ for odd $N$, may be }} 
 \newline \noindent
 {\it{found by using the hook rule and the application of the tying rules for that}} 
 \newline \noindent
 {\it{irrep's Young Tableau in $\mathfrak{su}(N)$, if that Young Tableau contains less than}}
 \newline \noindent
 {\it{or equal to three columns.}}
\end{quotation}

\noindent 
However, this still is not a replacement for a rigorous 
mathematical proof. 

In this work, we have presented preliminary evidence for the existence of
adynkrafields which have the potential to replace superfields as tools for the
study of dynamical systems that realize supersymmetry.  There are still some points
about this proposal that remain unclear and will require further study in the future.
One obvious consequence of the introduction of adynkrafields is their multiplication 
rules, following as a result of applying tensor product rules 
to the YT's that appear in the expansions of adinkrafields. This observation must have 
further implications for obtaining a deeper understanding of superspace integration theory.

Finally, in this work we have introduced an essentially coordinate-free ``generator of
supersymmetry multiplets'' in the form of the operator $\cal G$ that appears in
(\ref{equ:calGg}).  It operates on the space of fields $\{ {\cal F} \}$ introduced in (\ref{equ:Fspace}) 
to produce scalar supermultiplets via an ``inner product'' ${\cal G} \cdot \{ {\cal F} \}$.  Tensoring 
${\cal G}$ with any Young Tableaux $\lambda$ suitable to describe an irrep in the space of fields  
$\{ {\cal F} \}$ produces higher representation supermultiplets by tensoring first and then 
applying the inner product, i.e. $\left( \lambda \, {\cal G} \right) \cdot \{ {\cal F} \}$.  
It should also be clear that a very similar argument can be made for Salam-Strathdee superfields. Namely
a scalar superfield may be regarded as an ``inner product'' ${\cal K} \cdot \{ {\cal F} \}$.

With respect to supersymmetry, the operator ${\cal G}$ is reducible.  Understanding how to 
decompose it remains an unsolved, the one great unsolved, problem of supersymmetry representation theory.

\vspace{.05in}
 \begin{center}
\parbox{4in}{{\it ``By believing passionately in something that still does not $~\,$ exist,
we create it.'' \\ ${~}$
\\ ${~}$ }\,\,-\,\, Franz Kafka $~~~~~~~~~$}
 \parbox{4in}{
 $~~$}  
 \end{center}
 \noindent
{\bf Acknowledgements}\\[.1in] \indent

The research of S.\ J.\ G., Y.\ Hu, and S.-N.\ Mak is supported in part by the endowment
of the Ford Foundation Professorship of Physics at Brown University and they gratefully
acknowledge the support of the Brown Theoretical Physics Center.
SNHM would like to thank Kevin Iga and Caroline Klivans for conversations 
on combinatorics related to Young tableau methods.

\newpage
\appendix
\section{More Discussion of Index Notation \label{appen:example-notation}}
In Section \ref{subsec:BYTindex}, we introduced notational conventions to express the index
structure which corresponds to the irreducible bosonic Young Tableaux. The general
expression was shown in Figure~\ref{fig:index-notation0}.  In this appendix, we
will present more examples of this notation.

First, recall the general expression of the notation.
\begin{equation}
\includegraphics[width=1.0\textwidth]{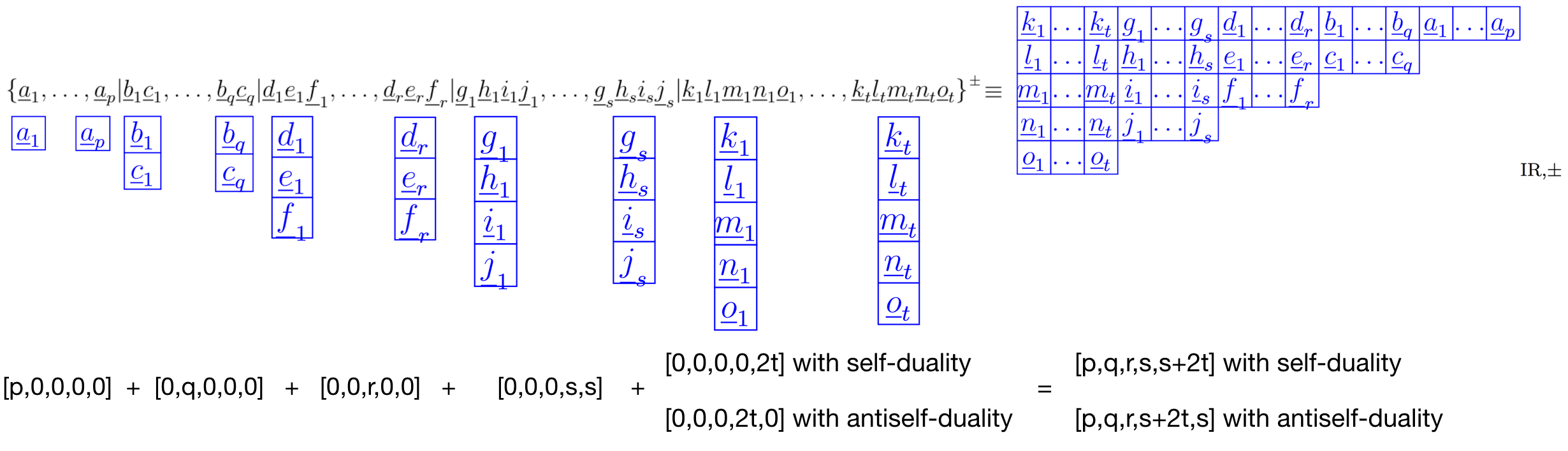}
\end{equation}

We start with some simple examples where only $p\neq 0$.
\begin{align}
\ytableausetup{boxsize=1.2em}
&\{{\un a}_1 \} ~~\equiv~~ { \CMTB{{\ytableaushort{\aone}} } }_{{\rm IR}} ~~\equiv~~
\CMTB{[1,0,0,0,0]}~~~,\\
&\{{\un a}_1 , {\un a}_2\} ~~\equiv~~ { \CMTB{{\ytableaushort{\aone \atwo}} } }_{{\rm IR
}} ~~\equiv~~ \CMTB{[2,0,0,0,0]}~~~,\\
&\{{\un a}_1 , {\un a}_2 , {\un a}_3\}~~\equiv~~ { \CMTB{{\ytableaushort{\aone \atwo
\athree}} } }_{{\rm IR}} ~~\equiv~~ \CMTB{[3,0,0,0,0]}~~~,\\
&\{{\un a}_1 , {\un a}_2 , {\un a}_3 , {\un a}_4\}~~\equiv~~ { \CMTB{{\ytableaushort{
\aone \atwo \athree \afour}} } }_{{\rm IR}} ~~\equiv~~ \CMTB{[4,0,0,0,0]}~~~.
\end{align}

Then we turn to some examples where $t=0$, which means we don't need to consider self
duality.
\begin{align}
&\{{\un a}_1 {\un b}_1\} ~~\equiv~~ { \CMTB{{\ytableaushort{\aone,\bone}} } }_{{\rm IR}}
~~\equiv~~ \CMTB{[0,1,0,0,0]}~~~,\\
&\{{\un a}_2 | {\un a}_1 {\un b}_1\} ~~\equiv~~ { \CMTB{{\ytableaushort{\aone \atwo,
\bone}} } }_{{\rm IR}} ~~\equiv~~ \CMTB{[1,1,0,0,0]}~~~,\\
&\{{\un a}_1 {\un b}_1 ,  {\un a}_2  {\un b}_2 \}~~\equiv~~ { \CMTB{{\ytableaushort{\aone
\atwo,\bone \btwo}} } }_{{\rm IR}} ~~\equiv~~ \CMTB{[0,2,0,0,0]}~~~,\\
&\{{\un a}_1  {\un b}_1   {\un c}_1\} ~~\equiv~~ { \CMTB{{\ytableaushort{\aone,\bone,
\cone}} } }_{{\rm IR}} ~~\equiv~~ \CMTB{[0,0,1,0,0]}~~~,\\
&\{{\un a}_2 | {\un a}_1   {\un b}_1   {\un c}_1\} ~~\equiv~~ { \CMTB{{\ytableaushort{
\aone \atwo,\bone,\cone}} } }_{{\rm IR}} ~~\equiv~~ \CMTB{[1,0,1,0,0]}~~~,\\
&\{{\un a}_2 , {\un a}_3| {\un a}_1   {\un b}_1   {\un c}_1\} ~~\equiv~~ { \CMTB{{
\ytableaushort{\aone \atwo \athree,\bone,\cone}} } }_{{\rm IR}} ~~\equiv~~
\CMTB{[2,0,1,0,0]}~~~,\\
&\{{\un a}_1 {\un b}_1 {\un c}_1 ,  {\un a}_2  {\un b}_2 {\un c}_2 \} ~~\equiv~~ { \CMTB{{
\ytableaushort{\aone \atwo ,\bone \btwo,\cone \ctwo}} } }_{{\rm IR}} ~~\equiv~~ \CMTB{
[0,0,2,0,0]}~~~,\\
&\{{\un a}_1 {\un b}_1 {\un c}_1 {\un d}_1 , {\un a}_2 {\un b}_2 {\un c}_2 {\un d}_2 , {\un
a}_3 {\un b}_3 {\un c}_3 {\un d}_3 , {\un a}_4 {\un b}_4 {\un c}_4 {\un d}_4 \} ~~\equiv~~{
\CMTB{{\ytableaushort{\aone \atwo \athree \afour,\bone \btwo \bthree \bfour,\cone \ctwo
\cthree \cfour,\done \dtwo \dthree \dfour}} } }_{{\rm IR}} ~~\equiv~~ \CMTB{[0,0,0,4,4]}~~~.
\end{align}

Finally, some examples where we have to take self duality into account are presented
as below.
\begin{align}
&\{{\un a}_1   {\un b}_1   {\un c}_1  {\un d}_1  {\un e}_1  \}{}^{+} ~~\equiv~~ {\CMTB{
\ytableaushort{\aone,\bone,\cone,\done,\eone}}}_{{\rm IR},+} ~~\equiv~~ \CMTB{
[0,0,0,0,2] }~~~,\\
&\{{\un a}_1  {\un b}_1  {\un c}_1 {\un d}_1 {\un e}_1 \}{}^{-} ~~\equiv~~ {\CMTB{
\ytableaushort{\aone,\bone,\cone,\done,\eone}}}_{{\rm IR},-} ~~\equiv~~ \CMTB{
[0,0,0,2,0] }~~~,\\
&\{{\un a}_2 | {\un a}_1    {\un b}_1   {\un c}_1   {\un d}_1  {\un e}_1  \}{}^{+} ~~\equiv~~
{\CMTB{\ytableaushort{\aone \atwo,\bone,\cone,\done,\eone}}}_{{\rm IR},+} ~~\equiv~~
\CMTB{[1,0,0,0,2]}~~~,\\
 &\{{\un a}_2| {\un a}_1 {\un b}_1 {\un c}_1 {\un d}_1 {\un e}_1  \}{}^{-} ~~\equiv~~
 {\CMTB{\ytableaushort{\aone \atwo,\bone,\cone,\done,\eone}}}_{{\rm IR},-} ~~\equiv
 ~~ \CMTB{[1,0,0,2,0]}~~~,\\
 &\{{\un a}_2 {\un b}_2   | {\un a}_1 {\un b}_1   {\un c}_1  {\un d}_1  {\un e}_1  \}{}^{+}
~~\equiv~~{\CMTB{\ytableaushort{\aone \atwo,\bone \btwo,\cone,\done,\eone}}}_{{
\rm IR,+}}~~\equiv~~ \CMTB{[0,1,0,0,2]}~~~,\\
&\{{\un a}_2 {\un b}_2   | {\un a}_1 {\un b}_1  {\un c}_1 {\un d}_1 {\un e}_1  \}{}^{-} ~~
\equiv~~{\CMTB{\ytableaushort{\aone \atwo,\bone \btwo,\cone,\done,\eone}}}_{{\rm
IR,-}}~~\equiv~~ \CMTB{[0,1,0,2,0]}~~~.  
 \end{align}

\newpage
\section{Explicit Examples of Spinorial Irrep Dimension Formulas}
\label{appen:SYTdimex}

In this appendix, we list explicit examples of spinorial irrep dimension formulas for completely antisymmetric, completely symmetric, two-equal-column and two-unequal-column BYTs attached with a $\CMTred{\{16\}}$ respectively. These are the types of spinorial irreps that appear in the 10D, $\mathcal{N}=1$ scalar superfield.

In the following sections, we list the irreps in Dynkin Label, mixed YT notation and dimensions on the left,
and how we calculate the dimensions graphically and numerically on the right. 
Note that on the right, we omit all the ``dim'' notation for compact presentation. 
Hence the ``$\times$'' just means multiplying the dimensions, 
but {\em not} any sort of tensor product or direct product; 
and the ``$-$'' just means subtracting the corresponding dimensions, 
but {\em not} any sort of complement.

\subsection{Completely antisymmetric BYTs attached with $\CMTred{\{16\}}$}

\begin{align}
\begin{split}
    ~ \CMTred{ [1,0,0,0,1] } ~=&~ {\CMTB{\ydiagram{1}}\CMTred{\ydiagram{1}}}_{{\rm IR
    }}  ~~~~~~~~~~~~~~  \left(~ {\CMTB{\ydiagram{1}}}_{\rm IR} -~ \CMTB{\cdot} ~\right)
    \times~ \CMTred{\ydiagram{1}}  \\
    ~=&~ \CMTred{\{\overline{144}\}}  ~~~~~~~~~~~~~~~~~~~~  (~ 10 ~-~ 1 ~) ~\times~ 16
\end{split} \label{eqn:anti-mixedYT1}  \\
\begin{split}\label{equ:560}
    ~ \CMTred{[0,1,0,0,1]} ~=&~ {\CMTB{\ydiagram{1,1}}\CMTred{\ydiagram{1,0}}}_{{\rm
    IR}}  ~~~~~~~~~~~~~  \left(~ {\CMTB{\ydiagram{1,1}}}_{\rm IR} -~ {\CMTB{\ydiagram{
    1}}}_{\rm IR} \right) \times~ \CMTred{\ydiagram{1}}  \\
    ~=&~ \CMTred{\{560\}}  ~~~~~~~~~~~~~~~~~~~~  (~ 45 ~-~ 10 ~) ~\times~ 16
\end{split}  \\
\begin{split}
    ~ \CMTred{[0,0,1,0,1]} ~=&~ {\CMTB{\ydiagram{1,1,1}}\CMTred{\ydiagram{1,0,0}}}_{{
    \rm IR}}  ~~~~~~~~~~~~~  \left(~ {\CMTB{\ydiagram{1,1,1}}}_{\rm IR} -~ {\CMTB{
    \ydiagram{1,1}}}_{\rm IR} \right) \times~ \CMTred{\ydiagram{1}}  \\
    ~=&~ \CMTred{\{\overline{1200}\}}  ~~~~~~~~~~~~~~~~~  (~ 120 ~-~ 45 ~) ~\times~
    16
\end{split}  \\
\begin{split}
    ~ \CMTred{[0,0,0,1,2]} ~=&~ {\CMTB{\ydiagram{1,1,1,1}}\CMTred{\ydiagram{1,0,0,0}}
    }_{{\rm IR}}  ~~~~~~~~~~~~~  \left(~ {\CMTB{\ydiagram{1,1,1,1}}}_{\rm IR} -~ {\CMTB
    {\ydiagram{1,1,1}}}_{\rm IR} \right) \times~ \CMTred{\ydiagram{1}}  \\
    ~=&~ \CMTred{\{1440\}}  ~~~~~~~~~~~~~~~~~  (~ 210 ~-~ 120 ~) ~\times~ 16
\end{split} \label{equ:1440dim} \\
\begin{split}
    ~ \CMTred{[0,0,0,0,3]} ~=&~ {\CMTB{\ydiagram{1,1,1,1,1}}\CMTred{\ydiagram{1,0,0,0,0
    }}}_{{\rm IR},+}  ~~~~~~~~~~~  \left(~ {\CMTB{\ydiagram{1,1,1,1,1}}}_{{\rm IR}} ~-~~
    {\CMTB{\ydiagram{1,1,1,1}}}_{\rm IR} \right) \times~ \CMTred{\ydiagram{1}}  \\
    ~=&~ \CMTred{\{\overline{672}\}}  ~~~~~~~~~~~~~  (~ ( 126 + 126 ) ~-~ 210 ~)
    ~\times~ 16
\end{split} \label{eqn:anti-mixedYT5} 
\end{align}

\subsection{Completely symmetric BYTs attached with $\CMTred{\{16\}}$}

\begin{align}
\begin{split}
    ~ \CMTred{[2,0,0,0,1]} ~=&~ {\CMTB{\ydiagram{2}}\CMTred{\ydiagram{1}}}_{{\rm
    IR}}  ~~~~~~~~~~  \left(~ {\CMTB{\ydiagram{2}}}_{\rm IR} -~ {\CMTB{\ydiagram{1
    }}}_{\rm IR} +~ \CMTB{\cdot} ~\right) \times~ \CMTred{\ydiagram{1}}  \\
    ~=&~ \CMTred{\{720\}}  ~~~~~~~~~~~~~~~~~~~  (~ ~~54~~ ~-~ ~10~ ~+~ 1 ~) ~
    \times~ 16
\end{split} \label{eqn:sym-mixedYT1} \\
\begin{split}
    ~ \CMTred{[3,0,0,0,1]} ~=&~ {\CMTB{\ydiagram{3}}\CMTred{\ydiagram{1}}}_{{\rm
    IR}}  ~~~~~~  \left(~ {\CMTB{\ydiagram{3}}}_{\rm IR} -~ {\CMTB{\ydiagram{2}}}_{\rm
    IR} +~ {\CMTB{\ydiagram{1}}}_{\rm IR} -~ \CMTB{\cdot} ~\right) \times~ \CMTred{
    \ydiagram{1}}  \\
    ~=&~ \CMTred{\{\overline{2640}\}}  ~~~~~~~~~~~~~~~~~  (~ ~~~~210~~~~ ~-~
    ~~54~~ ~+~ ~10~ ~-~ 1 ~) ~\times~ 16
\end{split}  \label{eqn:sym-mixedYT2} \\
\begin{split}
    ~ \CMTred{[4,0,0,0,1]} ~=&~ {\CMTB{\ydiagram{4}}\CMTred{\ydiagram{1}}}_{{\rm IR
    }}  ~~~  \left(~ {\CMTB{\ydiagram{4}}}_{\rm IR} -~ {\CMTB{\ydiagram{3}}}_{\rm IR} +~
    {\CMTB{\ydiagram{2}}}_{\rm IR} -~ {\CMTB{\ydiagram{1}}}_{\rm IR} +~ \CMTB{\cdot}
    ~\right) \times~ \CMTred{\ydiagram{1}}  \\
    ~=&~ \CMTred{\{7920\}}  ~~~~~~~~~~~~~~~~~  (~ ~~~~~~660~~~~~~ ~-~ ~~~~210
    ~~~~ ~+~ ~~54~~ ~-~ ~10~ ~+~ 1 ~) ~\times~ 16
\end{split} \label{eqn:sym-mixedYT3}
\end{align}

\subsection{Two-equal-column BYTs attached with $\CMTred{\{16\}}$}

\begin{align}
\begin{split}
    ~ \CMTred{[0,2,0,0,1]} ~=&~ {\CMTB{\ydiagram{2,2}}\CMTred{\ydiagram{1,0}}}_{
    {\rm IR}}  ~~~~~~~~~~~~~  \left(~ {\CMTB{\ydiagram{2,2}}}_{\rm IR} -~ {\CMTB{
    \ydiagram{2,1}}}_{\rm IR} +~ {\CMTB{\ydiagram{2}}}_{\rm IR} \right) \times~ \CMTred{
    \ydiagram{1}}  \\
    ~=&~ \CMTred{\{8064\}}  ~~~~~~~~~~~~~~~~~~~~  (~ ~~770~~ ~-~ ~~320~~ ~+~
    ~~~54~~~ ~) ~\times~ 16
\end{split} \label{eqn:twoeq-mixedYT1} \\
\begin{split}
    ~ \CMTred{[0,0,2,0,1]} ~=&~ {\CMTB{\ydiagram{2,2,2}}\CMTred{\ydiagram{1,0,0}}}_{{
    \rm IR}}  ~~~~~~~~~~~~~  \left(~ {\CMTB{\ydiagram{2,2,2}}}_{\rm IR} -~ {\CMTB{
    \ydiagram{2,2,1}}}_{\rm IR} +~ {\CMTB{\ydiagram{2,2}}}_{\rm IR} \right) \times~
    \CMTred{\ydiagram{1}}  \\
    ~=&~ \CMTred{\{30800\}}  ~~~~~~~~~~~~~~~~~~~  (~~ ~4125~ ~-~ ~2970~ ~+~
    ~~770~~ ~) ~\times~ 16
\end{split} \\
\begin{split}
    ~ \CMTred{[0,0,0,2,3]} ~=&~ {\CMTB{\ydiagram{2,2,2,2}}\CMTred{\ydiagram{1,0,0,0}}}_{{\rm IR}}  ~~~~~~~~~~~~~  \left(~ {\CMTB{\ydiagram{2,2,2,2}}}_{\rm IR} -~ {\CMTB{\ydiagram{2,2,2,1}}}_{\rm IR} +~ {\CMTB{\ydiagram{2,2,2}}}_{\rm IR} \right) \times~ \CMTred{\ydiagram{1}}  \\
    ~=&~ \CMTred{\{39600\}}  ~~~~~~~~~~~~~~~~~~~  (~~ ~8910~ ~-~ 10560 ~~+~~4125~ ~) ~\times~ 16
\end{split} \\
\begin{split}
    ~ \CMTred{[0,0,0,0,5]} ~=&~ {\CMTB{\ydiagram{2,2,2,2,2}}\CMTred{\ydiagram{1,0,0,0,0}}}_{{\rm IR},+}  ~~~~~~~~~~~  \left(~ {\CMTB{\ydiagram{2,2,2,2,2}}}_{{\rm IR}} ~-~~ {\CMTB{\ydiagram{2,2,2,2,1}}}_{{\rm IR}} ~+~~ {\CMTB{\ydiagram{2,2,2,2}}}_{\rm IR} \right) \times~ \CMTred{\ydiagram{1}}  \\
    ~=&~ \CMTred{\{9504\}}  ~~~~~~~~~~~~  (~ ( 2772 + 2772 ) ~-~ (6930 + 6930) ~+~ 8910 ~) ~\times~ 16
\end{split} \label{eqn:twoeq-mixedYT4}
\end{align}

\subsection{Two-unequal-column BYTs attached with $\CMTred{\{16\}}$}

\begin{align}
\begin{split}
    ~ \CMTred{[1,1,0,0,1]} ~=&~ {\CMTB{\ydiagram{2,1}}\CMTred{\ydiagram{1,0}}}_{{\rm IR}}  ~~~~~~~~~~  \left(~ {\CMTB{\ydiagram{2,1}}}_{{\rm IR}} -~ {\CMTB{\ydiagram{2}}}_{{\rm IR}} -~ {\CMTB{\ydiagram{1,1}}}_{{\rm IR}} +~ {\CMTB{\ydiagram{1}}}_{\rm IR} \right) \times~ \CMTred{\ydiagram{1}}  \\
    ~=&~ \CMTred{\{\overline{3696}\}}  ~~~~~~~~~~~~~~~~~  (~ ~~320~~ ~-~ ~~~54~~~ ~-~ ~45~ ~+~ ~10~ ~) ~\times~ 16
\end{split} \label{eqn:twouneq-mixedYT1} \\
\begin{split}
    ~ \CMTred{[0,1,1,0,1]} ~=&~ {\CMTB{\ydiagram{2,2,1}}\CMTred{\ydiagram{1,0,0}}}_{{\rm IR}}  ~~~~~~~~~~  \left(~ {\CMTB{\ydiagram{2,2,1}}}_{{\rm IR}} -~ {\CMTB{\ydiagram{2,2}}}_{{\rm IR}} -~ {\CMTB{\ydiagram{2,1,1}}}_{{\rm IR}} +~ {\CMTB{\ydiagram{2,1}}}_{\rm IR} \right) \times~ \CMTred{\ydiagram{1}}  \\
    ~=&~ \CMTred{\{\overline{25200}\}}  ~~~~~~~~~~~~~~~~  (~ ~2970~ ~-~ ~~770~~ ~-~ ~~945~~ ~+~ ~~320~~ ~) ~\times~ 16
\end{split}  \label{eqn:twouneq-mixedYT2} \\
\begin{split}
    ~ \CMTred{[0,0,1,0,3]} ~=&~ {\CMTB{\ydiagram{2,2,2,1,1}}\CMTred{\ydiagram{1,0,0,0,0}}}_{{\rm IR}}  ~~~~~~~~~~  \left(~ {\CMTB{\ydiagram{2,2,2,1,1}}}_{{\rm IR}} ~-~~ {\CMTB{\ydiagram{2,2,2,1}}}_{{\rm IR}} ~-~~ {\CMTB{\ydiagram{2,2,1,1,1}}}_{{\rm IR}} ~+~~ {\CMTB{\ydiagram{2,2,1,1}}}_{\rm IR} \right) \times~ \CMTred{\ydiagram{1}}  \\
    ~=&~ \CMTred{\{29568\}}  ~~~~~~~~~  (~ ( 6930 + 6930 ) ~-~ 10560 ~-~ ( 3696 + 3696 ) ~+~ 5940 ~) ~\times~ 16
\end{split} \label{eqn:twouneq-mixedYT3}
\end{align}

\newpage
\section{Spinorial Irreps of $\mathfrak{so}(10)$ in Field Theory Notation by Tensor Products
\label{appen:spinorial-tensorproduct}}


In Chapter \ref{sec:king-tensor}, we described the tensor product rule for a general bosonic irrep with the basic spinorial irrep, and explained how the Schur function series $Q$ serves as an inverse of the Schur function series $P$ by verifying the tensor product rule from the SYT dimension rule. In this appendix, we will turn to the spinorial irreps that appear in the 10D, $\mathcal{N}=1$ scalar superfield, i.e. 
\begin{equation}
    \CMTred{\ydiagram{1}} ~~~,~~~ 
    {\CMTB{\ydiagram{1,1}}\CMTred{\ydiagram{1,0}}}_{\rm IR} ~~~,~~~
    {\CMTB{\ydiagram{1,1,1,1,1}}\CMTred{\ydiagram{1,0,0,0,0}}}_{{\rm IR},\pm} ~~~,~~~
    {\CMTB{\ydiagram{3}}\CMTred{\ydiagram{1}}}_{\rm IR} ~~~,~~~
    {\CMTB{\ydiagram{2,1}}\CMTred{\ydiagram{1,0}}}_{\rm IR} ~~~,~~~
    {\CMTB{\ydiagram{2,1,1}}\CMTred{\ydiagram{1,0,0}}}_{\rm IR} ~~~.
\label{eqn:superfieldSYTs}
\end{equation}
In this appendix, we plan to achieve two goals:
\begin{enumerate}[label=(\alph*.)]
\item provide some explicit examples of tensor product rules; and
\item show that $Q$ is the inverse of $P$ through these examples by obtaining the formulas in Appendix \ref{appen:SYTdimex} from these tensor product rules.
\end{enumerate}
Note that (b.) is the converse of what we did in Section \ref{sec:king-tensor}. 

For the first SYT in (\ref{eqn:superfieldSYTs}), the relevant tensor product rule is 
\begingroup
\setlength{\tabcolsep}{2pt}
\renewcommand{\arraystretch}{1.2}
\begin{align}
\begin{split}
& \begin{tabular}{*{5}{c}}
    $\CMTB{\ytableaushort{{\none[\cdot]}}}$ & $\otimes$ & $\CMTred{\ydiagram{1}}$ & $=$ & $\CMTred{\ytableaushort{\sixteen}}$  \\
    $\CMTB{\{1\}}$ & $\otimes$ & $\CMTred{\{16\}}$ & $=$ & $\CMTred{\{16\}}$
\end{tabular}
\end{split} ~~~~~, \label{equ:tensor1} 
\end{align}
\endgroup
so that there is no irreducible condition, and we know that it translates to index notations as
\begin{align}
    \CMTred{\{16\}} ~=&~ \Psi^{\a} ~~~, \\
    \CMTred{\{\overline{16}\}} ~=&~ \Psi_{\a} ~~~.
\end{align}

For the second and the third SYTs in (\ref{eqn:superfieldSYTs}),
the relevant tensor product rules are those totally antisymmetric BYTs tensored with a basic spinorial irrep. They are listed as follows.
\begingroup
\setlength{\tabcolsep}{2pt}
\renewcommand{\arraystretch}{1.2}
\begin{align}
\begin{split}
& \hspace{5.05em} 
\begin{tabular}{*{7}{c}}
    ${\CMTB{\ydiagram{1}}}_{\rm IR}$ & $\otimes$ & $\CMTred{\ydiagram{1}}$ & $=$ & ${\CMTB{\ydiagram{1}}\CMTred{\ytableaushort{\sixteen}}}_{\rm IR}$ & $\oplus$ & $\CMTred{\ytableaushort{\sixteenbar}}$  \\
    $\CMTB{\{10\}}$ & $\otimes$ & $\CMTred{\{16\}}$ & $=$ & $\CMTred{\{\overline{144}\}}$ & $\oplus$ & $\CMTred{\{\overline{16}\}}$
\end{tabular}
\end{split} ~~~~~, \label{equ:tensor10} \\[5pt]
\begin{split}
& \hspace{5.05em} 
\begin{tabular}{*{9}{c}}
    ${\CMTB{\ydiagram{1,1}}}_{\rm IR}$ & $\otimes$ & $\CMTred{\ydiagram{1}}$ & $=$ & ${\CMTB{\ydiagram{1,1}}\CMTred{\ytableaushort{\sixteen,\none}}}_{\rm IR}$ & $\oplus$ & ${\CMTB{\ydiagram{1}}\CMTred{\ytableaushort{\sixteenbar}}}_{\rm IR}$ & $\oplus$ & $\CMTred{\ytableaushort{\sixteen}}$  \\
    $\CMTB{\{45\}}$ & $\otimes$ & $\CMTred{\{16\}}$ & $=$ & $\CMTred{\{560\}}$ & $\oplus$ & $\CMTred{\{144\}}$ & $\oplus$ & $\CMTred{\{16\}}$
\end{tabular}
\end{split} ~~~~~, \label{equ:tensor45} \\[5pt]
\begin{split}
& \hspace{4.6em} 
\begin{tabular}{*{11}{c}}
    ${\CMTB{\ydiagram{1,1,1}}}_{\rm IR}$ & $\otimes$ & $\CMTred{\ydiagram{1}}$ & $=$ & ${\CMTB{\ydiagram{1,1,1}}\CMTred{\ytableaushort{\sixteen,\none,\none}}}_{\rm IR}$ & $\oplus$ & ${\CMTB{\ydiagram{1,1}}\CMTred{\ytableaushort{\sixteenbar,\none}}}_{\rm IR}$ & $\oplus$ & ${\CMTB{\ydiagram{1}}\CMTred{\ytableaushort{\sixteen}}}_{\rm IR}$ & $\oplus$ & $\CMTred{\ytableaushort{\sixteenbar}}$  \\
    $\CMTB{\{120\}}$ & $\otimes$ & $\CMTred{\{16\}}$ & $=$ & $\CMTred{\{\overline{1200}\}}$ & $\oplus$ & $\CMTred{\{\overline{560}\}}$ & $\oplus$ & $\CMTred{\{\overline{144}\}}$ & $\oplus$ & $\CMTred{\{\overline{16}\}}$
\end{tabular}
\end{split} ~~~~~, \label{equ:tensor120} \\[5pt]
\begin{split}
& \hspace{4.6em} 
\begin{tabular}{*{13}{c}}
    ${\CMTB{\ydiagram{1,1,1,1}}}_{\rm IR}$ & $\otimes$ & $\CMTred{\ydiagram{1}}$ & $=$ & ${\CMTB{\ydiagram{1,1,1,1}}\CMTred{\ytableaushort{\sixteen,\none,\none,\none}}}_{\rm IR}$ & $\oplus$ & ${\CMTB{\ydiagram{1,1,1}}\CMTred{\ytableaushort{\sixteenbar,\none,\none}}}_{\rm IR}$ & $\oplus$ & ${\CMTB{\ydiagram{1,1}}\CMTred{\ytableaushort{\sixteen,\none}}}_{\rm IR}$ & $\oplus$ & ${\CMTB{\ydiagram{1}}\CMTred{\ytableaushort{\sixteenbar}}}_{\rm IR}$ & $\oplus$ & $\CMTred{\ytableaushort{\sixteen}}$  \\
    $\CMTB{\{210\}}$ & $\otimes$ & $\CMTred{\{16\}}$ & $=$ & $\CMTred{\{1440\}}$ & $\oplus$ & $\CMTred{\{1200\}}$ & $\oplus$ & $\CMTred{\{560\}}$ & $\oplus$ & $\CMTred{\{144\}}$ & $\oplus$ & $\CMTred{\{16\}}$
\end{tabular}
\end{split} ~~~~~, \label{equ:tensor210} \\[5pt]
\begin{split}
& \begin{tabular}{*{15}{c}}
    ${\CMTB{\ydiagram{1,1,1,1,1}}}_{\rm IR}$ & $\otimes$ & $\CMTred{\ydiagram{1}}$ & $=$ & ${\CMTB{\ydiagram{1,1,1,1,1}}\CMTred{\ytableaushort{\sixteen,\none,\none,\none,\none}}}_{{\rm IR},+}$ & $\oplus$ & ${\CMTB{\ydiagram{1,1,1,1}}\CMTred{\ytableaushort{\sixteenbar,\none,\none,\none}}}_{\rm IR}$ & $\oplus$ & ${\CMTB{\ydiagram{1,1,1}}\CMTred{\ytableaushort{\sixteen,\none,\none}}}_{\rm IR}$ & $\oplus$ & ${\CMTB{\ydiagram{1,1}}\CMTred{\ytableaushort{\sixteenbar,\none}}}_{\rm IR}$ & $\oplus$ & ${\CMTB{\ydiagram{1}}\CMTred{\ytableaushort{\sixteen}}}_{\rm IR}$ & $\oplus$ & $\CMTred{\ytableaushort{\sixteenbar}}$  \\
    $(\CMTB{\{126\}}\oplus\CMTB{\{\overline{126}\}})$ & $\otimes$ & $\CMTred{\{16\}}$ & $=$ & $\CMTred{\{\overline{672}\}}$ & $\oplus$ & $\CMTred{\{\overline{1440}\}}$ & $\oplus$ & $\CMTred{\{\overline{1200}\}}$ & $\oplus$ & $\CMTred{\{\overline{560}\}}$ & $\oplus$ & $\CMTred{\{\overline{144}\}}$ & $\oplus$ & $\CMTred{\{\overline{16}\}}$
\end{tabular}
\end{split} ~~~~~, \label{equ:tensor126126}
\end{align}
\endgroup
where Equation (\ref{equ:tensor126126}) can be split into
\begingroup
\setlength{\tabcolsep}{2pt}
\renewcommand{\arraystretch}{1.2}
\begin{align}
\begin{split}
& \begin{tabular}{*{9}{c}}
    ${\CMTB{\ydiagram{1,1,1,1,1}}}_{{\rm IR},+}$ & $\otimes$ & $\CMTred{\ydiagram{1}}$ & $=$ & ${\CMTB{\ydiagram{1,1,1,1,1}}\CMTred{\ytableaushort{\sixteen,\none,\none,\none,\none}}}_{{\rm IR},+}$ & $\oplus$ & ${\CMTB{\ydiagram{1,1,1}}\CMTred{\ytableaushort{\sixteen,\none,\none}}}_{\rm IR}$ & $\oplus$ & ${\CMTB{\ydiagram{1}}\CMTred{\ytableaushort{\sixteen}}}_{\rm IR}$ \\
    $\CMTB{\{\overline{126}\}}$ & $\otimes$ & $\CMTred{\{16\}}$ & $=$ & $\CMTred{\{\overline{672}\}}$ & $\oplus$ & $\CMTred{\{\overline{1200}\}}$ & $\oplus$ & $\CMTred{\{\overline{144}\}}$ 
\end{tabular}
\end{split} ~~~~~, \\[5pt]
\begin{split}
& \begin{tabular}{*{9}{c}}
    ${\CMTB{\ydiagram{1,1,1,1,1}}}_{{\rm IR},-}$ & $\otimes$ & $\CMTred{\ydiagram{1}}$ & $=$ &  ${\CMTB{\ydiagram{1,1,1,1}}\CMTred{\ytableaushort{\sixteenbar,\none,\none,\none}}}_{\rm IR}$ & $\oplus$ &  ${\CMTB{\ydiagram{1,1}}\CMTred{\ytableaushort{\sixteenbar,\none}}}_{\rm IR}$ & $\oplus$ & $\CMTred{\ytableaushort{\sixteenbar}}$  \\
    $\CMTB{\{126\}}$ & $\otimes$ & $\CMTred{\{16\}}$ & $=$ & $\CMTred{\{\overline{1440}\}}$ & $\oplus$ & $\CMTred{\{\overline{560}\}}$ & $\oplus$ & $\CMTred{\{\overline{16}\}}$
\end{tabular}
\end{split} ~~~~~. 
\end{align}
\endgroup
From the above tensor product rules, we obtain
\begingroup
\setlength{\tabcolsep}{2pt}
\renewcommand{\arraystretch}{1.2}
\begin{align}
\begin{split}
& \hspace{0.8em}
\begin{tabular}{*{9}{c}}
    ${\CMTB{\ydiagram{1,1}}\CMTred{\ytableaushort{\sixteen,\none}}}_{\rm IR}$ & $=$ & ${\CMTB{\ydiagram{1,1}}}_{\rm IR}$ & $\otimes$ & $\CMTred{\ytableaushort{\sixteen}}$ & $-$ & ${\CMTB{\ydiagram{1}}}_{\rm IR}$ & $\otimes$ & $\CMTred{\ytableaushort{\sixteenbar}}$  \\
    $\CMTred{\{560\}}$ & $=$ & $\CMTB{\{45\}}$ & $\otimes$ & $\CMTred{\{16\}}$ & $-$ & $\CMTB{\{10\}}$ & $\otimes$ & $\CMTred{\{\overline{16}\}}$
\end{tabular}
\end{split} ~~~~~, \\[5pt]
\begin{split}
& \begin{tabular}{*{9}{c}}
    ${\CMTB{\ydiagram{1,1,1,1,1}}\CMTred{\ytableaushort{\sixteen,\none,\none,\none,\none}}}_{{\rm IR},+}$ & $=$ & ${\CMTB{\ydiagram{1,1,1,1,1}}}_{\rm IR}$ & $\otimes$ & $\CMTred{\ytableaushort{\sixteen}}$ & $-$ & ${\CMTB{\ydiagram{1,1,1,1}}}_{\rm IR}$ & $\otimes$ & $\CMTred{\ytableaushort{\sixteenbar}}$  \\
    $\CMTred{\{\overline{672}\}}$ & $=$ & $(\CMTB{\{126\}}\oplus\CMTB{\{\overline{126}\}})$ & $\otimes$ & $\CMTred{\{16\}}$ & $-$ & $\CMTB{\{210\}}$ & $\otimes$ & $\CMTred{\{\overline{16}\}}$
\end{tabular}
\end{split} ~~~~~, 
\end{align}
\endgroup
and their conjugates. The ``$-$'' here denotes the removal of the terms in the direct sum. Note that the dimensions obtained from these two equations agree exactly with Equations (\ref{equ:560}) and (\ref{eqn:anti-mixedYT5}).

For the forth SYT in (\ref{eqn:superfieldSYTs}),
the relevant tensor product rules are Equations (\ref{equ:tensor1}) and (\ref{equ:tensor10}), and the totally symmetric BYTs tensored with a basic spinorial irrep listed as follows.
\begingroup
\setlength{\tabcolsep}{2pt}
\renewcommand{\arraystretch}{1.2}
\begin{align}
\begin{split}
& \hspace{1.25em}
\begin{tabular}{*{7}{c}}
    ${\CMTB{\ydiagram{2}}}_{\rm IR}$ & $\otimes$ & $\CMTred{\ydiagram{1}}$ & $=$ & ${\CMTB{\ydiagram{2}}\CMTred{\ytableaushort{\sixteen}}}_{\rm IR}$ & $\oplus$ & ${\CMTB{\ydiagram{1}}\CMTred{\ytableaushort{\sixteenbar}}}_{\rm IR}$  \\
    $\CMTB{\{54\}}$ & $\otimes$ & $\CMTred{\{16\}}$ & $=$ & $\CMTred{\{720\}}$ & $\oplus$ & $\CMTred{\{144\}}$
\end{tabular}
\end{split} ~~~~~, \label{equ:tensor54} \\[5pt]
\begin{split}
& \begin{tabular}{*{7}{c}}
    ${\CMTB{\ydiagram{3}}}_{\rm IR}$ & $\otimes$ & $\CMTred{\ydiagram{1}}$ & $=$ & ${\CMTB{\ydiagram{3}}\CMTred{\ytableaushort{\sixteen}}}_{\rm IR}$ & $\oplus$ & ${\CMTB{\ydiagram{2}}\CMTred{\ytableaushort{\sixteenbar}}}_{\rm IR}$  \\
    $\CMTB{\{210'\}}$ & $\otimes$ & $\CMTred{\{16\}}$ & $=$ & $\CMTred{\{\overline{2640}\}}$ & $\oplus$ & $\CMTred{\{\overline{720}\}}$
\end{tabular}
\end{split} ~~~~~. \label{equ:tensor210'}
\end{align}
\endgroup
From these tensor product rules we derive
\begingroup
\setlength{\tabcolsep}{2pt}
\renewcommand{\arraystretch}{1.2}
\begin{align}
\begin{split}
& \hspace{1.25em}
\begin{tabular}{*{17}{c}}
    ${\CMTB{\ydiagram{3}}\CMTred{\ytableaushort{\sixteen}}}_{\rm IR}$ & $=$ & ${\CMTB{\ydiagram{3}}}_{\rm IR}$ & $\otimes$ & $\CMTred{\ytableaushort{\sixteen}}$ & $-$ & ${\CMTB{\ydiagram{2}}}_{\rm IR}$ & $\otimes$ & $\CMTred{\ytableaushort{\sixteenbar}}$ & $\oplus$ & ${\CMTB{\ydiagram{1}}}_{\rm IR}$ & $\otimes$ & $\CMTred{\ytableaushort{\sixteen}}$ & $-$ & $\CMTB{\ytableaushort{{\none[\cdot]}}}$ & $\otimes$ & $\CMTred{\ytableaushort{\sixteenbar}}$   \\
    $\CMTred{\{\overline{2640}\}}$ & $=$ & $\CMTB{\{210'\}}$ & $\otimes$ & $\CMTred{\{16\}}$ & $-$ & $\CMTB{\{54\}}$ & $\otimes$ & $\CMTred{\{\overline{16}\}}$ & $\oplus$ & $\CMTB{\{10\}}$ & $\otimes$ & $\CMTred{\{16\}}$ & $-$ & $\CMTB{\{1\}}$ & $\otimes$ & $\CMTred{\{\overline{16}\}}$
\end{tabular}
\end{split} ~~~~~.
\end{align}
\endgroup
which is consistent with (\ref{eqn:sym-mixedYT2}).

For the last two SYTs in (\ref{eqn:superfieldSYTs}), the relevant tensor product rules are Equations (\ref{equ:tensor10}), (\ref{equ:tensor45}), (\ref{equ:tensor120}), (\ref{equ:tensor54}) and the two following equations,
\begingroup
\setlength{\tabcolsep}{2pt}
\renewcommand{\arraystretch}{1.2}
\begin{align}
\begin{split}
& \begin{tabular}{*{11}{c}}
    ${\CMTB{\ydiagram{2,1}}}_{\rm IR}$ & $\otimes$ & $\CMTred{\ydiagram{1}}$ & $=$ & ${\CMTB{\ydiagram{2,1}}\CMTred{\ytableaushort{\sixteen,\none}}}_{\rm IR}$ & $\oplus$ & ${\CMTB{\ydiagram{1,1}}\CMTred{\ytableaushort{\sixteenbar,\none}}}_{\rm IR}$ & $\oplus$ & ${\CMTB{\ydiagram{2}}\CMTred{\ytableaushort{\sixteenbar}}}_{\rm IR}$ & $\oplus$ & ${\CMTB{\ydiagram{1}}\CMTred{\ytableaushort{\sixteen}}}_{\rm IR}$  \\
    $\CMTB{\{320\}}$ & $\otimes$ & $\CMTred{\{16\}}$ & $=$ & $\CMTred{\{\overline{3696}\}}$ & $\oplus$ & $\CMTred{\{\overline{560}\}}$ & $\oplus$ & $\CMTred{\{\overline{720}\}}$ & $\oplus$ & $\CMTred{\{\overline{144}\}}$
\end{tabular}
\end{split} ~~~~~, \\[5pt]
\begin{split}
& \begin{tabular}{*{15}{c}}
    ${\CMTB{\ydiagram{2,1,1}}}_{\rm IR}$ & $\otimes$ & $\CMTred{\ydiagram{1}}$ & $=$ & ${\CMTB{\ydiagram{2,1,1}}\CMTred{\ytableaushort{\sixteen,\none,\none}}}_{\rm IR}$ & $\oplus$ & ${\CMTB{\ydiagram{1,1,1}}\CMTred{\ytableaushort{\sixteenbar,\none,\none}}}_{\rm IR}$ & $\oplus$ & ${\CMTB{\ydiagram{2,1}}\CMTred{\ytableaushort{\sixteenbar,\none}}}_{\rm IR}$ & $\oplus$ & ${\CMTB{\ydiagram{1,1}}\CMTred{\ytableaushort{\sixteen,\none}}}_{\rm IR}$ & $\oplus$ & ${\CMTB{\ydiagram{2}}\CMTred{\ytableaushort{\sixteen}}}_{\rm IR}$ & $\oplus$ & ${\CMTB{\ydiagram{1}}\CMTred{\ytableaushort{\sixteenbar}}}_{\rm IR}$  \\
    $\CMTB{\{945\}}$ & $\otimes$ & $\CMTred{\{16\}}$ & $=$ & $\CMTred{\{8800\}}$ & $\oplus$ & $\CMTred{\{1200\}}$ & $\oplus$ & $\CMTred{\{3696\}}$ & $\oplus$ & $\CMTred{\{560\}}$ & $\oplus$ & $\CMTred{\{720\}}$ & $\oplus$ & $\CMTred{\{144\}}$
\end{tabular}
\end{split} ~~~~~.
\end{align}
\endgroup
Then we have
\begingroup
\setlength{\tabcolsep}{2pt}
\renewcommand{\arraystretch}{1.2}
\begin{align}
\begin{split}
& \begin{tabular}{*{17}{c}}
    ${\CMTB{\ydiagram{2,1}}\CMTred{\ytableaushort{\sixteen,\none}}}_{\rm IR}$ & $=$ & ${\CMTB{\ydiagram{2,1}}}_{\rm IR}$ & $\otimes$ & $\CMTred{\ytableaushort{\sixteen}}$ & $-$ & $\CMTB{\ydiagram{1,1}}$ & $\otimes$ & $\CMTred{\ytableaushort{\sixteenbar}}$ & $-$ & ${\CMTB{\ydiagram{2}}}_{\rm IR}$ & $\otimes$ & $\CMTred{\ytableaushort{\sixteenbar}}$ & $\oplus$ & ${\CMTB{\ydiagram{1}}}_{\rm IR}$ & $\otimes$ & $\CMTred{\ytableaushort{\sixteen}}$   \\
    $\CMTred{\{\overline{3696}\}}$ & $=$ & $\CMTB{\{320\}}$ & $\otimes$ & $\CMTred{\{16\}}$ & $-$ & $\CMTB{\{45\}}$ & $\otimes$ & $\CMTred{\{\overline{16}\}}$ & $-$ & $\CMTB{\{54\}}$ & $\otimes$ & $\CMTred{\{\overline{16}\}}$ & $\oplus$ & $\CMTB{\{10\}}$ & $\otimes$ & $\CMTred{\{16\}}$ 
\end{tabular}
\end{split} ~~~~~, \label{eqn:3696bar} \\[5pt]
\begin{split}
& \begin{tabular}{*{17}{c}}
    ${\CMTB{\ydiagram{2,1,1}}\CMTred{\ytableaushort{\sixteen,\none,\none}}}_{\rm IR}$ & $=$ & ${\CMTB{\ydiagram{2,1,1}}}_{\rm IR}$ & $\otimes$ & $\CMTred{\ytableaushort{\sixteen}}$ & $-$ & $\CMTB{\ydiagram{1,1,1}}$ & $\otimes$ & $\CMTred{\ytableaushort{\sixteenbar}}$ & $-$ & ${\CMTB{\ydiagram{2,1}}}_{\rm IR}$ & $\otimes$ & $\CMTred{\ytableaushort{\sixteenbar}}$ & $\oplus$ & ${\CMTB{\ydiagram{1,1}}}_{\rm IR}$ & $\otimes$ & $\CMTred{\ytableaushort{\sixteen}}$   \\
    $\CMTred{\{8800\}}$ & $=$ & $\CMTB{\{945\}}$ & $\otimes$ & $\CMTred{\{16\}}$ & $-$ & $\CMTB{\{120\}}$ & $\otimes$ & $\CMTred{\{\overline{16}\}}$ & $-$ & $\CMTB{\{320\}}$ & $\otimes$ & $\CMTred{\{\overline{16}\}}$ & $\oplus$ & $\CMTB{\{45\}}$ & $\otimes$ & $\CMTred{\{16\}}$ 
\end{tabular}
\end{split} ~~~~~,
\end{align}
\endgroup
where Equation (\ref{eqn:3696bar}) agrees with Equation (\ref{eqn:twouneq-mixedYT1}).

\newpage
$$~~$$

\end{document}